\documentstyle[12pt,epsf,epsfig]{article}
\newcount\timecount
\newcount\hours \newcount\minutes  \newcount\temp \newcount\pmhours
\hours = \time
\divide\hours by 60
\temp = \hours
\multiply\temp by 60
\minutes = \time
\advance\minutes by -\temp
\def\hour{\the\hours}
\def\minute{\ifnum\minutes<10 0\the\minutes
            \else\the\minutes\fi}
\def\clock{
\ifnum\hours=0 12:\minute\ AM
\else\ifnum\hours<12 \hour:\minute\ AM
      \else\ifnum\hours=12 12:\minute\ PM
            \else\ifnum\hours>12
                 \pmhours=\hours
                 \advance\pmhours by -12
                 \the\pmhours:\minute\ PM
                 \fi
            \fi
      \fi
\fi
}

\def\monthname{\relax\ifcase\month 0/\or January\or February\or
   March\or April\or May\or June\or July\or August\or September\or
   October\or November\or December\else\number\month/\fi}

\def\bold#1{\setbox0=\hbox{$#1$}%
     \kern-.025em\copy0\kern-\wd0
     \kern.05em\copy0\kern-\wd0
     \kern-.025em\raise.0433em\box0 }

\def\beq{\begin{equation}}
\def\eeq{\end{equation}}

\def\st{\scriptstyle}
\def\ss{\scriptscriptstyle}
\def\ga{\mathrel{\raise.3ex\hbox{$>$\kern-.75em\lower1ex\hbox{$\sim$}}}}
\def\la{\mathrel{\raise.3ex\hbox{$<$\kern-.75em\lower1ex\hbox{$\sim$}}}}
\def\gev{{\rm \, Ge\kern-0.125em V}}
\def\tev{{\rm \, Te\kern-0.125em V}}
\def\gyr{{\rm \, G\kern-0.125em yr}}
\def\ohsq{\Omega_{\chi} h^2}

\def\tsq{|{\cal T}|^2}
\def\nl{\hfill\nonumber\\&&}
\def\nnl{\hfill\nonumber\\}
\def\thw{\theta_{\ss W}}
\def\thell{\theta_{\st \ell}}
\def\thf{\theta_{\st \ell}}

\def\ttbt{\tan^2 \beta}
\def\Atau{A_{\st \ell}}
\def\thA{\theta_{\st A}}
\def\thB{\theta_{\st B}}
\def\gappeq{\mathrel{\rlap {\raise.5ex\hbox{$>$}}
{\lower.5ex\hbox{$\sim$}}}}
\def\lappeq{\mathrel{\rlap{\raise.5ex\hbox{$<$}}
{\lower.5ex\hbox{$\sim$}}}}
\def\Toprel#1\over#2{\mathrel{\mathop{#2}\limits^{#1}}}

\def\schi{\widetilde \chi}        %\def\ch{{\widetilde \chi}}
\def\slept{\widetilde \ell} 

\def\sel{{\widetilde e}}

\def\stau{\widetilde \tau}

\def\snu{\widetilde \nu}

\def\mchi{m_{\tilde \chi}}
\def\msn{m_{\tilde\nu}}
\def\m12{m_{1\!/2}}
\def\mst{m_{\tilde{\ell}_1}}
\def\mstwo{m_{\tilde{\ell}_2}}
\def\msti{m_{\tilde{\ell}_i}}
\def\mstj{m_{\tilde{\ell}_j}}
\def\msei{m_{\tilde{e}_i}}
\def\msej{m_{\tilde{e}_j}}

\def\msl{m_{\tilde{\ell}_1}}
\def\mxi{m_{\tilde{\chi}_i^+}}
\def\mxj{m_{\tilde{\chi}_j^+}}
\def\mchar{m_{\tilde{\chi}_1^+}}

\def\mt{m_{t}}

\def\mw{m_{\ss W}}
\def\mz{m_{\ss Z}}
\def\mA{m_{\ss A}}

\def\mstau{m_{\tilde{\ell}_1}}
\def\mell{m_{\st \ell}}
\def\mtau{m_{\st \ell}}
\def\nevalsj{m_{\tilde{\chi}_j^0}}
\def\nevalsi{m_{\tilde{\chi}_i^0}}
\def\msn{m_{\tilde{\nu}_\ell}}
\def\msnu{m_{\tilde{\nu}}}
\def\mHp{m_{H^+}}
\def\mla{m_A}
\def\mlb{m_B}
\def\msa{m_{\widetilde{A}}}

\def\mselL{m_{\tilde{e}_L}}
\begin{document}
\begin{titlepage}
\pagestyle{empty}
\baselineskip=21pt
\rightline{hep-ph/0210205}
\rightline{CERN--TH/2002-238}
\rightline{UMN--TH--2113/02}
\rightline{TPI--MINN--02/42}
\vskip 0.2in
\begin{center}
{\large{\bf Exploration of the MSSM with Non-Universal Higgs Masses}}
\end{center}
\begin{center}
\vskip 0.2in
{{\bf John Ellis}$^1$, {\bf Toby Falk}$^{2}$, {\bf Keith
A.~Olive}$^{2}$ and {\bf Yudi Santoso}$^{2}$}\\
\vskip 0.1in
{\it
$^1${TH Division, CERN, Geneva, Switzerland}\\
$^2${Theoretical Physics Institute,
University of Minnesota, Minneapolis, MN 55455, USA}}\\
\vskip 0.2in
{\bf Abstract}
\end{center}
\baselineskip=18pt \noindent
%%%%%%%%%%%%%%%%%%%%%%%%%%%%%%%%%%%%%%%%%%%%%%%%%%%%%%%%%%%%%%%%%%%%%

We explore the parameter space of the minimal supersymmetric extension of
the Standard Model (MSSM), allowing the soft supersymmetry-breaking masses
of the Higgs multiplets, $m_{1,2}$, to be non-universal (NUHM). Compared
with the constrained MSSM (CMSSM) in which $m_{1,2}$ are required to be
equal to the soft supersymmetry-breaking masses $m_0$ of the squark and
slepton masses, the Higgs mixing parameter $\mu$ and the pseudoscalar
Higgs mass
$m_A$, which are calculated in the CMSSM, are free in the NUHM model. We
incorporate accelerator and dark matter constraints in determining
allowed regions of the $(\mu, m_A)$,
$(\mu, M_2)$ and
$(m_{1/2}, m_0)$ planes for selected choices of the other NUHM parameters.
In the examples studied, we find that the LSP mass cannot be reduced far
below its limit in the CMSSM, whereas $m_A$ may be as small as allowed by
LEP for large $\tan \beta$. We present in Appendices details of the
calculations of neutralino-slepton, chargino-slepton and
neutralino-sneutrino coannihilation needed in our exploration of the NUHM.

%%%%%%%%%%%%%%%%%%%%%%%%%%%%%%%%%%%%%%%%%%%%%%%%%%%%%%%%%%%%%%%%%%%%%
\vfill
\leftline{CERN--TH/2002-238}
\leftline{October 2002}
\end{titlepage}
\baselineskip=18pt
%%%%%%%%%%%%%%%%%%%%%%%%%%%%%%%%%%%%%%%%%%%%%%%%%%%%%%%%%%%%%%%%%%%%%

\section{Introduction}

The hierarchy of mass scales in physics is preserved in a natural way if
supersymmetric particles weigh less than about a TeV. Many supersymmetric
models conserve the quantity $R = (-1)^{3 B + L + 2 S}$, where $B$ is the
baryon number, $L$ the lepton number and $S$ the spin. If this $R$ parity
is conserved, the lightest supersymmetric particle (LSP) is expected to be
absolutely stable. The most plausible candidate for the LSP is the
lightest neutralino $\chi$, which is a good candidate~\cite{EHNOS} for the
cold dark matter (CDM) that is thought to dominate over baryonic and hot
dark matter.

In this paper, we refine and extend the many previous calculations of the
relic LSP density in the framework of the minimal supersymmetric extension
of the Standard Model (MSSM). In particular, we expand the recent analysis
of the MSSM parameter space in~\cite{Ellis:2002wv}, where we allowed
non-universal input soft supersymmetry-breaking scalar masses for the
Higgs multiplets. Here we explore in more detail the constraints imposed
by accelerator experiments - including searches at LEP, $b \to s \gamma$
and $g_\mu - 2$ - and the cosmological bound on the LSP relic density.

Before discussing our calculations in more detail, we first review the
range of the relic LSP density that we prefer in our calculations. An
important new constraint on this is provided by data on the cosmic
microwave background (CMB), which have recently been used to obtained the
following preferred 95\% confidence range: $\Omega_{\rm  CDM} h^2 = 0.12 \pm
0.04$~\cite{MS}. Values much smaller than $\Omega_{\rm CDM} h^2 = 0.10$ seem
to be disfavoured by earlier analyses of structure formation in the CDM
framework, so we restrict our attention to $\Omega_{\rm CDM} h^2 > 0.1$.
However, one should note that the LSP may not constitute all the CDM, in
which case
$\Omega_{\rm LSP}$ could be reduced below this value. On the upper side,
we prefer to remain very conservative, in particular because the upper
limit on
$\Omega_{\rm LSP}$ sets the upper limit for the sparticle mass scale. In
this paper, we use
$\Omega_{\rm CDM} h^2 < 0.3$, while being aware that the lower part of this
range currently appears the most plausible.

The parameter space of the MSSM with non-universal soft
supersymmetry-breaking masses for the two Higgs multiplets has two
additional dimensions, beyond those in the constrained MSSM (CMSSM), in
which all the soft supersymmetry-breaking scalar masses $m_0$ are assumed
to be universal. In the CMSSM, the underlying parameters may be taken
as $m_0$, the soft supersymmetry-breaking gaugino mass $m_{1/2}$ that is
also assumed to be universal, the trilinear supersymmetry-breaking
parameters $A_0$ that we set to zero at the GUT scale in this paper, the
ratio
$\tan
\beta$ of Higgs vacuum expectation values, the Higgs superpotential
coupling
$\mu$ and the pseudoscalar Higgs boson mass $m_A$. Two relations between
these parameters follow from the electroweak symmetry-breaking vacuum
conditions, which are normally used in the CMSSM to fix the values of
$\mu$ (up to a sign ambiguity) and $m_A$ in terms of the other parameters
$(m_0, m_{1/2}, A_0, \tan \beta)$.

In the more general MSSM with non-universal Higgs masses (NUHM), the
parameters $\mu$ and $m_A$ become independent again
\cite{nonu,oldnuhm,EFGOS}. Thus one may use the parameters $(m_0, m_{1/2},
\mu, m_A, A_0$, $\tan
\beta)$ to parametrize this more general NUHM.
The underlying theory is likely to specify the non-universalities of the
Higgs masses: ${\hat m}_i \equiv {\rm sign}(m_i^2) |{m_i / m_0}|: i =
1,2$, so it is important to know how the different values of the ${\hat
m}_i$ map into the $(m_0, m_{1/2}, \mu, m_A, A_0, \tan \beta)$ parameter
space, a point we discuss in Section 2. Furthermore, this non-universality 
leads
to new coannihilation processes becoming important, which are discussed in
Section  3.

We review and update in Section 4 the experimental and phenomenological
constraints on the MSSM parameter space that we use, applying them to the
CMSSM. 
Then, in Section 5, we explore the NUHM parameter space. 
Previously, we gave priority to a first scan of the extra dimensions of 
the parameter
space, and postponed a complete discussion of the NUHM at large $\tan
\beta$. 
Here we also show how our  results in the
$(\mu, m_A)$, $(\mu, M_2)$ and $(m_{1/2}, m_0)$ planes for $\tan
\beta  = 10$ change at larger $\tan \beta$, concentrating on the
behaviour of the relic LSP density, but also incorporating constraints on
the NUHM from accelerators. In our discussions of these planes, we
emphasize the novel features not present in the CMSSM, such as the forms
of the regions in which the LSP is charged, e.g., the lighter
${\widetilde \tau}$,  regions where the LSP is a sneutrino ${\widetilde
\nu}$ (thus necessitating the inclusion of additional $\chi -
{\widetilde\nu}$ coannihilation processes), and the potential importance
of Higgsino coannihilation processes. These are not usually relevant in
the CMSSM, where the relic LSP is usually mainly a ${\widetilde B}$.
Section 6 summarizes some conclusions  from our analysis, including
comments on the range of LSP and  pseudoscalar Higgs masses allowed in
the NUHM.

The Appendices provide the information needed to reproduce our
calculations of coannihilation processes relevant to this NUHM analysis.
In particular, Appendix A lists the MSSM couplings we use, Appendix B
extends previous results on neutralino-slepton coannihilation to include
left-right (L-R) mixing, Appendix C discusses chargino-slepton
coannihilation processes, and Appendix D concerns neutralino-sneutrino
coannihilation.

\section{Vacuum Conditions for Non-Universal Higgs Masses}

We assume that the soft supersymmetry-breaking parameters are specified at
some large input scale $M_X$, that may be identified with the supergravity
or grand unification scale. The low levels of flavour-changing neutral
interactions provide good reasons to think that sparticles with the same
Standard Model quantum numbers have universal soft scalar masses, e.g.,
for the ${\widetilde e}_L$, ${\widetilde \mu}_L$ and ${\widetilde \tau}_L$. 
Specific
grand unification models may equate the soft scalar masses of matter
sparticles with different Standard Model quantum numbers, e.g., $({\widetilde
d},{\widetilde s},{\widetilde b})_L$ and $({\widetilde e}, {\widetilde \mu},
{\widetilde
\tau})_L$ in SU(5), and all the Standard Model matter sparticles in SO(10).
However, there are no particularly good reasons to expect that the soft
supersymmetry-breaking scalar masses of the Higgs multiplets should be
equal to those of the matter sparticles. This is, however, the assumption
made in the CMSSM, which we relax in the NUHM studied here~\footnote{For
models with non-universality also in the sfermion masses,
see~\cite{nonu,Arnowitt:2001yh}.}.

One of the attractive features of the CMSSM is that it provides a
mechanism for generating electroweak symmetry breaking via the running of
the effective Higgs masses-squared $m^2_1$ and $m^2_2$ from $M_X$ down to
low energies. We use this mechanism also in the NUHM, which enables us to
relate $m^2_1(M_X)$ and $m^2_2(M_X)$ to the Higgs supermultiplet mixing
parameter $\mu$ and the pseudoscalar Higgs mass $m_A$. Therefore, we can and do
choose as our independent parameters $\mu(\mz) \equiv \mu$ and $m_A(Q) \equiv
m_A$,
where $Q\equiv (m_{\widetilde{t}_R} m_{\widetilde{t}_L})^{1/2}$; as well
as the CMSSM parameters $(m_0(M_X), m_{1/2}(M_X), A_0, \tan \beta)$. In
fact, in this paper we set $A_0 = 0$ for definiteness.
 
The electroweak symmetry breaking conditions may be written in the form:
\begin{equation}
m_A^2 (Q) = m_1^2(Q) + m_2^2(Q) + 2 \mu^2(Q) + \Delta_A(Q)
\end{equation}
and
\begin{equation}
\mu^2 = \frac{m_1^2 - m_2^2 \tan^2 \beta + \frac{1}{2} \mz^2 (1 - \tan^2 \beta)
+ \Delta_\mu^{(1)}}{\tan^2 \beta - 1 + \Delta_\mu^{(2)}},
\end{equation}
where $\Delta_A$ and $\Delta_\mu^{(1,2)}$ are loop
corrections~\cite{Barger:1993gh,deBoer:1994he,Carena:2001fw} and $m_{1,2} \equiv
m_{1,2}(\mz)$.  
We incorporate the known radiative
corrections~\cite{Barger:1993gh,IL,Martin:1993zk} 
$c_1, c_2$ and $c_\mu$ relating the values of the
NUHM parameters at $Q$ to their values at $\mz$: 
\begin{eqnarray}
m_1^2(Q) &=& m_1^2 + c_1 \nnl
m_2^2(Q) &=& m_2^2 + c_2 \nnl
\mu^2(Q) &=& \mu^2 + c_\mu \, .
\end{eqnarray}
Solving for $m^2_1$ and $m^2_2$, one has
\begin{eqnarray}
m_1^2(1+ \tan^2 \beta) &=& m_A^2(Q) \tan^2 \beta - \mu^2 (\tan^2 \beta + 1 -
\Delta_\mu^{(2)} ) 
- (c_1 + c_2 + 2 c_\mu) \ttbt \nl - \Delta_A(Q) \ttbt 
- \frac{1}{2} \mz^2 (1 - \ttbt) - \Delta_\mu^{(1)} 
\label{m1}
\end{eqnarray}
and 
\begin{eqnarray}
m_2^2(1+ \tan^2 \beta) &=& m_A^2(Q) - \mu^2 (\tan^2 \beta + 1 +
\Delta_\mu^{(2)} )
- (c_1 + c_2 + 2 c_\mu) \nl
- \Delta_A(Q) + \frac{1}{2} \mz^2 (1 - \ttbt) + \Delta_\mu^{(1)},
\label{m2}
\end{eqnarray}
which we use to perform our numerical calculations. 

It can be seen from (\ref{m1}) and (\ref{m2}) that, if $m_A$ is too small
or $\mu$ is too large, then $m_1^2$ and/or $m_2^2$ can become negative and
large. This could lead to $m_1^2(M_X) + \mu^2(M_X) < 0$ and/or $m_2^2(M_X)
+ \mu^2(M_X) < 0$, thus triggering
electroweak symmetry breaking at the GUT scale. The requirement that
electroweak symmetry breaking occurs far below the GUT scale forces us to
impose the conditions $m_1^2(M_X)+ \mu(M_X), m_2^2(M_X)+ \mu(M_X) > 0$ as
extra constraints, which we call the GUT
stability constraint~\footnote{For a different point of view, however,
see~\cite{fors}.}.

Specific models for the origin of supersymmetry breaking should be able to
predict the amounts by which universality is violated in $m^2_{1,2}$,
which can be read off immediately from (\ref{m1}, \ref{m2}).
Alternatively, for a given amount of universality breaking, these
equations may easily be inverted to yield the corresponding values of
$\mu$ and $m_A$. In this paper, we plot quantities in terms of $\mu$ and
$m_A$.

In the CMSSM, 
to obtain a consistent low energy model given GUT scale inputs, we must
run down the full set of renormalization group equations (RGE's) from the
GUT scale and use the electroweak symmetry breaking constraints which fix
$\mu$ and $m_A$.  Consistency requires the RGE's to be run back up to the
GUT scale, where the input parameters are reset and the RGE's are run
back down.  Many models require running this cycle about 3 times, though
in some cases convergence may be much slower, particularly at large $\tan
\beta$. Indeed, there are  no solutions for $\mu < 0$ when $\tan \beta$ is
large ($\ga 40$) because of diverging Yukawas. In the NUHM case considered
here,  we have boundary conditions at both the GUT and low-energy scales.
Once again, the numerical calculations of the RGE's must be iterated until
they converge. However, in this case, it is not always possible to arrive
at a solution, especially for large $\tan
\beta$. In our subsequent calculations, we start by making guesses for
the values of
$m_{1,2}(M_X)$ for use in the first run from $M_X$ down to $\mz$, and it
can happen that the iteration pushes the solution away from the
convergence point instead of towards it. Therefore, the first few
iterations must be monitored for any potential blow-ups.  

\section{Renormalization and Coannihilations in the NUHM Model}

The RGEs for the NUHM have additional terms beyond those appearing in the 
CMSSM, and the resulting sparticle spectrum may exhibit some novel 
features, as we now discuss.

The new terms in the RGEs which vanish in the CMSSM involve the
following combination of soft 
supersymmetry-breaking parameters~\cite{Martin:1993zk}: 
\begin{eqnarray}
S &\equiv& \frac{g_1^2}{4} ( m_2^2 - m_1^2 +
        2 ( m_{\widetilde{Q}_L}^2  - m_{\widetilde{L}_L}^2 - 2
	m_{\widetilde{u}_R}^2 +
	m_{\widetilde{d}_R}^2 + m_{\widetilde{e}_R}^2 ) \nonumber \\ && \, + \, 
          ( m_{\widetilde{Q}_{3L}}^2 - m_{\widetilde{L}_{3L}}^2 - 2
	  m_{\widetilde{t}_R}^2 
	  + m_{\widetilde{b}_R}^2 + m_{\widetilde{\tau}_R}^2 )) 
\label{defS}
\end{eqnarray}
Here $\widetilde{Q}_L$, $\widetilde{L}_L$ are the first two generations
left-handed sfermions, and $\widetilde{Q}_{3L}$, $\widetilde{L}_{3L}$ are the
third-generation sfermions. These new terms appear as follows in the RGEs
for  the NUHM:
\begin{eqnarray}
 \frac{d m_1^2}{dt} &=& \frac{1}{8 \pi^2} (-3 g_2^2 M_2^2 - g_1^2 M_1^2 +
          h_\tau^2 ( m_{\widetilde{\tau}_L}^2 + m_{\widetilde{\tau}_R}^2 +
	  m_1^2 + A_\tau^2) \nl +
          3 h_b^2 ( m_{\widetilde{b}_L}^2 + m_{\widetilde{b}_R}^2 + m_1^2 +
	  A_b^2) - 2 S) \nnl
 \frac{d m_2^2}{dt} &=& \frac{1}{8 \pi^2} (-3 g_2^2 M_2^2 - g_1^2 M_1^2 +
          3 h_t^2 ( m_{\widetilde{t}_L}^2 + m_{\widetilde{t}_R}^2 + m_2^2 +
	  A_t^2) + 2 S) \nnl
 \frac{d m_{\widetilde{L}_L}^2}{dt} &=& \frac{1}{8 \pi^2} (-3 g_2^2 M_2^2 -
          g_1^2  M_1^2 - 2 S)  \nnl
 \frac{d m_{\widetilde{e}_R}^2}{dt} &=& \frac{1}{8 \pi^2} (-4 g_1^2 M_1^2 + 4 S) 
         \nnl 
 \frac{d m_{\widetilde{L}_{3L}}^2}{dt} &=& \frac{1}{8 \pi^2} (-3 g_2^2 M_2^2 -
          g_1^2 M_1^2 +
          h_\tau^2 ( m_{\widetilde{L}_{3L}}^2 + m_{\widetilde{\tau}_R}^2 + 
	  m_1^2 + A_\tau^2 ) - 2 S) \nnl
 \frac{d m_{\widetilde{\tau}_R}^2}{dt} &=& \frac{1}{8 \pi^2} (-4 g_1^2 M_1^2 +
          2 h_\tau^2 ( m_{\widetilde{L}_{3L}}^2 + m_{\widetilde{\tau}_R}^2 +
	  m_1^2 +  A_\tau^2 ) + 4 S) \nnl 
 \frac{d m_{\widetilde{Q}_L}^2}{dt} &=& \frac{1}{8 \pi^2} (- \frac{16}{3} g_3^2
          M_3^2  - 3 g_2^2 M_2^2 -
          \frac{1}{9} g_1^2 M_1^2 + \frac{2}{3} S) \nnl 
 \frac{d m_{\widetilde{u}_R}^2}{dt} &=& \frac{1}{8 \pi^2} (- \frac{16}{3} g_3^2
          M_3^2 - \frac{16}{9} g_1^2 M_1^2 - \frac{8}{3} S) \nnl 
 \frac{d m_{\widetilde{d}_R}^2}{dt} &=& \frac{1}{8 \pi^2} (- \frac{16}{3} g_3^2
          M_3^2 - \frac{4}{9} g_1^2 M_1^2 + \frac{4}{3} S) \nnl 
 \frac{d m_{\widetilde{Q}_{3L}}^2}{dt} &=& \frac{1}{8 \pi^2} (- \frac{16}{3}
         g_3^2 M_3^2 - 3 g_2^2 M_2^2 
      - \frac{1}{9} g_1^2 M_1^2 +
          h_b^2 ( m_{\widetilde{Q}_{3L}}^2 + m_{\widetilde{b}_R}^2 + m_1^2 +
	  A_b^2 )  \nl +
          h_t^2 ( m_{\widetilde{Q}_{3L}}^2 + m_{\widetilde{t}_R}^2 + m_2^2 +
	  A_t^2) + 
	  \frac{2}{3} S) \nnl 
 \frac{d m_{\widetilde{t}_R}^2}{dt} &=& \frac{1}{8 \pi^2} (- \frac{16}{3} g_3^2
       M_3^2 - \frac{16}{9} g_1^2 M_1^2 
         + 2 h_t^2 ( m_{\widetilde{Q}_{3L}}^2 + m_{\widetilde{t}_R}^2 + m_2^2 +
	 A_t^2 ) - \frac{8}{3} S) \nnl 
 \frac{d m_{\widetilde{b}_R}^2}{dt} &=& \frac{1}{8 \pi^2} (- \frac{16}{3} g_3^2
       M_3^2 - \frac{4}{9} g_1^2 M_1^2 
       + 2 h_b^2 ( m_{\widetilde{Q}_{3L}}^2 + m_{\widetilde{b}_R}^2 + m_1^2 +
       A_b^2) + \frac{4}{3} S)
\label{rges}
\end{eqnarray}
where the $M_{1,2,3}$ are gaugino masses that we assume to be universal 
at the GUT scale.

In the CMSSM, with all scalar masses set equal to $m_0$ at the GUT scale,
$S = 0$ initially and remains zero at any
scale~\cite{Sterms}, since $S=0$ is a fixed point of the RGEs at the 
one-loop level. However, in the NUHM, with $m_1 \ne m_2$, as seen in Eq.
(\ref{defS})  $S \ne 0$ and
can cause the low-energy NUHM spectrum to differ
significantly from that in the CMSSM. For example, if $S < 0$ the left-handed
slepton can be lighter than the right-handed one. Also, $m_1^2$ and $m_2^2$
appear in the Yukawa parts of the RGEs for the third generation (Eq.
(\ref{rges})),  so NUHM initial conditions may cause their spectrum
to differ from that in the CMSSM. 
In the NUHM case, depending on the
parameters, we may find the LSP to be either (i) the lightest neutralino
$\chi$, (ii) the lighter stau $\widetilde{\tau}_1$, (iii) the
right-handed selectron
$\widetilde{e}_R$ and smuon $\widetilde{\mu}_R$~\footnote{We neglect
left-right (L-R) mixing for the first two generations of sfermions, so the
right-handed selectron and smuon are degenerate. Here and elsewhere, by
`right-handed' sfermion we mean the superpartner of right-handed
fermion.}, (iv) the left-handed selectron $\widetilde{e}_L$ and smuon
$\widetilde{\mu}_L$, (v) the electron and muon sneutrinos
$\widetilde{\nu}_{e,\mu}$, (vi) the tau sneutrino $\widetilde{\nu}_\tau$,
or (vii) one of the squarks, especially the stop and the sbottom~\footnote{
A squark LSP is possible only if $|A_0|$ is large, a
possibility we do not study in this paper.}. Note that in the cases that
we  consider here, the $\widetilde{\nu}_{e,\mu}$ are generally lighter
than  the $\widetilde{\nu}_{\tau}$ in the regions in which they are the
LSP (that is, when $m_1^2 < 0$, cf. Eq. (\ref{rges}) and Fig. 2 of ref.
\cite{Ellis:2002wv}), unless
$m_A$ is very large ($\gappeq 1000$~GeV).

We assume that $R$ parity is conserved, so that the LSP is stable and is
present in the Universe today as a relic from the Big Bang. Searches for
anomalous heavy isotopes tell us that the dark matter should be
weakly-interacting and neutral, and therefore eliminate all but the
neutralino and the sneutrinos as possible LSPs. LEP and direct dark-matter
searches together exclude a sneutrino LSP~\cite{Falk:1994es}, at least if
the majority of the CDM is the LSP. Thus we require in our analysis that the lightest neutralino be
the LSP.

Nevertheless there are new coannihilation processes to be considered when
one or more of these `wannabe' LSPs is almost degenerate with the lightest
neutralino. These include $\chi - \widetilde{\tau}_1$, $\chi
- \widetilde{e}_L - \widetilde{\mu}_L$, $\chi - \widetilde{e}_R -
\widetilde{\mu}_R$, $\chi - \widetilde{\nu}_{e,\mu}$,  $\chi -
\chi^\prime - \chi^\pm$ coannihilations and all possible
combinations~\footnote{Again, because we set $A_0=0$ here, squark
coannihilations are not important, but see~\cite{stopco} for a
calculation of neutralino-stop coannihilation. }.  
However, not all of these combinations 
are important as they are significant only in very small regions for a
particular set of parameters. For this reason we do not include, for example,
the
sneutrino-slepton coannihilation in our calculations. 

We include in our subsequent calculations neutralino-slepton $\chi -
\widetilde{\ell}$~\cite{Ellis:1999mm,Arnowitt:2001yh,coann}, $\chi -
\chi^\prime - \chi^\pm$~\cite{oldcoann}, $\chi - \snu_{e,\mu}$,
$\chi^\prime - \widetilde{\ell}$ and $\chi^\pm - \widetilde{\ell}$
coannihilations\footnote{See \cite{all} for recent work which includes
all coannihilation channels.}. The
$\chi^\prime$ (co)annihilation rates can be derived from the
corresponding $\chi$ (co)annihilations by appropriate mass and coupling
replacements. Details of our calculations are given in the Appendices.
Following a summary of the relevant couplings in Appendix A, in Appendix
B we update the neutralino-slepton coannihilation calculation
of~\cite{Ellis:1999mm} to include L-R mixing. These are not very
important at relatively low
$\tan \beta$  but are potentially important for large values of
$\tan \beta$. These improved coannihilation calculations
were in fact already used in~\cite{efgosi}, 
but no details were given there. Appendix C provides
chargino-slepton coannihilation processes, whilst Appendix D deals with
neutralino-sneutrino coannihilation processes.

\section{Summary of Constraints and Review of the CMSSM Parameter Space}

We impose in our analysis the constraints on the MSSM parameter space that
are provided by direct sparticle searches at LEP, including that on the
lightest chargino $\chi^\pm$: $m_{\chi^\pm} \gappeq$ 103.5
GeV~\cite{LEPsusy}, and that on the selectron $\tilde e$: $m_{\tilde
e}\gappeq$ 99 GeV \cite{LEPSUSYWG_0101}. Another important constraint is
provided by the LEP lower limit on the Higgs mass: $m_H > 114.4$~GeV
\cite{LEPHiggs} in the Standard Model\footnote{In view of the
theoretical uncertainty in calculating $m_h$, we apply this bound with
three significant digits, i.e., our figures use the
constraint $m_h > 114$ GeV.}. The lightest Higgs boson
$h$ in the general MSSM must obey a similar limit, which may in principle
be relaxed for larger $\tan \beta$. However, as we discussed in our
previous analysis of the NUHM~\cite{Ellis:2002wv}, the relaxation in the
LEP limit is not relevant in
the regions of MSSM parameter space of interest to us. We recall that
$m_h$ is sensitive to sparticle masses, particularly $m_{\tilde t}$, via
loop corrections~\cite{radcorrH,FeynHiggs}, implying that the LEP Higgs
limit constrains the MSSM parameters. 

We also impose the constraint imposed by measurements of $b\rightarrow
s\gamma$~\cite{bsg}, ${\rm BR}(B \rightarrow X_s \gamma) = (3.54 \pm 0.41
\pm 0.26) \times 10^{-4}$, which agree with the Standard Model calculation
${\rm BR}(B \rightarrow X_s \gamma)_{\rm SM} = (3.60 \pm 0.30) \times
10^{-4}$~\cite{bsgSM}. We recall that the $b\rightarrow s\gamma$
constraint is more important for $\mu < 0$, but it is also relevant for
$\mu > 0$, particularly when $\tan\beta$ is large, as we see again in this
paper.

We also take into account the latest value of the anomalous magnetic
moment of the muon reported~\cite{newBNL} by the BNL E821 experiment. The
world average of $a_\mu\equiv {1\over 2} (g_\mu -2)$ now deviates by
$(33.9 \pm 11.2) \times 10^{-10}$ from the Standard Model calculation
of~\cite{Davier} using $e^+ e^-$ data, and by $(17 \pm 11) \times
10^{-10}$ from the Standard Model calculation of~\cite{Davier} based on
$\tau$ decay data. Other recent analyses of the $e^+ e^-$ data yield
similar results. On some of the subsequent plots, we display the formal
2-$\sigma$ range $11.5 \times 10^{-10} < \delta a_\mu < 56.3 \times
10^{-10}$. However, in view of the chequered theoretical history of the
Standard Model calculations of $a_\mu$, we do not impose this as an 
absolute constraint on the supersymmetric parameter space.

\begin{figure}
\vspace*{-0.75in}
%\hspace*{-.70in}
\begin{minipage}{8in}
\epsfig{file=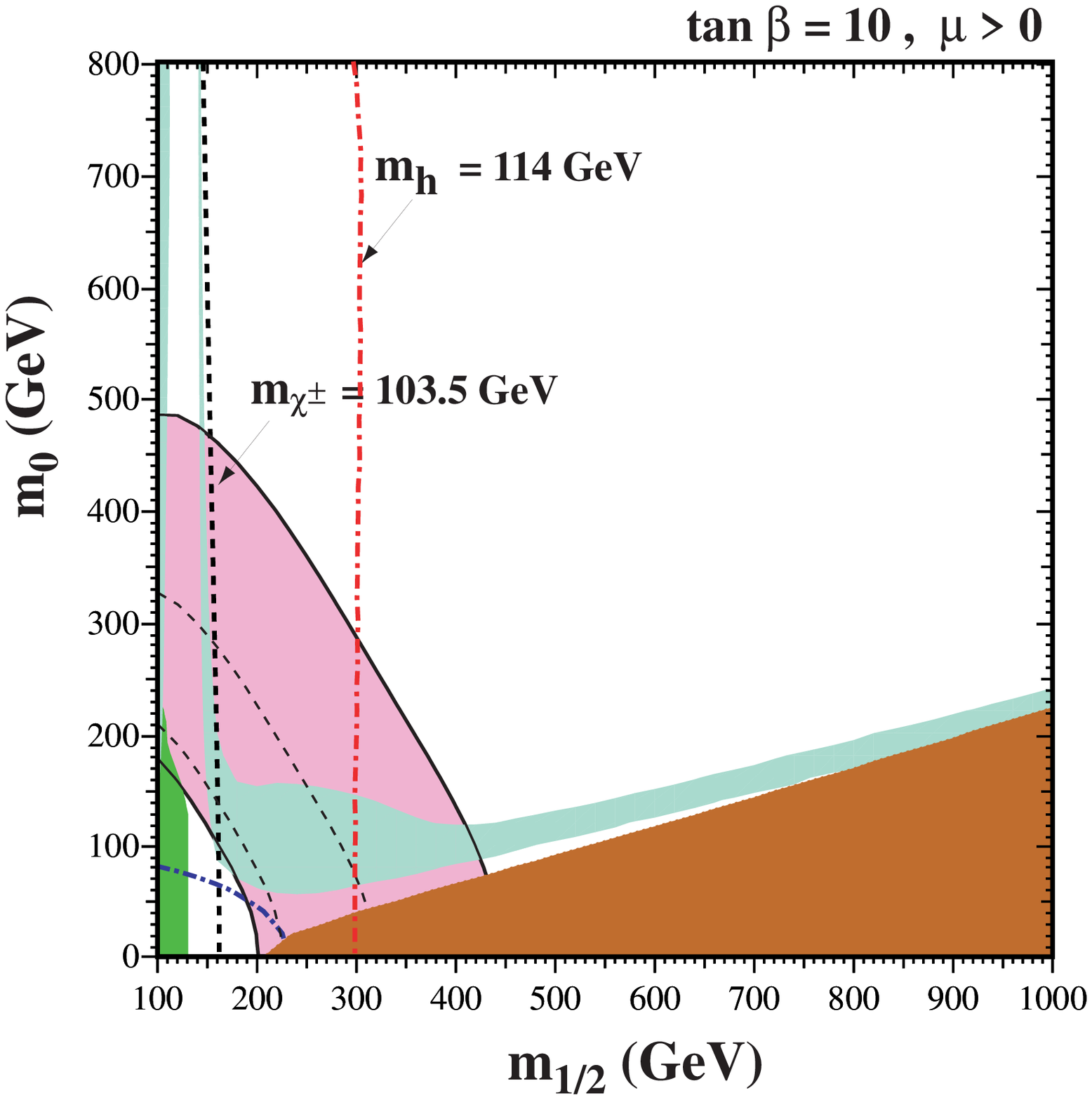,height=3.2in}
%\hspace*{-0.17in}
\epsfig{file=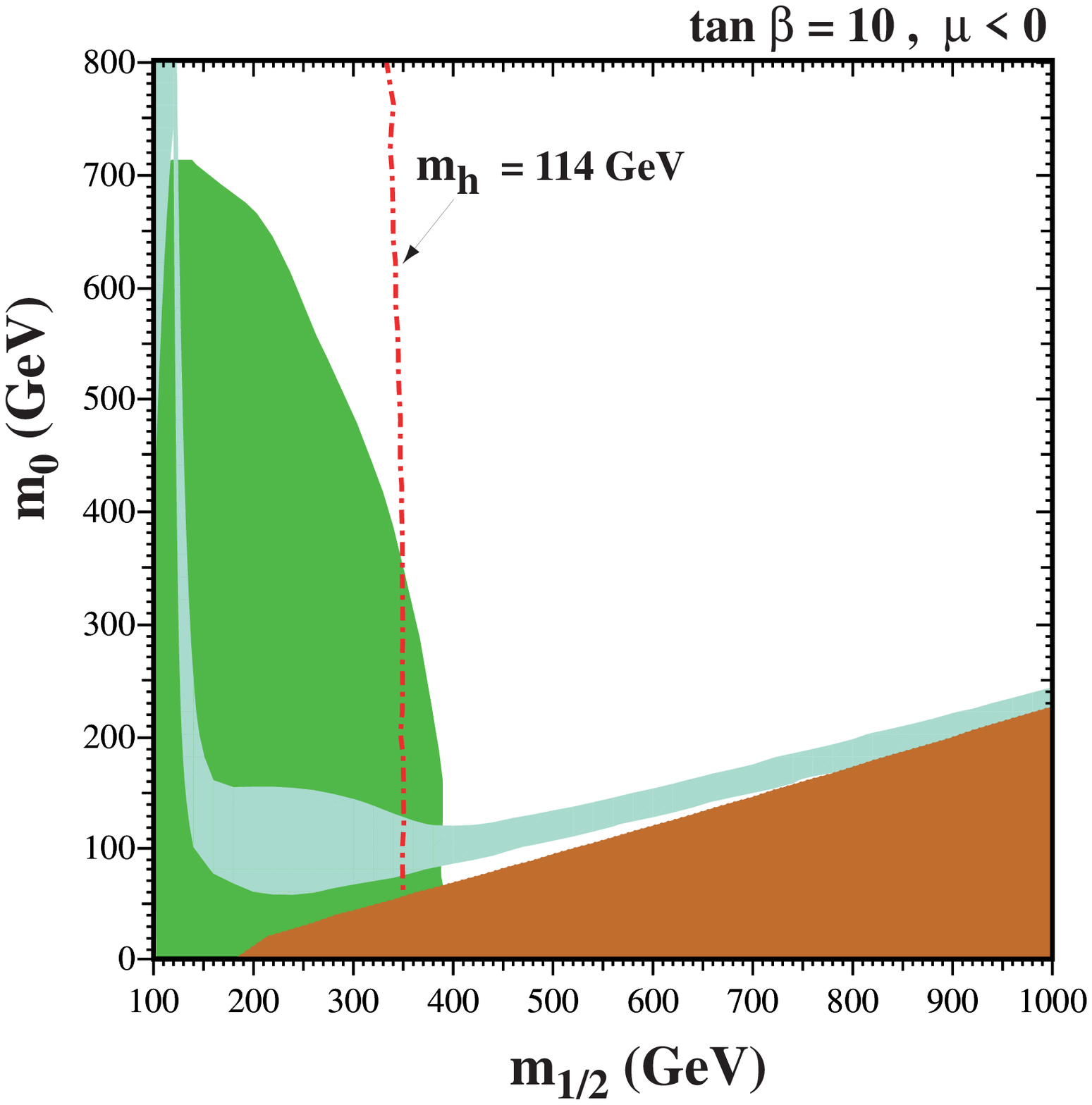,height=3.2in} \hfill
\end{minipage}
%\vspace*{-3in}
%\hspace*{-.70in}
\begin{minipage}{8in}
%\hskip -1.40in
%\vskip -.75in
\epsfig{file=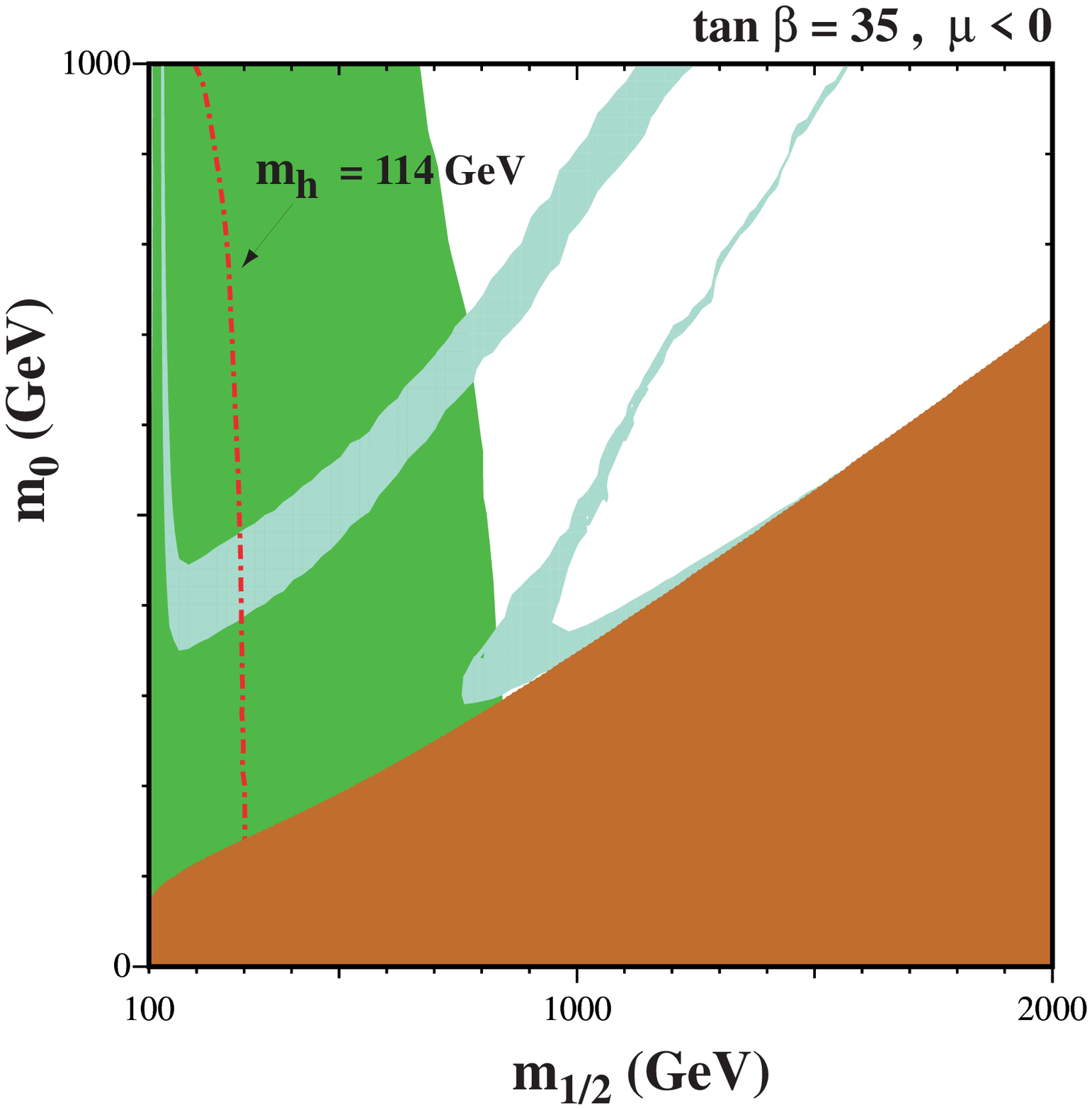,height=3.2in}
%\hspace*{-0.2in}
\epsfig{file=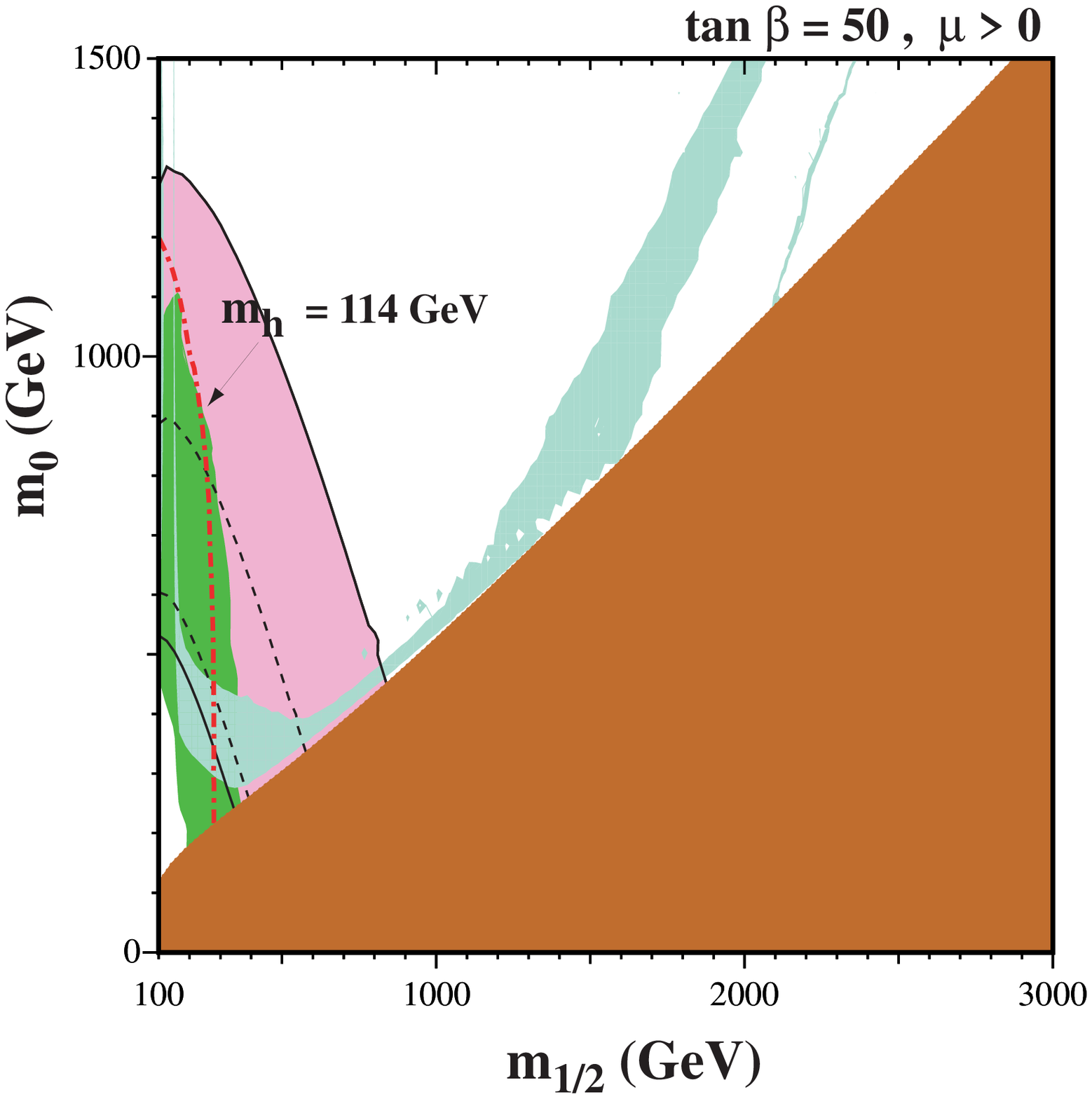,height=3.2in} \hfill
\end{minipage}
%\vskip 2.5in 
\caption{\label{fig:UHM}
{\it 
The CMSSM $(m_{1/2}, m_0)$ planes for (a) $\tan \beta = 10$ and $\mu > 0$,
(b) $\tan \beta = 10$ and $\mu < 0$, (c) $\tan \beta = 35$ and $\mu < 0$
and (d) $\tan \beta = 50$ and $\mu > 0$, assuming $A_0 = 0, m_t =
175$~GeV and
$m_b(m_b)^{\overline {MS}}_{SM} = 4.25$~GeV. The near-vertical (red)
dot-dashed lines are the contours $m_h = 114$~GeV as calculated using
{\tt FeynHiggs}~\cite{FeynHiggs}, and the near-vertical (black) dashed
line in panel (a) is the contour $m_{\chi^\pm} = 103.5$~GeV. The medium
(dark green) shaded regions are excluded by $b \to s \gamma$, and the
light (turquoise) shaded areas are the cosmologically preferred regions
with \protect\mbox{$0.1\leq\ohsq\leq 0.3$}. In the dark (brick red) shaded
regions, the LSP is the charged ${\tilde \tau}_1$, so these regions are
excluded. In panels (a) and (d), the regions allowed by the E821
measurement of $a_\mu$ at the 2-$\sigma$ level, as discussed in the text,
are shaded (pink) and bounded by solid black lines, with dashed lines
indicating the 1-$\sigma$ ranges.}}
\end{figure}

As a standard of comparison for our NUHM analysis, we first consider the
impacts of the above constraints on the parameter space of the CMSSM, in
which the soft supersymmetry-breaking Higgs scalar masses are assumed to
be universal at the input scale. In this case, as mentioned in the
Introduction, one may use the parameters $(m_{1/2}, m_0, A_0, \tan \beta)$
and the sign of $\mu$. We assume for simplicity that $A_0 = 0$, and plot
in Fig.~\ref{fig:UHM} the $(m_{1/2}, m_0)$ planes for certain choices of
$\tan \beta$ and the sign of $\mu$. These plots are similar to those
published previously~\cite{Ellis:2002rp}, but differ in using the latest
version of {\tt FeynHiggs}~\cite{FeynHiggs}  and  the latest
information on $a_\mu$ discussed above.

The shadings and lines in Fig. 1 are as follows. The dark (brick red) shaded
regions have a charged LSP, i.e. ${\tilde \tau}_1$, so these regions are
excluded. The $b \to s \gamma$ exclusion is presented by the medium
(dark green) shaded regions. The
light (turquoise) shaded areas are the cosmologically preferred regions
with $0.1\leq\ohsq\leq 0.3$. The regions allowed by the E821
measurement of $a_\mu$ at the 2-$\sigma$ level, $11.5 \times 10^{-10} < \delta
a_\mu < 56.3 \times
10^{-10}$,
are shaded (pink) and bounded by solid black lines. Only panel (a) and (d) have
regions allowed by $a_\mu$. The near-vertical (red)
dot-dashed lines are the contours $m_h = 114$~GeV, and the near-vertical
(black) dashed line in panel (a) is the contour $m_{\chi^\pm} = 103.5$~GeV
(though we do not plot this constraint in panels (b,c,d), the position of
the chargino contour would be very similar). Regions on the left of these
lines are excluded. 

We see in panel (a) of Fig.~\ref{fig:UHM} for $\tan \beta = 10$ and $\mu >
0$ that all the experimental constraints are compatible with the CMSSM for
$m_{1/2} \sim 300$ to 400~GeV and $m_0 \sim 100$~GeV, with larger values
of $m_{1/2}$ also being allowed if one relaxes the $a_\mu$ condition. In
the case of $\tan \beta = 10$ and $\mu < 0$ shown in panel (b) of
Fig.~\ref{fig:UHM}, valid only if one discards the $a_\mu$ condition, the
$m_h$ and $b \to s \gamma$ constraints both require $m_{1/2} \ga 400$~GeV
and $m_0 \ga 100$~GeV. In the case $\tan \beta = 35$ and $\mu < 0$ shown
in panel (c) of Fig.~\ref{fig:UHM}, the $b \to s \gamma$ constraint is
much stronger than the $m_h$ constraint, and imposes $m_{1/2} \ga
700$~GeV, with the $\Omega_\chi h^2$ constraint then allowing bands of
parameter space emanating from $m_0 \sim 600, 300$~GeV. Finally, in panel
(d) for $\tan \beta = 50$ and $\mu > 0$, we see again that the $m_h$ and
$b \to s \gamma$ constraints are almost equally important, imposing
$m_{1/2} \ga 300$~GeV for $m_0 \sim 400$~GeV. As in the case of panel (a),
there is again a region compatible with the $a_\mu$ constraint, extending
in this case as far as $m_{1/2} \sim 800$~GeV and $m_0 \sim 500$~GeV.

\section{Exploration of the NUHM Parameter Space}

Following our discussion of the CMSSM parameter space in
the previous Section, we now discuss how that analysis changes in the
NUHM. We extend our previous analysis~\cite{Ellis:2002wv} in two ways: 
(i) fixing
$\tan \beta = 10$ and $\mu > 0$, but choosing different values of $\mu$
and $m_A$, rather than assuming the CMSSM values, and (ii) varying $\tan
\beta$ for representative fixed values of $\mu$ and $m_A$. We make such 
selections for three projections of the NUHM, onto the $(m_{1/2}, m_0)$ 
plane, the $(\mu, m_A)$ plane and the $(\mu, M_2)$ plane.

\subsection{The $(m_{1/2}, m_0)$ Plane}

\begin{figure}
\vspace*{-0.75in}
%\hspace*{-.70in}
\begin{minipage}{8in}
\epsfig{file=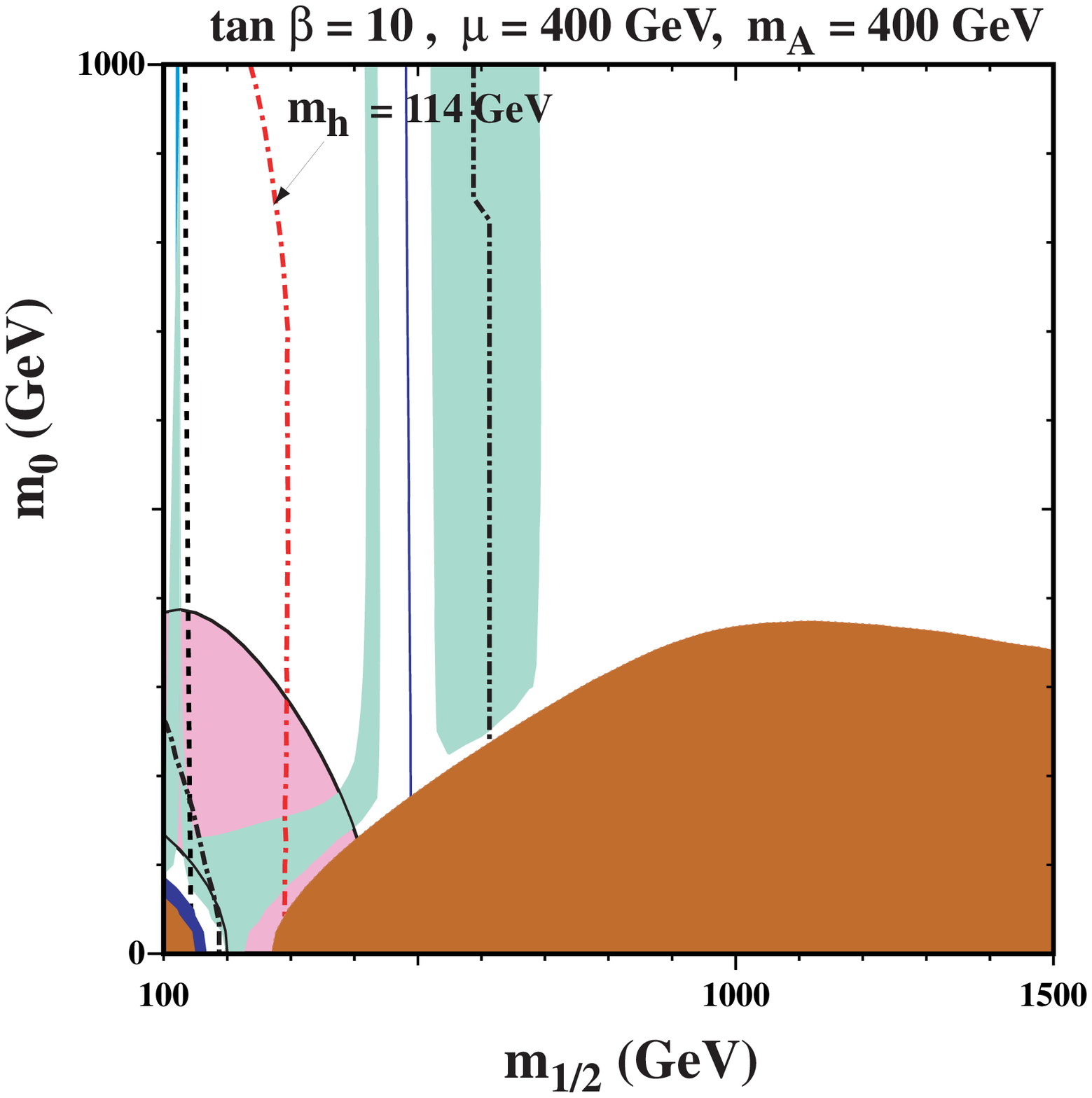,height=3.2in}
%\hspace*{-0.17in}
\epsfig{file=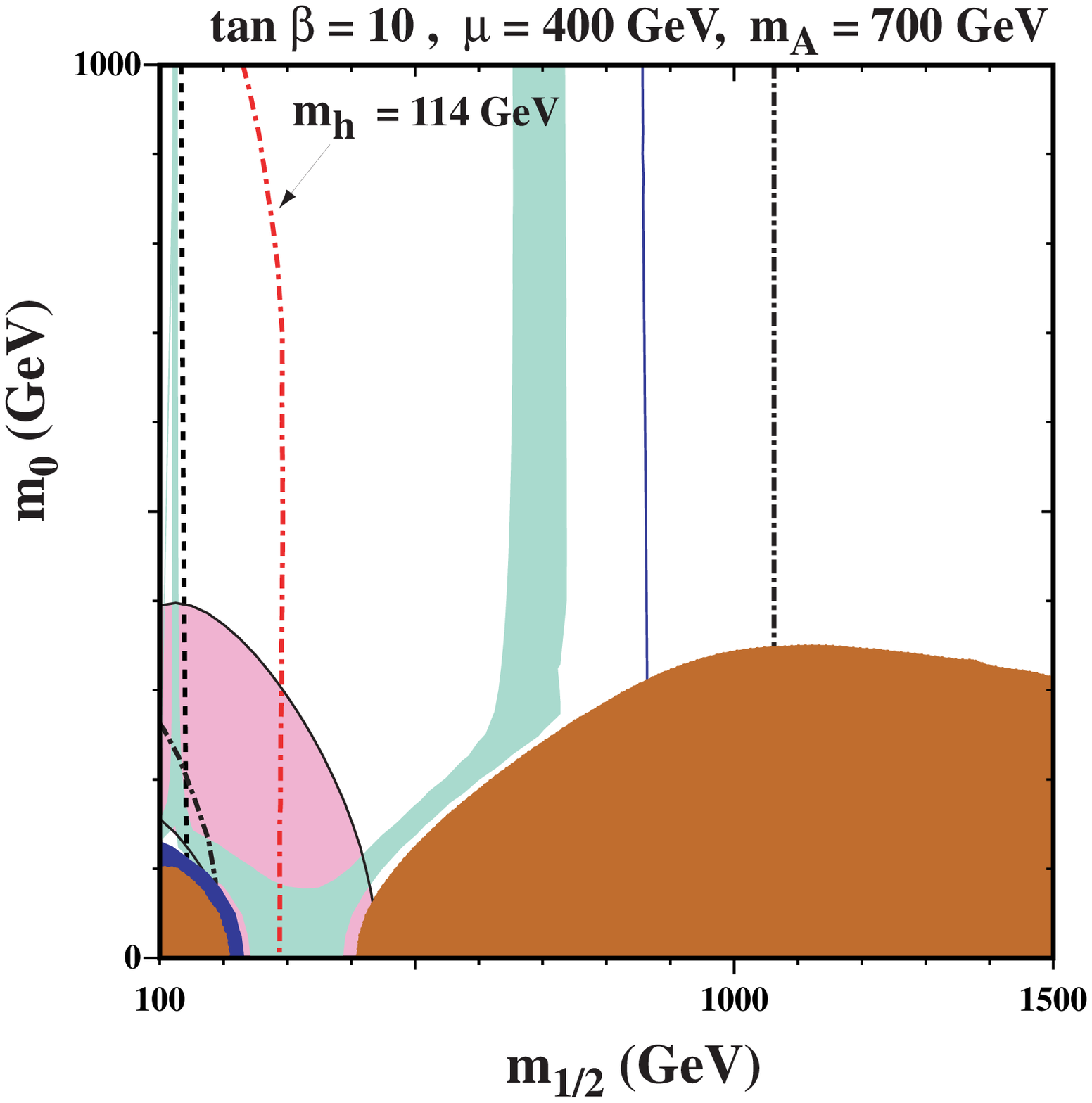,height=3.2in} \hfill
\end{minipage}
%\vspace*{1in}
%\hspace*{-.70in}
\begin{minipage}{8in}
%\hskip -1.40in
%\vskip -.75in
\epsfig{file=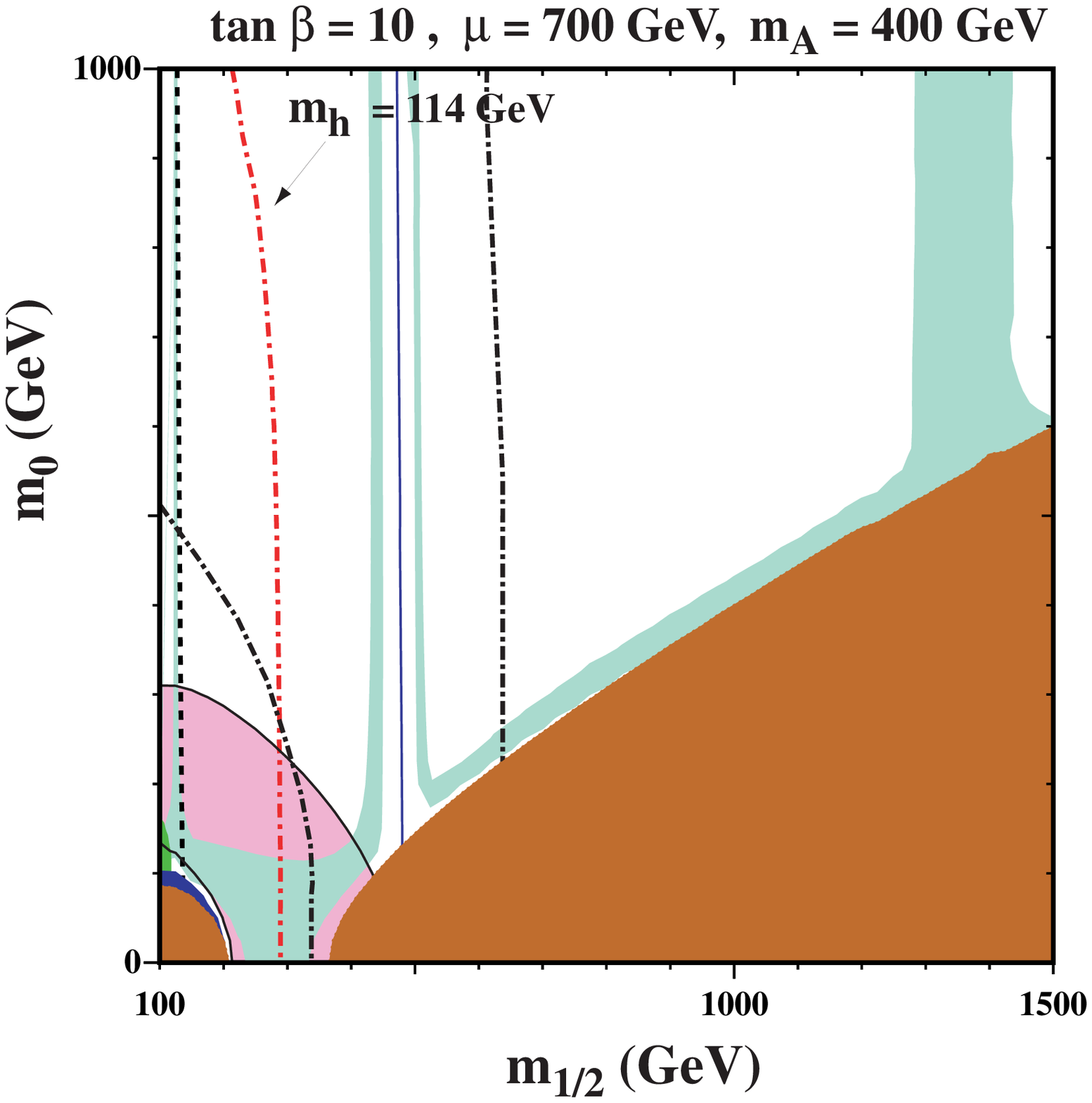,height=3.2in}
%\hspace*{-0.2in}
\epsfig{file=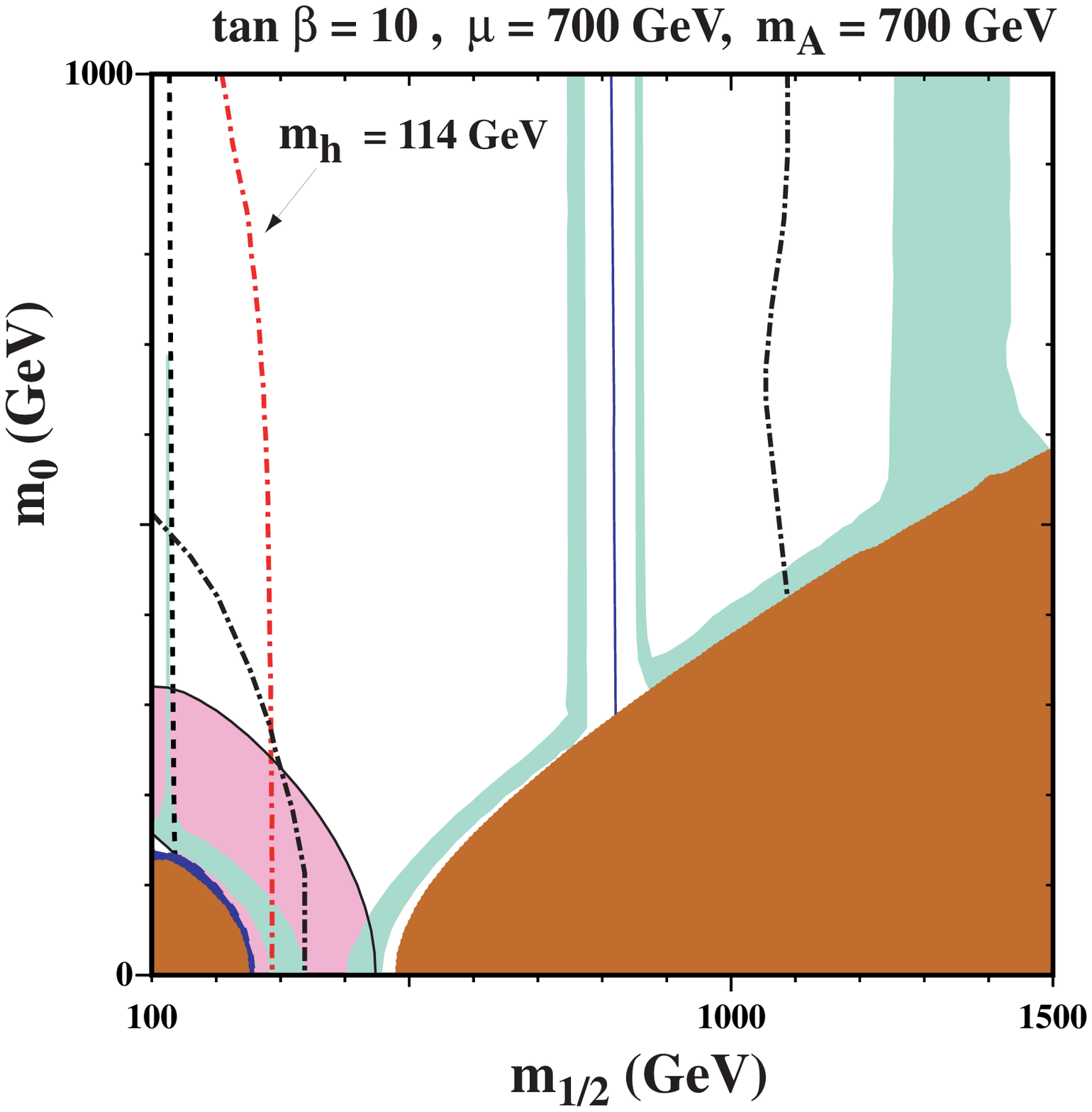,height=3.2in} \hfill
\end{minipage}
%\vskip 2.5in 
\caption{\label{fig:2}
{\it 
Projections of the NUHM model on the $(m_{1/2}, m_0)$ planes for $\tan
\beta =  10$ and (a) $\mu =
400$~GeV and $\mA = 400$~GeV, (b) $\mu = 400$~GeV and $\mA = 700$~GeV, 
(c) $\mu = 700$~GeV and $\mA
= 400$~GeV and (d)
$\mu = 700$~GeV and $\mA = 700$~GeV, assuming $A_0 = 0, m_t = 175$~GeV and
$m_b(m_b)^{\overline {MS}}_{SM} = 4.25$~GeV. The near-vertical (red)
dot-dashed lines are the contours $m_h = 114$~GeV as calculated using
{\tt FeynHiggs}~\cite{FeynHiggs}, and the near-vertical (black) dashed
lines are the contours $m_{\chi^\pm} = 103.5$~GeV. The dark (black) dot-dashed
lines indicate the GUT stability constraint. There are two such lines for each
panel and only the areas in between are allowed by this constraint. The
light (turquoise) shaded areas are the cosmologically preferred regions
with \protect\mbox{$0.1\leq\ohsq\leq 0.3$}. The dark (brick red) shaded
regions is excluded because a charged particle is lighter than the 
neutralino,
and the darker (dark blue) shaded regions is excluded because the LSP is a
sneutrino. In panel (c) there is a very small medium
(green) shaded region excluded by $b \to s \gamma$, at small $m_{1/2}$.
The regions allowed by the E821
measurement of $a_\mu$ at the 2-$\sigma$ level, as discussed in the text,
are shaded (pink) and bounded by solid black lines.}}
\end{figure}

Panel (a) of Fig.~\ref{fig:2} shows the $(m_{1/2}, m_0)$ plane for $\tan
\beta = 10$ and the particular choice $\mu = 400$~GeV and $\mA = 400$~GeV,
assuming $A_0 = 0, m_t = 175$~GeV and $m_b(m_b)^{\overline {MS}}_{SM} =
4.25$~GeV as usual. Again as usual, the light (turquoise) shaded area is
the cosmologically preferred region with $0.1 \leq \Omega_{\chi} h^2 \leq
0.3$. There is a bulk region satisfying this preference at $m_{1/2} \sim 50$~GeV
to 350~GeV, $m_0 \sim 50$~GeV to 150~GeV. 
The dark (red) shaded regions are excluded because a charged
sparticle is lighter than the neutralino.  As in the CMSSM shown in
Fig.~\ref{fig:UHM}, the ${\tilde \tau_1}$ is the LSP in the bigger area at
larger $m_{1/2}$, and there are light (turquoise) shaded strips close to
these forbidden regions where coannihilation suppresses the relic density
sufficiently to be cosmologically acceptable. Further away from these
regions, the relic density is generally too high. However, for larger
$m_{1/2}$ there is another suppression, discussed below, which makes the
relic density too low.  At small $m_{1/2}$ and $m_0$ the left handed
sleptons, and also the sneutrinos, become lighter than the neutralino. The
darker (dark blue) shaded area is where a sneutrino is the LSP. Within
these excluded regions there are also areas with tachyonic sparticles.

The near-vertical dark (black) dashed and light (red) dot-dashed lines in
Fig.~\ref{fig:2} are the LEP exclusion contours $m_{\chi^\pm} > 104$~GeV
and $m_h > 114$~GeV respectively. As in the CMSSM case, they exclude low values of
$m_{1/2}$, and hence rule out rapid relic annihilation via direct-channel
$h$ and $Z^0$ poles. The solid lines curved around small values of
$m_{1/2}$ and $m_0$ bound the light (pink) shaded region favoured by
$a_\mu$ and recent analyses of the $e^+ e^-$ data.

A striking feature in Fig.~\ref{fig:2}(a) when $m_{1/2} \sim 500$~GeV is a
strip with low $\ohsq$, which has bands with acceptable relic density on
either side.  The low-$\ohsq$ strip is due to rapid annihilation via the
direct-channel $A, H$ poles which occur when $m_\chi = m_A / 2 = 200$~GeV,
indicated by the near-vertical solid (blue) line. Analogous
rapid-annihilation strips have been noticed previously in the CMSSM
\cite{funnel,efgosi}, but at larger $\tan \beta$ as seen in
Fig.~\ref{fig:UHM}. There, they are diagonal in the $(m_{1/2}, m_0)$
plane, reflecting a CMSSM link between $m_0$ and $m_A$ that is absent in
our implementation of the NUHM. The right-hand band in Fig.~\ref{fig:2}(a)
with acceptable $\ohsq$ is broadened because the neutralino acquires
significant Higgsino content, and the relic density is suppressed by the
increased $W^+ W^-$ production. Hereafter, we will call this the
`transition' band, which in this case is incidentally coincident with the
right-hand rapid annihilation band~\footnote{As the
neutralino acquires more Higgsino content, annihilation to $W^+W^-$ 
production increases, whilst fermion-pair production decreases (except for 
$t\bar t$). Around $m_{1/2} \sim$ 625 GeV, there is a threshold for $hA, 
hH$ production, which decreases
$\ohsq$ to $\sim 0.08$, not far below the preferred range. However, 
the decrease in fermion production quickly raises $\ohsq$ again for 
larger $m_{1/2}$. There is a very narrow stripe with $\ohsq < 0.1$ of 
width $\delta m_{1/2} \sim 2$ GeV, which is not
shown in the figure due to problems of resolution.}. As
$m_{1/2}$ increases, the neutralino  becomes almost degenerate with the
second lightest neutralino and the lighter chargino, and the $\chi
- \chi^\prime - \chi^\pm$  coannihilation processes
eventually push $\ohsq < 0.1$ when $m_{1/2} \ga 700$~GeV. We note that
chargino-slepton coannihilation processes become important at the junction
between the vertical bands in Fig.~\ref{fig:2}(a)  and the
neutralino-slepton coannihilation strip that parallels the $m_\chi =
m_{\widetilde{\tau}_1}$ boundary of the forbidden (red) charged-LSP region.  

There are two dark (black) dash-dotted lines in Fig.~\ref{fig:2}(a) that
indicate where scalar squared masses become negative at the input GUT
scale for one of the Higgs multiplets, specifically when either $(m_1(M_X)^2 +
\mu(M_X)^2) < 0$ or $(m_2(M_X)^2 + \mu(M_X)^2) < 0$.  One of these GUT
stability lines is
near-vertical at $m_{1/2} \sim 600$~GeV, and the other is a curved line at
$m_{1/2} \sim 150$~GeV, $m_0 \sim 200$~GeV. We take the point of view that
regions outside either of these lines are excluded, because the preferred
electroweak vacuum should be energetically favoured and not bypassed early
in the evolution of the Universe, but a different point of view is argued
in~\cite{fors}.

Thus, combining all the constraints, the allowed regions are those between
the $m_h$ line at $m_{1/2} \sim 300$~GeV and the stability line at
$m_{1/2} \sim 600$~GeV, which include two rapid-annihilation bands, some of the
transition band and the
junction between the bulk and coannihilation regions around $m_{1/2} \sim
350$~GeV, $m_0 \sim 150$~GeV. If one incorporates also the putative
$a_\mu$ constraint, only the latter region survives. We note however,
that if the $\tau$ data were used in the $g-2$ analysis, the constraint
from $a_\mu$ only excludes the lower left corner of the plane and large
values of $m_{1/2}$ and $m_0$ survive at the 2$\sigma$ level. 

Panel (b) of Fig.~\ref{fig:2} is for $\mu = 400$~GeV and $\mA = 700$~GeV.
We notice immediately that the heavy Higgs pole and the right-hand
boundary of the GUT stability region move out to larger $m_{1/2} \sim 850,
1050$~GeV, respectively, as one would expect for larger $m_A$. At this
value of $m_A$, the transition strip and and the rapid annihilation
(`funnel') strip are separate. However the latter would be to the right of
the transition strip and hence the $\ohsq$ bands on both sides of the
rapid-annihilation strip that was prominent in panel (a) have disappeared,
due to enhanced chargino-neutralino coannihilation effects. Panel (b) has
the interesting feature that there is a region of $m_{1/2} \sim 300$ to
400~GeV where $m_0 = 0$ is allowed. As discussed in~\cite{ENO} this
possibility, which would be favoured in some specific no-scale models of
supersymmetry breaking, is {\it disallowed} in the CMSSM. The small-$m_0$
region is even favoured in this variant of the NUHM by the putative
$a_\mu$ constraint.

Panel (c) of Fig.~\ref{fig:2} is for $\mu = 700$~GeV and $\mA
= 400$~GeV. In this case, we see that the rapid-annihilation strip is back 
to $m_{1/2} \sim 500$~GeV, reflecting the smaller value of $\mA$, whereas 
the transition band has separated off to large 
$m_{1/2}$, reflecting the larger value of $\mu$. However, this band is 
excluded in this case by the GUT stability requirement. GUT stability also 
excludes the possibility that $m_0 =0$. In this case, the putative $a_\mu$ 
constraint would restrict one to around $m_{1/2} \sim 400$~GeV and $m_0 
\sim 100$~GeV.

Finally, panel (d) of  Fig.~\ref{fig:2} is for $\mu = 700$~GeV and $\mA = 
700$~GeV. In this case, the rapid-annihilation strip has again moved  
to larger $m_{1/2}$, related to the larger value of $\mA$, and the 
transition band  at large $m_{1/2}$ is again 
excluded by the GUT stability requirement. The bulk region has disappeared in
this panel, reflecting the fact that the values of $\mu$ and $\mA$ here has
strayed away from their values in the bulk region for the CMSSM. 
GUT stability no longer 
excludes $m_0 =0$, and this possibility would be selected by the 
putative $a_\mu$ constraint.

\begin{figure}
\vspace*{-0.75in}
%\hspace*{-.70in}
\begin{minipage}{8in}
\epsfig{file=m0M_10_0_400_700.eps,height=3.2in}
%\hspace*{-0.17in}
\epsfig{file=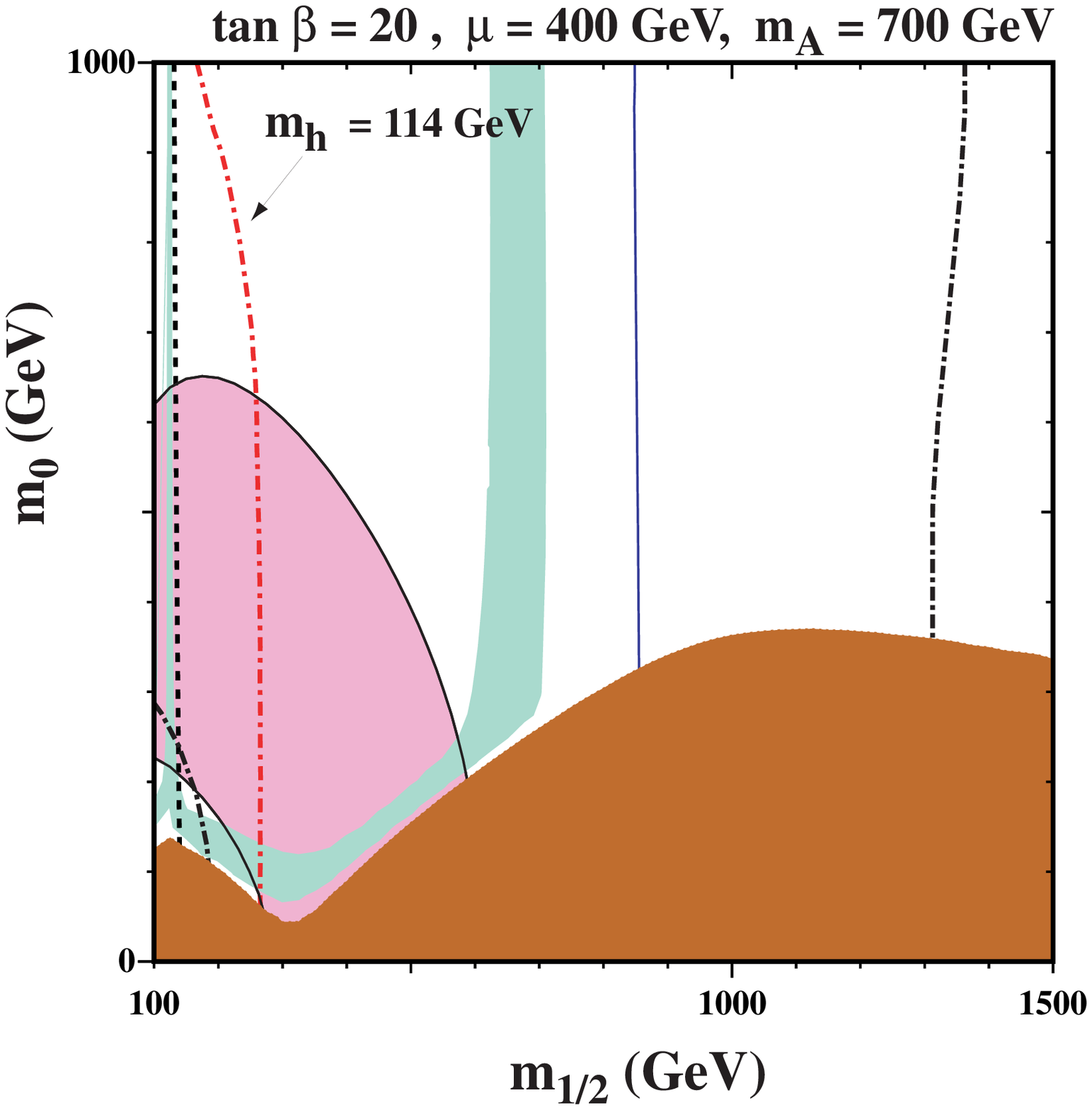,height=3.2in} \hfill
\end{minipage}
%\vspace*{-3in}
%\hspace*{-.70in}
\begin{minipage}{8in}
%\hskip -1.40in
%\vskip -.75in
\epsfig{file=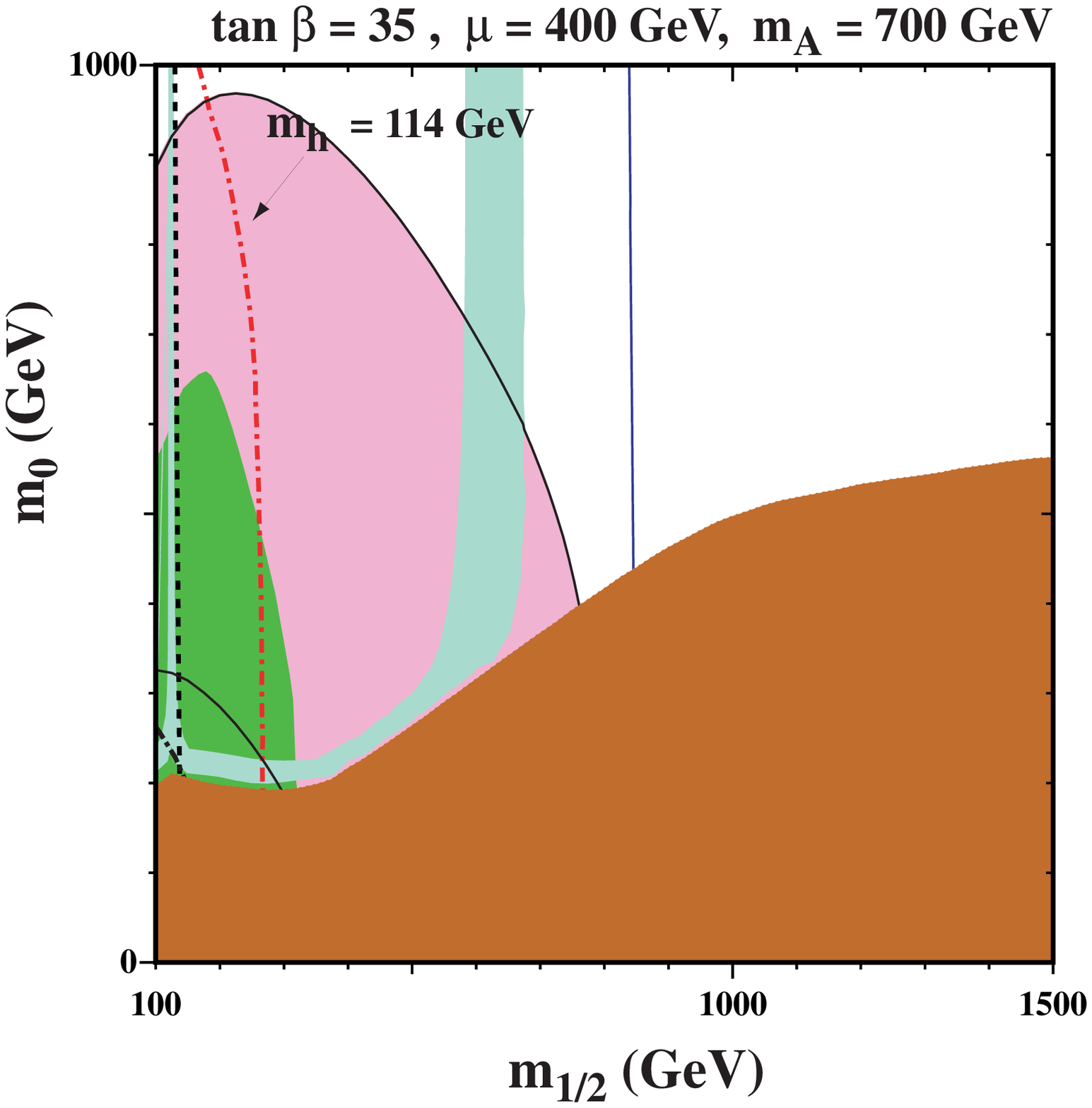,height=3.2in}
%\hspace*{-0.2in}
\epsfig{file=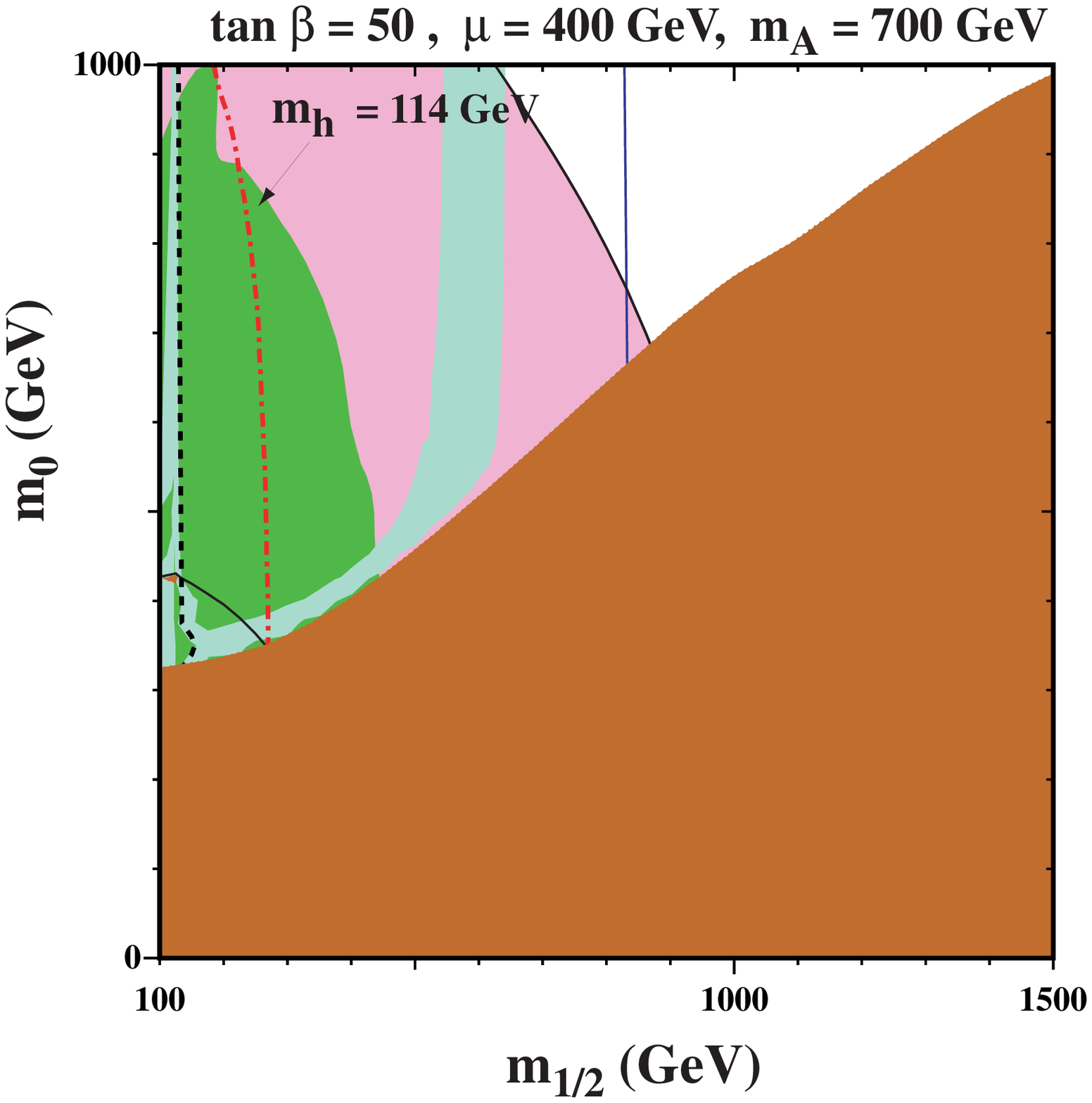,height=3.2in} \hfill
\end{minipage}
%\vskip 2.5in 
\caption{\label{fig:3}
{\it 
The NUHM $(m_{1/2}, m_0)$ planes for (a) $\tan \beta = 10$ , (b) $\tan
\beta = 20$, (c) $\tan \beta = 35$ and (d)
$\tan \beta = 50$, for $\mu = 400$~GeV, $\mA = 700$~GeV, assuming $A_0 =
0, m_t = 175$~GeV and $m_b(m_b)^{\overline {MS}}_{SM} = 4.25$~GeV. The
shadings and  line styles are the same as in Fig.~\ref{fig:2}.}}
\end{figure}

We now turn, in Fig.~\ref{fig:3}, to the impact of varying $\tan \beta$,
keeping $\mu = 400$~GeV and $\mA = 700$~GeV. For convenience, panel (a)  
reproduces Fig.~\ref{fig:2}(b) with $\tan \beta = 10$. In general as $\tan
\beta$ is increased, two major trends are visible. One is for the region
excluded by the requirement that the LSP be neutral to spread up to larger
values of $m_0$, and the other is for the $a_\mu$ constraint to move out
to larger values of $m_{1/2}$ and $m_0$.

Specifically, we see in panel (b) of Fig.~\ref{fig:3} for $\tan \beta =
20$ that, whereas the heavy Higgs pole at $m_{1/2} \sim 850$~GeV
essentially does not move, the ${\tilde \tau}_1$ LSP and coannihilation 
strip lying above the
excluded charged-LSP region rise to larger $m_0$. This has the effect of
excluding the $m_0 = 0$ option that was present in panel (a). At low
$m_{1/2}$, the $m_h$ constraint is stronger than the GUT stability and
other constraints. The $a_\mu$ constraint would allow a larger range of
$m_{1/2}$ than in panel (a), extending up to $\sim 550$~GeV.

Continuing in panel (c) of Fig.~\ref{fig:3} to $\tan \beta = 35$, we see
that the minimum value of $m_0$ has now risen to $\sim 200$~GeV. We also
see that the $b \to s \gamma$ constraint is now important, enforcing
$m_{1/2} > 300$~GeV in the region preferred by the relic density. Because
of this and the $m_h$ constraint, the GUT stability constraint is now
irrelevant at low $m_{1/2}$, whereas at high $m_{1/2}$ it has vanished off
the screen, and is in any case also irrelevant because of
chargino-neutralino coannihilation.  The $a_\mu$ constraint would now
allow part of the cosmological band on the left side of the
rapid-annihilation strip. These trends are strengthened in panel (d) of
Fig.~\ref{fig:3} for $\tan \beta = 50$, where we see that $m_0, m_{1/2}
\ga 400$~GeV because of the $b \to s \gamma$ constraint, and $a_\mu$ would
allow $m_0 \la 1100$~GeV along the transition band.

\subsection{The $(\mu, m_A)$ Plane}

\begin{figure}
\vspace*{-0.75in}
%\hspace*{-.70in}
\begin{minipage}{8in}
\epsfig{file=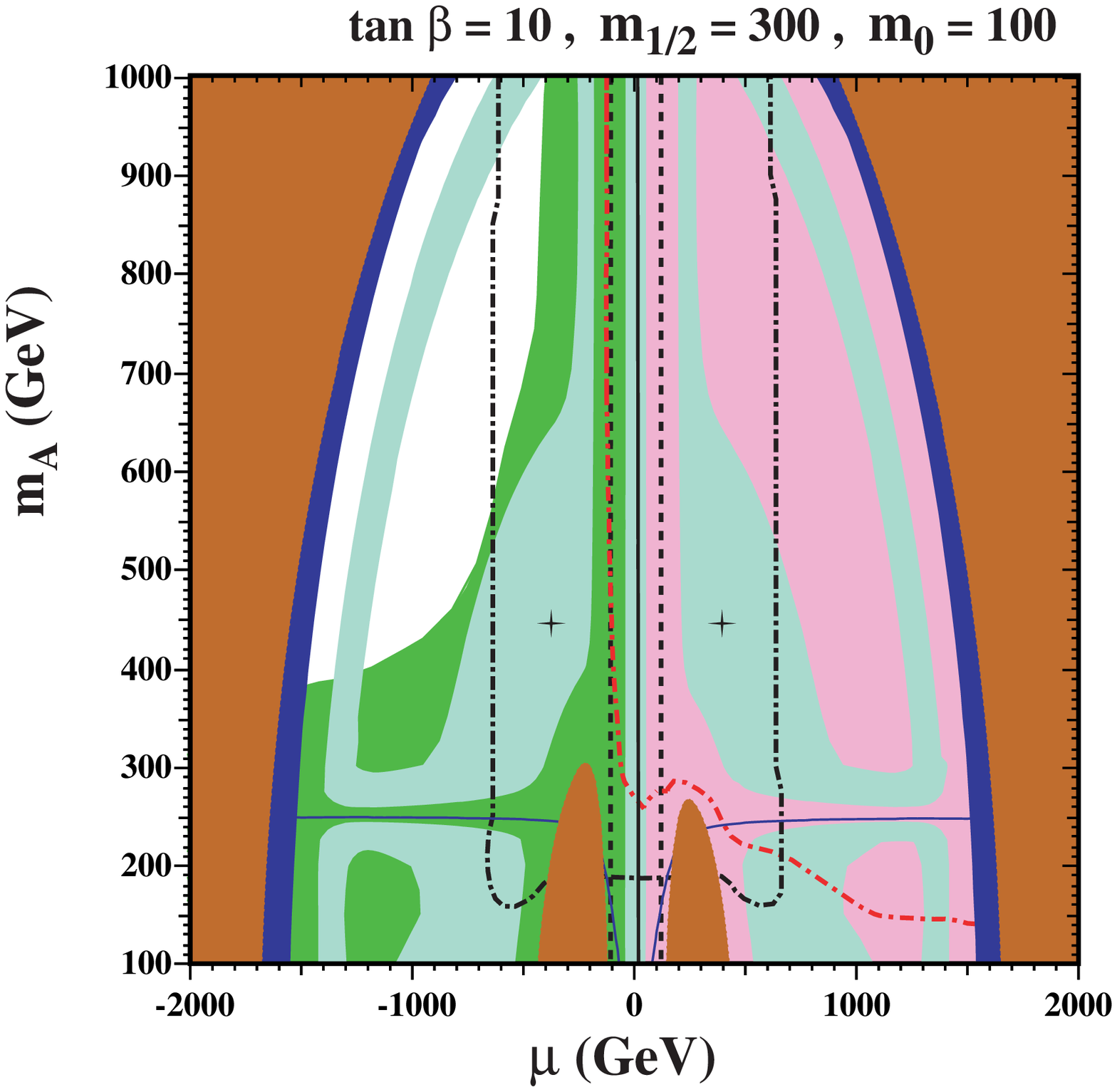,height=3.2in}
%\hspace*{-0.17in}
\epsfig{file=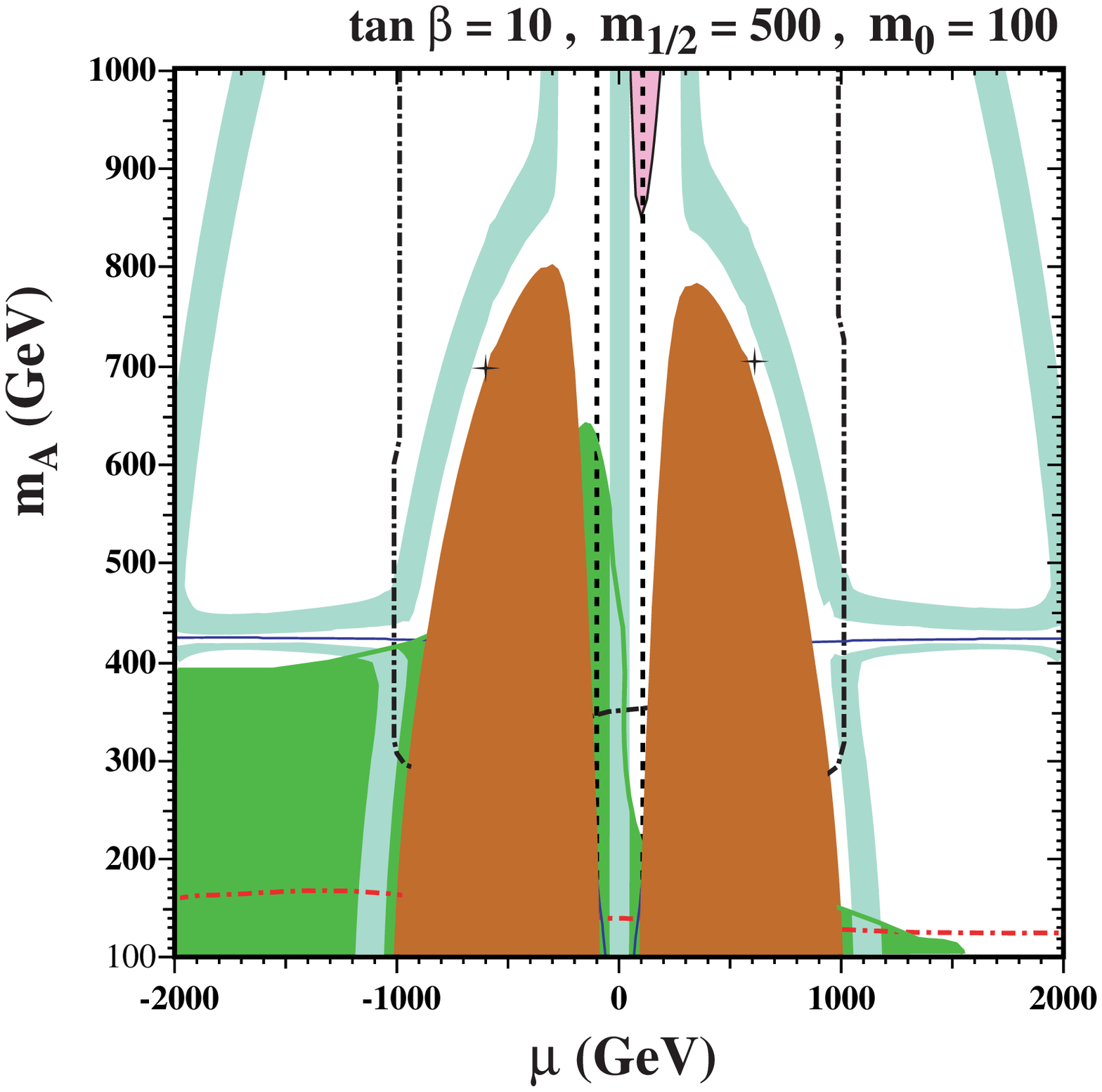,height=3.2in} \hfill
\end{minipage}
%\vspace*{-3in}
%\hspace*{-.70in}
\begin{minipage}{8in}
%\hskip -1.40in
%\vskip -.75in
\epsfig{file=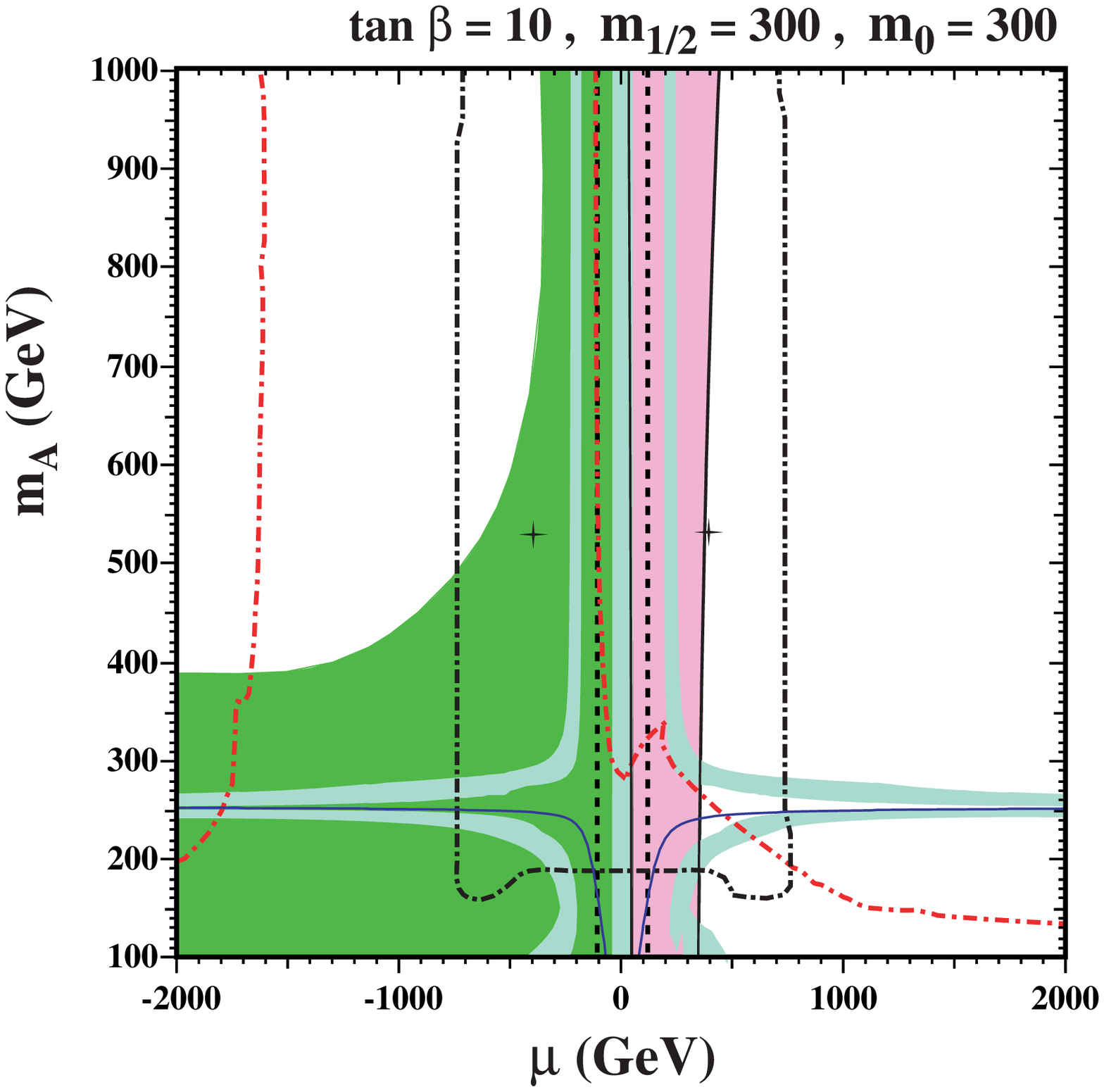,height=3.2in}
%\hspace*{-0.2in}
\epsfig{file=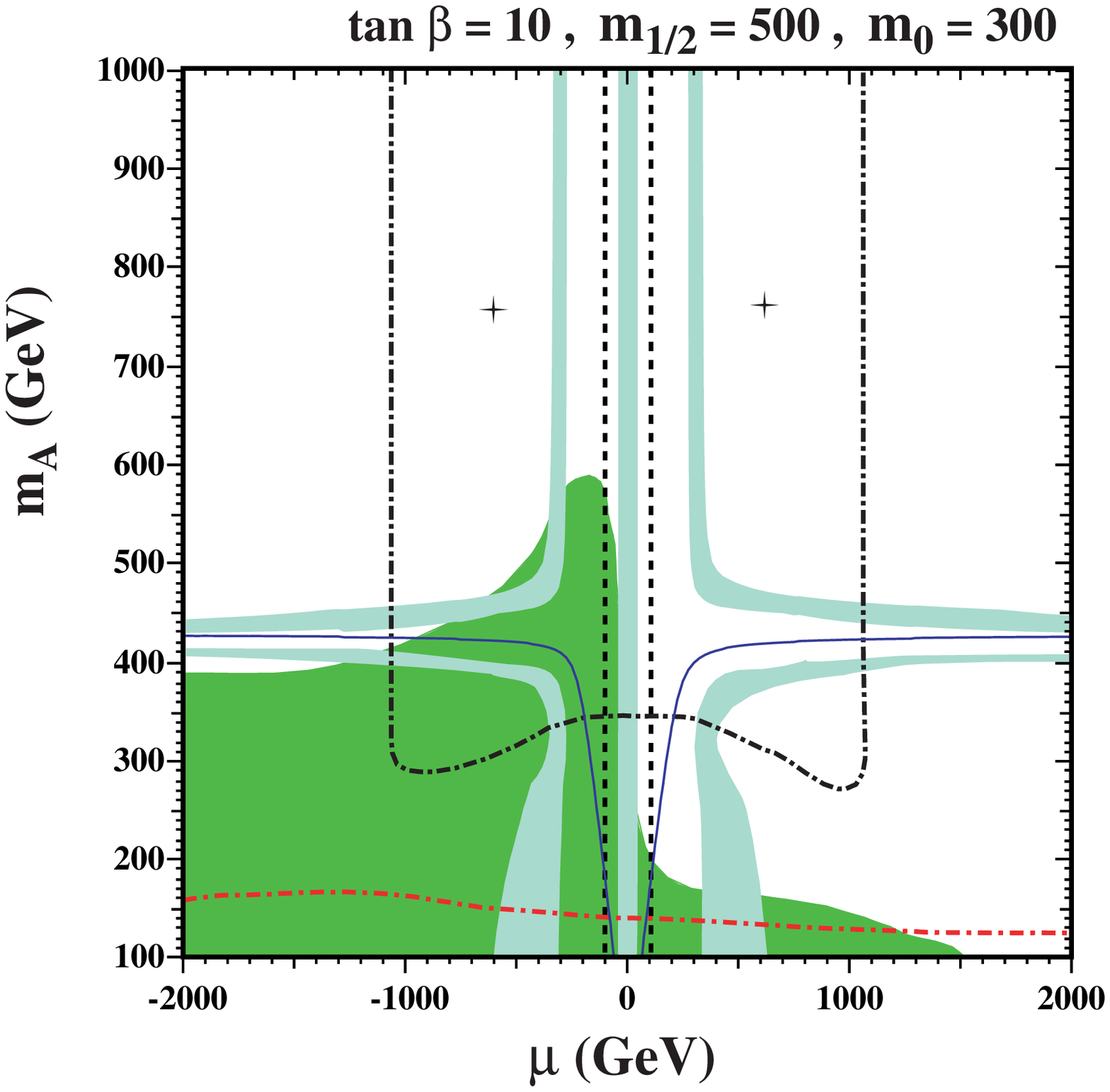,height=3.2in} \hfill
\end{minipage}
%\vskip 2.5in 
\caption{\label{fig:4}
{\it 
The NUHM $(\mu, \mA)$ planes for $\tan \beta = 10$, (a) $m_0 = 100$~GeV and
$m_{1/2} = 300$~GeV, (b) $m_0 = 100$~GeV and $m_{1/2} = 500$~GeV, (c) $m_0 =
300$~GeV and $m_{1/2} = 300$~GeV and (d)
$m_0 = 300$~GeV and $m_{1/2} = 500$~GeV, assuming $A_0 = 0, m_t = 175$~GeV
and
$m_b(m_b)^{\overline {MS}}_{SM} = 4.25$~GeV. The shadings and
line styles are the same as in Fig.~\ref{fig:2}.}}
\end{figure}

We now analyze the range of possibilities in the $(\mu, m_A)$ plane for
various fixed choices of $m_{1/2}$ and $m_0$, first choosing $\tan \beta 
= 10$.
Panel (a) of Fig.~\ref{fig:4} displays the $(\mu, m_A)$ plane for $m_{1/2}
= 300$~GeV, $m_0 = 100$~GeV. As we saw in Fig.~\ref{fig:UHM}(a),
there is a CMSSM point with $\mu > 0$ that is compatible with all the
constraints for these values of $m_{1/2}$ and $m_0$. The corresponding
CMSSM point for $\mu < 0$ in Fig.~\ref{fig:UHM}(b) is, however,
incompatible with the $m_h$ and $b \to s \gamma$ constraints, as well as
the putative $a_\mu$ constraint. The CMSSM equivalent points are shown as
crosses in panel (a) of Fig.~\ref{fig:4}, and have $(\mu, m_A) \simeq
(\pm 390, 450)$ GeV.

As usual, there are dark (red) regions where there is one or more charged
sparticle lighter than the neutralino $\chi$ so that $\chi$ is no longer the
LSP. First, there are `shark's teeth' at $|\mu| \sim 300$~GeV,
$m_A \lappeq 300$~GeV in panel (a) of Fig.~\ref{fig:4} where the ${\tilde
\tau}_1$ is the LSP. At small
$|\mu|$, particularly at small $m_A$ when the mass difference $m_2^2 -
m_1^2$ is small, the ${\widetilde \tau}_R$ mass 
is driven small, making the ${\widetilde \tau}_1$ the LSP~\footnote{Note that
the ${\widetilde e}_R$ and ${\widetilde \mu}_R$ masses are also driven
to small values, along
with the ${\widetilde \tau}_R$, and in fact there are small regions where the
degenerate ${\widetilde e}_R$ and ${\widetilde \mu}_R$ become the LSP.}.  
At even smaller
$|\mu|$, however, the lightest neutralino gets lighter again, 
since $m_\chi \simeq \mu$ when $\mu < M_1 \simeq 0.4 \, 
m_{1/2}$~\footnote{We note that regions in Figs.~\ref{fig:4}(a,b) and 
\ref{fig:6}(b) at small $|\mu|$ and $m_A$ are 
excluded by the LEP slectron search~\cite{LEPSUSYWG_0101}. However, these 
regions are all excluded by other limits, such as GUT stability, the LEP 
chargino limit~\cite{LEPsusy} and $b \to s \gamma$, and so are not shown 
separately.}. In addition, there are 
extended dark (red) shaded regions at large $|\mu|$ where left-handed sleptons
become lighter than the neutralino. However, the electron sneutrino $\snu_e$
and the muon sneutrino $\snu_\mu$ (which are degenerate within the
approximations used here) have become joint LSPs at a slightly smaller $|\mu|$. 
Since the possibility of
sneutrino dark matter has been essentially excluded by a combination of
`$\nu$ counting' at LEP, which excludes $m_{\tilde \nu} < 43$~GeV
\cite{efos}, and searches for cold dark matter, which exclude heavier
${\tilde \nu}$ weighing $\la 1$~TeV \cite{Falk:1994es}, we still demand
that the LSP be a neutralino $\chi$. The darker (dark blue) shaded
regions are where the sneutrinos are the LSPs, and therefore excluded. 

\begin{figure}
\vspace*{-0.75in}
\hspace*{.70in}
\begin{minipage}{8in}
\epsfig{file=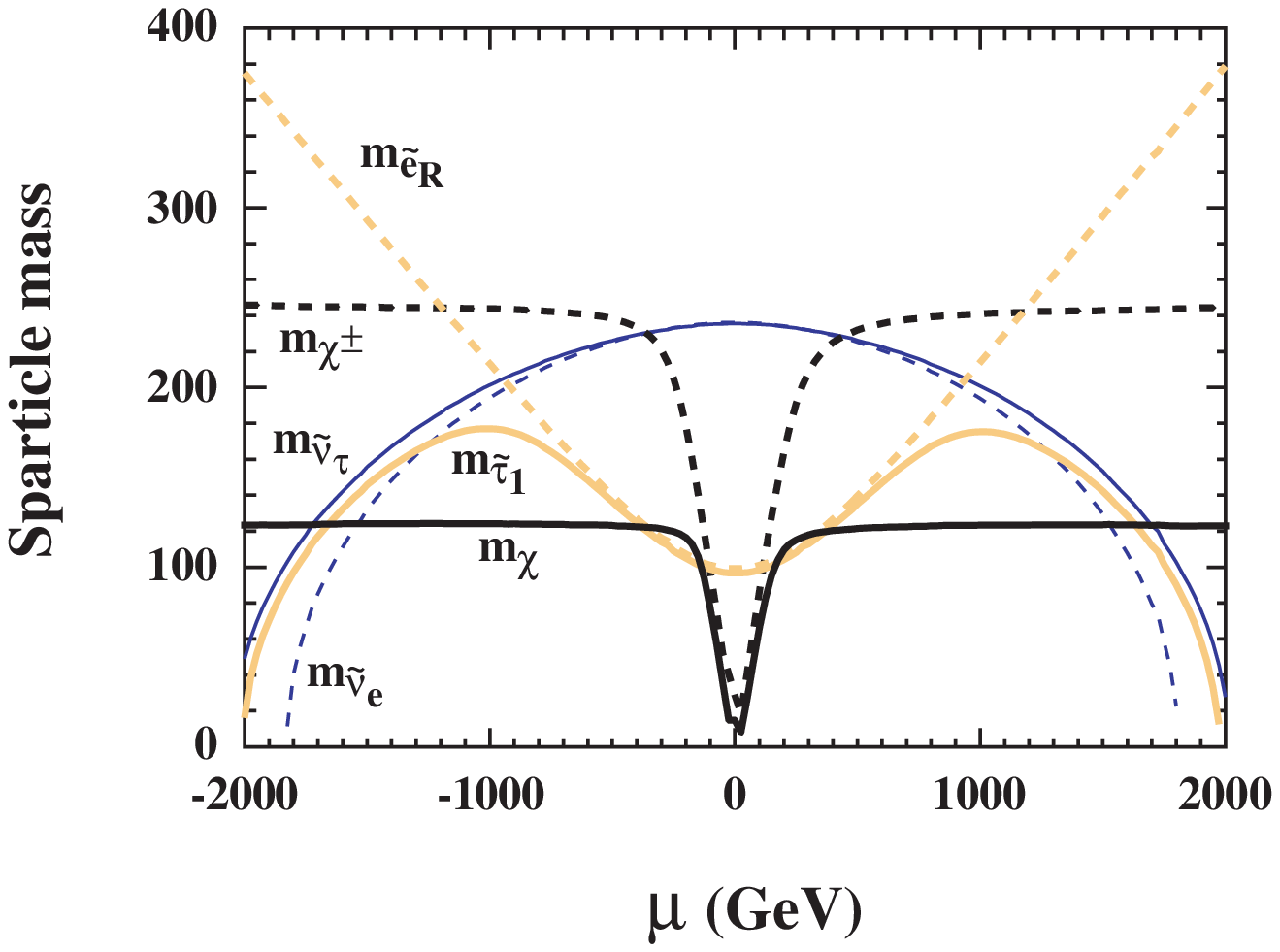,height=3.2in} \hfill
\end{minipage}
%\vspace*{-3in}
\hspace*{.70in}
\begin{minipage}{8in}
%\hskip -1.40in
%\vskip -.75in
\epsfig{file=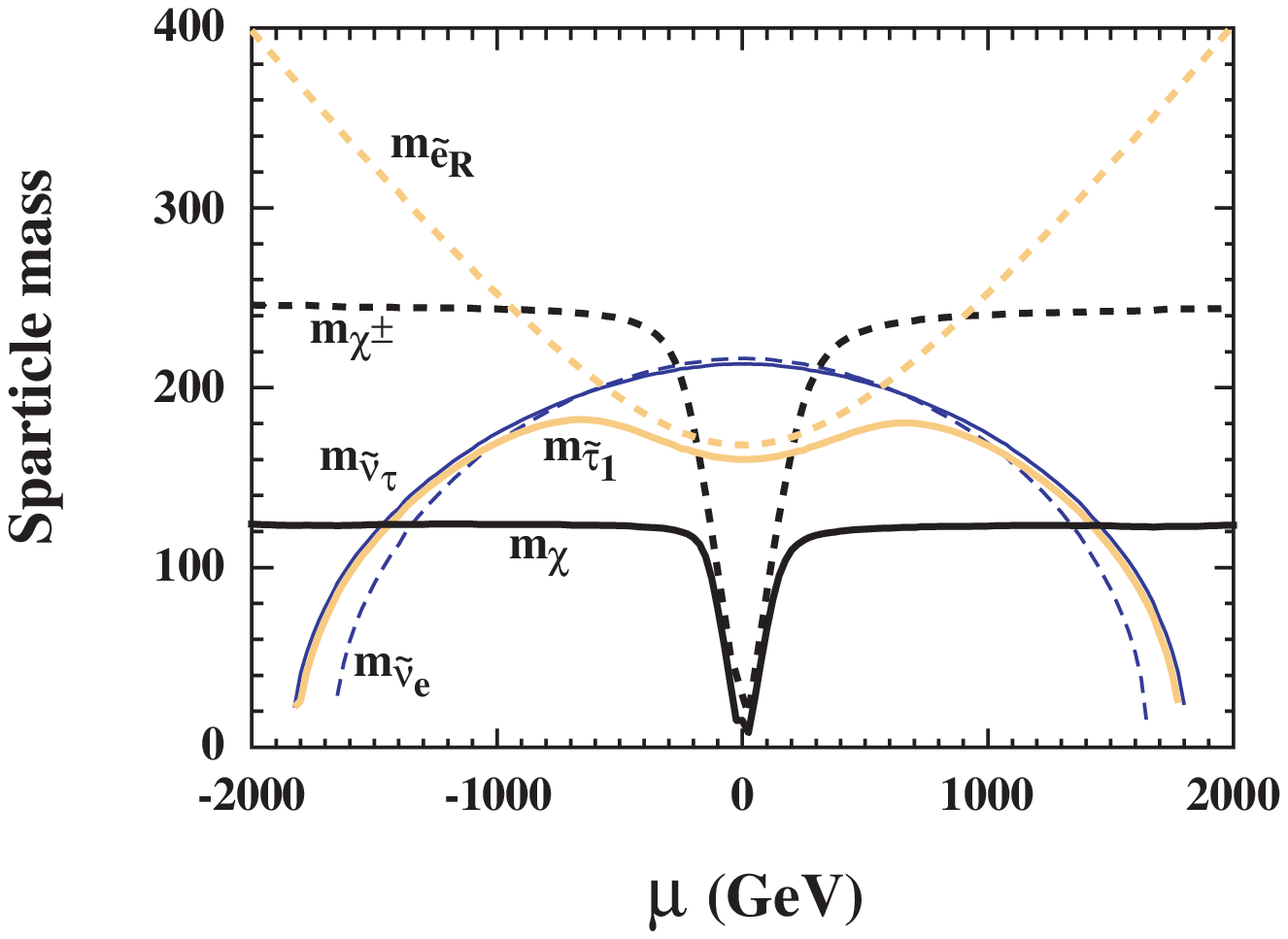,height=3.2in} \hfill
\end{minipage}
%\vskip 2.5in 
\caption{\label{fig:mass}
{\it 
The masses $m_\chi$ (dark solid), $m_{\chi^\pm}$ (dark dashed), 
$m_{\widetilde{\tau}_1}$ (light solid),
$m_{\widetilde{e}_R}$ (light dashed),
$m_{\widetilde{\nu}_\tau}$ (thin solid) and $m_{\widetilde{\nu}_e}$
(thin dashed) as functions of
$\mu$ for $\tan
\beta = 10$, $m_{1/2} = 300$~GeV, $m_0 = 100$~GeV for (a) $\mA = 200$~GeV and
(b) $\mA = 600$~GeV, assuming $A_0 = 0, m_t = 175$~GeV and
$m_b(m_b)^{\overline {MS}}_{SM} = 4.25$~GeV. }}
\end{figure}

To explain the possible spectra better, we plot in Fig.~\ref{fig:mass}
some sparticle masses as functions of $\mu$, with other parameters fixed.
In particular, we plot the neutralino mass $m_\chi$ (dark solid curve),
the chargino mass $m_{\chi^\pm}$ (dark dashed curve), the lighter stau
mass $m_{\widetilde \tau_1}$ (light sold curve), the right-handed
selectron mass $m_{\widetilde e_R}$ (light dashed curve), and the
sneutrino masses $m_{\widetilde {\nu_\tau}}$ and $m_{\widetilde {\nu_e}}$
(thin solid and dashed curves respectively). We have omitted the curve for
$m_{\widetilde e_L}$, which is very similar to the sneutrino curves, for
reasons of clarity. In both panels of Fig.~\ref{fig:mass}, we have fixed
most parameters as in Fig.~\ref{fig:4}(a).  Fig.~\ref{fig:mass}(a) is for
$m_A = 200$ GeV, and we see that the lighter stau and the sneutrinos are
lighter than the neutralino for very large values of $|\mu|$.  As $|\mu|$
is decreased, there is a small window in which ${\widetilde \nu_e}$ is the
LSP. The curious behaviour of the stau mass curve at large $|\mu|$ is
due to a large and negative value for $S$ which drives $m_{\widetilde
\tau_L}$ down while driving $m_{\widetilde \tau_R}$ up (cf. Eq.
(\ref{rges})).  For large values of
$|\mu|$, the lighter stau is mostly left-handed, whilst it is mostly
right-handed for smaller
$|\mu|
\la 1000$ GeV. When $|\mu| \la 400$ GeV, both the stau and the
right-handed selectron are lighter than the neutralino until $|\mu|$ is
very small and the neutralino becomes Higgsino-like, with its mass scaling
as $\mu$. The `shark's teeth' in Fig.~\ref{fig:4}(a)  correspond to the
range around $|\mu| \sim 200$~GeV where the lighter stau is the LSP, with
the right-handed selectron slightly heavier.

In Fig.~\ref{fig:mass}(b), we have fixed $m_A = 600$~GeV. Here, one sees a
similar pattern of masses at large $|\mu|$. However, at small $|\mu|$ the
lighter stau and right-handed selectron are much heavier, so the
neutralino remains the LSP. This reflects the fact that $m_A = 600$~GeV
is above the tips of the `shark's teeth' in Fig.~\ref{fig:4}(a). It is
clear that the  sizes of these `shark's teeth' must depend sensitively on
the NUHM  parameters, as seen in the other panels of Fig.~\ref{fig:4} and 
subsequent figures.

Returning to panel (a) of Fig.~\ref{fig:4}, we see strips adjacent to the
${\widetilde \nu}_{e,\mu}$ LSP regions, where neutralino-sneutrino
coannihilation is important in suppressing the relic density to an
acceptable level.  Appendix D presents details of our calculations of
${\tilde \nu} - \chi$ coannihilation channels. We consider here only
$\snu_e$ and $\snu_\mu$, for which the effects of $m_e$, $m_\mu$ and L-R
mixing in the slepton mass matrix are negligible, leading to some
simplifications compared to the $\snu_\tau$ (whose inclusion would have 
little effect on our figures). The inclusion of
neutralino-sneutrino coannihilation in panel (a) squeezes inwards slightly
the coannihilation strips. As we see later, the effect of
neutralino-sneutrino coannihilation is more noticeable at larger $\tan
\beta$. Again as usual, the light (turquoise) shaded regions are those for
which $0.1 < \Omega_\chi h^2 < 0.3$.

The thick cosmological region at smaller $\mu$ in panel (a) corresponds to
the `bulk' region familiar from CMSSM studies. Note that this region is
squeezed outward
near the `shark's teeth' by the neutralino-slepton coannihilation. 
Extending upward in $m_A$
from this bulk region, there is another light (turquoise) shaded band at
smaller $|\mu|$. This is the transition band, where the neutralino gets more
Higgsino-content and the
annihilation to $W^+ W^-$ becomes important, yielding a relic density in
the allowed range, as happens in the focus-point region~\cite{focus} in
the CMSSM. For smaller $|\mu|$, the relic density becomes too small due to
$\chi - \chi^{\prime} - \chi^+$ coannihilations, and the chargino-slepton
annihilations described in Appendix C must be taken into account where
this strip meets the neutralino-slepton coannihilation strip discussed
earlier. The LEP limit on the chargino mass excludes a strip at even
smaller $|\mu|$ where the relic density would again have come into the
cosmologically preferred region.
There are also horizontal bands of acceptable relic density when $m_A \sim
250$~GeV, that are separated by strips of low relic density, due to rapid
annihilation through the $A$ (indicated by solid (blue) lines) and $H$
poles.

Underlying the cosmological regions are dark (green) shaded
regions excluded by $b \to s \gamma$, which are more important for $\mu <
0$, as expected from previous analyses. Also important is the $m_h$
constraint, calculated using the {\tt FeynHiggs}
programme~\cite{FeynHiggs}, which excludes the option $\mu < 0$ in panel
(a) of Fig.~\ref{fig:4}. The putative $a_\mu$ constraint would also
exclude the $\mu < 0$ half-plane, while allowing all of the $\mu > 0$
parameter space for this particular choice of $\tan \beta$, $m_{1/2}$ and
$m_0$.

The darker (black) dot-dashed lines in Fig.~\ref{fig:4}(a) indicates where
one or the other of the Higgs mass-squared becomes negative at the input
GUT scale. We see that these constraints exclude much of the cosmological
region still permitted, apart from the `bulk' region and part of the
`transition' region for small $\mu > 0$.

We conclude from Fig.~\ref{fig:4}(a) that moderate positive values of $\mu
< 700$~GeV are favoured and $m_A$ is unlikely to be very small, though 
there is a very small allowed region below the rapid $A,H$ annihilation 
strip where
$m_A \sim 230$~GeV. The cosmological relic density lies within the range
favoured by astrophysics and cosmology in a large fraction of the
remaining NUHM parameter space for $\mu > 0$, generalizing the CMSSM point
that is indicated by the cross.

The notations used for the constraints illustrated in the other panels of
Fig.~\ref{fig:4} are the same, but the constraints interplay in different
ways. In panel (b), we have chosen a larger value of $m_{1/2}$. In this
case, the dark (red) shaded `shark's teeth' at moderate $|\mu|$ and small
$m_A$ where the LSP is charged have expanded greatly, and are flanked by
pale (turquoise)  shaded regions where neutralino-slepton coannihilation
produces an acceptable relic density. On the other hand, the dark (red) shaded 
and the darker (dark blue) shaded regions at large $|\mu|$ have moved out of
the panel displayed, and one sees
only parts of the adjacent coannihilation strips. The relatively large
value of $m_{1/2}$ keeps the rate of $b \to s \gamma$ under control unless
$m_A$ is small and $\mu < 0$. The chargino constraint is similar to that
in panel (a), whereas the $m_h$ constraint is irrelevant due to the large
value of $m_{1/2}$. The putative $a_\mu$ constraint would allow only a very
small region in this panel, but without any cosmological preferred region.
Finally, we observe that the GUT stability constraint now allows larger values
of $|\mu| \lappeq 1000$~GeV
and $m_A \gappeq 300$~GeV.

We now turn to panel (c) of Fig.~\ref{fig:4}, which is for $\tan \beta =
10, m_{1/2} = 300$~GeV and $m_0 = 300$~GeV. In this case, the `shark's
teeth' have disappeared, as have the neutralino-slepton strips at large
$|\mu|$ (due to the large value of $m_0$). Negative values of $\mu$ are
excluded partially by
$m_h$ and by $b
\to s \gamma$, as well as by the putative $a_\mu$ constraint, which would
permit a strip with $\mu > 0$. GUT stability enforces $\mu \lappeq
800$~GeV, and also provides a lower limit on $m_A$ that is irrelevant
because of the other constraints. In contrast, in panel (d) of
Fig.~\ref{fig:4} for $m_{1/2} = 500$~GeV and $m_0 = 300$~GeV, we see a
similar `cruciform' pattern for the regions allowed by cosmology, but $b
\to s \gamma$ has only rather limited impact for $\mu < 0$ and $m_h$ is
irrelevant: there is no region consistent with the putative $a_\mu$
constraint. In this case, $m_A$ could be as small as the $\sim 300$~GeV
allowed by the GUT stability constraint, and $|\mu|$ could be as large as
$\sim 1100$~GeV.

\begin{figure}
\vspace*{-0.75in}
%\hspace*{-.70in}
\begin{minipage}{8in}
\epsfig{file=mumA_10_0_100_300.eps,height=3.2in}
%\hspace*{-0.17in}
\epsfig{file=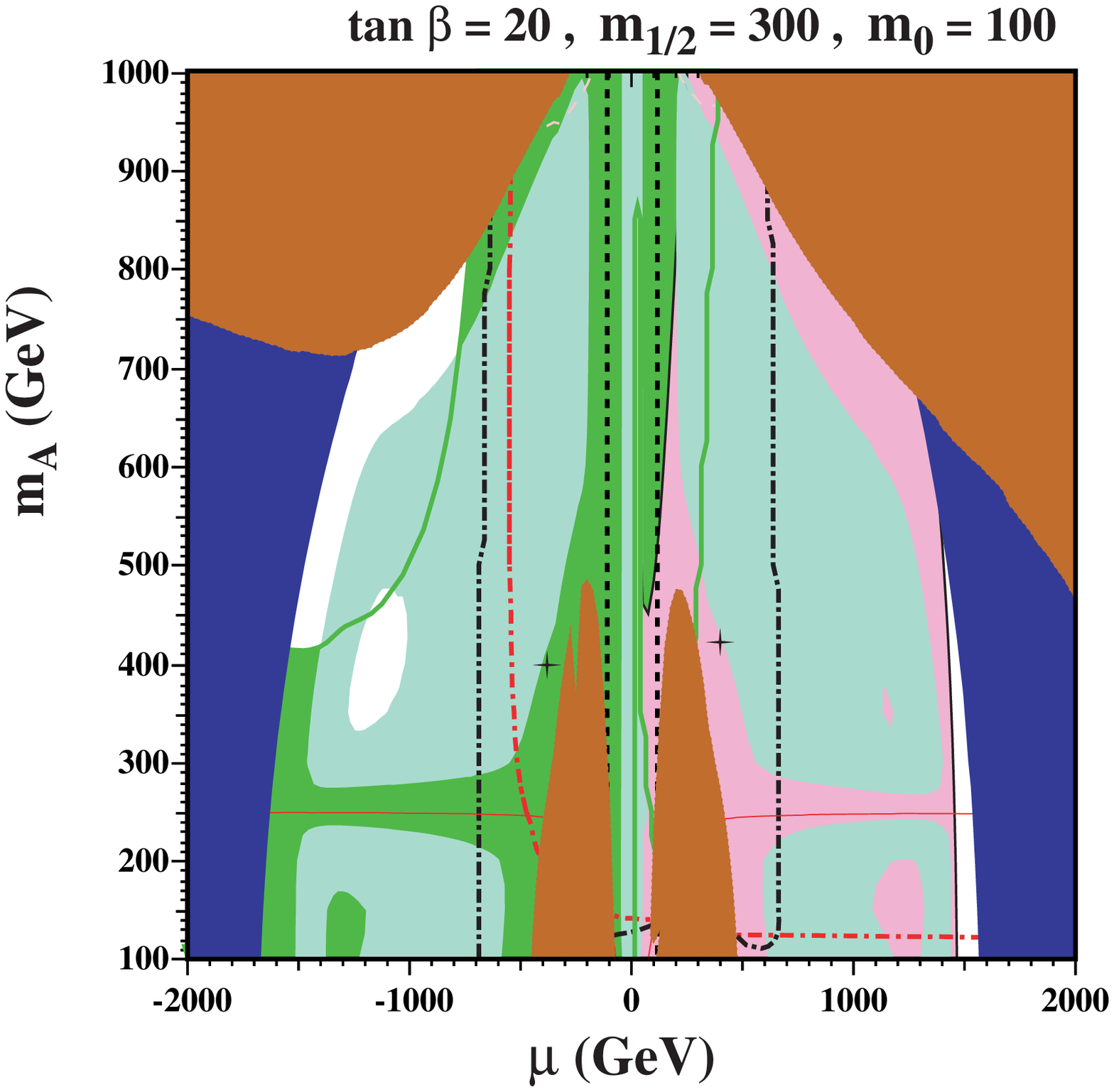,height=3.2in} \hfill
\end{minipage}
%\vspace*{-3in}
%\hspace*{-.70in}
\begin{minipage}{8in}
%\hskip -1.40in
%\vskip -.75in
\epsfig{file=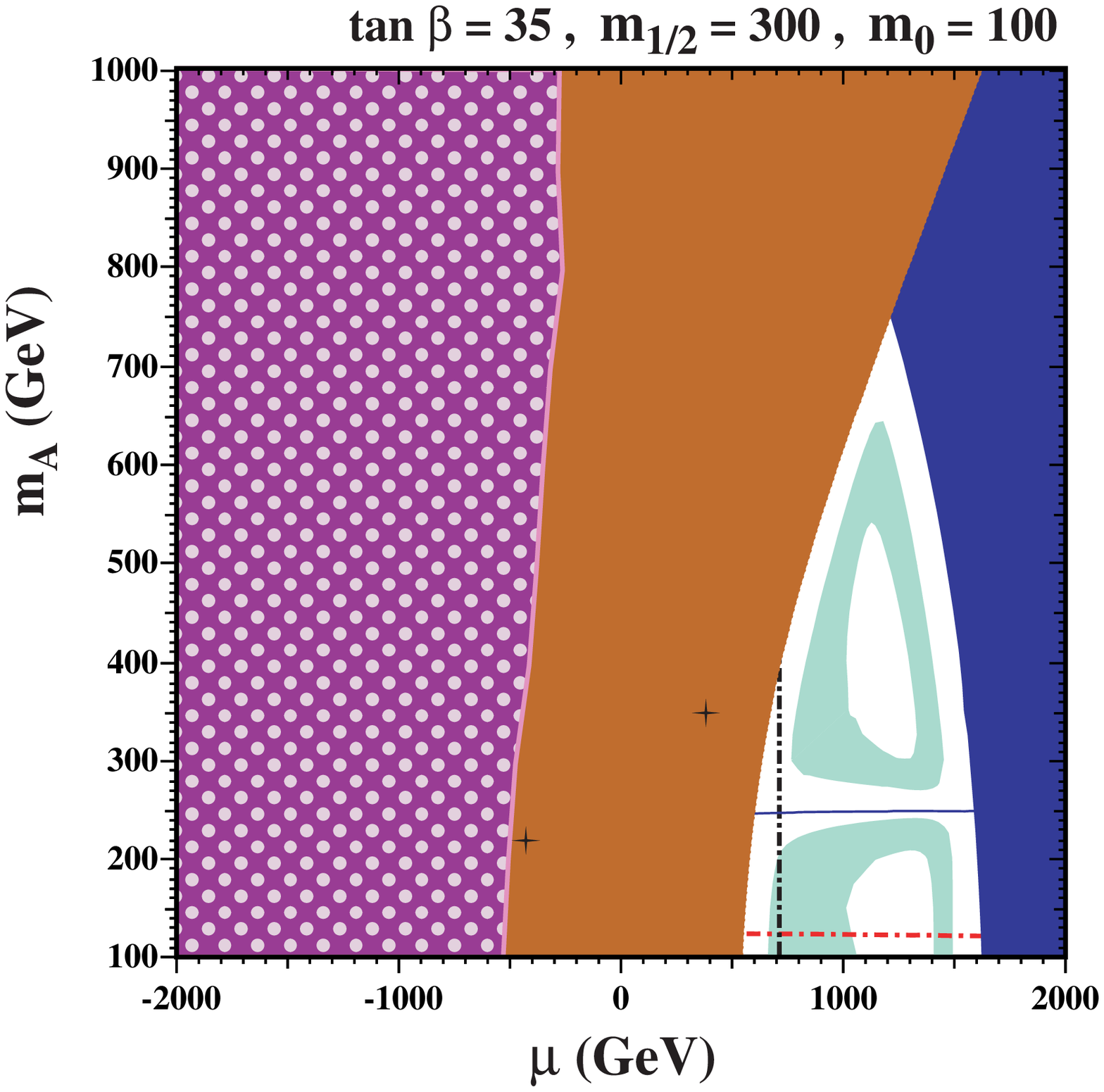,height=3.2in}
%\hspace*{-0.2in}
\epsfig{file=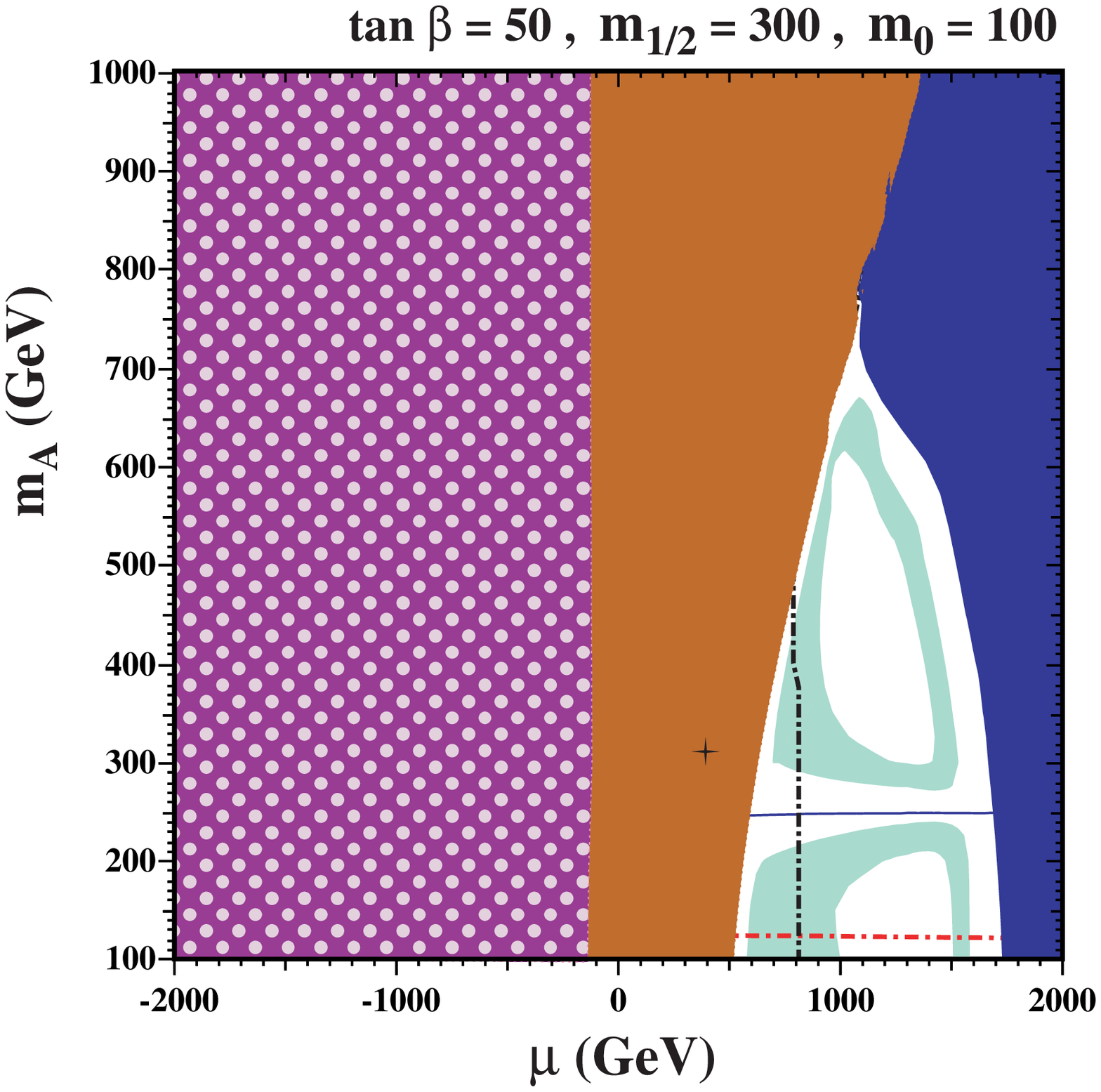,height=3.2in} \hfill
\end{minipage}
%\vskip 2.5in 
\caption{\label{fig:5}
{\it 
The NUHM $(\mu, \mA)$ planes for $m_0 = 100$~GeV and
$m_{1/2} = 300$~GeV, for (a) $\tan \beta = 10$, (b) $\tan \beta = 20$, (c) $\tan
\beta = 35$ and (d) $\tan \beta = 50$, assuming $A_0 = 0, m_t = 175$~GeV and
$m_b(m_b)^{\overline {MS}}_{SM} = 4.25$~GeV. The shadings and
line styles are the same as in Fig.~\ref{fig:2}, and 
there is no consistent electroweak vacuum in the polka-dotted region.}}
\end{figure}

\begin{figure}
\vspace*{-0.75in}
%\hspace*{-.70in}
\begin{minipage}{8in}
\epsfig{file=mumA_10_0_300_500.eps,height=3.2in}
%\hspace*{-0.17in}
\epsfig{file=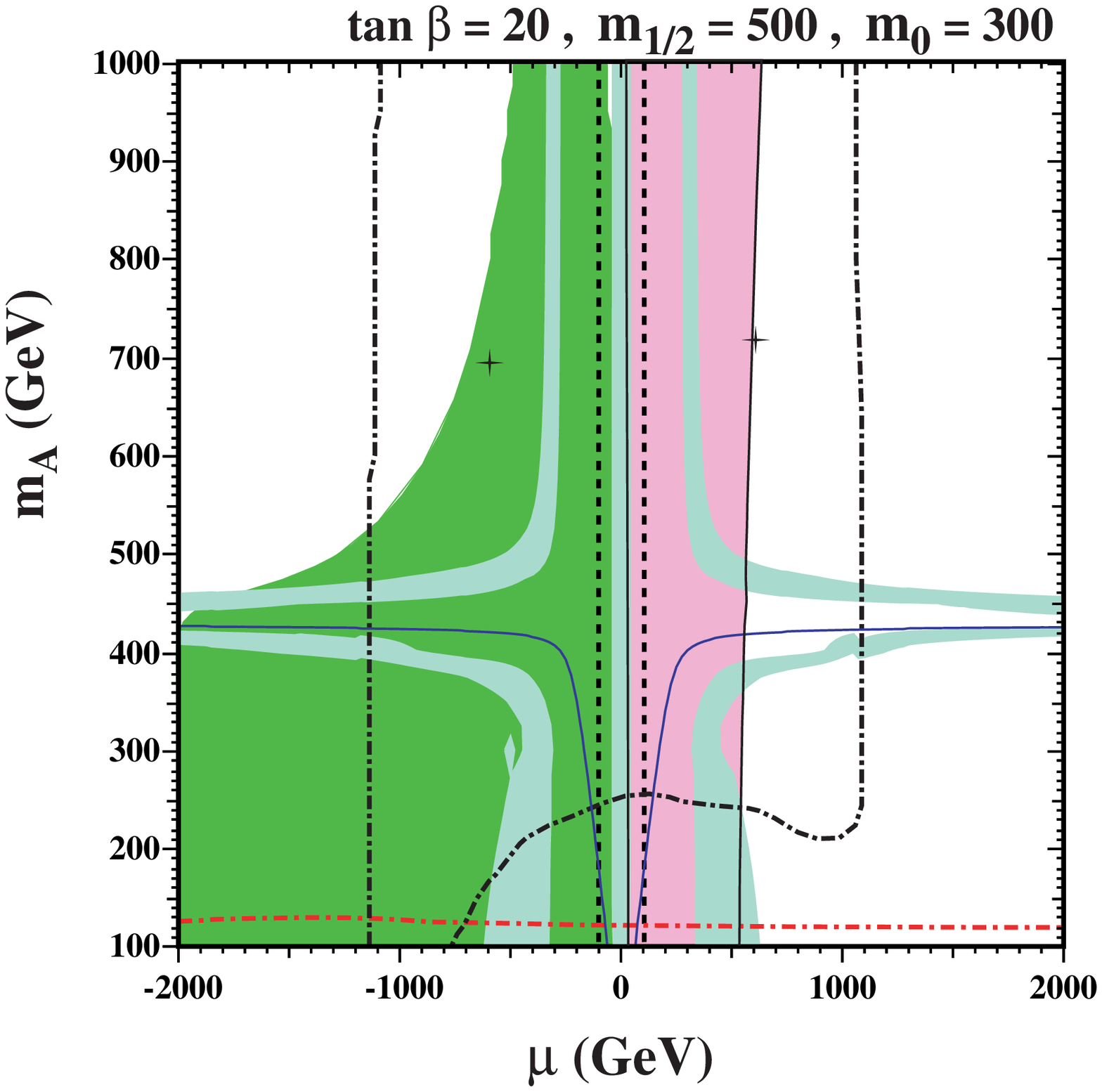,height=3.2in} \hfill
\end{minipage}
%\vspace*{-3in}
%\hspace*{-.70in}
\begin{minipage}{8in}
%\hskip -1.40in
%\vskip -.75in
\epsfig{file=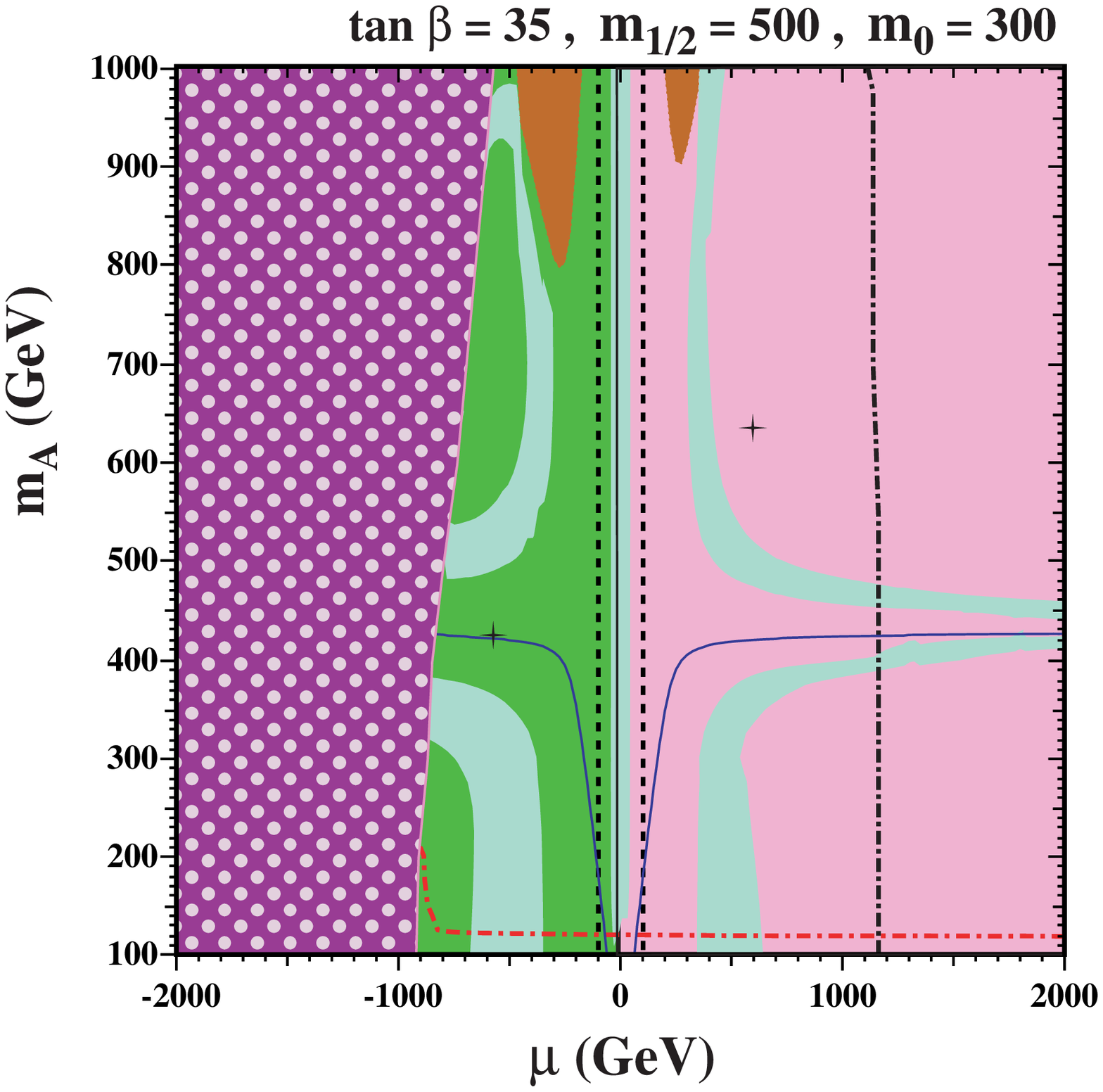,height=3.2in}
%\hspace*{-0.2in}
\epsfig{file=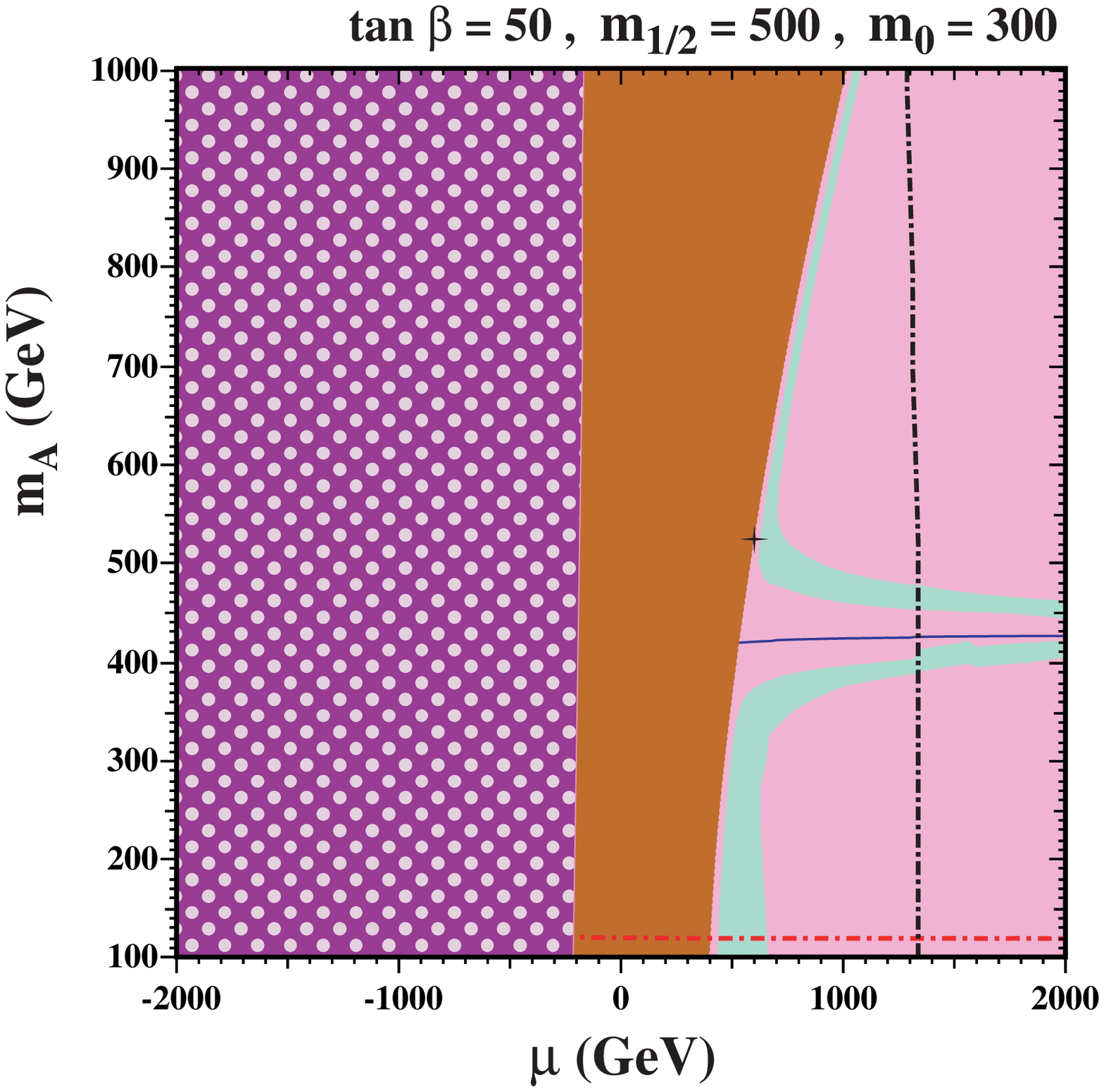,height=3.2in} \hfill
\end{minipage}
%\vskip 2.5in 
\caption{\label{fig:6}
{\it 
The NUHM $(\mu, \mA)$ planes for $m_0 = 300$~GeV and
$m_{1/2} = 500$~GeV, for (a) $\tan \beta = 10$, (b) $\tan \beta = 20$, (c) $\tan
\beta = 35$ and (d) $\tan \beta = 50$, assuming $A_0 = 0, m_t = 175$~GeV and
$m_b(m_b)^{\overline {MS}}_{SM} = 4.25$~GeV. The remaining shadings and
line styles of Fig.~\ref{fig:2} are used here, and 
there is no consistent electroweak vacuum in the polka-dotted region.}}
\end{figure}

We now discuss the variation with $\tan \beta$ shown in Fig.~\ref{fig:5},
keeping fixed $m_{1/2} = 300$~GeV and $m_0 = 100$~GeV, and starting (for
convenience) in panel (a) with the case $\tan \beta = 10$ shown previously
in panel (a) of Fig.~\ref{fig:4}. Increasing $\tan \beta$ to 20 in panel
(b), we note that the `shark's teeth' are somewhat expanded, whereas the
regions at large $|\mu|$ and/or $m_A$ where the lightest neutralino is no
longer the LSP change in shape. The allowed cosmological region is bounded
at larger $m_A$ by one where the LSP is the ${\tilde \tau_1}$, and at
larger $|\mu|$ by one where the LSP is a sneutrino: these different
regions had shapes similar to each other in panel (a). As usual,
neutralino-slepton coannihilations - as calculated in Appendix B - and
neutralino-sneutrino coannihilations - as calculated in Appendix D -
suppress the relic density close to these boundaries, and the relic 
density is within the preferred range over most of the region outside the 
`shark's teeth' and the chargino exclusion for $m_A > 2 m_\chi$.
As in previous cases, $b \to s \gamma$ and $m_h$ together exclude the
option $\mu < 0$, which would also be disfavoured by $a_\mu$. The $\mu >
0$ region surviving all the constraints when $\tan \beta = 20$ lies
between the GUT stability constraint at $\mu \sim 600$~GeV and $b \to s \gamma$
constraint around $\mu \sim 350$~GeV, and has
130~GeV$ \lappeq m_A \lappeq $900~GeV.

When $\tan \beta = 35$, in panel (c) of Fig.~\ref{fig:5}, we find no
consistent electroweak vacuum for a large (polka-dotted) region with $\mu 
< 0$, 
and the
condition that the LSP not be the ${\tilde \tau_1}$ excludes a broad
swathe with small $\mu > 0$. The condition that the LSP not be a sneutrino
provides an upper bound $\mu \la 1500$~GeV. The $m_h$ constraint imposes
$m_A \ga 120$~GeV and the $b \to s \gamma$ constraint is irrelevant. 
The GUT stability constraint allows only a
narrow sliver of $\mu \sim 700$~GeV for $120$~GeV $ \la m_A \la 200$~GeV, where
the lower bound comes from the LEP Higgs search. 
This reach is not compatible with $a_\mu$.
Here, the predicted discrepancy with the Standard Model is in excess of
that allowed at the 2$\sigma$ level by the $g-2$ experiment.

In the case of $\tan \beta = 50$, shown in panel (d) of Fig.~\ref{fig:5},
the problem with the non-existence of a consistent electroweak vacuum
extends to most of the $\mu < 0$ half-plane, as shown by the polka dots. 
The 
regions excluded because
the LSP is either the ${\tilde \tau}_1$ or a sneutrino hem in a small
region with $\mu > 0$. The region that survives all the experimental and
cosmological constraints is similar to that for $\tan \beta = 35$, but
extends to somewhat larger $m_A$ and broader $\mu$. Once again, these
regions are not compatible with the $g-2$ result.

Another series of plots for different values of $\tan \beta$ is shown in 
Fig.~\ref{fig:6}, this time for the fixed values $m_{1/2} = 500$~GeV and 
$m_0 = 300$~GeV. All of these have the `cruciform shape' of 
cosmological region familiar from 
panels (c, d) of Fig.~\ref{fig:4}. The $m_h$ constraint is irrelevant for 
this larger value of $m_{1/2}$, but much of the parameter space for $\mu < 
0$ is excluded by $b \to s \gamma$, particularly for larger $\tan \beta$. 
Likewise, $a_\mu$ consistently favours $\mu > 0$. One of the GUT stability 
constraints requires $|\mu| \lappeq 1200$~GeV, the exact value increasing 
slightly with $\tan \beta$. The other GUT stability constraint requires 
$m_A \gappeq 300$~GeV for $\tan \beta = 10$, but weakens for larger $\tan 
\beta$, becoming irrelevant when $\tan \beta \ge 35$. As in 
Fig.~\ref{fig:5}, the absence of a consistent electroweak vacuum also 
becomes a problem for $\mu < 0$ when $\tan \beta \ge 35$. Finally, we see 
in panel (d) of Fig.~\ref{fig:6} that small values of positive $\mu$ are 
disallowed because the LSP becomes the ${\tilde \tau_1}$. As in 
Fig.~\ref{fig:5}, values of $m_A$ as low as allowed by the Higgs search 
are consistent with all the constraints for $\tan \beta \ge 35$, as long 
as $\mu$ has a suitable positive value. 
Unlike the case shown, in Fig.~\ref{fig:5}, here the entire $\mu > 0 $
half-plane is within the $2\sigma$ range for $a_\mu$ when $\tan \beta =
35$ and 50.

\subsection{The $(\mu, M_2)$ Plane}

\begin{figure}
\vspace*{-0.75in}
%\hspace*{-.70in}
\begin{minipage}{8in}
\epsfig{file=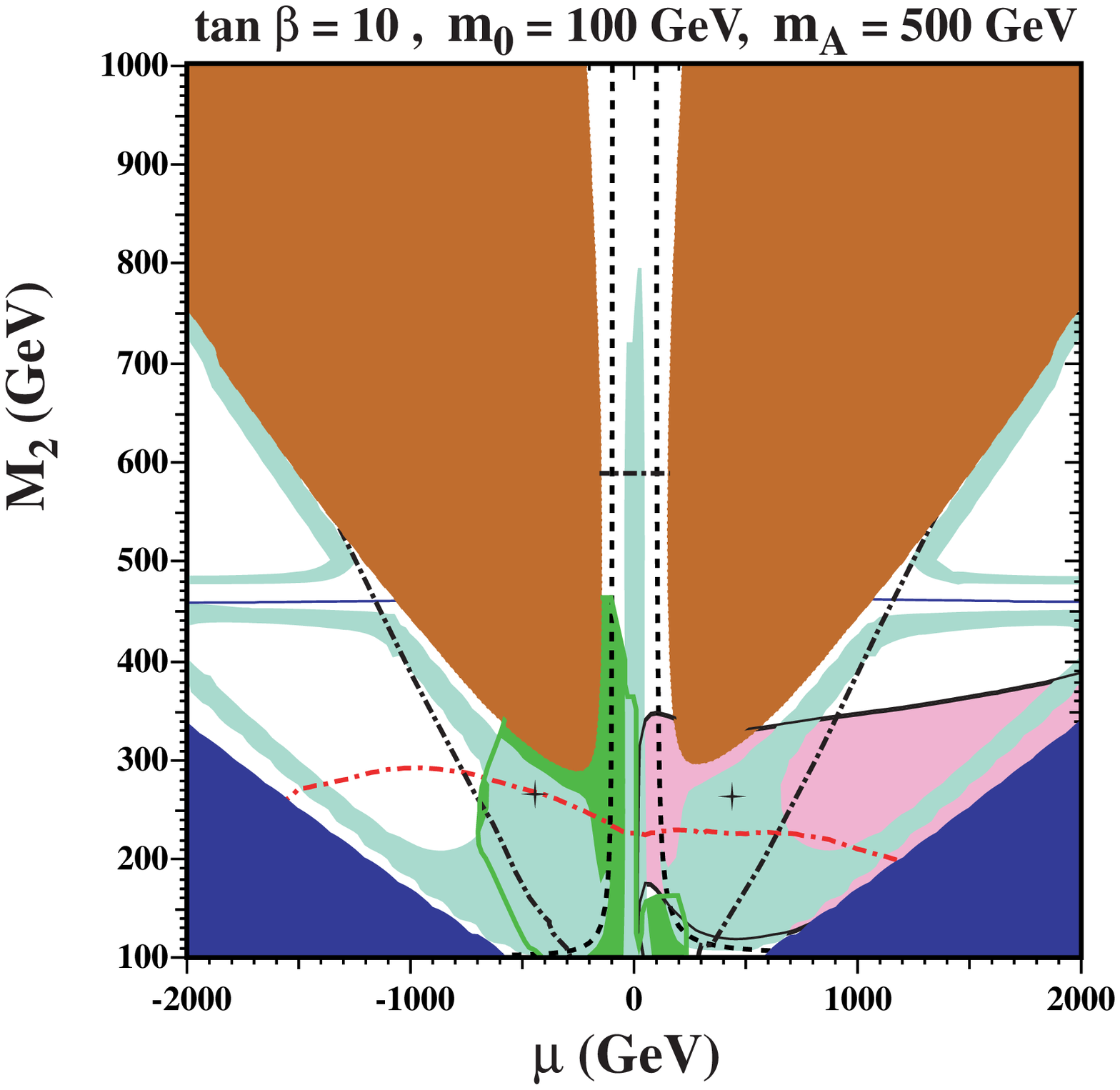,height=3.2in}
%\hspace*{-0.17in}
\epsfig{file=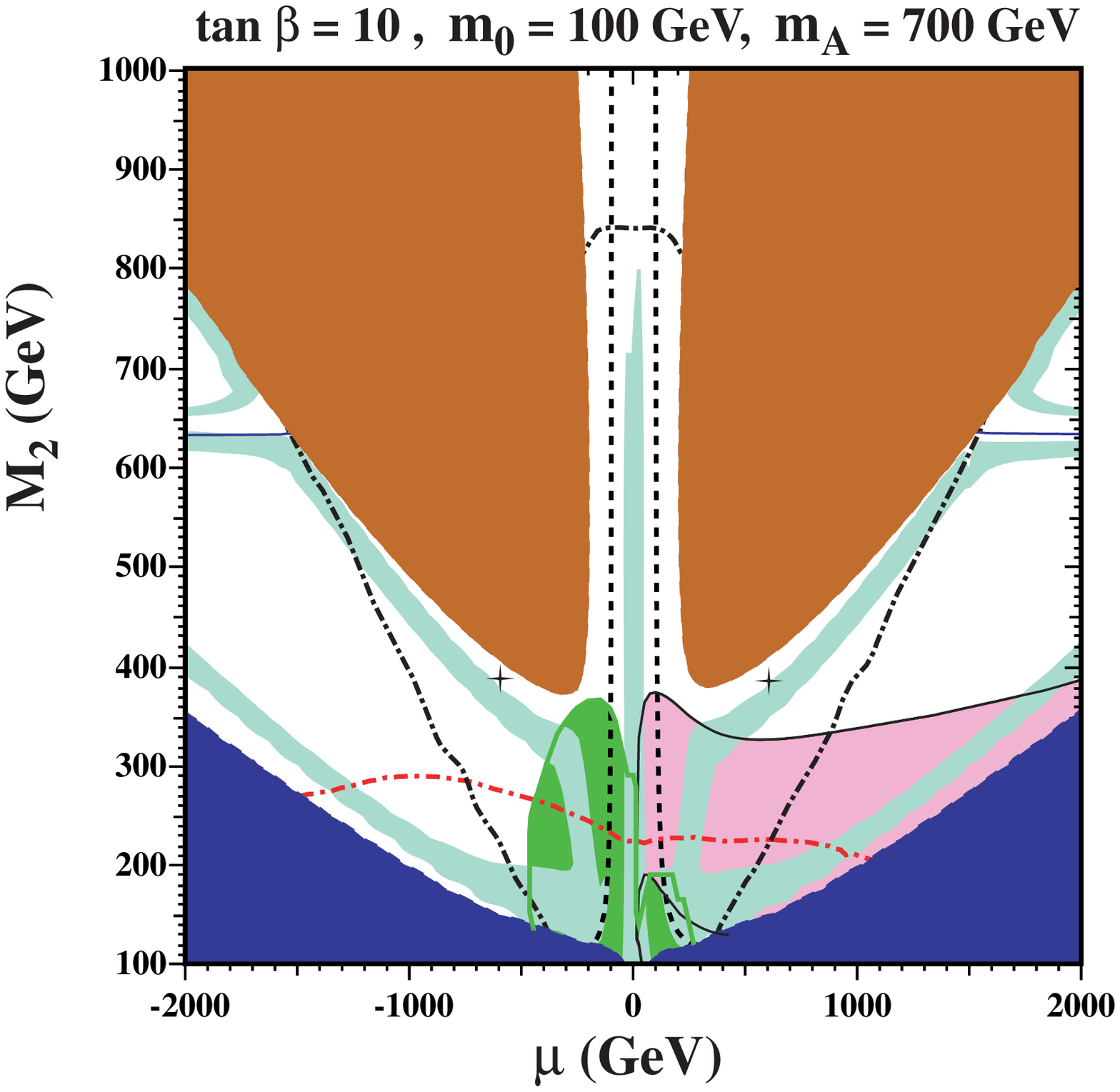,height=3.2in} \hfill
\end{minipage}
%\vspace*{-3in}
%\hspace*{-.70in}
\begin{minipage}{8in}
%\hskip -1.40in
%\vskip -.75in
\epsfig{file=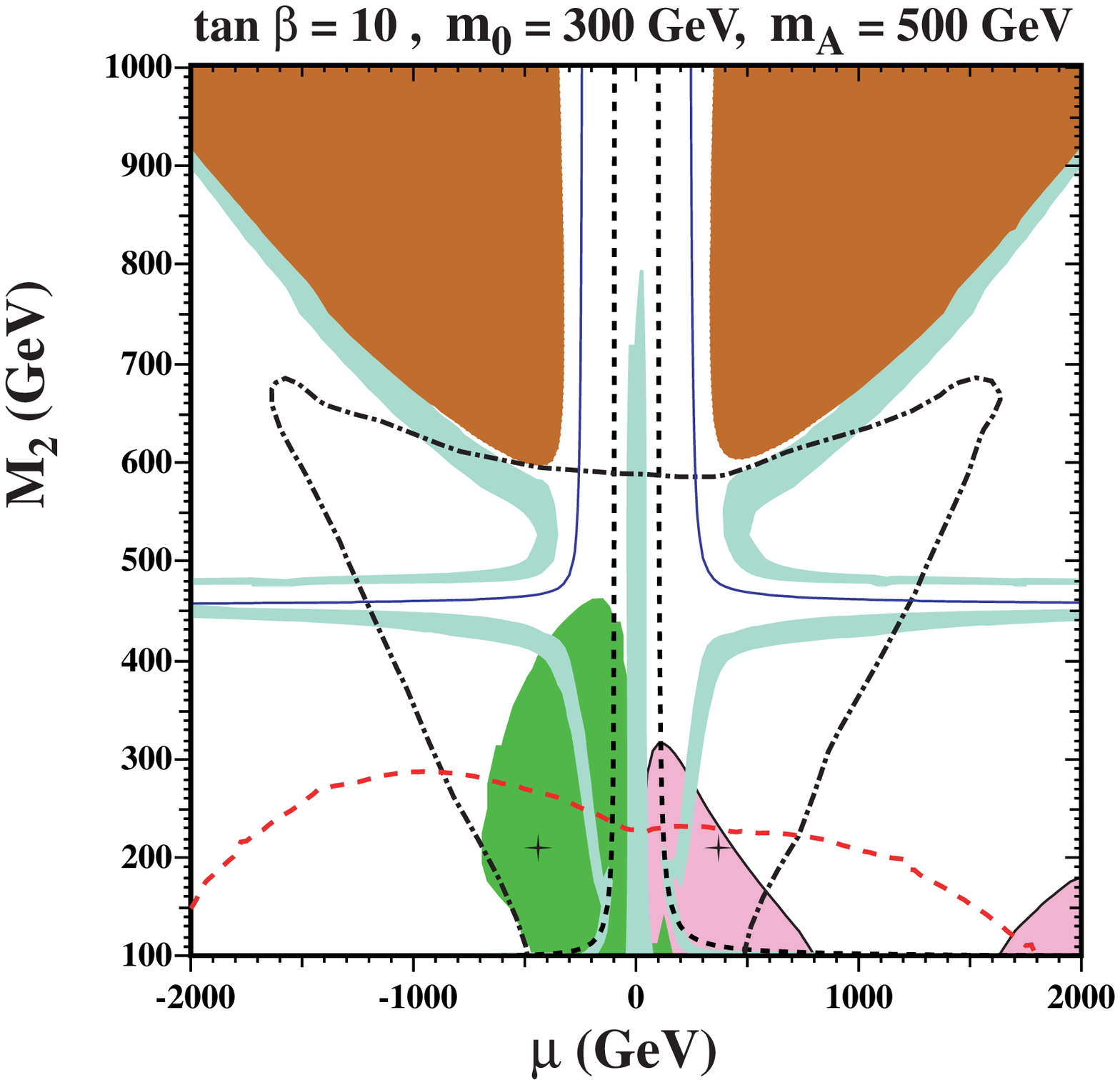,height=3.2in}
%\hspace*{-0.2in}
\epsfig{file=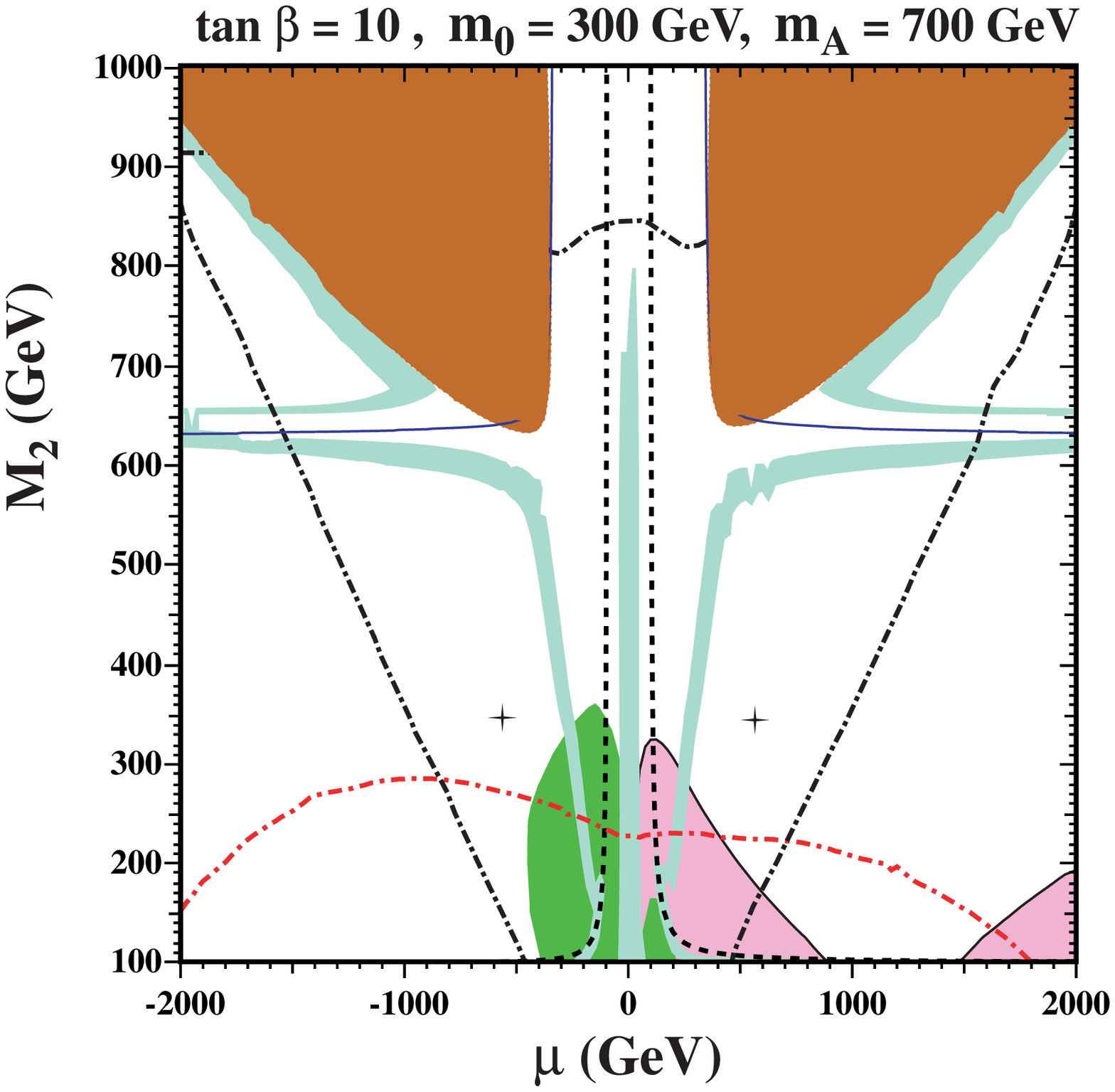,height=3.2in} \hfill
\end{minipage}
%\vskip 2.5in 
\caption{\label{fig:7}
{\it 
The NUHM $(\mu, M_2)$ planes for $\tan \beta = 10$, (a) $m_0 = 100$~GeV and
$\mA = 500$~GeV, (b) $m_0 = 100$~GeV and $\mA = 700$~GeV, (c) $m_0 =
300$~GeV and $\mA = 500$~GeV and (d)
$m_0 = 300$~GeV and $\mA = 700$~GeV, assuming $A_0 = 0, m_t = 175$~GeV and
$m_b(m_b)^{\overline {MS}}_{SM} = 4.25$~GeV. The shadings and
line styles are the same as in Fig.~\ref{fig:6}.}}
\end{figure}

Panel (a) of Fig.~\ref{fig:7} displays the $(\mu, M_2)$ plane for the
choices $\tan \beta = 10$, $m_0 = 100$~GeV and $m_A = 500~{\rm
GeV}$. We restrict our attention to the region allowed by the GUT
stability constraints, which is roughly triangular, extending from the
origin to vertices at $(\mu, M_2) = (\pm 1500, 600)$~GeV. Within this
region, there are two large triangular regions with $M_2 \gappeq 300$~GeV
and either sign of $\mu$ that are excluded because the LSP is not the
lightest neutralino. Also, a band around the $\mu = 0$ axis is excluded by
the LEP chargino constraint, regions with $M_2 \lappeq 250 (300)$~GeV for
$\mu > (<) 0$ are excluded by the LEP Higgs constraint, and most of the
surviving
$\mu < 0$ region is eroded by the $b \to s \gamma$ constraint. In the
lower corners (for both signs of $\mu$) the LSP is a sneutrino and we
see the effects of $\chi - \widetilde \nu$ coannihilation running
alongside these regions. However, these regions are in conflict with the
GUT stability constraint.

Much of the remaining area of the plane is consistent with the
cosmological relic density constraint, mainly along strips where
neutralino-slepton 
is important.  These
terminate around $M_2 \sim 450$~GeV, where rapid annihilations through the
direct-channel $A, H$ resonances cut down the relic density. The
putative $a_\mu$ constraint would favour $\mu > 0$ and $M_2 \lappeq
350$~GeV.

The above constraints interplay analogously in panel (b) of 
Fig.~\ref{fig:7}, for $m_0 = 100$~GeV and $m_A = 700$~GeV. We note, 
however, that the GUT stability limit has risen with $m_A$ to above
800~GeV, and that the tips of the non-neutralino LSP triangles have also 
moved up slightly. 
The neutralino-slepton coannihilation strip now 
extends up to $M_2 \sim 640$~GeV, where it is cut off by rapid $A, H$ 
annihilations. We also see the `transition' band at lower $M_2$. In fact, this
region with $\mu > 0$ is the one favoured by all constraints including $a_\mu$.

In panel (c) of Fig.~\ref{fig:7}, for $m_0 = 300$~GeV and $m_A = 500$~GeV,
the triangular GUT stability region is very similar to that in panel (a),
whereas the non-neutralino LSP triangles have moved to significantly
higher $M_2$, reflecting the larger value of $m_0$. The regions where the
cosmological relic density falls within the preferred range are now narrow
strips in the neutralino-slepton coannihilation regions, on either side of
the rapid $A, H$ annihilation strips, and the `transition' bands. 
This tendency towards `skinnier' cosmological regions is also
apparent in panel (d), where $m_0 = 300$~GeV and $m_A = 700$~GeV are
assumed. In both these panels, the $b \to s \gamma$ constraints disfavours
$\mu < 0$ and small $m_A$, whilst the putative $a_\mu$ constraint would
favour $\mu > 0$ and small $m_A$.

\begin{figure}
\vspace*{-0.75in}
%\hspace*{-.70in}
\begin{minipage}{8in}
\epsfig{file=M2mu_10_0_100_500.eps,height=3.2in}
%\hspace*{-0.17in}
\epsfig{file=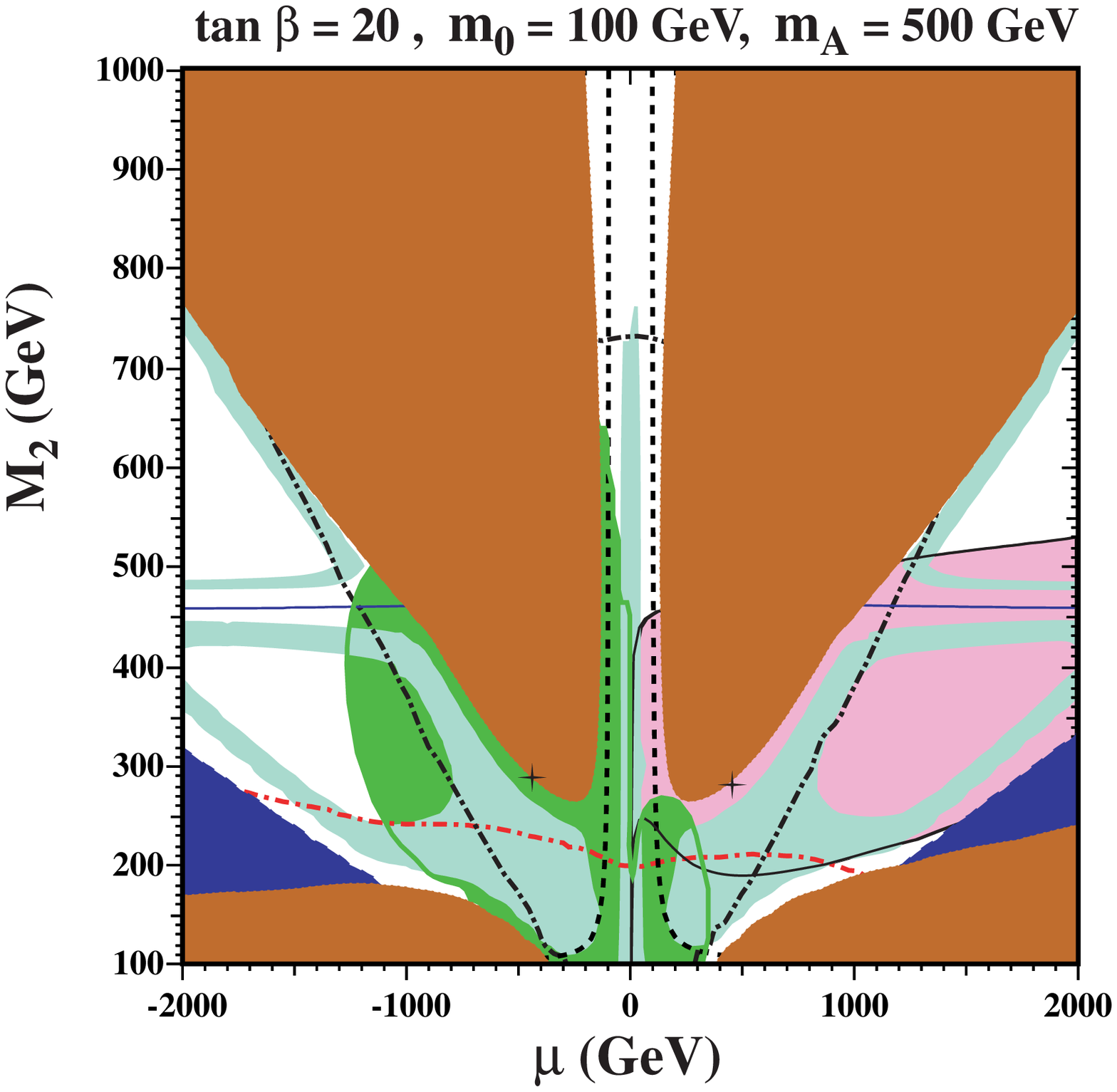,height=3.2in} \hfill
\end{minipage}
%\vspace*{-3in}
%\hspace*{-.70in}
\begin{minipage}{8in}
%\hskip -1.40in
%\vskip -.75in
\epsfig{file=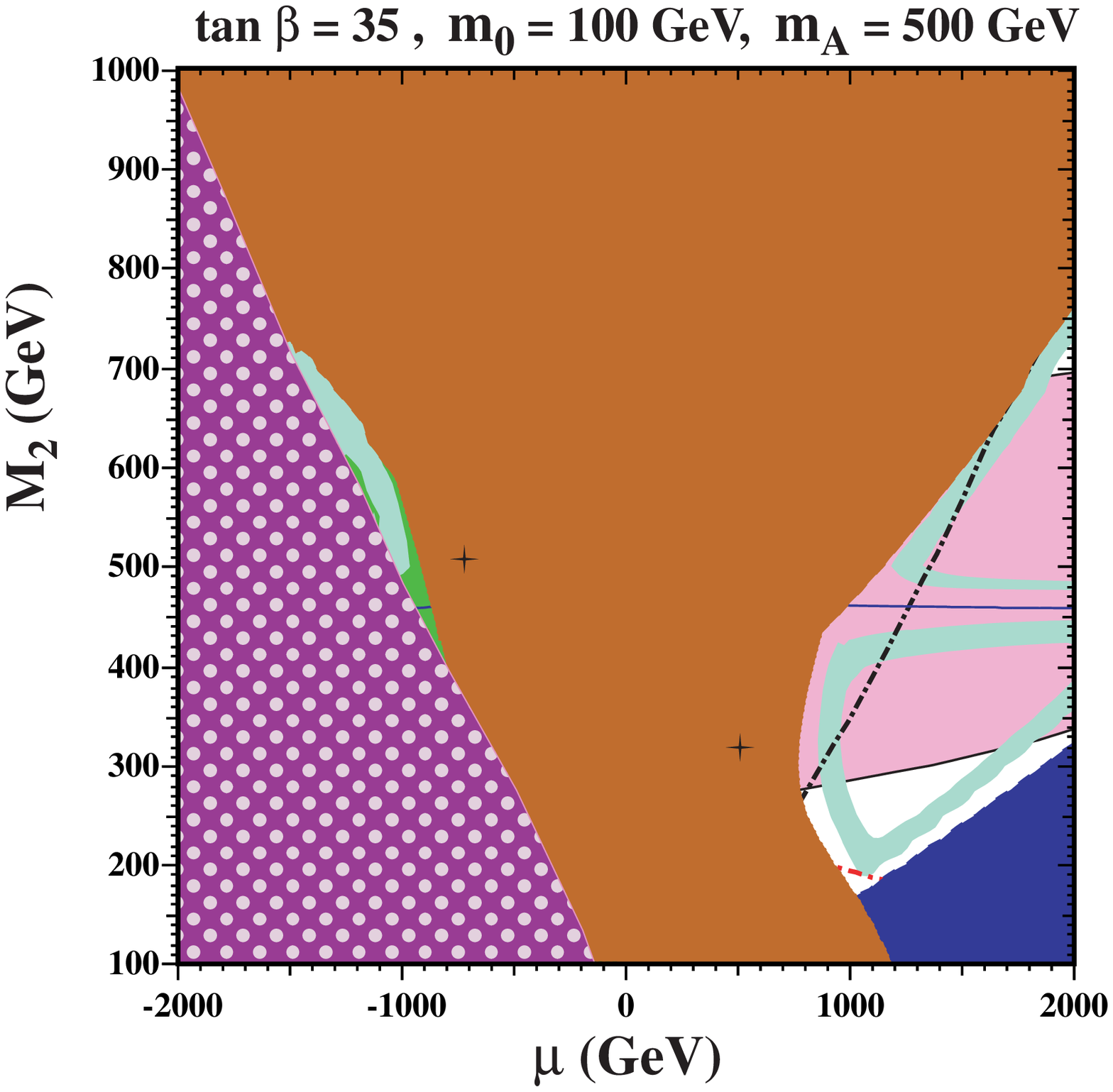,height=3.2in}
%\hspace*{-0.2in}
\epsfig{file=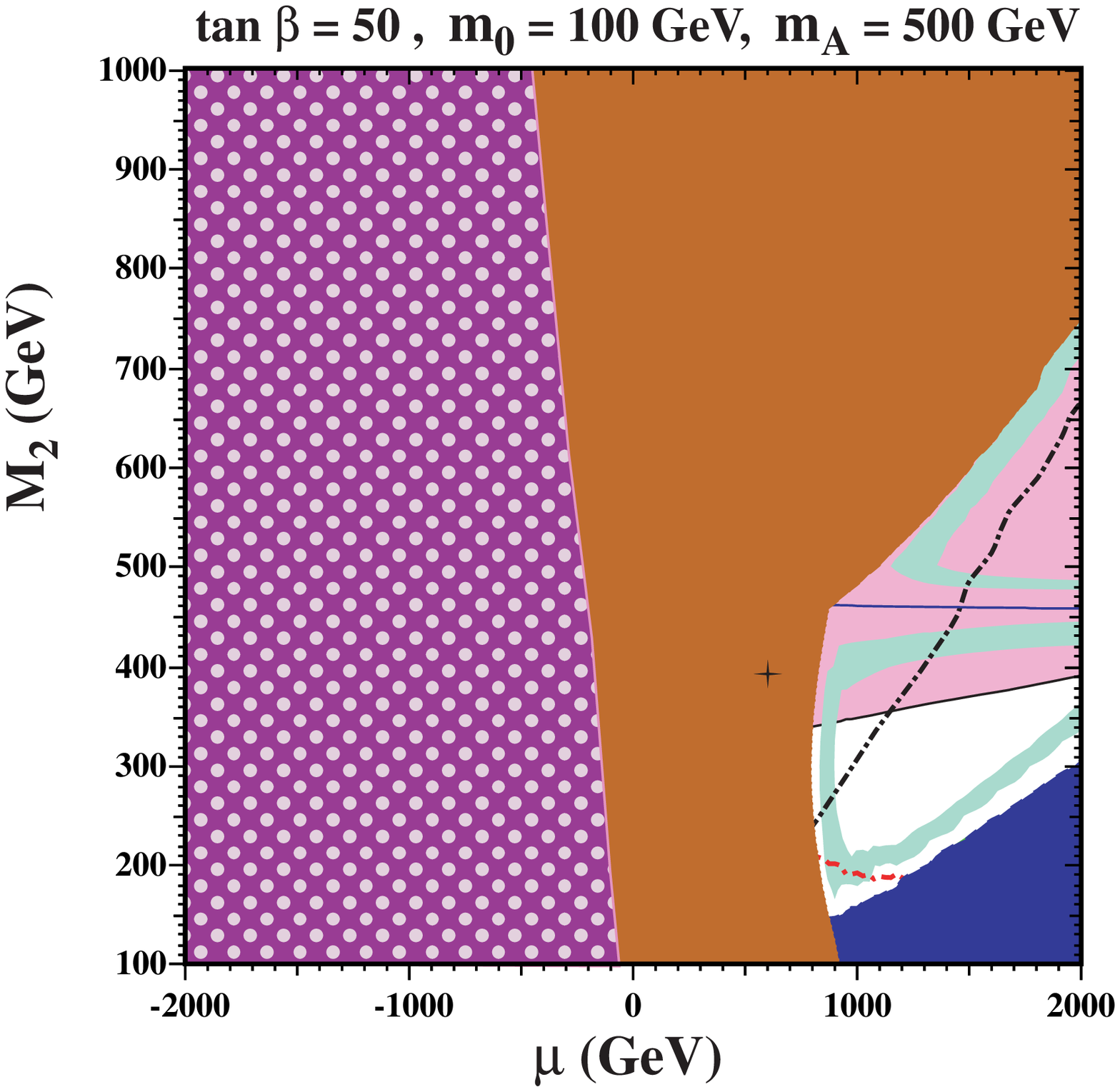,height=3.2in} \hfill
\end{minipage}
%\vskip 2.5in 
\caption{\label{fig:8}
{\it 
The NUHM $(\mu, M_2)$ planes for $m_0 = 100$~GeV and
$\mA = 500$~GeV, for (a) $\tan \beta = 10$, (b) $\tan \beta = 20$, (c) $\tan
\beta = 35$ and (d) $\tan \beta = 50$, assuming $A_0 = 0, m_t = 175$~GeV and
$m_b(m_b)^{\overline {MS}}_{SM} = 4.25$~GeV. The shadings and
line styles are the same as in Fig.~\ref{fig:6}.}}
\end{figure}

We now explore the variation of these results with $\tan \beta$, as
displayed in Fig.~\ref{fig:8} for the choices $m_0 = 100$~Gev and $m_A =
500$~GeV. Panel (a) reproduces for convenience the case $\tan \beta 
= 10$ that was shown also in panel (a) of Fig.~\ref{fig:7}. When $\tan 
\beta = 20$, as seen in panel (b) of Fig.~\ref{fig:8}, we first notice 
that the GUT stability region now extends to larger $M_2$. Next, we 
observe that the triangular non-neutralino LSP regions have extended down 
to lower $M_2$. As one would expect on general grounds and from earlier 
plots, the $m_h$ constraint at low $M_2$ is weaker than in panel (a), 
whereas the $b \to s \gamma$ constraint is stronger, ruling out much of 
the otherwise allowed region with $\mu < 0$. The allowed region with $\mu 
> 0$ is generally compatible with the putative $a_\mu$ constraint.

Panel (c) for $\tan \beta = 35$ again shows the feature that no consistent
electroweak vacuum is found over much of the half-plane with $\mu < 0$.
The GUT stability constraint now allows $M_2 \lappeq 1000$~GeV. The
non-neutralino LSP region is no longer triangular in shape, but now
requires $\mu \gappeq 800$~GeV. This happens as the regions with light
left-handed slepton at low $M_2$ meet with the ones with light right-handed
slepton at higher $M_2$. There is a minuscule allowed region for
$\mu < 0$. The residual regions with relic density in the preferred range
for $\mu > 0$ are limited to strips in the neutralino-slepton
coannihilation regions, and on either side of the rapid $A, H$
annihilation strip. All this preferred region is compatible with the
putative $a_\mu$ constraint.

A rather similar pattern is visible in panel (d) of Fig.~\ref{fig:8} for 
$\tan \beta = 50$. The most noticeable differences are that the absence of 
a consistent electroweak vacuum is even more marked for $\mu > 0$, and 
that the putative $a_\mu$ constraint would suggest a lower limit $M_2 
\gappeq 350$~GeV, whereas lower values would have been permitted in panel 
(c) for $\tan \beta = 35$.

\section{Conclusions and Open Issues}

We have provided in this paper the tools needed for a detailed study of
the NUHM, in the form of complete calculations of the most important
coannihilation processes. However, in this paper we have only been able to
scratch the surface of the phenomenology of the NUHM. Even this
exploratory study has shown that many new features appear compared with
the CMSSM, as results of the two additional parameters in the NUHM, but
much more remains to be studied. For example, we have not studied the NUHM
with a nonzero trilinear coupling $A_0$. Nevertheless, some interesting
preliminary conclusions about the NUHM can be drawn, though many questions
remain open.

\begin{figure}
\vspace*{-0.75in}
\hspace*{-.70in}
\begin{minipage}{8in}
\begin{center}
\epsfig{file=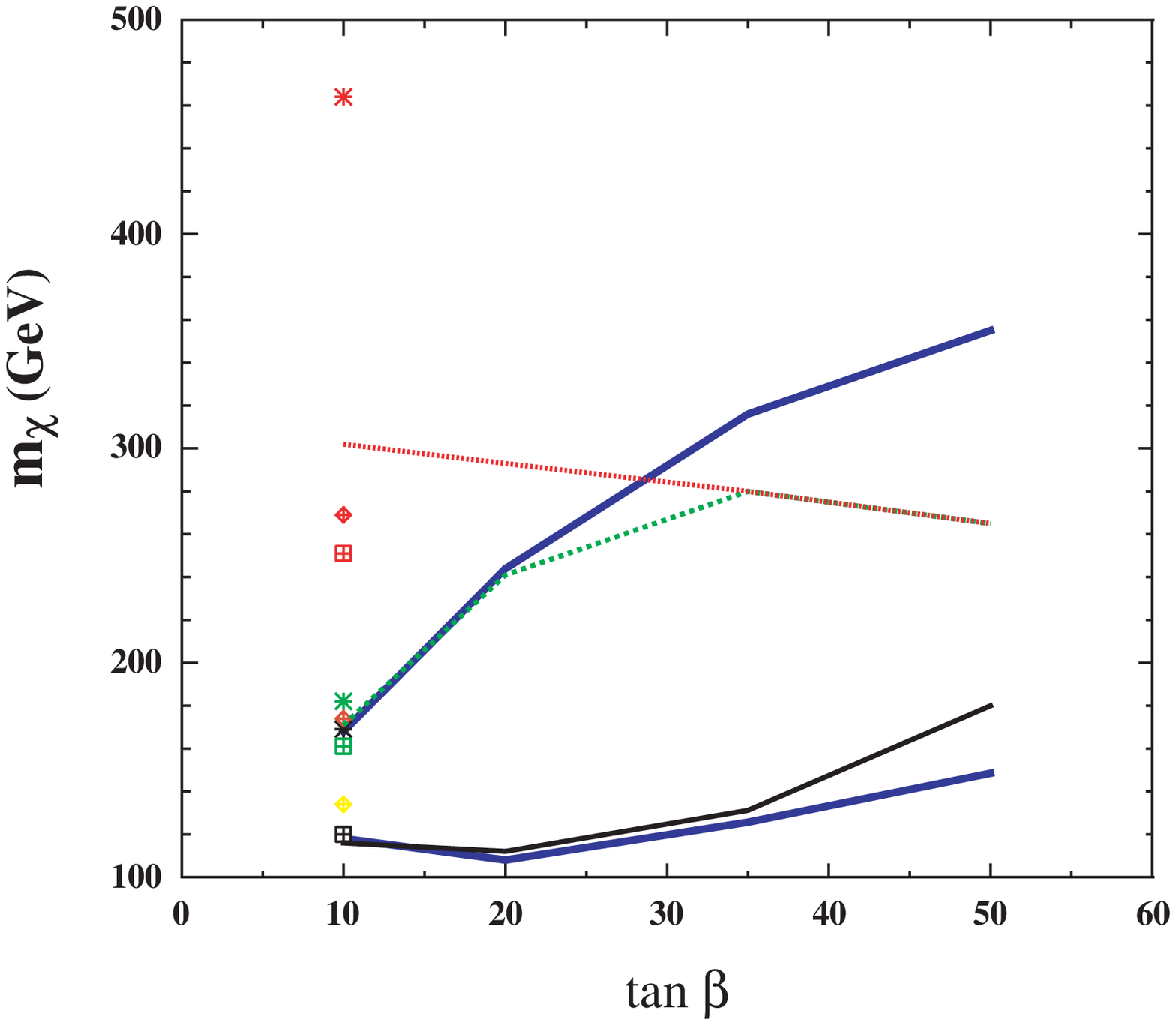,height=3.2in} \hfill
\end{center}
\end{minipage}
%\vspace*{-3in}
\hspace*{-.70in}
\begin{minipage}{8in}
\hskip -1.40in
%\vskip -.75in
\begin{center}
\epsfig{file=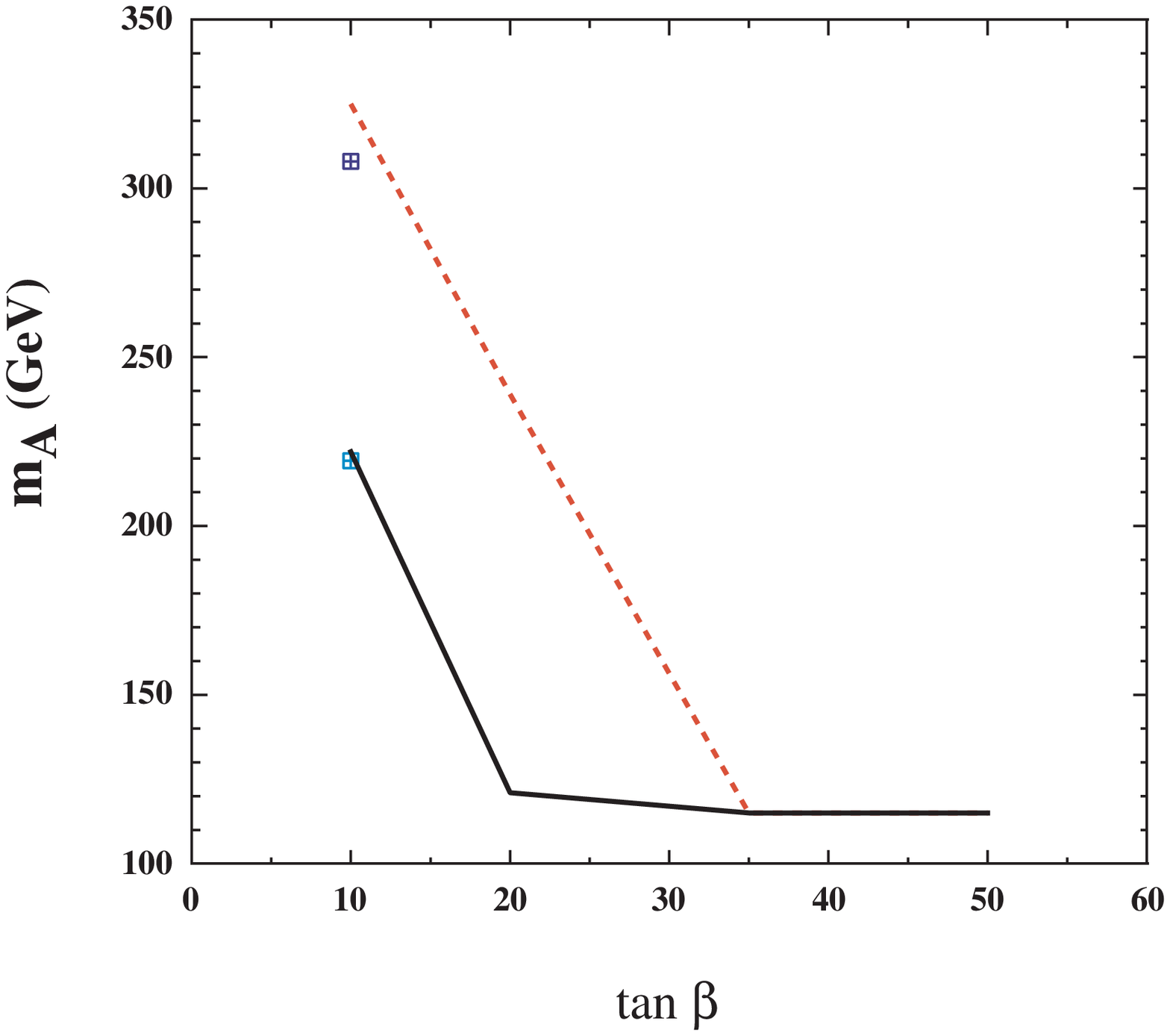,height=3.2in} \hfill
\end{center}
\end{minipage}
%\vskip 2.5in 
\caption{\label{fig:sum}
{\it Top panel: Bounds on the LSP mass $m_\chi$ found in the NUHM for the
particular parameter choices displayed in Figs.~\ref{fig:2} and
\ref{fig:3}. The solid (black) line is the lower limit on $m_\chi$ for 
$(\mu,
m_A) = (400, 700)$~GeV, The (red) dotted line is the upper bound for
the same values of $(\mu, m_A)$, but without imposing the putative $g_\mu 
- 2$ constraint, and the (green)  dashed line shows how this upper bound
would be strengthened if the $g_\mu - 2$ constraint were imposed. The
symbols shown for $\tan \beta = 10$ mark the lower ( $g_\mu - 2$ upper,
overall upper) limits on $m_\chi$ for $(\mu, m_A) = (400, 400)$~GeV
(squares), $(\mu, m_A) = (700, 400)$~GeV (diamonds) and $(\mu, m_A) =
(700, 700)$~GeV (stars). The thick (blue) lines correspond to CMSSM upper
and lower limits. Bottom panel: Lower bounds on
$m_A$ for
$(m_{1/2}, m_0) = (300, 100)$~GeV (solid black line) and $(m_{1/2}, m_0)  
= (500, 300)$~GeV (dashed red line). The extra symbols for $\tan \beta = 
10$ are lower bounds for $(m_{1/2}, m_0) = (500, 100)$~GeV (upper point)
and  $(300,  300)$~GeV (lower point). }}
\end{figure}

The lower limit on the LSP mass $m_\chi$ in the CMSSM has been much
discussed, and it is interesting to consider whether this could be greatly
reduced in the NUHM. The top panel of Fig.~\ref{fig:sum} compiles the
bounds on $m_\chi$ for the various sample parameter choices explored in
this paper. We focus on the case $\mu > 0$, which is favoured by $m_h$ and
$b \to s \gamma$ as well as the dubious $g_\mu - 2$ constraint. The solid
(black) line connects the lower limits on $m_\chi$ for the particular
choice $\mu = 400$~GeV and $m_A = 700$~GeV shown for the four choices of
$\tan \beta$ in Fig.~\ref{fig:3}. This lower limit is provided by $m_h$
for $\tan \beta = 10, 20$, but by $b \to s \gamma$ for $\tan \beta = 35,
50$. The CMSSM lower bound on $m_\chi$ is indicated by a thick (blue)
line. We see that this CMSSM lower bound on $m_\chi$ is similar to that in
the NUHM when $\tan \beta = 10$, but weaker when $\tan \beta = 50$ because
of the different behaviour in the CMSSM of the $b \to s \gamma$ constraint
in this case.

For $\tan \beta = 10$, we also indicate by different black symbols the
lower bounds on $m_\chi$ for the other choices of $(\mu, m_A)$ studied in
Fig.~\ref{fig:2}, namely $(\mu, m_A) = (400, 400)$~GeV (square), $(\mu,
m_A) = (700, 400)$~GeV (diamond) and $(\mu, m_A) = (700, 700)$~GeV (star).
The two latter lower bounds are significantly higher than in the default
case $(\mu, m_A) = (400, 700)$~GeV, due to the impacts of the GUT
stability condition and the $\ohsq$ constraint, respectively. In none of
the NUHM examples studied was the lower limit on $m_\chi$ relaxed compared
with the CMSSM, though this might be found possible in a more complete
survey of the NUHM parameter space.

We also show in the top panel of Fig.~\ref{fig:sum} the upper bounds on
$m_\chi$ found in the NUHM for the various parameter choices explored in
this paper. The (red) dot-dashed line is the upper bound for $\mu =
400$~GeV and $m_A = 400$~GeV, as obtained without imposing the putative
$g_\mu - 2$ constraint. The upper bound on $m_\chi$ is attained in the
rapid-annihilation strip where $m_\chi < m_A/2$. Also shown in
Fig.~\ref{fig:sum}, as a (green) dashed line, is the strengthening of this
upper bound that would be found if the $g_\mu - 2$ constraint were
imposed. This constraint would not strengthen the upper limit for $\tan
\beta \ge 35$. We also show as a thick (blue) line the CMSSM upper limit
on $m_\chi$, implementing the $g_\mu - 2$ constraint. We see that it is
similar to that in the NUHM for $\tan \beta = 10, 20$, but is weaker for
higher $\tan \beta$. The CMSSM upper bound would be much weaker still if
the $g_\mu - 2$ constraint were relaxed, because of the different
behaviours of the rapid-annihilation strips in the NUHM and CMSSM, as seen
by comparing Figs.~\ref{fig:UHM} and \ref{fig:3}.

For $\tan \beta = 10$, we also show in the top panel of Fig.~\ref{fig:sum}
the upper bounds that hold with and without $g_\mu - 2$ for the other
three choices of $(\mu, m_A)$ made in Fig.~\ref{fig:2}, using the same
symbols as for the lower bounds (squares, diamonds and stars,
respectively). The largest upper bound is for the choice $(\mu, m_A) =
(700, 700)$~GeV, for which there is an extension of the allowed region
along the coannihilation tail, extending up to the GUT stability line.

In the second panel of Fig.~\ref{fig:sum}, we summarize the lower limits
on $m_A$ found in the NUHM models that we have studied. The solid (black)  
line is for the choice $(m_{1/2}, m_0) = (300, 100)$~GeV displayed
previously in Fig.~\ref{fig:6}. The lower bounds shown for $\tan \beta =
10, 20$ are compatible with $g_\mu - 2$, but we find that no region would
be allowed by $g_\mu - 2$ in the cases $\tan \beta = 35, 50$, as seen in
panels (c)  and (d) of Fig.~\ref{fig:6}. The (red) dashed line in the
second panel of Fig.~\ref{fig:sum} is for the choice $(m_{1/2}, m_0) =
(500, 300)$~GeV shown previously in Fig.~\ref{fig:5}. In this case, we
find that no region would be allowed by $g_\mu - 2$ for $\tan \beta = 10$,
whereas the lower bounds for $\tan \beta = 20, 35, 50$ are compatible with
$g_\mu - 2$. The extra symbols for $\tan \beta = 10$ correspond to the
other choices $(m_{1/2}, m_0) = (500, 100)$~GeV and $(300, 300)$~GeV shown
in panels (b)  and (c) of Fig.~\ref{fig:4}, at $m_A = 308$ GeV and 
$m_A = 220$ GeV, respectively. We note that
neither of these choices satisfies the $g-2$ constraint. 

An interesting feature of the second panel of Fig.~\ref{fig:sum} is the
fact that the lower bounds coincide for $\tan \beta \ge 35$, and
correspond to the lower limit established by direct searches at LEP.  For
comparison, we note that in the CMSSM for $(m_{1/2}, m_0) = (300,
100)$~GeV one would have $m_A = 449, 424, 377, 315$~GeV for $\tan \beta =
10, 20, 35, 50$, whilst for $(m_{1/2}, m_0) = (500, 300)$~GeV one would
have $m_A = 762, 720, 639, 526$~GeV for $\tan \beta = 10, 20, 35, 50$. We
conclude that the NUHM allows $m_A$ to be considerably smaller than in the
CMSSM, particularly at large $\tan \beta$.

This brief survey is no substitute for a detailed study of the NUHM.  
However, we have provided in the Appendices the technical tools required
for such a study, and in this Section we have presented some preliminary
observations based on a cursory exploration of the NUHM. This has already
provided some interesting indications, for example that it may prove
difficult to relax significantly the CMSSM lower bound on the LSP, but
that $m_A$ may be greatly reduced. In our view, it would be interesting to 
pursue these questions more deeply in a detailed study of the NUHM, a 
project that lies beyond the scope of this work.

\vskip 0.5in
\vbox{
\noindent{ {\bf Acknowledgments} } \\
\noindent 
We would like to thank M. Gomez and S. Profumo for helpful
comments.
The work of K.A.O. and Y.S. was supported in part by DOE grant
DE--FG02--94ER--40823.}
%\newpage

\newpage
\baselineskip=14pt
\appendix
\setcounter{equation}{0}
\renewcommand{\theequation}{A\arabic{equation}}
\section*{Appendix A: Couplings used in the Calculations}

Here we list the couplings used in the calculations. For clarity, we have
not written factors such as $i$, $\gamma$'s and momenta.  
These are taken into account in the calculations of the amplitudes squared
in the other Appendices.

Couplings for chargino-slepton coannihilation:
\baselineskip=20pt
\begin{eqnarray*}
C_{\slept_1-\nu_{\ss \ell}-\schi^-_i}^{L} &=& 0 \\
C_{\slept_1-\nu_{\ss \ell}-\schi^-_i}^{R} &=& (g_2 \, \mell/(\sqrt{2} \,
    \mw \cos \beta)) U_{i2} \sin \thell - g_2 \, U_{i1} \cos \thell \\
C_{\schi^+_1-\schi^+_i-Z}^{L} &=&  g_2/ \cos \thw (-V_{i1} V_{11} - V_{i2}
    V_{12}/2  + \delta_{i1} \sin^2 \thw)\\
C_{\schi^+_1-\schi^+_i-Z}^{R} &=&  g_2/ \cos \thw (-U_{i1} U_{11} - U_{i2}
    U_{12}/2  + \delta_{i1} \sin^2 \thw) \\
C_{\slept_1-Z-\slept^-_1} &=& g_2/(2 \cos \thw) ( \cos^2 \thell - 2
    \sin^2 \thw) \\
C_{\slept_1-Z-\slept^-_2} &=& g_2/(2 \cos \thw) \sin \thell \cos \thell \\
C_{\schi^-_1-\nu_{\ss \ell}-\slept_i}^{L} &=& 0 \\
C_{\schi^-_1-\nu_{\ss \ell}-\slept_1}^{R} &=& (g_2 \mell /(\sqrt{2} \mw \cos
    \beta )) U_{12} \sin \thell - g_2 U_{11} \cos \thell \\
C_{\schi^-_1-\nu_{\ss \ell}-\slept_2}^{R} &=& (g_2 \mell /(\sqrt{2} \mw \cos
    \beta )) U_{12} \cos \thell + g_2 U_{11} \sin \thell \\
C_{\schi^+_i-\schi^+_i-\gamma} &=& -e \\
C_{\slept_i-\slept_i-\gamma} &=& e \\
C_{\schi^+_1-\schi^+_i-H}^{L} &=& -g_2 (V_{11} U_{i2}/ \sqrt{2} \cos \alpha
    + V_{12} U_{i1}/ \sqrt{2} \sin \alpha ) \\
C_{\schi^+_1-\schi^+_i-H}^{R} &=& -g_2 (V_{i1} U_{12}/ \sqrt{2} \cos \alpha
    + V_{i2} U_{11}/ \sqrt{2} \sin \alpha ) \\
C_{\schi^+_1-\schi^+_i-h}^{L} &=&  g_2 (V_{11} U_{i2}/ \sqrt{2} \sin \alpha
    - V_{12} U_{i1}/ \sqrt{2} \cos \alpha )  \\
C_{\schi^+_1-\schi^+_i-h}^{R} &=&  g_2 (V_{i1} U_{12}/ \sqrt{2} \sin \alpha
    - V_{i2} U_{11}/ \sqrt{2} \cos \alpha )   \\ 
C_{\slept_1-H-\slept_1} &=&  g_2 \mz / \cos \thw (1/2 - \sin^2 \thw) \cos
    (\alpha + \beta) \cos^2 \thell \nl - g_2 \mz / \cos \thw (-1) \sin^2 \thw
    \cos (\alpha + \beta) \sin^2 \thell \nl - g_2 \mell^2 / (\mw \cos \beta)
    \cos \alpha \nl + g_2 \mell / (\mw \cos \beta) (\mu \sin \alpha + A_{\st \ell}
    \cos \alpha ) \sin \thell \cos \thell \\
C_{\slept_1-H-\slept_2} &=&  g_2 \mz / \cos \thw (1/2 - \sin^2 \thw ) \cos
    (\alpha + \beta) \cos \thell (-\sin \thell) \nl -g_2 \mz/ \cos \thw (-1) \sin^2
    \thw \cos (\alpha + \beta) \sin \thell \cos \thell \nl + g_2 \mell / (2 \mw \cos
    \beta) (\mu \sin \alpha + A_{\st \ell} \cos \alpha) (\cos^2 \thell - \sin^2
    \thell) \\
C_{\slept_1-h-\slept_1} &=& - g_2 \mz / \cos \thw (1/2 - \sin^2 \thw ) \sin
    (\alpha + \beta) \cos^2 \thell \nl + g_2 \mz / \cos \thw (-1) \sin^2 \thw
    \sin (\alpha + \beta) \sin^2 \thell \nl + g_2 \mell^2 /(\mw \cos \beta)
    \sin \alpha \nl + g_2 \mell /(\mw \cos \beta) (\mu \cos \alpha
    - A_{\st \ell} \sin \alpha ) \sin \thell \cos \thell \\
C_{\slept_1-h-\slept_2} &=& - g_2 \mz/ \cos \thw (1/2 - \sin^2 \thw ) \sin
    (\alpha + \beta) \cos \thell (-\sin \thell) \nl + g_2 \mz / \cos \thw (-1)
    \sin^2 \thw \sin (\alpha + \beta) \sin \thell \cos \thell \nl + g_2 \mell
    /(2 \mw \cos \beta) (\mu \cos \alpha  - A_{\st \ell} \sin \alpha) (\cos^2
    \thell  - \sin^2 \thell) \\
C_{\schi^+_1-\schi^+_i-A}^{L} &=&  g_2 (V_{11} U_{i2}/ \sqrt{2} \sin \beta
    + V_{12} U_{i1}/ \sqrt{2} \cos \beta) \\
C_{\schi^+_1-\schi^+_i-A}^{R} &=&  - g_2 (V_{i1} U_{12}/ \sqrt{2} \sin \beta
    + V_{i2} U_{11}/ \sqrt{2} \cos \beta)  \\
C_{\slept_1-A-\slept^-_2} &=& -g_2 \mell /(2 \mw) (\mu - A_{\st \ell} \tan
    \beta  ) \\
C_{\slept_1-\ell-\schi^0_i}^{L} &=&  \cos \thell (g_2/ \sqrt{2} (N_{i2} + \tan
   \thw N_{i1})) \nl +\sin \thell (-g_2/ \sqrt{2} \mell N_{i3}/(\mw \cos \beta))
   \\
C_{\slept_1-\ell-\schi^0_i}^{R} &=& \cos \thell (-g_2/ \sqrt{2} \mell N_{i3}
    /(\mw \cos \beta)) \nl + \sin \thell (-g_2/
   \sqrt{2} (2) 
   \tan \thw N_{i1} ) \\
C_{\nu_{\ss \ell}-\ell-W^+} &=& -g_2/ \sqrt{2} \\
C_{W-\schi^-_1-\schi^0_i}^{L} &=&  g_2 (N_{i2} V_{11} - N_{i4} V_{12}
    / \sqrt{2}) \\
C_{W-\schi^-_1-\schi^0_i}^{R} &=&  g_2 (N_{i2} U_{11} + N_{i3} U_{12}
    / \sqrt{2})  \\
C_{\nu_{\ss \ell}-\ell-H^+} &=&  g_2/ (\sqrt{2} \mw) \mell \tan \beta \\
C_{H^+-\schi^-_1-\schi^0_i}^{L} &=& -g_2 (N_{i4} V_{11} + (N_{i2}
     + N_{i1} \tan \thw) V_{12} / \sqrt{2}) \cos \beta \\
C_{H^+-\schi^-_1-\schi^0_i}^{R} &=& -g_2 (N_{i3} U_{11} -(N_{i2} +N_{i1}
     \tan \beta) U_{12}/ \sqrt{2})  \sin \beta  \\
C_{\ell-{\widetilde \nu}_{\ss \ell}-\schi^-_1}^{L} &=&  -g_2 V_{11} \\
C_{\ell-{\widetilde \nu}_{\ss \ell}-\schi^-_1}^{R} &=&   g_2 \mell/(\sqrt{2} \mw
    \cos \beta) U_{12} \\
C_{{\widetilde \nu}_{\ss \ell}-\slept_1-W} &=&   -g_2/ \sqrt{2} \cos \thell \\
C_{{\widetilde \nu}_{\ss \ell}-\slept_1-H^+} &=&  -g_2 \mw/ \sqrt{2} (\sin (2
   \beta) - \mell^2 \tan \beta / \mw^2) \cos \thell \nl + g_2 \mell/( \sqrt{2}
   \mw)   (\mu - A_{\st \ell} \tan \beta) \sin \thell  	
\end{eqnarray*}

Couplings for neutralino-sneutrino coannihilation:
\begin{eqnarray*}
C_{H-W-W} &=& g_2 \mw \cos (\beta - \alpha) \\
C_{h-W-W} &=& g_2 \mw \sin (\beta - \alpha) \\
C_{\snu-\snu-H} &=& -g_2 \mz/ \cos \thw (1/2) \cos (\beta + \alpha) \\
C_{\snu-\snu-h} &=& g_2 \mz/\cos \thw (1/2) \sin (\beta + \alpha)  \\
C_{\snu-\snu-Z} &=& - g_2 / \cos \thw (1/2) \\
C_{\snu-\snu-W-W} &=&  g_2^2/2 \\
C_{Z-W-W} &=& g_2 \cos \thw \\
C_{\snu-\sel -W} &=& - g_2/ \sqrt{2} \\
C_{H-Z-Z} &=&  g_2 \mz / \cos \thw \cos (\beta - \alpha) \\
C_{h-Z-Z} &=&  g_2 \mz / \cos \thw \sin (\beta - \alpha) \\
C_{\snu-\snu-Z-Z} &=& 2 g_2^2/(\cos^2 \thw) (1/2)^2 \\
C_{Z-f-f}^L &=& - g_2/ \cos \thw (T_{3f}  - Q_f \sin^2 \thw ) \\
C_{Z-f-f}^R &=& g_2/ \cos \thw Q_f \sin^2 \thw \\
C_{Z-\nu-\nu} &=& -g_2/ (2 \cos \thw ) \\
C_{H-f-f} &=& -g_2 m_f /(2 \mw) \sin \alpha / \sin \beta \\
C_{h-f-f} &=& - g_2 m_f /(2 \mw) \cos \alpha / \sin \beta \\
C_{\snu-\schi^+_i -e} &=& - g_2 V_{i1} \\
C_{\snu-\schi^0_i -\nu} &=& - g_2/ \sqrt{2} (N_{i2} - \tan \thw N_{i1} ) \\
C_{H - W^+ - H^-} &=& -g_2 \mz/ \cos \thw (1/2) \cos (\alpha + \beta) \\
C_{h - W^+ - H^-} &=& g_2 \mz/ \cos \thw (1/2) \sin (\alpha + \beta) \\
C_{Z - H^+ - H^-} &=& - g_2 \cos (2 \thw)/(2 \cos \thw ) \\
C_{\snu - \sel_L - H^+} &=& - g_2 \mw/ \sqrt{2} \\
C_{H - H^+ - H^-} &=& -g_2 (\mw \cos (\beta - \alpha) - \mz/(2 \cos \thw ) 
    \cos (2 \beta) \cos (\beta + \alpha)) \\
C_{h - H^+ - H^-} &=& -g_2 (\mw \sin (\beta-\alpha) + \mz/(2 \cos \thw) 
     \cos (2 \beta) \sin (\beta+\alpha))  \\
C_{\snu-\snu-H^+-H^-} &=& g_2^2/2 \cos (2 \beta) (-2 (1/2) + (1/2)/ 
        \cos^2 \thw )   \\
C_{H-H-H} &=& -3 g_2 \mz/(2 \cos \thw ) \cos (2 \alpha) \cos (\beta+\alpha) \\
C_{H-H-h} &=& g_2 \mz/(2 \cos \thw )(2 \sin (2 \alpha) \cos (\beta + \alpha)
             + \sin (\beta + \alpha) \cos (2 \alpha))  \\
C_{H-h-h} &=&	- g_2 \mz/(2 \cos \thw )(2 \sin (2 \alpha) \sin (\beta + \alpha)
            - \cos (\beta + \alpha) \cos (2 \alpha))    	\\
C_{h-h-h} &=& -3 g_2 \mz/(2 \cos \thw) \cos (2 \alpha) \sin (\beta + \alpha) \\
C_{\snu - \snu - H -H} &=& g_2^2/2 ((1/2)/ \cos^2 \thw (- \cos (2 \alpha))) \\
C_{\snu - \snu - h -h} &=& g_2^2/2 ((1/2)/ \cos^2 \thw \cos (2 \alpha))  \\
C_{\snu - \snu - H -h} &=& g_2^2/2 \sin (2 \alpha)((1/2)/ \cos^2 \thw ) \\
C_{H-A-A} &=& g_2 \mz/(2 \cos \thw ) \cos (2 \beta) \cos (\beta+\alpha) \\
C_{h-A-A} &=& -g_2 \mz/(2 \cos \thw) \cos(2 \beta) \sin(\beta+\alpha) \\
C_{\snu - \snu - A-A } &=& g_2^2/2 ((1/2)/ \cos^2 \thw \cos(2 \beta)) \\
C_{H-Z-A} &=& g_2 \sin(\beta-\alpha)/(2 \cos \thw ) \\
C_{h-Z-A} &=& -g_2 \cos(\beta-\alpha)/(2 \cos \thw ) \\
C_{\schi^0_1-\schi^0_i - Z}^L &=& g_2/(2 \cos \thw )(N_{i4} N_{14} - N_{i3}
            N_{13} )  \\
C_{\schi^0_1-\schi^0_i - Z}^R &=& -g_2/(2 \cos \thw)(N_{i4}N_{14} - N_{i3}
     N_{13})  \\
C_{\nu-e-W} &=& -g_2/ \sqrt{2} \\
C_{\snu-\sel-W} &=&  -g_2/ \sqrt{2} \\
C_{\sel_1-\schi^0_1 - e}^L &=& -g_2/ \sqrt{2} (\sin \theta_e 2 \tan \thw 
       N_{11})  \\
C_{\sel_2-\schi^0_1 - e}^L &=& -g_2/ \sqrt{2} (\cos \theta_e 2 \tan \thw 
     N_{11})  \\ 
C_{\sel_1-\schi^0_1 - e}^R &=& -g_2/ \sqrt{(2} (\cos \theta_e (-N_{12}
       - \tan \thw N_{11} )) \\
C_{\sel_2-\schi^0_1 - e}^R &=& -g_2/ \sqrt{2} (- \sin \theta_e (-N_{12}
      - \tan \thw N_{11} )) \\
C_{\schi^0_1 - \schi^+_i - W}^L  &=& g_2 (-1/\sqrt{2} N_{14} V_{i2} 
            + N_{12} V_{i1} )\\
C_{\schi^0_1 - \schi^+_i - W}^R  &=&  g_2 (1 /\sqrt{2} N_{13} U_{i2}
           + N_{12} U_{i1} )   \\
C_{\schi^0_1 - \schi^0_i - h}^L &=& g_2/2 ((N_{13} (N_{i2} - N_{i1}
         \tan \thw ) + N_{i3} (N_{12} - N_{11}
         \tan \thw ))  \sin \alpha \nl + (N_{14}
          (N_{i2}-
         N_{i1} \tan \thw ) N_{i4} (N_{12} -
          N_{11} \tan \thw )) \cos \alpha ) \\
C_{\schi^0_1 - \schi^0_i - h}^R &=& g_2/2 ((N_{i3} (N_{12} - N_{11}
         \tan \thw ) + N_{13} (N_{i2} - N_{i1}
         \tan \thw ))  \sin \alpha \nl + (N_{i4}
          (N_{12} -
         N_{11} \tan \thw )+ N_{14} (N_{i2} -
          N_{i1} \tan \thw )) \cos \alpha ) \\
C_{\schi^0_1 - \schi^0_i - H}^L &=& g_2/2 ((N_{13} (N_{i2} - N_{i1}
         \tan \thw ) + N_{i3} (N_{12} - N_{11}
         \tan \thw ))  (-\cos \alpha) \nl + (N_{14}
          (N_{i2}-
         N_{i1} \tan \thw ) N_{i4} (N_{12} -
          N_{11} \tan \thw )) \sin \alpha ) \\
C_{\schi^0_1 - \schi^0_i - H}^R &=& g_2/2 ((N_{i3} (N_{12} - N_{11}
         \tan \thw ) + N_{13} (N_{i2} - N_{i1}
         \tan \thw ))  (-\cos \alpha) \nl + (N_{i4}
          (N_{12} -
         N_{11} \tan \thw )+ N_{14} (N_{i2} -
          N_{i1} \tan \thw )) \sin \alpha ) \\	  	  	  	    
C_{\schi^0_1 - \schi^0_i - A}^L &=& g_2/2 ((N_{13} (N_{i2} - N_{i1}
         \tan \thw ) + N_{i3} (N_{12} - N_{11}
         \tan \thw ))  \sin \beta \nl + (N_{14}
          (N_{i2}-
         N_{i1} \tan \thw ) N_{i4} (N_{12} -
          N_{11} \tan \thw )) (-\cos \beta) ) \\
C_{\schi^0_1 - \schi^0_i - A}^R &=& g_2/2 ((N_{i3} (N_{12} - N_{11}
         \tan \thw ) + N_{13} (N_{i2} - N_{i1}
         \tan \thw ))  (-\sin \beta) \nl + (N_{i4}
          (N_{12} -
         N_{11} \tan \thw )+ N_{14} (N_{i2} -
          N_{i1} \tan \thw )) \cos \beta ) \\
C_{\snu - \sel - H^+} &=& -g_2/\sqrt{2} \sin (2 \beta) \\
C_{\schi^0_1 - \schi^-_i - H^+}^L &=& -g_2 (N_{14} V_{i1} \sqrt{1/2} (N_{12}
          + N_{11} \tan \thw ) V_{i2} ) \cos \beta \\
C_{\schi^0_1 - \schi^-_i - H^+}^R &=& -g_2(N_{13} U_{i1} - \sqrt{1/2} (N_{12}
        + N_{11} \tan \thw ) U_{i2})  \sin \beta	   	  	           
\end{eqnarray*}

\setcounter{equation}{0}
\renewcommand{\theequation}{B\arabic{equation}} 
\section*{Appendix B: Neutralino-Slepton Coannihilation with L-R 
Mixing}

We list below some modifications and additions to the slepton-slepton and
neutralino-slepton coannihilation channels previously given in the
Appendix of~\cite{Ellis:1999mm}. There the lighter sleptons were assumed
to be pure partners of the right-handed leptons, which is a very good
approximation for CMSSM when $\tan \beta$ is small. Here, we include L-R
mixing and denote the lighter sleptons by $\slept_1$, denoting $\ell
\equiv e, \mu, \tau$. The obvious change needed is then replacing
${\widetilde \tau}_{\scriptscriptstyle\rm R}$ in~\cite{Ellis:1999mm} with
$\slept_1$. Furthermore, the presence of the left component entails some
modifications in the couplings and opens up some new channels. We note
that L-R mixing had already been included in some of our previous
work~\cite{efgosi,Ellis:2002wv}. Our convention for the slepton mixing 
angle is 
\begin{equation}
\left( \begin{array}{c} \slept_1 \\ \slept_2 \end{array} \right) = 
\left( \begin{array}{cc} \cos \thf & \sin \thf \\ -\sin \thf & \cos \thf
\end{array} \right) \left(
\begin{array}{c} \slept_L \\ \slept_R \end{array} \right)
\end{equation}
The sign of $\nevalsi$ must be kept in all of the formulae below. 

\subsection*{$\slept_1 \slept^\ast_1 \longrightarrow W^+ W^-$}
There are two additional channels: \hfill \\
 III. $t$-channel $\snu_\ell$  exchange \hfill\\ 
 IV. point interaction \hfill\\
The couplings $f_1$, $f_2$ and $f_5$ are modified, while $f_6$ remains the same.
\begin{eqnarray}
f_1 &=& (-g_2 \mw \cos (\beta-\alpha))( g_2 \mz/\cos \thw  (
          (1/2-\sin^2 \thw) \cos (\beta + \alpha) \cos^2 \thell
        \nl  + \sin^2 \thw \cos (\beta + \alpha) \sin^2 \thell )
          + g_2 \mtau^2/(\mw \cos \beta) (- \cos \alpha)
       \nl   - g_2 \mtau/(\mw \cos \beta ) 
          (- \Atau \cos \alpha - \mu \sin \alpha) \sin \thell \cos \thf )
	  \nonumber \\
f_2 &=& (-g_2 \mw \sin (\beta - \alpha))(g_2 \mz/ \cos \thw  (
          (-1/2+ \sin^2 \thw ) \sin (\beta + \alpha) \cos^2 \thf
      \nl    - \sin^2 \thw  \sin ( \beta + \alpha) \sin^2 \thf )
          + g_2 \mtau^2/(\mw \cos \beta) \sin \alpha
      \nl    - g_2 \mtau/(\mw \cos \beta) 
        (\Atau \sin \alpha - \mu \cos \alpha) \sin \thf \cos \thf ) \nonumber \\
f_3 &=&  (-g_2 \cos \thf / \sqrt{2}  )^2 \nonumber \\
f_4 &=&  -g_2^2 \cos^2 \thf /2   \nonumber \\
f_5 &=&  (g_2 \cos \thw)  ( - g_2 / \cos \thw (\sin^2 \thw - \cos^2 \thf/2 )) 
        \nonumber \\
%f_6 &=&  e^2  \nonumber \\
{\cal T}_{\rm III}\!\!\times\!\!{\cal T}_{\rm III} &=&  (\msl^8 -
     4  \msl^6 \mw^2 + 6  \msl^4 \mw^4 - 
         4  \msl^2 \mw^6 + \mw^8 - 4  \msl^6 u + 
         4  \msl^4 \mw^2 u \nl + 4  \msl^2 \mw^4 u - 4  \mw^6 u + 
         6  \mst^4 u^2 + 4  \mst^2 \mw^2 u^2 + 
         6  \mw^4 u^2 - 4  \mst^2 u^3 \nl - 4  \mw^2 u^3 + u^4
         )/(\mw^4 (\msn^2 - u)^2)  \nonumber \\
{\cal T}_{\rm IV}\!\!\times\!\!{\cal T}_{\rm IV} &=& (12  \mw^4 - 4  \mw^2 s +
       s^2)/(4 \mw^4)   \nonumber \\
{\cal T}_{\rm I}\!\!\times\!\!{\cal T}_{\rm III} &=&   (-6 \mst^4 \mw^2 -
     20 \mst^2 \mw^4 - 6  \mw^6 + 
         \mst^4 s + 2  \mst^2 \mw^2 s + 5 \mw^4 s + 
         4  \mst^2 \mw^2 t \nl + 4  \mw^4 t + 8  \mst^2 \mw^2 u + 
         8  \mw^4 u - 2  \mst^2 s u - 2  \mw^2 s u - 
         4  \mw^2 t u - 2  \mw^2 u^2 \nl + s u^2)/
       (2 \mw^4 (m_H^2 - s) (\msn^2 - u))    \nonumber \\
{\cal T}_{\rm II}\!\!\times\!\!{\cal T}_{\rm III} &=&   (-6 \mst^4 \mw^2 -
      20 \mst^2 \mw^4 - 6  \mw^6 + 
         \mst^4 s + 2  \mst^2 \mw^2 s + 5 \mw^4 s + 
         4  \mst^2 \mw^2 t \nl + 4  \mw^4 t + 8  \mst^2 \mw^2 u + 
         8  \mw^4 u - 2  \mst^2 s u - 2  \mw^2 s u - 
         4  \mw^2 t u - 2  \mw^2 u^2 \nl + s u^2)/
       (2  \mw^4 (m_h^2 - s) (\msn^2 - u))    \nonumber \\
{\cal T}_{\rm III}\!\!\times\!\!{\cal T}_{\rm IV} &=&   (-6  \mst^4 \mw^2 -
     20 \mst^2 \mw^4 - 6  \mw^6 + 
         \mst^4 s + 2  \mst^2 \mw^2 s + 5 \mw^4 s + 
         4  \mst^2 \mw^2 t \nl + 4  \mw^4 t + 8  \mst^2 \mw^2 u + 
         8  \mw^4 u - 2  \mst^2 s u - 2  \mw^2 s u - 
         4  \mw^2 t u - 2  \mw^2 u^2 \nl + s u^2)/
       (2 \mw^4 (\msn^2 - u))    \nonumber \\
{\cal T}_{\rm III}\!\!\times\!\!{\cal T}_{\rm V} &=&   (64  \mst^4 \mw^4 \mz^2
       + 64  \mst^2 \mw^6 \mz^2- 
         16 \mst^4 \mw^2 \mz^2 s - 
         32  \mst^2 \mw^4 \mz^2 s \nl - 16 \mw^6 \mz^2 s + 
         4  \mst^2 \mw^2 \mz^2 s^2 + 4  \mw^4 \mz^2 s^2 - 
         2  \mst^4 \mw^2 \mz^2 t - 
         28 \mst^2 \mw^4 \mz^2 t \nl - 2  \mw^6 \mz^2 t + 
         \mst^4 \mz^2 s t + 2  \mst^2 \mw^2 \mz^2 s t + 
         5 \mw^4 \mz^2 s t + 2  \mst^4 \mw^2 \mz^2 u \nl - 
         36 \mst^2 \mw^4 \mz^2 u + 2  \mw^6 \mz^2 u - 
         \mst^4 \mz^2 s u + 14 \mst^2 \mw^2 \mz^2 s u + 
         11 \mw^4 \mz^2 s u \nl - 4  \mw^2 \mz^2 s^2 u + 
         4  \mst^2 \mw^2 \mz^2 t u + 4  \mw^4 \mz^2 t u - 
         2  \mst^2 \mz^2 s t u - 2  \mw^2 \mz^2 s t u \nl - 
         4  \mst^2 \mw^2 \mz^2 u^2 - 4  \mw^4 \mz^2 u^2 + 
         2  \mst^2 \mz^2 s u^2 + 2  \mw^2 \mz^2 s u^2 - 
         2  \mw^2 \mz^2 t u^2 \nl + \mz^2 s t u^2 + 
         2  \mw^2 \mz^2 u^3 - \mz^2 s u^3)/
       (2 \mw^4 \mz^2 (\mz^2 - s) (\msn^2 - u))    \nonumber \\
{\cal T}_{\rm III}\!\!\times\!\!{\cal T}_{\rm VI} &=&   (-64  \mst^4 \mw^4 -
      64  \mst^2 \mw^6 + 
         16 \mst^4 \mw^2 s + 32  \mst^2 \mw^4 s + 
         16 \mw^6 s  - 4  \mst^2 \mw^2 s^2 \nl - 4  \mw^4 s^2 + 
         2  \mst^4 \mw^2 t + 28 \mst^2 \mw^4 t + 
         2  \mw^6 t - \mst^4 s t - 2  \mst^2 \mw^2 s t - 
         5 \mw^4 s t \nl - 2  \mst^4 \mw^2 u + 
         36 \mst^2 \mw^4 u - 2  \mw^6 u + \mst^4 s u - 
         14 \mst^2 \mw^2 s u - 11 \mw^4 s u \nl + 
         4  \mw^2 s^2 u - 4  \mst^2 \mw^2 t u - 
         4  \mw^4 t u + 2  \mst^2 s t u + 2  \mw^2 s t u + 
         4  \mst^2 \mw^2 u^2 \nl + 4  \mw^4 u^2 - 
         2  \mst^2 s u^2 - 2  \mw^2 s u^2 + 
         2  \mw^2 t u^2 - s t u^2 - 2  \mw^2 u^3 + s u^3)
       \nl  /(2 \mw^4 s (\msn^2 - u))    \nonumber \\ 
{\cal T}_{\rm I}\!\!\times\!\!{\cal T}_{\rm IV} &=&   (12  \mw^4 - 4  \mw^2 s
     + s^2)/(4 \mw^4 (m_H^2 - s))    \nonumber \\
{\cal T}_{\rm II}\!\!\times\!\!{\cal T}_{\rm IV} &=&   (12  \mw^4 - 4  \mw^2 s
    + s^2)/(4 \mw^4 (m_h^2 - s))    \nonumber \\
{\cal T}_{\rm IV}\!\!\times\!\!{\cal T}_{\rm V} &=&   (-12  \mw^4 \mz^2 t +
          \mz^2 s^2 t + 
         12  \mw^4 \mz^2 u - \mz^2 s^2 u)/
       (4 \mw^4 \mz^2 (\mz^2 - s))    \nonumber \\
{\cal T}_{\rm IV}\!\!\times\!\!{\cal T}_{\rm VI} &=&   (12  \mw^4 t - s^2 t
       -12  \mw^4 u + s^2 u)/(4 \mw^4 s)    \nonumber \\ 
\tsq &=&  f_1^2 {\cal T}_{\rm I}\!\!\times\!\!{\cal T}_{\rm I} 
   +  f_2^2 {\cal T}_{\rm II}\!\!\times\!\!{\cal T}_{\rm II} 
   +  f_3^2 {\cal T}_{\rm III}\!\!\times\!\!{\cal T}_{\rm III} 
   +  f_4^2 {\cal T}_{\rm IV}\!\!\times\!\!{\cal T}_{\rm IV}  
   +  f_5^2 {\cal T}_{\rm V}\!\!\times\!\!{\cal T}_{\rm V} 
   +  f_6^2 {\cal T}_{\rm VI}\!\!\times\!\!{\cal T}_{\rm VI}  
  \nl + 2 f_1 f_2 {\cal T}_{\rm I}\!\!\times\!\!{\cal T}_{\rm II} 
   + 2 f_1 f_3  {\cal T}_{\rm I}\!\!\times\!\!{\cal T}_{\rm III} 
   + 2 f_1 f_4 {\cal T}_{\rm I}\!\!\times\!\!{\cal T}_{\rm IV}  
   + 2 f_1 f_5  {\cal T}_{\rm I}\!\!\times\!\!{\cal T}_{\rm V} 
   + 2 f_2 f_3 {\cal T}_{\rm II}\!\!\times\!\!{\cal T}_{\rm III} 
 \nl   + 2 f_2 f_4 {\cal T}_{\rm II}\!\!\times\!\!{\cal T}_{\rm IV} 
   + 2 f_2 f_5 {\cal T}_{\rm II}\!\!\times\!\!{\cal T}_{\rm V}  
   + 2 f_3 f_4 {\cal T}_{\rm III}\!\!\times\!\!{\cal T}_{\rm IV} 
   + 2 f_3 f_5 {\cal T}_{\rm III}\!\!\times\!\!{\cal T}_{\rm V} 
 \nl  + 2 f_4 f_5 {\cal T}_{\rm IV}\!\!\times\!\!{\cal T}_{\rm V} 
   + 2 f_1 f_6 {\cal T}_{\rm I}\!\!\times\!\!{\cal T}_{\rm VI}  
   + 2 f_2 f_6 {\cal T}_{\rm II}\!\!\times\!\!{\cal T}_{\rm VI} 
   + 2 f_3 f_6 {\cal T}_{\rm III}\!\!\times\!\!{\cal T}_{\rm VI} 
 \nl  + 2 f_4 f_6  {\cal T}_{\rm IV}\!\!\times\!\!{\cal T}_{\rm VI} 
   + 2 f_5 f_6 {\cal T}_{\rm V}\!\!\times\!\!{\cal T}_{\rm VI}  
\end{eqnarray}

\subsection*{$\slept_1 \slept^\ast_1 \longrightarrow Z Z$}
Besides the ${\widetilde \ell}_{\scriptscriptstyle\rm 1}$ exchange (III and IV
in~\cite{Ellis:1999mm}), we also have  ${\widetilde \ell}_{\scriptscriptstyle\rm
2}$ exchange, written here as VI and VII. \hfill \\
 VI. $t$-channel  ${\widetilde \ell}_{\scriptscriptstyle\rm 1}$ exchange \hfill \\
 VII.  $u$-channel  ${\widetilde \ell}_{\scriptscriptstyle\rm 1}$ exchange \hfill
 \\
The couplings $f_1, \ldots f_5$ are modified, and we have new $f_6$ and 
$f_7$ couplings.
\begin{eqnarray}
f_1 &=& (-g_2 \mz \cos(\beta-\alpha)/\cos \thw ) ( g_2 \mz/\cos \thw  (
          (1/2- \sin^2 \thw) \cos(\beta+\alpha)
	  \nl  \cos^2 \thf
          + \sin^2 \thw  \cos(\beta+\alpha) \sin^2 \thf )
          + g_2 \mtau^2/( \mw \cos \beta ) (-\cos \alpha )
         \nl  - g_2 \mtau/(\mw \cos \beta ) 
          (- \Atau \cos \alpha - \mu \sin \alpha ) \sin \thf \cos \thf )
	  \nonumber \\
f_2 &=&  (-g_2 \mz \sin(\beta-\alpha)/\cos \thw ) (g_2 \mz/\cos \thw (
          (-1/2+ \sin^2 \thw ) \sin(\beta+\alpha) 
	  \nl  \cos^2 \thf
          - \sin^2 \thw  \sin(\beta+\alpha)  \sin^2 \thf )
          + g_2 \mtau^2/(\mw \cos \beta ) \sin \alpha
         \nl - g_2 \mtau/(\mw \cos \beta ) 
          (\Atau \sin \alpha - \mu \cos \alpha ) \sin \thf \cos \thf )\nonumber
	  \\
f_3 &=&  ( -g_2 / \cos \thw ( \sin^2 \thw - \cos^2 \thf/2))^2 \nonumber \\
f_4 &=&  ( -g_2 /\cos \thw ( \sin^2 \thw - \cos^2 \thf/2))^2 \nonumber \\
f_5 &=&   -2 g_2^2/\cos^2 \thw ( \sin^4 \thw +\cos^2 \thf 
          (1/4 - \sin^2 \thw))       \nonumber \\
f_6 &=&  ( -(g_2 /\cos \thw) \cos \thf \sin \thf/2)^2 \nonumber \\
f_7 &=&  ( -(g_2 /\cos \thw )\cos \thf \sin \thf/2)^2 \nonumber \\
{\cal T}_{\rm VI}\!\!\times\!\!{\cal T}_{\rm VI} &=&  (\mst^8 - 4 \mst^6 \mz^2
        + 6 \mst^4 \mz^4 - 
         4 \mst^2 \mz^6 + \mz^8 - 4 \mst^6 t + 
         4 \mst^4 \mz^2 t \nl + 4 \mst^2 \mz^4 t - 4 \mz^6 t + 
         6 \mst^4 t^2 + 4 \mst^2 \mz^2 t^2 + 
         6 \mz^4 t^2 - 4 \mst^2 t^3 - 4 \mz^2 t^3 \nl + t^4
         )/(\mz^4 (\mstwo^2 - t)^2)     \nonumber \\
{\cal T}_{\rm VII}\!\!\times\!\!{\cal T}_{\rm VII} &=&  (\mst^8 - 4 \mst^6
        \mz^2 + 6 \mst^4 \mz^4 - 
         4 \mst^2 \mz^6 + \mz^8 - 4 \mst^6 u + 
         4 \mst^4 \mz^2 u \nl + 4 \mst^2 \mz^4 u - 4 \mz^6 u + 
         6 \mst^4 u^2 + 4 \mst^2 \mz^2 u^2 + 
         6 \mz^4 u^2 - 4 \mst^2 u^3 - 4 \mz^2 u^3 \nl + u^4
         )/(\mz^4 (\mstwo^2 - u)^2)     \nonumber \\
{\cal T}_{\rm VI}\!\!\times\!\!{\cal T}_{\rm VII} &=&   (\mst^8 + 12 \mst^6
       \mz^2 + 38 \mst^4 \mz^4 + 
         12 \mst^2 \mz^6 + \mz^8 - 4 \mst^4 \mz^2 s - 
         24 \mst^2 \mz^4 s \nl - 4 \mz^6 s + 4 \mz^4 s^2 - 
         2 \mst^6 t - 14 \mst^4 \mz^2 t - 
         14 \mst^2 \mz^4 t - 2 \mz^6 t + 
         4 \mst^2 \mz^2 s t \nl + 4 \mz^4 s t + \mst^4 t^2 + 
         2 \mst^2 \mz^2 t^2 + \mz^4 t^2 - 2 \mst^6 u - 
         14 \mst^4 \mz^2 u - 14 \mst^2 \mz^4 u \nl - 
         2 \mz^6 u + 4 \mst^2 \mz^2 s u + 4 \mz^4 s u + 
         4 \mst^4 t u + 16 \mst^2 \mz^2 t u + 
         4 \mz^4 t u \nl - 4 \mz^2 s t u - 2 \mst^2 t^2 u - 
         2 \mz^2 t^2 u + \mst^4 u^2 + 
         2 \mst^2 \mz^2 u^2 + \mz^4 u^2 - 
         2 \mst^2 t u^2 \nl - 2 \mz^2 t u^2 + t^2 u^2)/
       (\mz^4 (\mstwo^2 - t) (\mstwo^2 - u))    \nonumber \\
{\cal T}_{\rm I}\!\!\times\!\!{\cal T}_{\rm VI} &=&   (-6 \mst^4 \mz^2 - 20
        \mst^2 \mz^4 - 6 \mz^6 + 
         \mst^4 s + 2 \mst^2 \mz^2 s + 5 \mz^4 s + 
         8 \mst^2 \mz^2 t \nl + 8 \mz^4 t - 2 \mst^2 s t - 
         2 \mz^2 s t - 2 \mz^2 t^2 + s t^2 + 
         4 \mst^2 \mz^2 u + 4 \mz^4 u - 4 \mz^2 t u) \nl /
       (2 \mz^4 (m_H^2 - s) (\mstwo^2 - t))    \nonumber \\
{\cal T}_{\rm II}\!\!\times\!\!{\cal T}_{\rm VI} &=&   (-6 \mst^4 \mz^2 - 20
         \mst^2 \mz^4 - 6 \mz^6 + 
         \mst^4 s + 2 \mst^2 \mz^2 s + 5 \mz^4 s + 
         8 \mst^2 \mz^2 t \nl + 8 \mz^4 t - 2 \mst^2 s t - 
         2 \mz^2 s t - 2 \mz^2 t^2 + s t^2 + 
         4 \mst^2 \mz^2 u + 4 \mz^4 u - 4 \mz^2 t u) \nl /
       (2 \mz^4 (m_h^2 - s) (\mstwo^2 - t))    \nonumber \\
{\cal T}_{\rm III}\!\!\times\!\!{\cal T}_{\rm VI} &=&   -((\mst^4 + (\mz^2 -
           t)^2 - 2 \mst^2 (\mz^2 + t))^2/
          (\mz^4 (\mst^2 - t) (-\mstwo^2 + t)))    \nonumber \\ 
{\cal T}_{\rm IV}\!\!\times\!\!{\cal T}_{\rm VI} &=&   (\mst^4 + \mz^4 + \mst^2
          (6 \mz^2 - t - u) + t u -  \mz^2 (2 s + t + u))^2 
	  \nl /(\mz^4 (\mstwo^2 - t) (\mst^2 - u))    \nonumber \\
{\cal T}_{\rm V}\!\!\times\!\!{\cal T}_{\rm VI} &=&   (-6 \mst^4 \mz^2 - 20
         \mst^2 \mz^4 - 6 \mz^6 + 
         \mst^4 s + 2 \mst^2 \mz^2 s + 5 \mz^4 s + 
         8 \mst^2 \mz^2 t \nl + 8 \mz^4 t - 2 \mst^2 s t - 
         2 \mz^2 s t - 2 \mz^2 t^2 + s t^2 + 
         4 \mst^2 \mz^2 u + 4 \mz^4 u - 4 \mz^2 t u) \nl /
       (2 \mz^4 (\mstwo^2 - t))    \nonumber \\
{\cal T}_{\rm I}\!\!\times\!\!{\cal T}_{\rm VII} &=&   (-6 \mst^4 \mz^2 - 20
           \mst^2 \mz^4 - 6 \mz^6 + 
         \mst^4 s + 2 \mst^2 \mz^2 s + 5 \mz^4 s + 
         4 \mst^2 \mz^2 t \nl + 4 \mz^4 t + 8 \mst^2 \mz^2 u + 
         8 \mz^4 u - 2 \mst^2 s u - 2 \mz^2 s u - 
         4 \mz^2 t u - 2 \mz^2 u^2 + s u^2) \nl /
       (2 \mz^4 (m_H^2 - s) (\mstwo^2 - u))    \nonumber \\
{\cal T}_{\rm II}\!\!\times\!\!{\cal T}_{\rm VII} &=&   (-6 \mst^4 \mz^2 - 20
        \mst^2 \mz^4 - 6 \mz^6 + 
         \mst^4 s + 2 \mst^2 \mz^2 s + 5 \mz^4 s + 
         4 \mst^2 \mz^2 t \nl + 4 \mz^4 t + 8 \mst^2 \mz^2 u + 
         8 \mz^4 u - 2 \mst^2 s u - 2 \mz^2 s u - 
         4 \mz^2 t u - 2 \mz^2 u^2 + s u^2) \nl /
       (2 \mz^4 (m_h^2 - s) (\mstwo^2 - u))    \nonumber \\
{\cal T}_{\rm III}\!\!\times\!\!{\cal T}_{\rm VII} &=&  (\mst^4 + \mz^4 + \mst^2
          (6 \mz^2 - t - u) + t u - \mz^2 (2 s + t + u))^2 
	  \nl /(\mz^4 (\mst^2 - t) (\mstwo^2 - u))     \nonumber \\ 
{\cal T}_{\rm IV}\!\!\times\!\!{\cal T}_{\rm VII} &=&  -((\mst^4 + (\mz^2 - u)^2
           - 2 \mst^2 (\mz^2 + u))^2/
          (\mz^4 (\mst^2 - u) (-\mstwo^2 + u)))     \nonumber \\
{\cal T}_{\rm V}\!\!\times\!\!{\cal T}_{\rm VII} &=&   (-6 \mst^4 \mz^2 - 20
          \mst^2 \mz^4 - 6 \mz^6 + 
         \mst^4 s + 2 \mst^2 \mz^2 s + 5 \mz^4 s + 
         4 \mst^2 \mz^2 t \nl + 4 \mz^4 t + 8 \mst^2 \mz^2 u + 
         8 \mz^4 u - 2 \mst^2 s u - 2 \mz^2 s u - 
         4 \mz^2 t u - 2 \mz^2 u^2 + s u^2) \nl /
       (2 \mz^4 (\mstwo^2 - u))    \nonumber \\
\tsq &=&  f_1^2 {\cal T}_{\rm I}\!\!\times\!\!{\cal T}_{\rm I} 
   +  f_2^2 {\cal T}_{\rm II}\!\!\times\!\!{\cal T}_{\rm II} 
   +  f_3^2 {\cal T}_{\rm III}\!\!\times\!\!{\cal T}_{\rm III} 
   +  f_4^2 {\cal T}_{\rm IV}\!\!\times\!\!{\cal T}_{\rm IV}  
   +  f_5^2 {\cal T}_{\rm V}\!\!\times\!\!{\cal T}_{\rm V} 
   +  f_6^2 {\cal T}_{\rm VI}\!\!\times\!\!{\cal T}_{\rm VI} 
   \nl +  f_7^2 {\cal T}_{\rm VII}\!\!\times\!\!{\cal T}_{\rm VII}     
   + 2 f_1 f_2 {\cal T}_{\rm I}\!\!\times\!\!{\cal T}_{\rm II} 
   + 2 f_1 f_3  {\cal T}_{\rm I}\!\!\times\!\!{\cal T}_{\rm III} 
   + 2 f_1 f_4 {\cal T}_{\rm I}\!\!\times\!\!{\cal T}_{\rm IV}  
   + 2 f_1 f_5  {\cal T}_{\rm I}\!\!\times\!\!{\cal T}_{\rm V}
  \nl + 2 f_1 f_6  {\cal T}_{\rm I}\!\!\times\!\!{\cal T}_{\rm VI}
   + 2 f_1 f_7  {\cal T}_{\rm I}\!\!\times\!\!{\cal T}_{\rm VII}       
   + 2 f_2 f_3 {\cal T}_{\rm II}\!\!\times\!\!{\cal T}_{\rm III} 
    + 2 f_2 f_4 {\cal T}_{\rm II}\!\!\times\!\!{\cal T}_{\rm IV} 
   + 2 f_2 f_5 {\cal T}_{\rm II}\!\!\times\!\!{\cal T}_{\rm V}
  \nl  + 2 f_2 f_6 {\cal T}_{\rm II}\!\!\times\!\!{\cal T}_{\rm VI}   
   + 2 f_2 f_7 {\cal T}_{\rm II}\!\!\times\!\!{\cal T}_{\rm VII}     
   + 2 f_3 f_4 {\cal T}_{\rm III}\!\!\times\!\!{\cal T}_{\rm IV} 
   + 2 f_3 f_5 {\cal T}_{\rm III}\!\!\times\!\!{\cal T}_{\rm V}
  \nl  + 2 f_3 f_6 {\cal T}_{\rm III}\!\!\times\!\!{\cal T}_{\rm VI}   
    + 2 f_3 f_7 {\cal T}_{\rm III}\!\!\times\!\!{\cal T}_{\rm VII}    
   + 2 f_4 f_5 {\cal T}_{\rm IV}\!\!\times\!\!{\cal T}_{\rm V} 
   + 2 f_4 f_6  {\cal T}_{\rm IV}\!\!\times\!\!{\cal T}_{\rm VI}
  \nl + 2 f_4 f_7  {\cal T}_{\rm IV}\!\!\times\!\!{\cal T}_{\rm VII}   
   + 2 f_5 f_6 {\cal T}_{\rm V}\!\!\times\!\!{\cal T}_{\rm VI}
   + 2 f_5 f_7 {\cal T}_{\rm V}\!\!\times\!\!{\cal T}_{\rm VII}     
\end{eqnarray}

\subsection*{$\slept_1 \slept^\ast_1 \longrightarrow \gamma \gamma$}
There is no change for this channel, except that ${\widetilde
\tau}_{\scriptscriptstyle\rm R} \rightarrow {\widetilde
\ell}_{\scriptscriptstyle\rm 1}$. 

\subsection*{$\slept_1 \slept^\ast_1 \longrightarrow Z \gamma$}
There is no change for this channel, except that ${\widetilde
\tau}_{\scriptscriptstyle\rm R} \rightarrow {\widetilde
\ell}_{\scriptscriptstyle\rm 1}$.

\subsection*{$\slept_1 \slept^\ast_1 \longrightarrow Z h [H]$}
Besides the ${\widetilde \ell}_{\scriptscriptstyle\rm 1}$ exchange (I and II
in~\cite{Ellis:1999mm}), we also have  ${\widetilde \ell}_{\scriptscriptstyle\rm
2}$ exchanged, written here as IV and V. \hfill \\
 IV. $t$-channel  ${\widetilde \ell}_{\scriptscriptstyle\rm 2}$ exchange \hfill \\
 V.  $u$-channel  ${\widetilde \ell}_{\scriptscriptstyle\rm 2}$ exchange \hfill
 \\
The couplings are modified. 
\begin{eqnarray}
f_1 &=& (- g_2 /\cos \thw (\sin^2 \thw - \cos^2 \thf/2))(g_2 \mz/\cos \thw (
          (-1/2+ \sin^2 \thw ) \nl \sin [-\cos](\beta+\alpha) \cos^2 \thf
          - \sin^2 \thw  \sin [-\cos] (\beta+\alpha) \sin^2 \thf )
       \nl    + g_2 \mtau^2/( \mw \cos \beta) \sin [-\cos] \alpha
          - g_2 \mtau/(\mw \cos \beta ) 
          (\Atau \sin [-\cos] \alpha \nl - \mu \cos [\sin] \alpha ) \sin \thf
	  \cos \thf )	 
	  \nonumber \\
f_2 &=& - (- g_2 /\cos \thw (\sin^2 \thw - \cos^2 \thf/2))(g_2 \mz/\cos \thw (
          (-1/2+ \sin^2 \thw ) \nl \sin [-\cos](\beta+\alpha) \cos^2 \thf
          - \sin^2 \thw  \sin [-\cos] (\beta+\alpha) \sin^2 \thf )
       \nl    + g_2 \mtau^2/( \mw \cos \beta) \sin [-\cos] \alpha
          - g_2 \mtau/(\mw \cos \beta ) 
          (\Atau \sin [-\cos] \alpha \nl - \mu \cos [\sin] \alpha ) \sin \thf
	  \cos \thf ) \nonumber
	  \\
f_3 &=&  (- g_2 /\cos \thw (\sin^2 \thw - \cos^2 \thf/2))
           (-g_2 \mz \sin [\cos] (\beta-\alpha) /\cos \thw )\nonumber \\
f_4 &=& ( -g_2 \cos \thf \sin \thf /(2 \cos \thw) ) (g_2 \mz/\cos \thw (
          (-1/2+ \sin^2 \thw) \nl 
	  \sin [-\cos] (\beta+\alpha) (-\cos \thf \sin \thf)
          - \sin^2 \thw \sin [-\cos] (\beta+\alpha) \cos \thf \sin \thf )
        \nl   - g_2 \mtau/(2 \mw \cos \beta) 
          (\Atau \sin [-\cos] \alpha - \mu \cos [\sin ] \alpha ) \cos (2 \thf) )
	   \nonumber \\
f_5 &=&  (-)  ( -g_2 \cos \thf \sin \thf /(2 \cos \thw) ) (g_2 \mz/\cos \thw (
          (-1/2+ \sin^2 \thw) \nl 
	  \sin [-\cos] (\beta+\alpha) (-\cos \thf \sin \thf)
          - \sin^2 \thw \sin [-\cos] (\beta+\alpha) \cos \thf \sin \thf )
       \nl   - g_2 \mtau/(2 \mw \cos \beta) 
          (\Atau \sin [-\cos] \alpha - \mu \cos [\sin ] \alpha ) \cos (2 \thf) )
	        \nonumber \\
{\cal T}_{\rm IV}\!\!\times\!\!{\cal T}_{\rm IV} &=&  (\mst^4 + (\mz^2 - t)^2
          - 2 \mst^2 (\mz^2 + t))/
          (\mz^2 (\mstwo^2 - t)^2)  \nonumber \\
{\cal T}_{\rm V}\!\!\times\!\!{\cal T}_{\rm V} &=&  (\mst^4 + (\mz^2 - u)^2 -
           2 \mst^2 (\mz^2 + u))/
          (\mz^2 (\mstwo^2 - u)^2)  \nonumber \\
{\cal T}_{\rm I}\!\!\times\!\!{\cal T}_{\rm IV} &=& -((\mst^4 + (\mz^2 - t)^2
           - 2 \mst^2 (\mz^2 + t))/
          (\mz^2 (\mst^2 - t) (-\mstwo^2 + t)))      \nonumber \\
{\cal T}_{\rm I}\!\!\times\!\!{\cal T}_{\rm V} &=&   (\mst^4 + \mz^4 +
       \mst^2 (6 \mz^2 - t - u) + t u - 
          \mz^2 (2 s + t + u)) \nl /
          (\mz^2 (\mst^2 - t) (\mstwo^2 - u))   \nonumber \\ 
{\cal T}_{\rm II}\!\!\times\!\!{\cal T}_{\rm IV} &=&  (\mst^4 + \mz^4 +
         \mst^2 (6 \mz^2 - t - u) + t u - 
          \mz^2 (2 s + t + u)) \nl /
          (\mz^2 (\mstwo^2 - t) (\mst^2 - u))    \nonumber \\
{\cal T}_{\rm II}\!\!\times\!\!{\cal T}_{\rm V} &=&  -((\mst^4 + (\mz^2 - u)^2
        - 2 \mst^2 (\mz^2 + u))/
          (\mz^2 (\mst^2 - u) (-\mstwo^2 + u)))     \nonumber \\
{\cal T}_{\rm III}\!\!\times\!\!{\cal T}_{\rm IV} &=&   (t (t - u) +
           \mst^2 (-8 \mz^2 - t + u) + 
          \mz^2 (2 s - t + u)) \nl /
          (2 \mz^2 (\mz^2 - s) (\mstwo^2 - t))    \nonumber \\
{\cal T}_{\rm III}\!\!\times\!\!{\cal T}_{\rm V} &=&  ((t - u) u +
           \mst^2 (8 \mz^2 - t + u) + 
          \mz^2 (-2 s - t + u)) \nl /
          (2 \mz^2 (\mz^2 - s) (\mstwo^2 - u))     \nonumber \\ 
{\cal T}_{\rm IV}\!\!\times\!\!{\cal T}_{\rm V} &=&(\mst^4 + \mz^4 +
           \mst^2 (6 \mz^2 - t - u) + t u - 
          \mz^2 (2 s + t + u)) \nl /
          (\mz^2 (\mstwo^2 - t) (\mstwo^2 - u))	  \nonumber \\ 
\tsq &=&  f_1^2 {\cal T}_{\rm I}\!\!\times\!\!{\cal T}_{\rm I} 
   +  f_2^2 {\cal T}_{\rm II}\!\!\times\!\!{\cal T}_{\rm II} 
   +  f_3^2 {\cal T}_{\rm III}\!\!\times\!\!{\cal T}_{\rm III} 
   +  f_4^2 {\cal T}_{\rm IV}\!\!\times\!\!{\cal T}_{\rm IV}  
   +  f_5^2 {\cal T}_{\rm V}\!\!\times\!\!{\cal T}_{\rm V} 
  \nl + 2 f_1 f_2 {\cal T}_{\rm I}\!\!\times\!\!{\cal T}_{\rm II} 
   + 2 f_1 f_3  {\cal T}_{\rm I}\!\!\times\!\!{\cal T}_{\rm III} 
   + 2 f_1 f_4 {\cal T}_{\rm I}\!\!\times\!\!{\cal T}_{\rm IV}  
   + 2 f_1 f_5  {\cal T}_{\rm I}\!\!\times\!\!{\cal T}_{\rm V} 
 \nl  + 2 f_2 f_3 {\cal T}_{\rm II}\!\!\times\!\!{\cal T}_{\rm III} 
    + 2 f_2 f_4 {\cal T}_{\rm II}\!\!\times\!\!{\cal T}_{\rm IV} 
   + 2 f_2 f_5 {\cal T}_{\rm II}\!\!\times\!\!{\cal T}_{\rm V}  
   + 2 f_3 f_4 {\cal T}_{\rm III}\!\!\times\!\!{\cal T}_{\rm IV} 
 \nl  + 2 f_3 f_5 {\cal T}_{\rm III}\!\!\times\!\!{\cal T}_{\rm V} 
   + 2 f_4 f_5 {\cal T}_{\rm IV}\!\!\times\!\!{\cal T}_{\rm V} 
\end{eqnarray}

\subsection*{$\slept_1 \slept^\ast_1 \longrightarrow \gamma h [H]$}
There is no new channels, however the couplings are modified.
\begin{eqnarray}
f_1 &=& (e) (g_2 \mz/\cos \thw  (
          (-1/2+ \sin^2 \thw) \sin [-\cos] (\beta+\alpha) \cos^2 \thf
        \nl   - \sin^2 \thw  \sin [-\cos] (\beta+\alpha) \sin^2 \thf )
          + g_2 \mtau^2/(\mw \cos \beta ) \sin [-\cos] \alpha
       \nl    - g_2 \mtau/(\mw \cos \beta ) 
     (\Atau \sin [-\cos] \alpha -\mu \cos [\sin] \alpha ) \sin \thf \cos \thf )
	  \nonumber \\
f_2 &=&  -(e) (g_2 \mz/\cos \thw  (
          (-1/2+ \sin^2 \thw) \sin [-\cos] (\beta+\alpha) \cos^2 \thf
        \nl   - \sin^2 \thw  \sin [-\cos] (\beta+\alpha) \sin^2 \thf )
          + g_2 \mtau^2/(\mw \cos \beta ) \sin [-\cos] \alpha
       \nl    - g_2 \mtau/(\mw \cos \beta ) 
     (\Atau \sin [-\cos] \alpha -\mu \cos [\sin] \alpha ) \sin \thf \cos \thf ) 
\end{eqnarray}

\subsection*{$\slept_1 \slept^\ast_1 \longrightarrow Z A$}
We now have $t$- and $u$-channel $\slept_2$ exchanges, written here as III
and IV.
\hfill \\ 
 III. $t$-channel  ${\widetilde \ell}_{\scriptscriptstyle\rm 2}$ exchange \hfill
 \\
 IV. $u$-channel  ${\widetilde \ell}_{\scriptscriptstyle\rm 2}$ exchange \hfill
 \\
The couplings $f_1$ and $f_2$ are also modified. 
\begin{eqnarray}
f_1 &=& ( g_2 \mz/\cos \thw (
          (-1/2 + \sin^2 \thw ) \sin (\beta+\alpha) \cos^2 \thf 
          - \sin^2 \thw  \sin(\beta+\alpha) \sin^2 \thf )
        \nl  + g_2 \mtau^2/(\mw \cos \beta ) \sin \alpha
          - g_2 \mtau/(\mw \cos \beta ) 
          (\Atau \sin \alpha - \mu \cos \alpha ) \sin \thf \cos \thf )
	\nl  (-g_2 \cos(\alpha-\beta)/(2  \cos \thw )) \nonumber \\
f_2 &=&  (g_2 \mz/\cos \thw  (
          (1/2 - \sin^2 \thw ) \cos (\beta+\alpha) \cos^2 \thf 
          + \sin^2 \thw  \cos(\beta+\alpha) \sin^2 \thf  )
       \nl   - g_2 \mtau^2/( \mw \cos \beta ) \cos \alpha 
          - g_2 \mtau/(\mw \cos \beta ) 
          (-\Atau \cos \alpha -\mu \sin \alpha ) \sin \thf \cos \thf )
	\nl  (-g_2 \sin(\alpha-\beta)/(2  \cos \thw )) \nonumber \\
f_3 &=& (- g_2 \cos \thf \sin \thf /(2 \cos \thw)  )
           (- g_2 \mtau/(2 \mw)  (\Atau \tan \beta - \mu)) \nonumber \\
f_4 &=& (- g_2 \cos \thf \sin \thf /(2 \cos \thw)  )
           (- g_2 \mtau/(2 \mw)  (\Atau \tan \beta - \mu)) \nonumber \\
{\cal T}_{\rm III}\!\!\times\!\!{\cal T}_{\rm III} &=&  (\mst^4 + (\mz^2 -
          t)^2 - 2 \mst^2 (\mz^2 + t))/
          (\mz^2 (\mstwo^2 - t)^2)     \nonumber \\
{\cal T}_{\rm IV}\!\!\times\!\!{\cal T}_{\rm IV} &=&   (\mst^4 + (\mz^2 - u)^2
        - 2 \mst^2 (\mz^2 + u))/
          (\mz^2 (\mstwo^2 - u)^2)   \nonumber \\ 
{\cal T}_{\rm III}\!\!\times\!\!{\cal T}_{\rm IV} &=&	(\mst^4 + \mz^4 +
        \mst^2 (6 \mz^2 - t - u) + t u - 
          \mz^2 (2 s + t + u)) \nl /
          (\mz^2 (\mstwo^2 - t) (\mstwo^2 - u))    \nonumber \\
{\cal T}_{\rm I}\!\!\times\!\!{\cal T}_{\rm III} &=&  (-\mz^4 + \mz^2 s +
          \mst^2 (-3 \mz^2 + s) + 
         \mz^2 t - s t + \mA^2 (-\mst^2 - 3 \mz^2 + t) \nl + 
          2 \mz^2 u)/(\mz^2 (m_h^2 - s) (\mstwo^2 - t))     \nonumber \\
{\cal T}_{\rm II}\!\!\times\!\!{\cal T}_{\rm III} &=&  (-\mz^4 + \mz^2 s +
        \mst^2 (-3 \mz^2 + s) + 
          \mz^2 t - s t + \mA^2 (-\mst^2 - 3 \mz^2 + t) \nl + 
          2 \mz^2 u)/(\mz^2 (m_H^2 - s) (\mstwo^2 - t))     \nonumber \\
{\cal T}_{\rm I}\!\!\times\!\!{\cal T}_{\rm IV} &=&   (-\mz^4 + \mz^2 s +
           \mst^2 (-3 \mz^2 + s) + 
          2 \mz^2 t + \mz^2 u - s u + 
          \mA^2 (-\mst^2 - 3 \mz^2 \nl + u))/
          (\mz^2 (m_h^2 - s) (\mstwo^2 - u))    \nonumber \\
{\cal T}_{\rm II}\!\!\times\!\!{\cal T}_{\rm IV} &=&  (-\mz^4 + \mz^2 s +
            \mst^2 (-3 \mz^2 + s) + 
          2 \mz^2 t + \mz^2 u - s u + 
          \mA^2 (-\mst^2 - 3 \mz^2 \nl + u))/
          (\mz^2 (m_H^2 - s) (\mstwo^2 - u))     \nonumber \\ 
\tsq &=&  f_1^2 {\cal T}_{\rm I}\!\!\times\!\!{\cal T}_{\rm I} 
   +  f_2^2 {\cal T}_{\rm II}\!\!\times\!\!{\cal T}_{\rm II} 
   +  f_3^2 {\cal T}_{\rm III}\!\!\times\!\!{\cal T}_{\rm III} 
   +  f_4^2 {\cal T}_{\rm IV}\!\!\times\!\!{\cal T}_{\rm IV}  
   + 2 f_1 f_2 {\cal T}_{\rm I}\!\!\times\!\!{\cal T}_{\rm II} 
   + 2 f_1 f_3  {\cal T}_{\rm I}\!\!\times\!\!{\cal T}_{\rm III} 
 \nl  + 2 f_1 f_4 {\cal T}_{\rm I}\!\!\times\!\!{\cal T}_{\rm IV}  
   + 2 f_2 f_3 {\cal T}_{\rm II}\!\!\times\!\!{\cal T}_{\rm III} 
    + 2 f_2 f_4 {\cal T}_{\rm II}\!\!\times\!\!{\cal T}_{\rm IV} 
   + 2 f_3 f_4 {\cal T}_{\rm III}\!\!\times\!\!{\cal T}_{\rm IV} 
\end{eqnarray}

\subsection*{$\slept_1 \slept^\ast_1 \longrightarrow \ell \bar{\ell} $}
The $s$-channel $h$ and $H$ annihilations are neglected, due
to the small Yukawa couplings for leptons. The $t$-channel $\chi$ exchange is
modified, with more couplings introduced $(K,K^\prime) \rightarrow
(K_a,K_b,K_a^\prime,K_b^\prime)$. The terms with $m_\ell$ were neglected
in~\cite{Ellis:1999mm}, but not here.    
The couplings $f_{3c}$ and  $f_{3d}$ are modified while
$f_{4c}$ remains the same.
\begin{eqnarray}
f_{3c} &=&  ( - g_2 (\sin^2 \thw - \cos^2 \thf /2)/\cos \thw )
    (g_2 ( 1 - 4  \sin^2 \thw ) /(4 \cos \thw )) \nonumber \\
f_{3d} &=&  ( - g_2 (\sin^2 \thw - \cos^2 \thf /2)/\cos \thw )
     (- g_2/( 4 \cos \thw ) )  \nonumber \\ 
K_a &=&   -\sin \thf (g_2 \mtau/(2 \mw \cos \beta ) N_{i3} + g_1 N_{i1}
       )/ \sqrt{2}
          \nl    -\cos \thf (g_2 \mtau/(2 \mw \cos \beta ) N_{i3} 
	   - (  g_1 N_{i1}  + g_2/2 
                (N_{i2}-N_{i1} \tan \thw )))/ \sqrt{2}  \nonumber \\
K_b &=&  -\sin \thf (g_2 \mtau/(2 \mw \cos \beta ) N_{i3} - g_1 N_{i1} 
           )/ \sqrt{2}
         \nl     -\cos \thf (-g_2 \mtau/(2 \mw \cos \beta ) N_{i3} 
	  - (  g_1 N_{i1}  + g_2/2 
                (N_{i2}-N_{i1} \tan \thw )))/ \sqrt{2}   \nonumber \\
K_a^\prime &=&  -\sin \thf (g_2 \mtau/(2 \mw \cos \beta ) N_{j3} + g_1
       N_{j1} )/ \sqrt{2}
       \nl       -\cos \thf (g_2 \mtau/(2 \mw \cos \beta ) N_{j3} 
        - ( g_1 N_{j1}  + g_2/2 
                (N_{j2}-N_{j1} \tan \thw )))/ \sqrt{2}   \nonumber \\ 
K_b^\prime &=&  -\sin \thf (g_2 \mtau/(2 \mw \cos \beta ) N_{j3} - g_1
       N_{j1} )/ \sqrt{2}
        \nl      -\cos \thf (-g_2 \mtau/(2 \mw \cos \beta ) N_{j3} 
	 -( g_1 N_{j1}  + g_2/2 
                (N_{j2}-N_{j1} \tan \thw )))/ \sqrt{2}    \nonumber \\
{\cal T}_{\rm III}\!\!\times\!\!{\cal T}_{\rm V} &=& ( 2 f_{3d} K_b K_a ( -
        16 \mst^2 \mtau^2 + 4 \mst^2 s + 4 \mtau^2 s 
	- s^2 	+ t^2	- 2 t u + u^2	)
	\nl + f_{3c} ( K_a^2 ( - 4 \mst^2 s + s^2 - 4 \mtau^2 t 
       - 4 \mtau \nevalsi t   - t^2 	+ 4 \mtau^2 u 
	+ 4 \mtau \nevalsi u 	+ 2 t u	- u^2 )	
        \nl      + K_b^2 (	- 4 \mst^2 s + s^2 - 4 \mtau^2 t 
	+ 4 \mtau \nevalsi t 	- t^2 	+ 4 \mtau^2 u 	- 4 \mtau \nevalsi u 
	+ 2 t u	- u^2   )	
	) ) \nl /  ((\mz^2 - s) (\nevalsi^2 - t))   \nonumber \\ 
{\cal T}_{\rm IV}\!\!\times\!\!{\cal T}_{\rm V} &=& -( f_{4c} ( K_a^2 ( 
       -4 \mst^2  s    + s^2  - 4 \mtau^2 t   - 4 \mtau \nevalsi t 
	       - t^2        + 4 \mtau^2 u 
	       + 4 \mtau \nevalsi u       + 2 t u \nl  - u^2 )
    + K_b^2 (  - 4 \mst^2 s + s^2  - 4 \mtau^2 t  + 4 \mtau \nevalsi t
	   - t^2  + 4 \mtau^2 u  - 4 \mtau \nevalsi u  + 2 t u
	  \nl - u^2  ) )	   
	   /(s (\nevalsi^2 - t)))   \nonumber \\
{\cal T}_{\rm V}\!\!\times\!\!{\cal T}_{\rm V} &=&  (2 ( 
 K_a^{\prime 2} K_b^2  ( - \mst^4    - 2 \mst^2 \mtau^2 
	        + 3  \mtau^4       - \mtau^3  \nevalsi 
	    - 3 \mtau^3  \nevalsi    +  \mtau^3   \nevalsj 
	   + 3  \mtau^3    \nevalsj   \nl - 2  \mtau^2  \nevalsi \nevalsj 
	     - 2   \mtau^2   \nevalsi \nevalsj    +  \mst^2 s
         -  \mtau^2  s 	+ \mtau  \nevalsi s 	-  \mtau  \nevalsj s 
	\nl + \nevalsi \nevalsj s +  \mst^2 t
         + 2  \mtau^2  t +  \mtau^2 t - \mtau  \nevalsi t 
	    +  \mtau \nevalsj t  +  \mst^2 u -  \mtau^2 u 
	  \nl   + \mtau \nevalsi u  -  \mtau \nevalsj u  -  t u )
 +  K_a^\prime K_a K_b^\prime K_b (
         -   2  \mst^4       - 4 \mst^2 \mtau^2 
	    + 6  \mtau^4      +  2  \mst^2 s
       \nl    -  2  \mtau^2 s 	
	+ 2 \nevalsi \nevalsj s 
	+  2  \mst^2 t
           -4 \mtau^2  t 
	+  2  \mtau^2   t 
	 +  2  \mst^2 u
            -  2  \mtau^2   u 
	 -  2  t u  )
   \nl +  K_a^2 K_b^{\prime 2} (
         -   \mst^4     - 2  \mst^2  \mtau^2 
	 + 3  \mtau^4  +  4  \mtau^3    \nevalsi 
	   - 4 \mtau^3      \nevalsj 
  - 4 \mtau^2   \nevalsi \nevalsj 
	 \nl   +  \mst^2 s  -  \mtau^2   s -  \mtau  \nevalsi s 
	    + \mtau    \nevalsj s + \nevalsi \nevalsj     s +  \mst^2 t
         + 3  \mtau^2    t +  \mtau    \nevalsi t 
	  \nl  - \mtau    \nevalsj t  +  \mst^2 u  -  \mtau^2    u 
	    -  \mtau     \nevalsi u 	 + \mtau     \nevalsj u  -  t u )
\nl +(K_a K_a^\prime + K_b K_b^\prime) (K_a K_a^\prime + K_b K_b^\prime)(
       -  \mtau^2  s -  \nevalsi \nevalsj       s 
	+  \mst^2 t +  \mtau^2     t 
	 \nl    +  \mst^2 u -  \mtau^2   u  -  t u )
+(K_a K_a^\prime + K_b K_b^\prime) (K_a K_a^\prime - K_b K_b^\prime)(
      4  \mtau^3   \nevalsi  \nl
   + 4 \mtau^3  \nevalsj - \mtau \nevalsi s - \mtau \nevalsj s 
 + \mtau \nevalsi t  + \mtau \nevalsj t - \mtau \nevalsi u - \mtau \nevalsj u )
\nl +(K_a K_a^\prime - K_b K_b^\prime) (K_a K_a^\prime - K_b K_b^\prime)(
    2  \mtau^2   \nevalsi \nevalsj  + 2  \mtau^2   t )
  )) \nl /   ((\nevalsi^2 - t) (-\nevalsj^2 + t))  \nonumber \\
\tsq &=&    {\cal T}_{\rm III}\!\!\times\!\!{\cal T}_{\rm III} 
   +   {\cal T}_{\rm IV}\!\!\times\!\!{\cal T}_{\rm IV}  
   +  \sum_{i,j} {\cal T}_{\rm V}\!\!\times\!\!{\cal T}_{\rm V} 
   + 2  {\cal T}_{\rm III}\!\!\times\!\!{\cal T}_{\rm IV} 
   + 2 \sum_i ({\cal T}_{\rm III}\!\!\times\!\!{\cal T}_{\rm V} 
   +  {\cal T}_{\rm IV}\!\!\times\!\!{\cal T}_{\rm V}) 
\end{eqnarray}

\subsection*{$\slept_1 \slept^\ast_1 \longrightarrow f \bar{f} $}
When $f$ is a fermion other than a lepton or $t$ quark, again the 
$s$-channel $h$
and $H$ annihilations can be  neglected. We separate the channel $\slept_1
\slept^\ast_1 \longrightarrow\nu_\ell
\bar{\nu}_\ell$ because it has $t$-channel chargino exchange.
The couplings $f_{3c}$ and $f_{3d}$ are modified while $f_{4c}$
remains the same.
\begin{eqnarray}
f_{3c} &=& ( -g_2 / \cos \thw ( \sin^2 \thw - \cos^2 \thf /2))
   (g_2 ( -2 T_3^f + 4 Q_f \sin^2 \thw )/(4 \cos \thw))
  \nonumber \\
f_{3d} &=&  ( -g_2/ \cos \thw ( \sin^2 \thw - \cos^2 \thf /2))
  (g_2 (2 T_3^f )/(4 \cos \thw))
\end{eqnarray}

\subsection*{$\slept_1 \slept^\ast_1 \longrightarrow \nu_\ell \bar{\nu}_\ell $}
With $\slept_R$ we have $t$-channel chargino exchange. \hfill \\
 III. $s$-channel $Z$ annihilation \hfill \\
 IV. $t$-channel $\schi^-_{(1,2)}$ exchange
\begin{eqnarray}
f_{3} &=& ( - g_2 / \cos \thw (\sin^2 \thw - \cos^2 \thf /2 ))
(- g_2/(2 \cos \thw ))
  \nonumber \\
f_{4}(i) &=&  (g_2 m_\ell / (\sqrt{2} \mw \cos \beta ) U_{i2} \sin \thf
             - g_2 U_{i1} \cos \thf )^2  \nonumber \\
{\cal T}_{\rm III}\!\!\times\!\!{\cal T}_{\rm III} &=& (- 4 \mst^2 s + s^2 + 2
         t u - t^2 - u^2) /(s-\mz^2)^2 \nonumber \\
{\cal T}_{\rm IV}\!\!\times\!\!{\cal T}_{\rm IV} &=&  (- \mst^2 s
          - \mst^2 t - \mst^2 u + \mst^4    + t u)
	  /((t- \mxi^2) (t-\mxj^2)) \nonumber \\
{\cal T}_{\rm III}\!\!\times\!\!{\cal T}_{\rm IV} &=& - (- 2 \mst^2 s
               + 1/2 s^2 + t u 
                - 1/2 t^2 - 1/2 u^2)/((s-\mz^2) 
                (t- \mxi^2)) \nonumber \\
\tsq &=&  f_3^2 {\cal T}_{\rm III}\!\!\times\!\!{\cal T}_{\rm III} 
   +  \sum_{i,j} f_4(i) f_4(j) {\cal T}_{\rm IV}\!\!\times\!\!{\cal T}_{\rm IV}
      + 2 f_3 \sum_i f_4(i) {\cal T}_{\rm III}\!\!\times\!\!{\cal T}_{\rm IV} 
\end{eqnarray}

\subsection*{$\slept_1 \slept^\ast_1 \longrightarrow t \bar{t} $}
The couplings $f_{1a},f_{2a},f_{3c}$ and $f_{3d}$ are modified while $f_{4c}$
remains the same.
\begin{eqnarray}
f_{1a} &=& ( -g_2 m_t \sin \alpha/(2 \mw \sin \beta )) ( g_2 \mz ((1/2 - \sin^2
       \thw  ) \cos (\beta+ \alpha) \cos^2 \thf 
       \nl   + \sin^2 \thw \cos (\beta + \alpha ) \sin^2 \thf )/\cos \thw  
           - g_2 \mtau^2 \cos \alpha /( \mw \cos \beta ) 
     \nl     - g_2 \mtau (- \Atau \cos \alpha -\mu \sin \alpha ) \sin \thf \cos
	  \thf /(\mw \cos \beta )        )
	  \nonumber \\
f_{2a} &=&  ( -g_2 m_t \cos \alpha /(2 \mw \sin \beta )) ( g_2 \mz ( (-1/2 +
     \sin^2 \thw )  \sin (\beta + \alpha) \cos^2 \thf 
       \nl   - \sin^2 \thw  \sin (\beta + \alpha) \sin^2 \thf )/\cos \thw  
          + g_2 \mtau^2 \sin \alpha/(\mw \cos \beta ) 
    \nl   - g_2 \mtau (\Atau \sin \alpha -\mu \cos \alpha ) \sin \thf \cos \thf 
	  /(\mw \cos \beta )        ) \nonumber	  \\
f_{3c} &=&  ( - g_2 (\sin^2 \thw - \cos^2 \thf/2) /\cos \thw ) (g_2
   (- 1 + 8/3 \sin^2 \thw )/(4 \cos \thw))  \nonumber \\
f_{3d} &=&   ( - g_2 (\sin^2 \thw - \cos^2 \thf /2)/\cos \thw ) (g_2
  / (4 \cos \thw))
\end{eqnarray}

\subsection*{$\slept_1 \slept^\ast_1 \longrightarrow hh $}
The couplings  are modified. Channels IV ($t$-channel $\slept$ exchange) and V
($u$-channel $\slept$ exchange) are now summed over $\slept_1$ and $\slept_2$,
with appropriate propagators.
\begin{eqnarray}
f_1 &=&  (-3 g_2 \mz \cos (2 \alpha) \sin (\alpha + \beta) /( 2 \cos \thw ) ) 
           (g_2 \mz (
          (1/2- \sin^2 \thw )  \cos (\beta + \alpha) \nl \cos^2 \thf
          + \sin^2 \thw \cos (\beta + \alpha) \sin^2 \thf )/\cos \thw  
          - g_2 \mtau^2 \cos \alpha /(\mw \cos \beta )   
          \nl - g_2 \mtau (-\Atau \cos \alpha - \mu \sin \alpha ) \sin \thf \cos
	  \thf /(\mw \cos \beta ) 
           )
	  \nonumber \\
f_2 &=&  (g_2 \mz (\cos (2 \alpha) \cos (\alpha + \beta ) -
          2 \sin (2 \alpha) \sin (\alpha + \beta ))/(2 \cos \thw ) )
	\nl   ( g_2 \mz (
          (1/2- \sin^2 \thw ) \cos (\beta + \alpha ) \cos^2 \thf
          + \sin^2 \thw \cos (\beta + \alpha ) \sin^2 \thf )/\cos \thw  
         \nl - g_2 \mtau^2 \cos \alpha/( \mw \cos \beta ) 
          - g_2 \mtau (-\Atau \cos \alpha - \mu \sin \alpha ) \sin \thf \cos
	  \thf /(\mw \cos \beta ) 
           )
	  \nonumber \\
f_3 &=&  (g_2^2/2) ((-1/2 + \sin^2 \thw ) \cos (2 \alpha)/\cos^2 \thw 
          - \mtau^2 \sin^2 \alpha  /(\mw^2 \cos^2 \beta )) \cos^2 \thf
      \nl    + (g_2^2/2) (- \sin^2 \thw  \cos (2 \alpha)/\cos^2 \thw
          - \mtau^2 \sin^2 \alpha /(\mw^2 \cos^2 \beta )) \sin^2 \thf
	  \nonumber \\
f_4(1) &=&  (g_2 \mz (
          (1/2 - \sin^2 \thw ) \cos (\beta + \alpha ) \cos^2 \thf 
          +\sin^2 \thw  \cos (\beta + \alpha ) \sin^2 \thf )/\cos \thw  
       \nl   - g_2 \mtau^2 \cos \alpha/(\mw \cos \beta ) 
          - g_2 \mtau (-\Atau \cos \alpha -\mu \sin \alpha ) \sin \thf \cos
	  \thf /(\mw \cos \beta ) 
           )^2
	  \nonumber \\
f_4(2) &=& ( g_2 \mz \cos \thf \sin \thf (
          (1/2 - \sin^2 \thw ) \sin (\beta + \alpha) 
          - \sin^2 \thw \sin (\beta + \alpha)  ) /\cos \thw  
      \nl    - g_2 \mtau (\Atau \sin \alpha - \mu \cos \alpha ) \cos (2 \thf)
            /(2 \mw \cos \beta ) 
           )^2 \nonumber \\
f_5(1) &=& (g_2 \mz (
          (1/2 - \sin^2 \thw ) \cos (\beta + \alpha ) \cos^2 \thf 
          +\sin^2 \thw  \cos (\beta + \alpha ) \sin^2 \thf )/\cos \thw  
       \nl   - g_2 \mtau^2 \cos \alpha/(\mw \cos \beta ) 
          - g_2 \mtau (-\Atau \cos \alpha -\mu \sin \alpha ) \sin \thf \cos
	  \thf /(\mw \cos \beta ) 
           )^2
	  \nonumber \\
f_5(2) &=& ( g_2 \mz \cos \thf \sin \thf (
          (1/2 - \sin^2 \thw ) \sin (\beta + \alpha) 
          - \sin^2 \thw \sin (\beta + \alpha)  ) /\cos \thw  
      \nl    - g_2 \mtau (\Atau \sin \alpha - \mu \cos \alpha ) \cos (2 \thf)
            /(2 \mw \cos \beta ) 
           )^2 \nonumber \\
\tsq &=&  f_1^2 {\cal T}_{\rm I}\!\!\times\!\!{\cal T}_{\rm I} 
   +  f_2^2 {\cal T}_{\rm II}\!\!\times\!\!{\cal T}_{\rm II} 
   +  f_3^2 {\cal T}_{\rm III}\!\!\times\!\!{\cal T}_{\rm III} 
   +  \sum_{i,j} (f_4(i) f_4(j) {\cal T}_{\rm IV}\!\!\times\!\!{\cal T}_{\rm IV}
   +  f_5(i) f_5(j) {\cal T}_{\rm V}\!\!\times\!\!{\cal T}_{\rm V} )   
  \nl  + 2 f_1 f_2 {\cal T}_{\rm I}\!\!\times\!\!{\cal T}_{\rm II} 
   + 2 f_1 f_3  {\cal T}_{\rm I}\!\!\times\!\!{\cal T}_{\rm III} 
   + 2 \sum_i (f_1 f_4(i) {\cal T}_{\rm I}\!\!\times\!\!{\cal T}_{\rm IV}
   +  f_1 f_5(i) {\cal T}_{\rm I}\!\!\times\!\!{\cal T}_{\rm V} )
 \nl   + 2 f_2 f_3 {\cal T}_{\rm II}\!\!\times\!\!{\cal T}_{\rm III} 
     + 2 \sum_i ( f_2 f_4(i) {\cal T}_{\rm II}\!\!\times\!\!{\cal T}_{\rm IV}
   +  f_2 f_5(i) {\cal T}_{\rm II}\!\!\times\!\!{\cal T}_{\rm V} )  
 \nl  + 2 \sum_i (f_3 f_4(i) {\cal T}_{\rm III}\!\!\times\!\!{\cal T}_{\rm IV}
+  f_3 f_5(i) {\cal T}_{\rm III}\!\!\times\!\!{\cal T}_{\rm V} )
+  2 \sum_{i,j} f_4(i) f_5(j) {\cal T}_{\rm IV}\!\!\times\!\!{\cal T}_{\rm V}
\end{eqnarray}

\subsection*{$\slept_1 \slept^\ast_1 \longrightarrow h A [HA] $}
Besides the $s$-channel $Z$ annihilation, we have $t$- and $u$-channel $\slept_2$
exchanges. \hfill \\
 II. $t$-channel $\slept_2$ exchange \hfill \\
 III. $u$-channel $\slept_2$ exchange \hfill \\
The coupling $f_1$ is also modified.
\begin{eqnarray}
f_1 &=& (-g_2 \cos [\sin] (\alpha-\beta)/(2 \cos \thw ))(-g_2 (\sin^2 \thw 
          - \cos^2 \thf /2 )/\cos \thw )\nonumber \\
f_2 &=& (- g_2 \mtau (\Atau \tan \beta - \mu)/(2 \mw ) )
          ( g_2 \mz (
          (1/2 - \sin^2 \thw ) \sin [-\cos] (\beta+\alpha)  
 \nl - \sin^2 \thw  \sin [-\cos] (\beta + \alpha ))  \cos \thf \sin \thf  /\cos
	  \thw  
 \nl    - g_2 \mtau (\Atau \sin [-\cos] \alpha -\mu \cos [\sin] \alpha ) 
         \cos (2 \thf)
	   /(2 \mw \cos \beta ) 
          ) \nonumber \\
f_3 &=& (- g_2 \mtau (\Atau \tan \beta - \mu)/(2 \mw ) )
          ( g_2 \mz (
          (1/2 - \sin^2 \thw ) \sin [-\cos] (\beta+\alpha)  
\nl  - \sin^2 \thw  \sin [-\cos] (\beta + \alpha ))  \cos \thf \sin \thf  /\cos
	  \thw  
 \nl    - g_2 \mtau (\Atau \sin [-\cos] \alpha -\mu \cos [\sin] \alpha ) 
         \cos (2 \thf)
	   /(2 \mw \cos \beta ) 
          ) \nonumber \\
{\cal T}_{\rm II}\!\!\times\!\!{\cal T}_{\rm II} &=& 1/(\mstwo^2 - t)^2 
           \nonumber \\
{\cal T}_{\rm III}\!\!\times\!\!{\cal T}_{\rm III} &=& 1/(\mstwo^2 - u)^2 
           \nonumber \\
{\cal T}_{\rm I}\!\!\times\!\!{\cal T}_{\rm II} &=& (t - u)/((\mz^2 - s)
           (\mstwo^2 - t)) \nonumber \\
{\cal T}_{\rm I}\!\!\times\!\!{\cal T}_{\rm III} &=& (t - u)/((\mz^2 - s) 
           (\mstwo^2 - u)) \nonumber \\
{\cal T}_{\rm II}\!\!\times\!\!{\cal T}_{\rm III} &=& 1/((\mstwo^2 - t) 
           (\mstwo^2 - u)) \nonumber \\
\tsq &=&   f_1^2 {\cal T}_{\rm I}\!\!\times\!\!{\cal T}_{\rm I}
+ f_2^2 {\cal T}_{\rm II}\!\!\times\!\!{\cal T}_{\rm II}
+ f_3^2 {\cal T}_{\rm III}\!\!\times\!\!{\cal T}_{\rm III} 
    + 2 f_1  f_2 {\cal T}_{\rm I}\!\!\times\!\!{\cal T}_{\rm II}
    + 2 f_1  f_3 {\cal T}_{\rm I}\!\!\times\!\!{\cal T}_{\rm III}    
 \nl    + 2 f_2  f_3 {\cal T}_{\rm II}\!\!\times\!\!{\cal T}_{\rm III}
\end{eqnarray}

\subsection*{$\slept_1 \slept^\ast_1 \longrightarrow W^+ H^- $}
Besides the $s$-channel $H$ and $h$ annihilation, we have $t$-channel sneutrino
exchange. \hfill \\
 III. $t$-channel $\widetilde{\nu}_\ell$ exchange. \hfill \\
The couplings $f_1$ and $f_2$ are modified. 
\begin{eqnarray}
f_1 &=&  ( g_2 \mz (
          (1/2- \sin^2 \thw ) \cos (\beta+\alpha) \cos^2 \thf
          + \sin^2 \thw \cos (\beta+\alpha) \sin^2 \thf ) /\cos \thw 
      \nl    - g_2 \mtau^2 \cos \alpha /(\mw \cos \beta ) 
        - g_2 \mtau (-\Atau \cos \alpha -\mu \sin \alpha ) \sin \thf \cos \thf 
	  /(\mw \cos \beta ) 
          ) \nl (g_2 \sin (\alpha-\beta) /2 )
	  \nonumber \\
f_2 &=&  ( g_2 \mz (
          (-1/2+\sin^2 \thw ) \sin (\beta+\alpha) \cos^2 \thf
          - \sin^2 \thw \sin (\beta+\alpha) \sin^2 \thf )/\cos \thw  
       \nl   + g_2 \mtau^2 \sin \alpha/ (\mw \cos \beta ) 
          - g_2 \mtau (\Atau \sin \alpha -\mu \cos \alpha) \sin \thf \cos \thf
	  /(\mw \cos \beta) 
          ) \nl (g_2 \cos (\alpha-\beta)/2 ) \nonumber \\
f_3 &=& (-g_2/\sqrt{2} \cos \thf )(-g_2 \mw \cos \thf (-\mtau^2 \tan \beta
           /\mw^2 + 
           \sin (2 \beta))/ \sqrt{2}
          \nl  + g_2 \mtau/( \sqrt{2} \mw) \sin \thf 
             (\mu - \Atau \tan \beta )) \nonumber \\
{\cal T}_{\rm III}\!\!\times\!\!{\cal T}_{\rm III} &=& (\mst^4 + (\mw^2 - t)^2
         - 2 \mst^2 (\mw^2 + t))/
          (\mw^2 (\msn^2 - t)^2) \nonumber \\
{\cal T}_{\rm I}\!\!\times\!\!{\cal T}_{\rm III} &=&  (-\mw^4 + \mw^2 s +
          \mst^2 (-3 \mw^2 + s) + 
          \mw^2 t - s t + \mHp^2 (-\mst^2 - 3 \mw^2 + t) \nl + 
          2 \mw^2 u)/(\mw^2 (m_H^2 - s) (\msn^2 - t)) 
	  \nonumber \\
{\cal T}_{\rm II}\!\!\times\!\!{\cal T}_{\rm III} &=&  	  
	  (-\mw^4 + \mw^2 s + \mst^2
          (-3 \mw^2 + s) + 
          \mw^2 t - s t + \mHp^2 (-\mst^2 - 3 \mw^2 + t)  \nl + 
          2 \mw^2 u)/(\mw^2 (m_h^2 - s) (\msn^2 - t)) \nonumber \\
\tsq &=&  f_1^2 {\cal T}_{\rm I}\!\!\times\!\!{\cal T}_{\rm I}
+ f_2^2 {\cal T}_{\rm II}\!\!\times\!\!{\cal T}_{\rm II}
+ f_3^2 {\cal T}_{\rm III}\!\!\times\!\!{\cal T}_{\rm III} 
    + 2 f_1  f_2 {\cal T}_{\rm I}\!\!\times\!\!{\cal T}_{\rm II}
    + 2 f_1  f_3 {\cal T}_{\rm I}\!\!\times\!\!{\cal T}_{\rm III}    
 \nl    + 2 f_2  f_3 {\cal T}_{\rm II}\!\!\times\!\!{\cal T}_{\rm III}
\end{eqnarray} 

\subsection*{$\slept_1 \slept^\ast_1 \longrightarrow AA $}
The new channels are \hfill \\
 IV.$t$-channel $\slept_2$ exchange \hfill \\
 V. $u$-channel $\slept_2$ exchange \hfill \\
The couplings are modified. 
\begin{eqnarray}
f_1 &=& ( g_2 \mz (
          (1/2- \sin^2 \thw ) \cos (\beta +\alpha ) \cos^2 \thf
          + \sin^2 \thw \cos (\beta +\alpha ) \sin^2 \thf ) /\cos \thw 
       \nl   - g_2 \mtau^2 \cos \alpha /(\mw \cos \beta ) 
          - g_2 \mtau (-\Atau \cos \alpha -\mu \sin \alpha ) \sin \thf \cos \thf
	  /(\mw \cos \beta ) 
          )
	\nl (g_2 \mz \cos (2 \beta ) \cos (\alpha +\beta ) /(2 \cos \thw) ) 
	  \nonumber \\
f_2 &=&  ( g_2 \mz (
          (-1/2+ \sin^2 \thw ) \sin (\beta +\alpha ) \cos^2 \thf
          - \sin^2 \thw \sin (\beta +\alpha ) \sin^2 \thf ) /\cos \thw 
       \nl   + g_2 \mtau^2 \sin \alpha /(\mw \cos \beta ) 
          - g_2 \mtau (\Atau \sin \alpha -\mu \cos \alpha ) \sin \thf \cos \thf 
	  /(\mw \cos \beta ) 
          )
   \nl (-g_2 \mz \cos (2 \beta ) \sin (\alpha +\beta ) /(2 \cos \thw) ) 
	  \nonumber \\
f_3 &=& g_2^2/2 ((-1/2+\sin^2 \thw) \cos (2 \beta )/\cos^2 \thw
          - \mtau^2 \tan^2 \beta /\mw^2) \cos^2 \thf
      \nl    +g_2^2/2 (-\sin^2 \thw  \cos (2 \beta )/\cos^2 \thw
          - \mtau^2 \tan^2 \beta /\mw^2) \sin^2 \thf 
	  \nonumber \\
f_4 &=&  (- g_2 \mtau (\Atau \tan \beta -\mu) /(2 \mw ) )^2 \nonumber \\
f_5 &=& (- g_2 \mtau (\Atau \tan \beta -\mu) /(2 \mw ) )^2 \nonumber \\
{\cal T}_{\rm IV}\!\!\times\!\!{\cal T}_{\rm IV} &=&  1/(\mstwo^2 - t)^2 
          \nonumber \\
{\cal T}_{\rm V}\!\!\times\!\!{\cal T}_{\rm V} &=& 1/(\mstwo^2 - u)^2 
          \nonumber \\
{\cal T}_{\rm I}\!\!\times\!\!{\cal T}_{\rm IV} &=& 1/((m_H^2 - s)(\mstwo^2 -
          t)) \nonumber \\
{\cal T}_{\rm I}\!\!\times\!\!{\cal T}_{\rm V} &=& 1/((m_H^2 - s)(\mstwo^2 - u))
         \nonumber \\
{\cal T}_{\rm II}\!\!\times\!\!{\cal T}_{\rm IV} &=& 1/((m_h^2 - s)(\mstwo^2 -
          t)) \nonumber \\
{\cal T}_{\rm II}\!\!\times\!\!{\cal T}_{\rm V} &=& 1/((m_h^2 - s)(\mstwo^2 -
           u)) \nonumber \\
{\cal T}_{\rm III}\!\!\times\!\!{\cal T}_{\rm IV} &=& 1/(\mstwo^2 - t) 
           \nonumber \\
{\cal T}_{\rm III}\!\!\times\!\!{\cal T}_{\rm V} &=& 1/(\mstwo^2 - u) 
           \nonumber \\
\tsq &=&  f_1^2 {\cal T}_{\rm I}\!\!\times\!\!{\cal T}_{\rm I} 
   +  f_2^2 {\cal T}_{\rm II}\!\!\times\!\!{\cal T}_{\rm II} 
   +  f_3^2 {\cal T}_{\rm III}\!\!\times\!\!{\cal T}_{\rm III} 
   +  f_4^2 {\cal T}_{\rm IV}\!\!\times\!\!{\cal T}_{\rm IV}  
   +  f_5^2 {\cal T}_{\rm V}\!\!\times\!\!{\cal T}_{\rm V} 
  \nl + 2 f_1 f_2 {\cal T}_{\rm I}\!\!\times\!\!{\cal T}_{\rm II} 
   + 2 f_1 f_3  {\cal T}_{\rm I}\!\!\times\!\!{\cal T}_{\rm III} 
   + 2 f_1 f_4 {\cal T}_{\rm I}\!\!\times\!\!{\cal T}_{\rm IV}  
   + 2 f_1 f_5  {\cal T}_{\rm I}\!\!\times\!\!{\cal T}_{\rm V} 
 \nl  + 2 f_2 f_3 {\cal T}_{\rm II}\!\!\times\!\!{\cal T}_{\rm III} 
    + 2 f_2 f_4 {\cal T}_{\rm II}\!\!\times\!\!{\cal T}_{\rm IV} 
   + 2 f_2 f_5 {\cal T}_{\rm II}\!\!\times\!\!{\cal T}_{\rm V}  
   + 2 f_3 f_4 {\cal T}_{\rm III}\!\!\times\!\!{\cal T}_{\rm IV} 
 \nl  + 2 f_3 f_5 {\cal T}_{\rm III}\!\!\times\!\!{\cal T}_{\rm V} 
   + 2 f_4 f_5 {\cal T}_{\rm IV}\!\!\times\!\!{\cal T}_{\rm V}
\end{eqnarray}

\subsection*{$\slept_1 \slept^\ast_1 \longrightarrow h H $}
The $t$- and $u$-channels $\slept$ exchanges are now summed over $\slept_1$
and
$\slept_2$ with appropriate propagators. The couplings are 
\begin{eqnarray}
f_1 &=& ( g_2 \mz (
          (1/2- \sin^2 \thw ) \cos (\beta +\alpha ) \cos^2 \thf
          + \sin^2 \thw \cos (\beta +\alpha ) \sin^2 \thf ) /\cos \thw 
      \nl    - g_2 \mtau^2 \cos \alpha /(\mw \cos \beta ) 
          - g_2 \mtau (-\Atau \cos \alpha -\mu \sin \alpha ) \sin \thf \cos \thf 
	  /(\mw \cos \beta ) 
          )
	  \nl (g_2 \mz (\cos (2 \alpha ) \sin (\alpha  + \beta ) +
          2 \sin (2 \alpha ) \cos (\alpha  + \beta )) /(2 \cos \thw )) 
	  \nonumber \\
f_2 &=& ( g_2 \mz (
          (-1/2+ \sin^2 \thw ) \sin (\beta +\alpha ) \cos^2 \thf
          - \sin^2 \thw \sin (\beta +\alpha ) \sin^2 \thf ) /\cos \thw 
         \nl  + g_2 \mtau^2 \sin \alpha /(\mw \cos \beta )  
          - g_2 \mtau (\Atau \sin \alpha -\mu \cos \alpha ) \sin \thf \cos \thf 
	  /(\mw \cos \beta ) 
          ) \nl (g_2 \mz (\cos (2 \alpha ) \cos (\alpha  + \beta ) -
          2 \sin (2 \alpha ) \sin (\alpha  + \beta )) /(2 \cos \thw ) ) 
	  \nonumber \\
f_3 &=& g_2^2 \sin (2 \alpha ) (
            (-1/2 + \sin^2 \thw )/\cos^2 \thw
          + \mtau^2/(2 \mw^2 \cos^2 \beta )) \cos^2 \thf /2 
       \nl   + g_2^2 \sin (2 \alpha ) ( -\sin^2 \thw/\cos^2 \thw
          + \mtau^2 /(2 \mw^2 \cos^2 \beta ) ) \sin^2 \thf /2  \nonumber \\
f_4(1) &=& ( g_2 \mz (
          (1/2- \sin^2 \thw ) \cos (\beta +\alpha ) \cos^2 \thf
          + \sin^2 \thw \cos (\beta +\alpha ) \sin^2 \thf ) /\cos \thw 
       \nl   - g_2 \mtau^2 \cos \alpha /(\mw \cos \beta ) 
       - g_2 \mtau (-\Atau \cos \alpha - \mu \sin \alpha ) \sin \thf \cos \thf 
	  /(\mw \cos \beta ) 
          ) \nl ( g_2 \mz (
          (-1/2+ \sin^2 \thw ) \sin (\beta +\alpha ) \cos^2 \thf
          - \sin^2 \thw \sin (\beta +\alpha ) \sin^2 \thf ) /\cos \thw 
         \nl + g_2 \mtau^2 \sin \alpha /(\mw \cos \beta ) 
          - g_2 \mtau (\Atau \sin \alpha -\mu \cos \alpha ) \sin \thf \cos \thf 
	  /(\mw \cos \beta ) 
          ) \nonumber \\
f_4(2) &=& (g_2 \mz (
          (-1/2 + \sin^2 \thw ) \cos (\beta +\alpha ) 
          + \sin^2 \thw \cos (\beta +\alpha ) ) \cos \thf \sin \thf  /\cos \thw 
        \nl  - g_2 \mtau (-\Atau \cos \alpha -\mu \sin \alpha ) \cos (2 \thf) 
	  /(2 \mw \cos \beta ) 
          ) \nl ( g_2 \mz (
          (1/2 - \sin^2 \thw ) \sin (\beta +\alpha ) 
          - \sin^2 \thw \sin (\beta +\alpha ) ) \cos \thf \sin \thf /\cos \thw 
     \nl     - g_2 \mtau (\Atau \sin \alpha -\mu \cos \alpha ) \cos (2 \thf) 
	  /(2 \mw \cos \beta ) 
          ) \nonumber \\
f_5(1) &=& ( g_2 \mz (
          (1/2- \sin^2 \thw ) \cos (\beta +\alpha ) \cos^2 \thf
          + \sin^2 \thw \cos (\beta +\alpha ) \sin^2 \thf ) /\cos \thw 
       \nl   - g_2 \mtau^2 \cos \alpha /(\mw \cos \beta ) 
       - g_2 \mtau (-\Atau \cos \alpha - \mu \sin \alpha ) \sin \thf \cos \thf 
	  /(\mw \cos \beta ) 
          ) \nl ( g_2 \mz (
          (-1/2+ \sin^2 \thw ) \sin (\beta +\alpha ) \cos^2 \thf
          - \sin^2 \thw \sin (\beta +\alpha ) \sin^2 \thf ) /\cos \thw 
         \nl + g_2 \mtau^2 \sin \alpha /(\mw \cos \beta ) 
          - g_2 \mtau (\Atau \sin \alpha -\mu \cos \alpha ) \sin \thf \cos \thf 
	  /(\mw \cos \beta ) 
          ) \nonumber \\
f_5(2) &=& (g_2 \mz (
          (-1/2 + \sin^2 \thw ) \cos (\beta +\alpha ) 
          + \sin^2 \thw \cos (\beta +\alpha ) ) \cos \thf \sin \thf  /\cos \thw 
        \nl  - g_2 \mtau (-\Atau \cos \alpha -\mu \sin \alpha ) \cos (2 \thf) 
	  /(2 \mw \cos \beta ) 
          ) \nl ( g_2 \mz (
          (1/2 - \sin^2 \thw ) \sin (\beta +\alpha ) 
          - \sin^2 \thw \sin (\beta +\alpha ) ) \cos \thf \sin \thf /\cos \thw 
     \nl     - g_2 \mtau (\Atau \sin \alpha -\mu \cos \alpha ) \cos (2 \thf) 
	  /(2 \mw \cos \beta ) 
          ) \nonumber \\
\tsq &=&  f_1^2 {\cal T}_{\rm I}\!\!\times\!\!{\cal T}_{\rm I} 
   +  f_2^2 {\cal T}_{\rm II}\!\!\times\!\!{\cal T}_{\rm II} 
   +  f_3^2 {\cal T}_{\rm III}\!\!\times\!\!{\cal T}_{\rm III} 
+  \sum_{i,j} ( f_4(i) f_4(j) {\cal T}_{\rm IV}\!\!\times\!\!{\cal T}_{\rm IV}  
   +  f_5(i) f_5(j) {\cal T}_{\rm V}\!\!\times\!\!{\cal T}_{\rm V} )
  \nl + 2 f_1 f_2 {\cal T}_{\rm I}\!\!\times\!\!{\cal T}_{\rm II} 
   + 2 f_1 f_3  {\cal T}_{\rm I}\!\!\times\!\!{\cal T}_{\rm III} 
   + 2 \sum_i f_1 (f_4(i) {\cal T}_{\rm I}\!\!\times\!\!{\cal T}_{\rm IV}  
   +  f_5(i)  {\cal T}_{\rm I}\!\!\times\!\!{\cal T}_{\rm V} )
 \nl  + 2 f_2 f_3 {\cal T}_{\rm II}\!\!\times\!\!{\cal T}_{\rm III} 
    + 2 \sum_i f_2 (f_4(i) {\cal T}_{\rm II}\!\!\times\!\!{\cal T}_{\rm IV} 
   +  f_5(i) {\cal T}_{\rm II}\!\!\times\!\!{\cal T}_{\rm V} )  
 \nl  + 2 f_3 \sum_i (f_4(i) {\cal T}_{\rm III}\!\!\times\!\!{\cal T}_{\rm IV} 
   +  f_5(i) {\cal T}_{\rm III}\!\!\times\!\!{\cal T}_{\rm V} ) 
   + 2 \sum_{i,j} f_4(i) f_5(j) {\cal T}_{\rm IV}\!\!\times\!\!{\cal T}_{\rm V}
\end{eqnarray}

\subsection*{$\slept_1 \slept^\ast_1 \longrightarrow HH $}
The $t$- and $u$-channels $\slept$ exchanges are now summed over $\slept_1$
and
$\slept_2$ with appropriate propagators. The couplings are 
\begin{eqnarray}
f_1 &=& (-3 g_2 \mz \cos (2 \alpha ) \cos (\alpha  + \beta ) /( 2 \cos \thw) ) 
          (g_2 \mz (
          (1/2- \sin^2 \thw ) \cos (\beta +\alpha ) \cos^2 \thf
       \nl   + \sin^2 \thw \cos (\beta +\alpha ) \sin^2 \thf ) /\cos \thw 
          - g_2 \mtau^2 \cos \alpha /(\mw \cos \beta ) 
   \nl   - g_2 \mtau (-\Atau \cos \alpha -\mu \sin \alpha ) \sin \thf \cos \thf 
	  /(\mw \cos \beta ) 
          ) \nonumber \\
f_2 &=& ( g_2 \mz (\cos (2 \alpha ) \sin (\alpha  + \beta ) +
          2 \sin (2 \alpha ) \cos (\alpha  + \beta )) /(2 \cos \thw) ) 
   \nl	  ( g_2 \mz (
          (-1/2+ \sin^2 \thw ) \sin (\beta +\alpha ) \cos^2 \thf
          - \sin^2 \thw \sin (\beta +\alpha ) \sin^2 \thf ) /\cos \thw 
   \nl       + g_2 \mtau^2 \sin \alpha /(\mw \cos \beta ) 
          - g_2 \mtau  (\Atau \sin \alpha -\mu \cos \alpha ) \sin \thf \cos \thf
	  /(\mw \cos \beta ) 
          ) \nonumber \\
f_3 &=& g_2^2/2 (-(-1/2+\sin^2 \thw ) \cos (2 \alpha )/\cos^2 \thw 
          - \mtau^2 \cos^2 \alpha /( \mw^2 \cos^2 \beta )) \cos^2 \thf
      \nl    + g_2^2/2 (\sin^2 \thw \cos (2 \alpha )/(\cos^2 \thw)
          - \mtau^2 (\cos^2 \alpha )/(\mw^2 \cos^2 \beta )) \sin^2 \thf
      \nonumber \\
f_4(1) &=& (g_2 \mz (
          (1/2- \sin^2 \thw ) \cos (\beta +\alpha ) \cos^2 \thf
          + \sin^2 \thw \cos (\beta +\alpha ) \sin^2 \thf ) /\cos \thw 
    \nl      - g_2 \mtau^2 \cos \alpha /(\mw \cos \beta ) 
          - g_2 \mtau (-\Atau \cos \alpha -\mu \sin \alpha ) \sin \thf \cos \thf 
	  /(\mw \cos \beta ) 
          )^2  \nonumber \\
f_4(2) &=& ( g_2 \mz (
          (1/2- \sin^2 \thw ) \cos (\beta +\alpha ) (-\cos \thf \sin \thf)
          + \sin^2 \thw \cos (\beta +\alpha ) \cos \thf \sin \thf ) 
	  \nl /\cos \thw 
          - g_2 \mtau (-\Atau \cos \alpha -\mu \sin \alpha ) \cos (2 \thf) 
	  /(2 \mw \cos \beta ) 
          )^2  \nonumber \\
f_5(1) &=&  (g_2 \mz (
          (1/2- \sin^2 \thw ) \cos (\beta +\alpha ) \cos^2 \thf
          + \sin^2 \thw \cos (\beta +\alpha ) \sin^2 \thf ) /\cos \thw 
    \nl      - g_2 \mtau^2 \cos \alpha /(\mw \cos \beta ) 
          - g_2 \mtau (-\Atau \cos \alpha -\mu \sin \alpha ) \sin \thf \cos \thf 
	  /(\mw \cos \beta ) 
          )^2 \nonumber \\
f_5(2) &=&  ( g_2 \mz (
          (1/2- \sin^2 \thw ) \cos (\beta +\alpha ) (-\cos \thf \sin \thf)
          + \sin^2 \thw \cos (\beta +\alpha ) \cos \thf \sin \thf ) 
	  \nl /\cos \thw 
          - g_2 \mtau (-\Atau \cos \alpha -\mu \sin \alpha ) \cos (2 \thf) 
	  /(2 \mw \cos \beta ) 
          )^2 \nonumber \\
\tsq &=&  f_1^2 {\cal T}_{\rm I}\!\!\times\!\!{\cal T}_{\rm I} 
   +  f_2^2 {\cal T}_{\rm II}\!\!\times\!\!{\cal T}_{\rm II} 
   +  f_3^2 {\cal T}_{\rm III}\!\!\times\!\!{\cal T}_{\rm III} 
   +  \sum_{i,j} (f_4(i) f_4(j) {\cal T}_{\rm IV}\!\!\times\!\!{\cal T}_{\rm IV}
   +  f_5(i) f_5(j) {\cal T}_{\rm V}\!\!\times\!\!{\cal T}_{\rm V} )   
  \nl  + 2 f_1 f_2 {\cal T}_{\rm I}\!\!\times\!\!{\cal T}_{\rm II} 
   + 2 f_1 f_3  {\cal T}_{\rm I}\!\!\times\!\!{\cal T}_{\rm III} 
   + 2 \sum_i (f_1 f_4(i) {\cal T}_{\rm I}\!\!\times\!\!{\cal T}_{\rm IV}
   +  f_1 f_5(i) {\cal T}_{\rm I}\!\!\times\!\!{\cal T}_{\rm V} )
 \nl   + 2 f_2 f_3 {\cal T}_{\rm II}\!\!\times\!\!{\cal T}_{\rm III} 
     + 2 \sum_i ( f_2 f_4(i) {\cal T}_{\rm II}\!\!\times\!\!{\cal T}_{\rm IV}
   +  f_2 f_5(i) {\cal T}_{\rm II}\!\!\times\!\!{\cal T}_{\rm V} )  
 \nl  + 2 \sum_i (f_3 f_4(i) {\cal T}_{\rm III}\!\!\times\!\!{\cal T}_{\rm IV}
+  f_3 f_5(i) {\cal T}_{\rm III}\!\!\times\!\!{\cal T}_{\rm V} )
+  2 \sum_{i,j} f_4(i) f_5(j) {\cal T}_{\rm IV}\!\!\times\!\!{\cal T}_{\rm V}
\end{eqnarray}

\subsection*{$\slept_1 \slept^\ast_1 \longrightarrow H^+ H^- $}
We have one new channel, \hfill \\
 VI. $t$-channel $\widetilde{\nu}_\ell$ exchange \hfill \\
The couplings, except $f_4$, are modified
\begin{eqnarray}
f_1&=& ( g_2 \mz (
          (1/2- \sin^2 \thw ) \cos (\beta +\alpha ) \cos^2 \thf 
          +\sin^2 \thw  \cos (\beta +\alpha ) \sin^2 \thf ) /\cos \thw 
   \nl       - g_2 \mtau^2 \cos \alpha /(\mw \cos \beta ) 
       - g_2 \mtau (-\Atau \cos \alpha -\mu \sin \alpha ) \sin \thf \cos \thf
	  /(\mw \cos \beta ) 
        )\nl (-g_2 (\mw \cos (\beta -\alpha )
          - \mz  \cos (2 \beta ) \cos (\beta +\alpha )/(2 \cos \thw ) )) 
     \nonumber \\
f_2 &=& ( g_2 \mz  (
          (-1/2+\sin^2 \thw ) \sin (\beta +\alpha ) \cos^2 \thf
          - \sin^2 \thw \sin (\beta +\alpha ) \sin^2 \thf ) /\cos \thw 
     \nl     + g_2 \mtau^2 \sin \alpha /(\mw \cos \beta ) 
          - g_2 \mtau (\Atau \sin \alpha -\mu \cos \alpha ) \sin \thf \cos \thf 
	  /(\mw \cos \beta ) 
         ) \nl (-g_2 (\mw \sin (\beta -\alpha )
          + \mz /(2 \cos \thw ) \cos (2 \beta ) \sin (\beta +\alpha ))) 
     \nonumber \\
f_3 &=& ( -g_2 (\sin^2 \thw - \cos^2 \thf/2) /\cos \thw ) (g_2 \cos (2 \thw)/(2
       \cos \thw )) \nonumber \\
f_5 &=& g_2^2 \cos (2 \beta ) (1 +(-1+2 \sin^2 \thw)/
          (2 \cos^2 \thw)) \cos^2 \thf /2 
      \nl    - ( g_2^2 \cos (2 \beta ) \sin^2 \thw /2  +
              g_2^2 \mtau^2 \tan^2 \beta /(2 \mw^2)  ) \sin^2 \thf
     \nonumber \\
f_6 &=& (-g_2 \mw \cos \thf (-\mtau^2 \tan \beta /\mw^2 + 
           \sin (2 \beta ))/\sqrt{2}
       \nl     + g_2 \mtau \sin \thf 
             (\mu-\Atau \tan \beta  ) /(\sqrt{2} \mw)  )^2 \nonumber \\
{\cal T}_{\rm VI}\!\!\times\!\!{\cal T}_{\rm VI} &=& 1/(\msn^2 - t)^2 
         \nonumber \\
{\cal T}_{\rm I}\!\!\times\!\!{\cal T}_{\rm VI} &=& 1/((m_H^2 - s)(\msn^2 - t)) 
         \nonumber \\
{\cal T}_{\rm II}\!\!\times\!\!{\cal T}_{\rm VI} &=& 1/((m_h^2 - t)(\msn^2 - t))
        \nonumber \\
{\cal T}_{\rm III}\!\!\times\!\!{\cal T}_{\rm VI} &=& (t - u)/((\mz^2 -
        s)(\msn^2 - t)) \nonumber \\
{\cal T}_{\rm IV}\!\!\times\!\!{\cal T}_{\rm VI} &=& (-t + u)/(s(\msn^2 - t)) 
        \nonumber \\
{\cal T}_{\rm V}\!\!\times\!\!{\cal T}_{\rm VI} &=& 1/(\msn^2 - t) \nonumber \\
\tsq &=&  f_1^2 {\cal T}_{\rm I}\!\!\times\!\!{\cal T}_{\rm I} 
   +  f_2^2 {\cal T}_{\rm II}\!\!\times\!\!{\cal T}_{\rm II} 
   +  f_3^2 {\cal T}_{\rm III}\!\!\times\!\!{\cal T}_{\rm III} 
   +  f_4^2 {\cal T}_{\rm IV}\!\!\times\!\!{\cal T}_{\rm IV}
   +  f_5^2 {\cal T}_{\rm V}\!\!\times\!\!{\cal T}_{\rm V}
   +  f_6^2 {\cal T}_{\rm VI}\!\!\times\!\!{\cal T}_{\rm VI}    
  \nl  + 2 f_1 f_2 {\cal T}_{\rm I}\!\!\times\!\!{\cal T}_{\rm II} 
   + 2 f_1 f_3  {\cal T}_{\rm I}\!\!\times\!\!{\cal T}_{\rm III} 
   + 2 f_1 f_4 {\cal T}_{\rm I}\!\!\times\!\!{\cal T}_{\rm IV}
   + 2 f_1 f_5 {\cal T}_{\rm I}\!\!\times\!\!{\cal T}_{\rm V} 
   + 2 f_1 f_6 {\cal T}_{\rm I}\!\!\times\!\!{\cal T}_{\rm VI}    
 \nl   + 2 f_2 f_3 {\cal T}_{\rm II}\!\!\times\!\!{\cal T}_{\rm III} 
     + 2  f_2 f_4 {\cal T}_{\rm II}\!\!\times\!\!{\cal T}_{\rm IV}
   + 2 f_2 f_5 {\cal T}_{\rm II}\!\!\times\!\!{\cal T}_{\rm V} 
   + 2 f_2 f_6 {\cal T}_{\rm II}\!\!\times\!\!{\cal T}_{\rm VI}     
 \nl  + 2 f_3 f_4 {\cal T}_{\rm III}\!\!\times\!\!{\cal T}_{\rm IV}
+ 2 f_3 f_5 {\cal T}_{\rm III}\!\!\times\!\!{\cal T}_{\rm V} 
+ 2 f_3 f_6 {\cal T}_{\rm III}\!\!\times\!\!{\cal T}_{\rm VI} 
+  2  f_4 f_5 {\cal T}_{\rm IV}\!\!\times\!\!{\cal T}_{\rm V}
\nl +  2  f_4 f_6 {\cal T}_{\rm IV}\!\!\times\!\!{\cal T}_{\rm VI}
+  2  f_5 f_6 {\cal T}_{\rm V}\!\!\times\!\!{\cal T}_{\rm VI}
\end{eqnarray}

\subsection*{$\slept_1 \slept_1 \longrightarrow \ell \ell $}
The channels 
 I. $t$-channel $\chi$ exchange
 II. $u$-channel $\chi$ exchange
are modified as follows.
\begin{eqnarray}
K_a &=& -\sin \thf  (g_2 \mtau/(2 \mw \cos \beta ) N_{i3} 
      + g_1 N_{i1} )/\sqrt{2}
  \nl   -\cos \thf  (g_2 \mtau/(2 \mw \cos \beta ) N_{i3} 
      +( - g_1 N_{i1} -g_2/2 
                (N_{i2}-N_{i1} \tan \thw )))/\sqrt{2}
     \nonumber \\
K_b &=& -\sin \thf  (g_2 \mtau/(2 \mw \cos \beta ) N_{i3} 
      - g_1 N_{i1} )/\sqrt{2}
 \nl   -\cos \thf  (-g_2 \mtau/(2 \mw \cos \beta ) N_{i3} 
      +( - g_1 N_{i1}-g_2/2 
                (N_{i2}-N_{i1} \tan \thw )))/\sqrt{2}
     \nonumber \\
K_a^\prime &=& -\sin \thf  (g_2 \mtau/(2 \mw \cos \beta ) N_{j3} 
      + g_1 N_{j1} )/\sqrt{2}
 \nl  -\cos \thf  (g_2 \mtau/(2 \mw \cos \beta ) N_{j3} 
      +( - g_1 N_{j1} - g_2/2 
                (N_{j2}-N_{j1} \tan \thw )))/\sqrt{2}
     \nonumber \\
K_b^\prime &=& -\sin \thf  (g_2 \mtau/(2 \mw \cos \beta ) N_{j3} 
      - g_1 N_{j1} )/\sqrt{2}
 \nl   -\cos \thf  (-g_2 \mtau/(2 \mw \cos \beta ) N_{j3} 
      +( - g_1 N_{j1}-g_2/2 
                (N_{j2}-N_{j1} \tan \thw )))/\sqrt{2}
     \nonumber \\
{\cal T}_{\rm I}\!\!\times\!\!{\cal T}_{\rm I} &=& (-2 (
         4 K_a K_a^\prime K_b K_b^\prime \nevalsi \nevalsj s 
 \nl   +    K_b^2 K_a^{\prime 2} (-\mst^4 - 2 \mst^2 \mtau^2 + 3 \mtau^4 - 
       4 \mtau^3 \nevalsi + 4 \mtau^3 \nevalsj - 4 \mtau^2 \nevalsi \nevalsj 
       \nl + 
              \mst^2 s - \mtau^2 s + \mtau \nevalsi s - \mtau \nevalsj s + 
              \nevalsi \nevalsj s + \mst^2 t + 3 \mtau^2 t - \mtau \nevalsi t 
	     \nl + 
              \mtau \nevalsj t + \mst^2 u - \mtau^2 u + \mtau \nevalsi u - 
              \mtau \nevalsj u - t u) 
\nl   +     K_b^2 K_b^{\prime 2} (\mst^4 + 2 \mst^2 \mtau^2 - 3 \mtau^4 + 
        4 \mtau^3 \nevalsi + 4 \mtau^3 \nevalsj - 4 \mtau^2 \nevalsi \nevalsj 
	\nl - 
              \mst^2 s + \mtau^2 s - \mtau \nevalsi s - \mtau \nevalsj s + 
              \nevalsi \nevalsj s - \mst^2 t - 3 \mtau^2 t + \mtau \nevalsi t 
	      \nl + 
              \mtau \nevalsj t - \mst^2 u + \mtau^2 u - \mtau \nevalsi u - 
              \mtau \nevalsj u + t u) 
  \nl +     K_a^2 K_b^{\prime 2} (-\mst^4 - 2 \mst^2 \mtau^2 + 3 \mtau^4 + 
        4 \mtau^3 \nevalsi - 4 \mtau^3 \nevalsj - 4 \mtau^2 \nevalsi \nevalsj 
	\nl + 
              \mst^2 s - \mtau^2 s - \mtau \nevalsi s + \mtau \nevalsj s + 
              \nevalsi \nevalsj s + \mst^2 t + 3 \mtau^2 t + \mtau \nevalsi t 
	      \nl - 
              \mtau \nevalsj t + \mst^2 u - \mtau^2 u - \mtau \nevalsi u + 
              \mtau \nevalsj u - t u) 
 \nl  +     K_a^2 K_a^{\prime 2} (\mst^4 + 2 \mst^2 \mtau^2 - 3 \mtau^4 - 
        4 \mtau^3 \nevalsi - 4 \mtau^3 \nevalsj - 4 \mtau^2 \nevalsi \nevalsj 
	\nl - 
              \mst^2 s + \mtau^2 s + \mtau \nevalsi s + \mtau \nevalsj s + 
              \nevalsi \nevalsj s - \mst^2 t - 3 \mtau^2 t - \mtau \nevalsi t 
	      \nl - 
              \mtau \nevalsj t - \mst^2 u + \mtau^2 u + \mtau \nevalsi u + 
              \mtau \nevalsj u + t u)
		 )	 )
     \nl	  /((\nevalsi^2 - t) (-\nevalsj^2 + t)) \nonumber \\
{\cal T}_{\rm II}\!\!\times\!\!{\cal T}_{\rm II} &=& (-2 (
    4 K_a K_a^\prime K_b K_b^\prime \nevalsi \nevalsj s 
  \nl  + K_a^2 K_b^{\prime 2} (-\mst^4 - 2 \mst^2 \mtau^2 + 3 \mtau^4 + 
        4 \mtau^3 \nevalsi - 4 \mtau^3 \nevalsj - 4 \mtau^2 \nevalsi \nevalsj 
	\nl + 
         \mst^2 s - \mtau^2 s - \mtau \nevalsi s + \mtau \nevalsj s + 
         \nevalsi \nevalsj s + \mst^2 t - \mtau^2 t - \mtau \nevalsi t 
	 \nl + 
         \mtau \nevalsj t + \mst^2 u + 3 \mtau^2 u + \mtau \nevalsi u - 
         \mtau \nevalsj u - t u) 
  \nl +  K_a^2 K_a^{\prime 2} (\mst^4 + 2 \mst^2 \mtau^2 - 3 \mtau^4 - 
        4 \mtau^3 \nevalsi - 4 \mtau^3 \nevalsj - 4 \mtau^2 \nevalsi \nevalsj 
	\nl - 
         \mst^2 s + \mtau^2 s + \mtau \nevalsi s + \mtau \nevalsj s + 
         \nevalsi \nevalsj s - \mst^2 t + \mtau^2 t + \mtau \nevalsi t 
	 \nl + 
         \mtau \nevalsj t - \mst^2 u - 3 \mtau^2 u - \mtau \nevalsi u - 
         \mtau \nevalsj u + t u) 
  \nl +  K_b^2 K_a^{\prime 2} (-\mst^4 - 2 \mst^2 \mtau^2 + 3 \mtau^4 - 
         4 \mtau^3 \nevalsi + 4 \mtau^3 \nevalsj - 4 \mtau^2 \nevalsi \nevalsj 
	 \nl + 
          \mst^2 s - \mtau^2 s + \mtau \nevalsi s - \mtau \nevalsj s + 
          \nevalsi \nevalsj s + \mst^2 t - \mtau^2 t + \mtau \nevalsi t 
	  \nl - 
          \mtau \nevalsj t + \mst^2 u + 3 \mtau^2 u - \mtau \nevalsi u + 
          \mtau \nevalsj u - t u) 
  \nl +   K_b^2 K_b^{\prime 2} (\mst^4 + 2 \mst^2 \mtau^2 - 3 \mtau^4 + 
         4 \mtau^3 \nevalsi + 4 \mtau^3 \nevalsj - 4 \mtau^2 \nevalsi \nevalsj 
	 \nl - 
          \mst^2 s + \mtau^2 s - \mtau \nevalsi s - \mtau \nevalsj s + 
          \nevalsi \nevalsj s - \mst^2 t + \mtau^2 t - \mtau \nevalsi t 
	  \nl - 
          \mtau \nevalsj t - \mst^2 u - 3 \mtau^2 u + \mtau \nevalsi u + 
          \mtau \nevalsj u + t u)
    )) \nl /((\nevalsi^2 - u) (-\nevalsj^2 + u)) \nonumber \\
{\cal T}_{\rm I}\!\!\times\!\!{\cal T}_{\rm II} &=& (
   8 K_a K_a^\prime K_b K_b^\prime \nevalsi \nevalsj s + 
 \nl    K_a^2 (-K_b^{\prime 2} (2 \mst^4 + 4 \mst^2 \mtau^2 - 6 \mtau^4 - 
        8 \mtau^3 \nevalsi + 8 \mtau^3 \nevalsj + 8 \mtau^2 \nevalsi \nevalsj 
	\nl + 
          2 \mst^2 s + 2 \mtau^2 s + 2 \mtau \nevalsi s - 
          2 \mtau \nevalsj s - 2 \nevalsi \nevalsj s - s^2 - 2 \mst^2 t - 
          2 \mtau^2 t 
	  \nl + 2 \mtau \nevalsi t + 2 \mtau \nevalsj t + t^2 - 
          2 \mst^2 u - 2 \mtau^2 u - 2 \mtau \nevalsi u - 
          2 \mtau \nevalsj u + u^2)) + 
  \nl  K_a^2 K_a^{\prime 2} (2 \mst^4 + 4 \mst^2 \mtau^2 - 6 \mtau^4 - 
        8 \mtau^3 \nevalsi - 8 \mtau^3 \nevalsj - 8 \mtau^2 \nevalsi \nevalsj 
	\nl + 
          2 \mst^2 s + 2 \mtau^2 s + 2 \mtau \nevalsi s + 
          2 \mtau \nevalsj s + 2 \nevalsi \nevalsj s - s^2 - 2 \mst^2 t - 
          2 \mtau^2 t 
	  \nl + 2 \mtau \nevalsi t - 2 \mtau \nevalsj t + t^2 - 
          2 \mst^2 u - 2 \mtau^2 u - 2 \mtau \nevalsi u + 
          2 \mtau \nevalsj u + u^2) + 
   \nl  K_b^2 K_b^{\prime 2} (2 \mst^4 + 4 \mst^2 \mtau^2 - 6 \mtau^4 + 
        8 \mtau^3 \nevalsi + 8 \mtau^3 \nevalsj - 8 \mtau^2 \nevalsi \nevalsj 
	\nl + 
          2 \mst^2 s + 2 \mtau^2 s - 2 \mtau \nevalsi s - 
          2 \mtau \nevalsj s + 2 \nevalsi \nevalsj s - s^2 - 2 \mst^2 t - 
          2 \mtau^2 t 
	  \nl - 2 \mtau \nevalsi t + 2 \mtau \nevalsj t + t^2 - 
          2 \mst^2 u - 2 \mtau^2 u + 2 \mtau \nevalsi u - 
          2 \mtau \nevalsj u + u^2) - 
   \nl K_b^2 K_a^{\prime 2} (2 \mst^4 + 4 \mst^2 \mtau^2 - 6 \mtau^4 + 
        8 \mtau^3 \nevalsi - 8 \mtau^3 \nevalsj + 8 \mtau^2 \nevalsi \nevalsj 
	\nl + 
          2 \mst^2 s + 2 \mtau^2 s - 2 \mtau \nevalsi s + 
          2 \mtau \nevalsj s - 2 \nevalsi \nevalsj s - s^2 - 2 \mst^2 t - 
          2 \mtau^2 t 
	  \nl - 2 \mtau \nevalsi t - 2 \mtau \nevalsj t + t^2 - 
          2 \mst^2 u - 2 \mtau^2 u + 2 \mtau \nevalsi u + 
          2 \mtau \nevalsj u + u^2)
   ) \nl /((\nevalsi^2 - t) (\nevalsj^2 - u)) \nonumber \\
\tsq &=&  \sum_{i,j} ( {\cal T}_{\rm I}\!\!\times\!\!{\cal T}_{\rm I} 
   +    {\cal T}_{\rm II}\!\!\times\!\!{\cal T}_{\rm II}
      + 2  {\cal T}_{\rm I}\!\!\times\!\!{\cal T}_{\rm II} )
\end{eqnarray}

\subsection*{$\slept^A_1 \slept^{B \ast}_1 \longrightarrow \ell^A \bar{\ell^B}$}
 I. $t$-channel $\chi$ exchange \hfill \\
Here $m_{A,B} \equiv m_{\ell^{A,B}}$ and $m_{\widetilde{A},\widetilde{B}} \equiv
       m_{\tilde{\ell}_1^{A,B}} $.    
\begin{eqnarray}
A(i) &=& -g_2/\sqrt{2} (\cos \thA (- N_{i2} - \tan \thw N_{i1} )
              + \sin \thA m_A /(\mw \cos \beta ) N_{i3} )             
	      \nonumber \\
B(i) &=& -g_2/\sqrt{2} (\cos \thA m_A /(\mw \cos \beta ) N_{i3} 
              + 2 \sin \thA \tan \thw N_{i1} ) 
	      \nonumber \\
C(i) &=& -g_2/\sqrt{2} (\cos \thB m_B /(\mw \cos \beta ) N_{i3} 
              + 2 \sin \thB \tan \thw N_{i1} ) 
	      \nonumber \\
D(i) &=& -g_2/\sqrt{2} (\cos \thB (-N_{i2} - \tan \thw N_{i1} )
              + \sin \thB m_B /(\mw \cos \beta ) N_{i3} )          
	         \nonumber \\
{\cal T}_{\rm I}\!\!\times\!\!{\cal T}_{\rm I} &=&  ( A(i) C(i) A(j) C(j)( 
   - \mla^2 \nevalsi \nevalsj - \mlb^2 \nevalsi \nevalsj + \nevalsi \nevalsj s )
   \nl    + A(i) C(i) A(j) D(j) ( - \msa^2 \mlb \nevalsi + \mla^2 \mlb \nevalsi 
                           + \mlb \nevalsi t )
  \nl     + A(i) C(i) B(j) C(j) ( \msa^2 \mla \nevalsi + 2 \mla \mlb^2 \nevalsi 
                      - \mla \nevalsi s - \mla \nevalsi u + \mla^3 \nevalsi )
  \nl     + A(i) C(i) B(j) D(j) ( - 2 \mla \mlb \nevalsi \nevalsj   )          
  \nl     + A(i) D(i) A(j) C(j) ( - \msa^2 \mlb \nevalsj + \mla^2 \mlb \nevalsj 
                       + \mlb \nevalsj t )
  \nl     + A(i) D(i) A(j) D(j) (  2 \msa^2 \mlb^2 - \msa^2 s - \msa^2 t 
                      - \msa^2 u + \msa^4 - 2 \mla^2 \mlb^2 
                  + \mla^2 s \nl + \mla^2 u - \mla^4 - \mlb^2 t + t u )
  \nl     + A(i) D(i) B(j) C(j) ( - 2 \mla \mlb t )
  \nl    + A(i) D(i) B(j) D(j) (  \msa^2 \mla \nevalsj + 2 \mla \mlb^2 \nevalsj 
           - \mla \nevalsj s - \mla \nevalsj u + \mla^3 \nevalsj )             
  \nl   + B(i) C(i) A(j) C(j) (  \msa^2 \mla \nevalsj + 2 \mla \mlb^2 \nevalsj 
               - \mla \nevalsj s - \mla \nevalsj u + \mla^3 \nevalsj )
  \nl     + B(i) C(i) A(j) D(j) ( - 2 \mla \mlb t )
  \nl     + B(i) C(i) B(j) C(j) (  2 \msa^2 \mlb^2 - \msa^2 s - \msa^2 t 
                   - \msa^2 u + \msa^4 - 2 \mla^2 \mlb^2 + \mla^2 s 
                  \nl  + \mla^2 u - \mla^4 - \mlb^2 t + t u )
  \nl     + B(i) C(i) B(j) D(j) ( - \msa^2 \mlb \nevalsj + \mla^2 \mlb \nevalsj 
                   + \mlb \nevalsj t )          
  \nl        + B(i) D(i) A(j) C(j) ( - 2 \mla \mlb \nevalsi \nevalsj  )
 \nl    + B(i) D(i) A(j) D(j) (  \msa^2 \mla \nevalsi + 2 \mla \mlb^2 \nevalsi 
                    - \mla \nevalsi s - \mla \nevalsi u + \mla^3 \nevalsi )
  \nl     + B(i) D(i) B(j) C(j) ( - \msa^2 \mlb \nevalsi + \mla^2 \mlb \nevalsi 
                  + \mlb \nevalsi t )
       \nl + B(i) D(i) B(j) D(j) ( - \mla^2 \nevalsi \nevalsj 
             - \mlb^2 \nevalsi \nevalsj 
                  + \nevalsi \nevalsj s )  )    
    \nl      / ((t - \nevalsi^2) (t - \nevalsj^2))  \nonumber \\	      
\tsq &=&  \sum_{i,j} {\cal T}_{\rm I}\!\!\times\!\!{\cal T}_{\rm I} 
\end{eqnarray}

\subsection*{$\slept^A_1 \slept^B_1 \longrightarrow \ell^A \ell^B $}
 I. $t$-channel $\chi$ exchange
\begin{eqnarray}
A(i) &=& -g_2/\sqrt{2} (\cos \thA (- N_{i2} - \tan \thw N_{i1} )
              + \sin \thA m_A /(\mw \cos \beta ) N_{i3} )             
	      \nonumber \\
B(i) &=& -g_2/\sqrt{2} (\cos \thA m_A /(\mw \cos \beta ) N_{i3} 
              + 2 \sin \thA \tan \thw N_{i1} ) 
	      \nonumber \\
C(i) &=& -g_2/\sqrt{2} (\cos \thB m_B /(\mw \cos \beta ) N_{i3} 
              + 2 \sin \thB \tan \thw N_{i1} ) 
	      \nonumber \\
D(i) &=& -g_2/\sqrt{2} (\cos \thB (-N_{i2} - \tan \thw N_{i1} )
              + \sin \thB m_B /(\mw \cos \beta ) N_{i3} )          
	         \nonumber \\
{\cal T}_{\rm I}\!\!\times\!\!{\cal T}_{\rm I} &=& (-A(i) (-B(j) \mla (D(i) 
          \nevalsi (2 C(j) \mlb \nevalsj + 
                      D(j) (\mla^2 + 2 \mlb^2 + \msa^2 - s - u)) \nl + 
                C(i) (2 D(j) \mlb t + 
                      C(j) \nevalsj (\mla^2 + 2 \mlb^2 + \msa^2 - s - 
                            u))) \nl + 
          A(j) (D(i) \nevalsi (D(j) \nevalsj (\mla^2 + \mlb^2 - s) + 
                      C(j) \mlb (\mla^2 - \msa^2 + t)) \nl + 
                C(i) (D(j) \mlb \nevalsj (\mla^2 - \msa^2 + t)  + 
                      C(j) (\mla^4 - \msa^4 + \msa^2 s + \msa^2 t \nl + 
                            \mlb^2 (-2 \msa^2 + t)  + 
                            \mla^2 (2 \mlb^2 - s - u) + \msa^2 u - 
                            t u))))  \nl - 
         B(i) (C(i) \nevalsi (B(j) (C(j) \nevalsj (\mla^2 + \mlb^2 - s) + 
                      D(j) \mlb (\mla^2 - \msa^2 + t)) \nl - 
                A(j) \mla (2 D(j) \mlb \nevalsj + 
                      C(j) (\mla^2 + 2 \mlb^2 + \msa^2 - s - u))) \nl + 
          D(i) (C(j) \mlb (-2 A(j) \mla t + 
                      B(j) \nevalsj (\mla^2 - \msa^2 + t)) \nl + 
                D(j) (A(j) \mla \nevalsj (-\mla^2 - 2 \mlb^2 - \msa^2 + s + 
                            u) + 
                      B(j) (\mla^4 - \msa^4 + \msa^2 s + \msa^2 t \nl + 
                            \mlb^2 (-2 \msa^2 + t) + 
                            \mla^2 (2 \mlb^2 - s - u) + \msa^2 u - 
                            t u))))) \nl /
                ((t - \nevalsi^2 ) (t - \nevalsj^2 )) \nonumber \\
\tsq &=&  \sum_{i,j}  {\cal T}_{\rm I}\!\!\times\!\!{\cal T}_{\rm I} 
\end{eqnarray}

\subsection*{$\slept_1 \chi \longrightarrow Z  \ell $}
Channel II (the $t$-channel $\slept$ exchange) is now summed over $\slept_1$ and
$\slept_2$, with appropriate propagators. Also added is the neutralino
$u$-channel exchange. 
\begin{eqnarray}
f_A(1,i) &=& - g_2/ \sqrt{2} (\cos \thf (-N_{i2} - \tan \thw 
            N_{i1} ) + \sin \thf \mtau/(\mw \cos \beta ) 
            N_{i3} ) \nonumber \\
f_A(2,i) &=&  - g_2/\sqrt{2} (- \sin \thf (- N_{i2} - \tan \thw 
            N_{i1} ) + \cos \thf \mtau/(\mw \cos \beta ) 
            N_{i3} )\nonumber \\
f_B(1,i) &=& - g_2/\sqrt{2} (\cos \thf \mtau/(\mw \cos \beta ) 
            N_{i3}  + \sin \thf (2 \tan \thw 
            N_{i1}  ) ) \nonumber \\
f_B(2,i) &=& - g_2/\sqrt{2} (- \sin \thf \mtau/(\mw \cos \beta ) 
            N_{i3}  + \cos \thf (2 \tan \thw 
            N_{i1}  ) ) \nonumber \\
f_C(1) &=& - g_2/\cos \thw ((-1/2) \cos^2 \thf 
             - (-1) \sin^2 \thw ) \nonumber \\
f_C(2) &=& g_2/\cos \thw ((-1/2) \cos \thf \sin \thf ) \nonumber \\
f_L &=& -g_2/\cos \thw ((-1/2) - (-1) \sin^2 \thw )  \nonumber \\
f_R &=& g_2/ \cos \thw ((-1) \sin^2 \thw )  \nonumber \\
f_{OL}(i) &=& g_2/(2 \cos \thw ) (N_{i4} N_{14} -
            N_{i3} N_{13} )  \nonumber \\
f_{OR}(i) &=& -g_2/(2 \cos \thw ) (N_{i4} N_{14} -
            N_{i3} N_{13} ) \nonumber \\	    
{\cal T}_{\rm I}\!\!\times\!\!{\cal T}_{\rm I} &=& (1/2)(2 f_A(1,1) f_B(1,1) 
          \mchi \mtau 
           (2 \mz^4 f_L^2 + 
            6 \mtau^2 \mz^2 f_L f_R + 2 \mz^4 f_R^2 \nl + 
            \mchi^2 (f_L^2 + f_R^2) (\mtau^2 - s)  + 
            \mst^2 (f_L^2 + f_R^2) (\mtau^2 - s) - 
            2 \mz^2 f_L^2 s + 6 \mz^2 f_L f_R s \nl - 
            2 \mz^2 f_R^2 s - \mtau^2 f_L^2 t - \mtau^2 f_R^2 t + 
            f_L^2 s t + f_R^2 s t - \mtau^2 f_L^2 u - 
            \mtau^2 f_R^2 u + f_L^2 s u + f_R^2 s u) \nl + 
      f_A(1,1)^2 (-\mtau^2 \mz^4 f_R^2  + 
            \mst^4 f_L^2 (\mtau^2 - s) - \mz^4 f_L^2 s - 
            6 \mtau^2 \mz^2 f_L f_R s \nl + \mtau^2 \mz^2 f_R^2 s + 
            \mz^2 f_L^2 s^2  + \mchi^4 f_L^2 (-\mtau^2 + s) - 
            \mtau^2 \mz^2 f_R^2 t + \mtau^2 f_L^2 s t \nl + 
            \mz^2 f_L^2 s t - f_L^2 s^2 t + \mtau^4 f_R^2 u  + 
            \mtau^2 \mz^2 f_R^2 u - \mz^2 f_L^2 s u - 
            \mtau^2 f_R^2 s u \nl + 
            \mst^2 f_L (2 \mz^4 f_L + 
                  \mz^2 (6 \mtau^2 f_R - 2 f_L s)  - 
                  f_L (\mtau^2 - s) (s + t + u)) \nl + 
            \mchi^2 (-2 \mz^4 f_L^2 + 
                  2 \mz^2 f_L (-3 \mtau^2 f_R + 
                        f_L s)  - (\mtau^2 - s) (\mtau^2 f_R^2 - 
                        f_L^2 (t + u)))) \nl - 
      f_B(1,1)^2 (\mchi^4 f_R^2 (\mtau^2 - s) - \mtau^4 f_L^2 u + 
            \mtau^2 (\mz^4 f_L^2  - \mst^4 f_R^2 - f_R^2 s t + 
                  f_L^2 s u \nl + \mst^2 f_R^2 (s + t + u) - 
                  \mz^2 f_L (6 \mst^2 f_R - 6 f_R s  + 
                        f_L (s - t + u))) \nl + 
            \mchi^2 (\mtau^4 f_L^2 + 
                  \mtau^2 (6 \mz^2 f_L f_R - f_L^2 s - 
                        f_R^2 (t + u)) + 
                  f_R^2 (2 \mz^4 \nl - 2 \mz^2 s + 
                        s (t + u))) - 
            f_R^2 (-\mst^4 s - 
                  s (\mz^4 + s t - \mz^2 (s + t - u)) \nl + 
                  \mst^2 (2 \mz^4 - 2 \mz^2 s + 
                        s (s + t + u)))))
            /(\mz^2(s - \mtau^2 )^2) \nonumber \\
{\cal T}_{\rm II}\!\!\times\!\!{\cal T}_{\rm II} &=& (1/2)((f_A(i,1) 
        (-2 f_B(j,1) 
       \mchi  \mtau + 
                f_A(j,1) (\mchi^2 + \mtau^2 - t)) \nl + 
          f_B(i,1) (-2 f_A(j,1) \mchi \mtau \nl + 
                f_B(j,1) (\mchi^2 + \mtau^2 - t))) (\mst^4 + \mz^4 - 
          2 \mz^2 t + t^2 - 2 \mst^2 (\mz^2 + t))) \nl /
          (\mz^2 (t - \msti^2 )(t-\mstj^2)) \nonumber \\
{\cal T}_{\rm III}\!\!\times\!\!{\cal T}_{\rm III} &=& 1/2 ((1/\mz^2) (f_B(1,i) 
        (f_A(1,j) \mtau (
            \mchi^4 (\nevalsj f_{OL}(i) f_{OL}(j) + \nevalsi f_{OR}(i) 
	       f_{OR}(j)) \nl + 
                 \mchi^2 (\nevalsj f_{OL}(i) f_{OL}(j) + 
                       \nevalsi f_{OR}(i) f_{OR}(j) ) (\mz^2 - 2 u) \nl - 
                 6 \mchi \mz^2 (\nevalsi \nevalsj f_{OR}(j) f_{OR}(i) + 
                       f_{OL}(i) f_{OR}(j) u) - (\nevalsj f_{OL}(i) f_{OL}(j)
		       \nl + 
                       \nevalsi f_{OR}(i) f_{OR}(j)) (2 \mz^4 - \mz^2 u - 
                       u^2)) + 
        f_B(1,j) (\mchi^6 f_{OL}(i) f_{OL}(j) \nl - 4 \mtau^2 \mz^4 f_{OL}(i)
	f_{OL}(j) - 
              2 \mz^6 f_{OL}(i) f_{OL}(j) - \nevalsi \nevalsj \mz^4 f_{OR}(i)
	      f_{OR}(j) \nl - 
                 3 \mchi^3 \mz^2 (\nevalsi f_{OL}(j) f_{OR}(i) + 
                     \nevalsj f_{OL}(i) f_{OR}(j)) + 2 \mz^4 f_{OL}(i) f_{OL}(j)
		      s \nl + 
                 \nevalsi \nevalsj \mz^2 f_{OR}(i) f_{OR}(j) s - 
                 3 \mchi \mz^2 (\nevalsi f_{OL}(j) f_{OR}(i) \nl + 
                     \nevalsj f_{OL}(i) f_{OR}(j)) (2 \mtau^2 + \mz^2 - s - 
                       t) + 2 \mz^4 f_{OL}(i) f_{OL}(j) t \nl - 
                 \nevalsi \nevalsj \mz^2 f_{OR}(i) f_{OR}(j) t + 
            \mchi^4 f_{OL}(i) f_{OL}(j) (2 \mtau^2 + 2 \mz^2 - s - t - 
                       2 u) \nl + 2 \mtau^2 \mz^2 f_{OL}(i) f_{OL}(j) u + 
                 2 \mz^4 f_{OL}(i) f_{OL}(j) u + 
                 \mtau^2 \nevalsi \nevalsj f_{OR}(i) f_{OR}(j) u \nl + 
                 \nevalsi \nevalsj \mz^2 f_{OR}(i) f_{OR}(j) u - 
             2 \mz^2 f_{OL}(i) f_{OL}(j) s u 
	     - \nevalsi \nevalsj f_{OR}(i) f_{OR}(j) s u \nl + 
                 \mtau^2 f_{OL}(i) f_{OL}(j) u^2 - f_{OL}(i) f_{OL}(j) t u^2 + 
                 \mchi^2 (-\mz^4 f_{OL}(i) f_{OL}(j) \nl + 
                       \nevalsi \nevalsj f_{OR}(i) f_{OR}(j) s 
		       + f_{OL}(i) f_{OL}(j) s u + 
                     2 f_{OL}(i) f_{OL}(j) t u \nl + f_{OL}(i) f_{OL}(j) u^2 - 
                       \mz^2 f_{OL}(i) f_{OL}(j) (s + t + u) + 
                       \mtau^2 (2 \mz^2 f_{OL}(i) f_{OL}(j) \nl - 
                             \nevalsi \nevalsj f_{OR}(i) f_{OR}(j) - 
                             3 f_{OL}(i) f_{OL}(j) u)))) \nl + 
        f_A(1,i) (f_B(1,j) \mtau (\mchi^4 (\nevalsi f_{OL}(i) f_{OL}(j) + 
                       \nevalsj f_{OR}(i) f_{OR}(j)) \nl + 
                 \mchi^2 (\nevalsi f_{OL}(i) f_{OL}(j) + 
                       \nevalsj f_{OR}(i) f_{OR}(j)) (\mz^2 - 2 u) \nl - 
                 6 \mchi \mz^2 (\nevalsi \nevalsj f_{OL}(i) f_{OR}(j) + 
                       f_{OL}(j) f_{OR}(i) u) - (\nevalsi f_{OL}(i) f_{OL}(j)
		       \nl + 
                       \nevalsj f_{OR}(i) f_{OR}(j)) (2 \mz^4 - \mz^2 u - 
                       u^2)) + 
           f_A(1,j) (\mchi^6 f_{OR}(i) f_{OR}(j) \nl - 
                 3 \mchi^3 \mz^2 (\nevalsj f_{OL}(j) f_{OR}(i) + 
                       \nevalsi f_{OL}(i) f_{OR}(j)) - 
                 3 \mchi \mz^2 (\nevalsj f_{OL}(j) f_{OR}(i) \nl + 
                  \nevalsi f_{OL}(i) f_{OR}(j)) (2 \mtau^2 + \mz^2 - s - 
                       t) \nl + 
              \mchi^4 f_{OR}(i) f_{OR}(j) (2 \mtau^2 + 2 \mz^2 - s - t - 
                       2 u) \nl + 
                 \nevalsi \nevalsj f_{OL}(i) f_{OL}(j) (-\mz^4 + (\mtau^2 - 
                             s) u + \mz^2 (s - t + u)) \nl + 
                 f_{OR}(i) f_{OR}(j) (-2 \mz^6 - 2 \mz^2 s u - t u^2 + 
                       2 \mz^4 (s + t + u) \nl + 
                       \mtau^2 (-4 \mz^4 + 2 \mz^2 u + 
                             u^2))  - 
          \mchi^2 (\mz^4 f_{OR}(i) f_{OR}(j) \nl - \nevalsi \nevalsj 
	  f_{OL}(i) f_{OL}(j) s - 
                     f_{OR}(i) f_{OR}(j) s u - 2 f_{OR}(i) f_{OR}(j) t u \nl - 
                       f_{OR}(i) f_{OR}(j) u^2 + 
                       \mz^2 f_{OR}(i) f_{OR}(j) (s + t + u) + 
                       \mtau^2 (\nevalsi \nevalsj f_{OL}(i) f_{OL}(j) \nl + 
                   f_{OR}(i) f_{OR}(j) (-2 \mz^2 + 3 u)))))))
             /((u-\nevalsi^2) (u-\nevalsj^2))\nonumber \\
{\cal T}_{\rm I}\!\!\times\!\!{\cal T}_{\rm II} &=& (1/2)(-f_A(1,1) (2 f_B(j,1) 
                \mchi 
                 \mtau (\mst^4 f_L + 
                  2 \mz^4 f_L - 3 \mtau^2 \mz^2 f_R - \mz^4 f_R \nl - 
                  2 \mz^2 f_L s  + \mz^2 f_R s + 
                  \mchi^2 f_L (\mst^2 + 3 \mz^2 - t) - 
                  3 \mz^2 f_L t + \mtau^2 f_R t + \mz^2 f_R t \nl - 
                  f_R s t  + f_L t^2  - \mz^2 f_L u + 2 \mz^2 f_R u + 
                  f_L t u + 
                  \mst^2 (\mz^2 (f_L - 3 f_R) - \mtau^2 f_R + 
                        f_R s \nl - 2 f_L t - f_L u)) + 
            f_A(j,1) (-2 \mtau^2 \mz^4 f_R + \mtau^2 \mz^2 f_L s + 
                  \mz^4 f_L s + 4 \mtau^2 \mz^2 f_R s \nl - 
                  \mz^2 f_L s^2 + \mtau^2 \mz^2 f_L t + 
                  \mz^4 f_L t + 2 \mtau^2 \mz^2 f_R t - 
                  \mtau^2 f_L s t - \mz^2 f_L s t + f_L s^2 t \nl - 
                  \mtau^2 f_L t^2 - 2 \mz^2 f_L t^2 + f_L t^3 + 
                  \mst^4 f_L (-2 \mtau^2 + s + t - u) - 
                  \mtau^2 \mz^2 f_L u \nl - \mz^4 f_L u + 
                  2 \mtau^2 \mz^2 f_R u + \mtau^2 f_L t u + 
                  \mz^2 f_L t u - 2 \mtau^2 f_R t u + 
                  \mz^2 f_L u^2 - f_L t u^2 \nl + 
                  \mst^2 (\mtau^2 (2 \mz^2 (f_L - 3 f_R) + 
                              f_L (s + 3 t - u) + 2 f_R u) + 
                        f_L (-s^2 - s t - 2 t^2 \nl + 
                              2 \mz^2 (s - t - u) + t u + 
                              u^2)) + 
                  \mchi^2 (2 \mtau^2 (f_L - f_R) (3 \mz^2 - 
                              t) \nl + 
                        \mst^2 (2 \mtau^2 (f_L - f_R) + 
                              f_L (8 \mz^2 - s + t - u)) - 
                        f_L (t (-s + t - u) \nl + 
                              \mz^2 (s - t + 5 u))))) + 
      f_B(1,1) (-2 f_A(j,1) \mchi \mtau (-\mz^4 f_L + \mst^4 f_R + 
                  3 \mchi^2 \mz^2 f_R \nl + 2 \mz^4 f_R + \mz^2 f_L s - 
                  2 \mz^2 f_R s + \mz^2 f_L t - \mchi^2 f_R t - 
                  3 \mz^2 f_R t - f_L s t + f_R t^2 \nl + 
                  \mtau^2 f_L (-3 \mz^2 + t) + 
                  2 \mz^2 f_L u - \mz^2 f_R u + f_R t u + 
                  \mst^2 (-\mtau^2 f_L + \mchi^2 f_R \nl + 
                        \mz^2 (-3 f_L + f_R) + f_L s - 2 f_R t - 
                        f_R u)) + 
            f_B(j,1) (2 \mtau^2 \mz^4 f_L - 4 \mtau^2 \mz^2 f_L s \nl - 
                  \mtau^2 \mz^2 f_R s - \mz^4 f_R s + 
                  \mz^2 f_R s^2 - 2 \mtau^2 \mz^2 f_L t - 
                  \mtau^2 \mz^2 f_R t - \mz^4 f_R t \nl + 
                  \mtau^2 f_R s t + \mz^2 f_R s t - f_R s^2 t + 
                  \mtau^2 f_R t^2 + 2 \mz^2 f_R t^2 - f_R t^3 - 
                  2 \mtau^2 \mz^2 f_L u \nl + \mtau^2 \mz^2 f_R u + 
                  \mz^4 f_R u + 2 \mtau^2 f_L t u - \mtau^2 f_R t u - 
                  \mz^2 f_R t u - \mz^2 f_R u^2 + f_R t u^2 \nl + 
                  \mst^4 f_R (2 \mtau^2 - s - t + u) + 
                  \mst^2 (-f_R (-s^2 - s t - 2 t^2 + 
                              2 \mz^2 (s - t - u) + t u \nl + u^2) + 
                        \mtau^2 (\mz^2 (6 f_L - 2 f_R) - 2 f_L u + 
                              f_R (-s - 3 t + u))) + 
                  \mchi^2 (2 \mtau^2 (f_L \nl - f_R) (3 \mz^2 - 
                              t) + 
                        \mst^2 (2 \mtau^2 (f_L - f_R) + 
                              f_R (-8 \mz^2 + s - t + u)) + 
                        f_R (t (-s \nl + t - u) + 
                              \mz^2 (s - t + 5 u))))))/
           (2 \mz^2 (s - \mtau^2 )(t - \mstj^2 )) \nonumber \\
{\cal T}_{\rm I}\!\!\times\!\!{\cal T}_{\rm III} &=& 1/2 ((1/(2 \mz^2))
        (f_A(1,i) 
(2 f_B(1,1) \mtau (
           \mchi^5 f_R f_{OR}(i) + 
                 \mchi^2 \nevalsi f_{OL}(i) (-\mtau^2 f_L \nl - 3 \mz^2 f_R + 
                       f_L s) + 
                 \mchi^3 f_{OR}(i) (\mz^2 (-3 f_L + f_R) + 
                       f_R (\mst^2 - t - 2 u)) \nl + 
                 \mchi f_{OR}(i) (-6 \mtau^2 \mz^2 f_L - 
                       \mz^4 (3 f_L + 4 f_R) + 
                       \mz^2 (-3 \mst^2 f_R + 3 f_L (s + t) \nl + 
                             f_R (s + 2 t)) + 
                       f_R u (-\mst^2 + t + u)) + 
                 \nevalsi f_{OL}(i) (-\mz^4 f_L + f_L (\mtau^2 - s) u \nl + 
                       \mz^2 (3 \mst^2 f_R - 3 f_R s + 
                             f_L (s - t + u)))) + 
           f_A(1,1) (2 \mchi^3 \nevalsi f_L f_{OL}(i) (-\mtau^2 + s) \nl + 
                \mchi^4 f_{OR}(i) (2 \mtau^2 f_R - f_L (s + t - u)) - 
                 2 \mchi \nevalsi f_{OL}(i) (2 \mz^4 f_L + 
                       \mst^2 f_L (\mtau^2 - s) \nl + 
                       \mz^2 (6 \mtau^2 f_R - 2 f_L s) - 
                       f_L (\mtau^2 - s) (t + u)) + 
                 \mchi^2 f_{OR}(i) (\mst^2 f_L (4 \mtau^2 + 8 \mz^2 \nl - 
                             3 s - t + u) + 
                       \mtau^2 (2 \mz^2 f_R - 4 f_R u + 
                             f_L (-s - 3 t + u)) + 
                       f_L (s^2 + t^2 + t u \nl - 2 u^2 + 
                             s (2 t + u) - 
                             \mz^2 (5 s + t + 3 u))) + 
                 f_{OR}(i) (\mst^2 f_L (4 \mz^4 - 4 \mz^2 (s + t) \nl + 
                             \mtau^2 (8 \mz^2 - 2 u) + (s + t - 
                                   u) u) + 
                       \mtau^2 (-4 \mz^4 f_R + 
                             \mz^2 (-8 f_L s + 2 f_R u) \nl + 
                             u (f_L (s + t - u) + 2 f_R u)) + 
                       f_L (-4 \mz^4 s - 
                             u (s^2 + t^2 - u^2) \nl + 
                             \mz^2 (4 s^2 + (t - u) u + 
                                   s (4 t + u)))))) + 
            f_B(1,i) (2 f_A(1,1) \mtau (\mchi^5 f_L f_{OL}(i) \nl + 
                 \mchi^2 \nevalsi f_{OR}(i) (-3 \mz^2 f_L + 
                       f_R (-\mtau^2 + s)) + 
                 \mchi f_{OL}(i) (-\mz^4 (4 f_L + 3 f_R) \nl + 
                       \mz^2 (-6 \mtau^2 f_R + 3 f_R (s + t) + 
                            f_L (s + 2 t)) - 
                       \mst^2 f_L (3 \mz^2 + u) + 
                       f_L u (t + u)) \nl + 
                 \mchi^3 f_{OL}(i) (\mst^2 f_L + \mz^2 (f_L - 3 f_R) - 
                       f_L (t + 2 u)) + 
                 \nevalsi f_{OR}(i) (3 \mst^2 \mz^2 f_L \nl - \mz^4 f_R + 
                       f_R (\mtau^2 - s) u + 
                       \mz^2 (-3 f_L s + 
                             f_R (s - t + u)))) \nl + 
        f_B(1,1) (2 \mchi^3 \nevalsi f_R f_{OR}(i) (-\mtau^2 + s) + 
              \mchi^4 f_{OL}(i) (2 \mtau^2 f_L - f_R (s + t - u)) \nl - 
                 2 \mchi \nevalsi f_{OR}(i) (\mtau^2 (6 \mz^2 f_L + 
                             f_R (\mst^2 - t - u)) + 
                       f_R (2 \mz^4 - 2 \mz^2 s + 
                             s (-\mst^2 \nl + t + u))) + 
                 \mchi^2 f_{OL}(i) (\mtau^2 (2 \mz^2 f_L + 
                4 \mst^2 f_R - f_R s - 3 f_R t - 4 f_L u + 
                             f_R u) \nl + 
                       f_R (s^2 + 2 s t + t^2 + s u + t u - 
                             2 u^2 + 
                             \mst^2 (8 \mz^2 - 3 s - t + u) - 
                             \mz^2 (5 s + t \nl + 3 u))) + 
                 f_{OL}(i) (\mtau^2 (-4 \mz^4 f_L + 
                       2 \mz^2 (4 \mst^2 f_R - 4 f_R s + 
                                   f_L u) + 
                             u (-2 \mst^2 f_R \nl + 
                           f_R (s + t - u) + 2 f_L u)) + 
                       f_R (-4 \mz^4 s + 
                             \mst^2 (4 \mz^4 - 
                            4 \mz^2 (s + t) + (s + t \nl - 
                                       u) u) - 
                             u (s^2 + t^2 - u^2) + 
                             \mz^2 (4 s^2 + (t - u) u + 
                                   s (4 t + u))))))))
          \nl  /((u-\nevalsi^2) (s-\mtau^2))\nonumber \\
{\cal T}_{\rm II}\!\!\times\!\!{\cal T}_{\rm III} &=& 1/2 ((1/\mz^2) (f_B(j,1) 
(f_A(1,i) \mtau (\nevalsi f_{OL}(i) + 
         \mchi f_{OR}(i)) (\mz^4 - 2 \mz^2 s \nl + \mchi^2 (\mst^2 + 
         3 \mz^2 - t) - \mz^2 t + \mst^2 (3 \mz^2 - u) - 
         \mz^2 u + t u) \nl + 
        f_B(1,i) (\mchi^4 f_{OL}(i) (\mst^2 + 3 \mz^2 - t) + 
         \mchi \nevalsi f_{OR}(i) (-\mz^4 + \mz^2 s \nl + 
         \mst^2 (-\mtau^2 - 3 \mz^2 + s) + 
         \mz^2 t - s t + \mtau^2 (-3 \mz^2 + t) + 2 \mz^2 u) \nl + 
        \mchi^2 f_{OL}(i) (3 \mz^4 - 4 \mz^2 s + 
         \mtau^2 (6 \mz^2 - 2 t) - 5 \mz^2 t + s t + t^2 + 
         \mst^2 (2 \mtau^2 \nl + 5 \mz^2 - s - t - u) - 
              2 \mz^2 u + t u) + 
           f_{OL}(i) (-\mz^4 s + \mz^2 s^2 - \mz^4 t + 
              2 \mz^2 s t \nl + \mz^2 t^2 + \mz^4 u + 
                  \mz^2 t u - t^2 u - \mz^2 u^2 + 
                  \mtau^2 (-\mz^2 (3 s + t) + t u) \nl + 
                  \mst^2 (\mtau^2 (4 \mz^2 - u) + t u + 
                              \mz^2 (-s - 3 t + u))))) \nl + 
     f_A(j,1) (f_B(1,i) \mtau (\mchi f_{OL}(i) + \nevalsi f_{OR}(i)) (\mz^4 - 
           2 \mz^2 s + \mchi^2 (\mst^2 \nl + 3 \mz^2 - t) - 
           \mz^2 t + \mst^2 (3 \mz^2 - u) - \mz^2 u + t u) + 
       f_A(1,i) (\mchi^4 f_{OR}(i) (\mst^2 \nl + 3 \mz^2 - t) + 
           \mchi \nevalsi f_{OL}(i) (-\mz^4 + \mz^2 s + 
           \mst^2 (-\mtau^2 - 3 \mz^2 + s) + 
           \mz^2 t \nl - s t + \mtau^2 (-3 \mz^2 + t) + 2 \mz^2 u) + 
           \mchi^2 f_{OR}(i) (3 \mz^4 - 4 \mz^2 s + 
           \mtau^2 (6 \mz^2 - 2 t) \nl - 5 \mz^2 t + s t + t^2 + 
           \mst^2 (2 \mtau^2 + 5 \mz^2 - s - t - u) - 
                        2 \mz^2 u + t u) \nl + 
         f_{OR}(i) (-\mz^4 s + \mz^2 s^2 - \mz^4 t + 
           2 \mz^2 s t + \mz^2 t^2 + \mz^4 u + 
              \mz^2 t u - t^2 u  - \mz^2 u^2 \nl + 
          \mtau^2 (-\mz^2 (3 s + t) + t u) + 
         \mst^2 (\mtau^2 (4 \mz^2 - u) + t u + \mz^2 (-s - 3 t + 
                        u)))))))
       \nl  /((u-\nevalsi^2) (t-\mstj^2)) \nonumber \\	   
\tsq &=&   {\cal T}_{\rm I}\!\!\times\!\!{\cal T}_{\rm I} 
  +  \sum_{i,j} f_C(i) f_C(j) {\cal T}_{\rm II}\!\!\times\!\!{\cal T}_{\rm II}
  +  \sum_{i,j}  {\cal T}_{\rm III}\!\!\times\!\!{\cal T}_{\rm III}
      + 2 \sum_j f_C(j) {\cal T}_{\rm I}\!\!\times\!\!{\cal T}_{\rm II}
    \nl  + 2 \sum_i  {\cal T}_{\rm I}\!\!\times\!\!{\cal T}_{\rm III}
      + 2 \sum_{i,j} f_C(j) {\cal T}_{\rm II}\!\!\times\!\!{\cal T}_{\rm III}
\end{eqnarray}

\subsection*{$\slept_1 \chi \longrightarrow \gamma  \ell $}
%Unchanged, except that $\stau_R \rightarrow \slept_1$
The couplings are modified.
\begin{eqnarray}
f_A &=& - g_2/ \sqrt{2} ( \cos \thf (- N_{12} - \tan \thw
            N_{11} ) + \sin \thf \mtau/( \mw \cos \beta )
            N_{13} ) (e) \nonumber \\
f_B &=&  - g_2/ \sqrt{2} ( \cos \thf \mtau/(\mw \cos \beta )
            N_{13}  + \sin \thf (2 \tan \thw
            N_{11}  ) )  (e) \nonumber \\
{\cal T}_{\rm I}\!\!\times\!\!{\cal T}_{\rm I} &=& (1/2)(2 (4 f_A f_B \mchi 
         \mtau
         (-\mchi^2 - \mst^2 + 2 s + t + u) + 
        f_A^2 (\mchi^4 - \mst^4 + \mtau^4 - 3 \mtau^2 s \nl - \mtau^2 t - 
              s u - \mchi^2 (\mtau^2 + t + u) + 
              \mst^2 (2 \mtau^2 + s + t + u)) + 
        f_B^2 (\mchi^4 - \mst^4 + \mtau^4 \nl - 3 \mtau^2 s - \mtau^2 t - 
              s u - \mchi^2 (\mtau^2 + t + u) + 
              \mst^2 (2 \mtau^2 + s + t + u))))/(s-\mtau^2)^2 
		    \nonumber \\
{\cal T}_{\rm II}\!\!\times\!\!{\cal T}_{\rm II} &=& (1/2)(-2 (-4 f_A f_B \mchi 
       \mtau + 
        f_A^2 (\mchi^2 + \mtau^2 - t) + 
        f_B^2 (\mchi^2 + \mtau^2 - t)) (\mst^2 + t)) \nl /(t-\mst^2)^2 
	\nonumber \\
{\cal T}_{\rm I}\!\!\times\!\!{\cal T}_{\rm II} &=& (1/2)(1/2 (-4 f_A f_B 
            \mchi \mtau
          (3 \mchi^2 - 3 \mst^2 - 
              3 \mtau^2 - s - t + u) + 
        f_A^2 (-5 \mtau^2 s + s^2 \nl - \mtau^2 t  + t^2 - 
              \mtau^2 u - u^2 + \mst^2 (2 \mtau^2 - s + 3 t + u) + 
              \mchi^2 (-8 \mst^2 + s - t + 5 u)) \nl + 
        f_B^2 (-5 \mtau^2 s + s^2 - \mtau^2 t + t^2 - 
              \mtau^2 u - u^2 + \mst^2 (2 \mtau^2 - s + 3 t + u) \nl + 
       \mchi^2 (-8 \mst^2 + s - t + 5 u))))/((s-\mtau^2)(t-\mst^2))
       \nonumber \\
\tsq &=&   {\cal T}_{\rm I}\!\!\times\!\!{\cal T}_{\rm I} 
   +   {\cal T}_{\rm II}\!\!\times\!\!{\cal T}_{\rm II}
      + 2  {\cal T}_{\rm I}\!\!\times\!\!{\cal T}_{\rm II}
\end{eqnarray}

\subsection*{$\slept_1 \chi \longrightarrow   \ell h [H]  $}
The $s$-channel $\tau$ annihilation was neglected in ~\cite{Ellis:1999mm} due to
the small Yukawa coupling. However, at large $\tan \beta$, these
couplings are  enhanced particularly for the $h$ final state. The
$t$-channel
$\slept$ exchange is now summed over
$\slept_1$ and $\slept_2$. Also added is the neutralino $u$-channel
exchange.  
\hfill \\
 I. $s$-channel $\ell$ annihilation  \hfill \\
 II. $t$-channel $\slept_{1,2}$ exchange  \hfill \\ 
 III. $u$-channel $\chi_{1,2,3,4}$ exchange 
\begin{eqnarray}
f_1 &=& -g_2 \mtau \cos \alpha /(2 \mw \cos \beta ) \nonumber \\
f_2(1) &=&  g_2 \mz (
          (-1/2+ \sin^2 \thw ) \sin [-\cos](\beta + \alpha ) \cos^2 \thf
       \nl  - \sin^2 \thw  \sin [-\cos](\beta + \alpha )  \sin^2 \thf ) 
	  /\cos \thw 
       \nl   + g_2 \mtau^2 \sin [-\cos] \alpha /(\mw \cos \beta ) 
     \nl  - g_2 \mtau (\Atau \sin [-\cos] \alpha   -\mu \cos [\sin] \alpha ) 
       \sin \thf \cos \thf
	  /(\mw \cos \beta ) 
              \nonumber \\
f_2(2) &=&  g_2 \mz (
          (1/2 - \sin^2 \thw ) \sin [-\cos] (\beta + \alpha ) 
      \nl - \sin^2 \thw  \sin [-\cos] (\beta + \alpha ) ) \cos \thf \sin \thf  
	  /\cos \thw 
      \nl  - g_2 \mtau (\Atau \sin [-\cos] \alpha -\mu \cos [\sin] \alpha ) 
	  \cos (2 \thf) 
	  /(2 \mw \cos \beta) 
               \nonumber \\
f_A(1,i) &=& - g_2/ \sqrt{2} (\cos \thf (-N_{i2} - \tan \thw 
            N_{i1} ) + \sin \thf \mtau/(\mw \cos \beta ) 
            N_{i3} ) \nonumber \\
f_A(2,i) &=&  - g_2/\sqrt{2} (- \sin \thf (- N_{i2} - \tan \thw 
            N_{i1} ) + \cos \thf \mtau/(\mw \cos \beta ) 
            N_{i3} )\nonumber \\
f_B(1,i) &=& - g_2/\sqrt{2} (\cos \thf \mtau/(\mw \cos \beta ) 
            N_{i3}  + \sin \thf (2 \tan \thw 
            N_{i1}  ) ) \nonumber \\
f_B(2,i) &=& - g_2/\sqrt{2} (- \sin \thf \mtau/(\mw \cos \beta ) 
            N_{i3}  + \cos \thf (2 \tan \thw 
            N_{i1}  ) ) \nonumber \\
f_{CL}(i) &=&  g_2/ 2 (
            (N_{13} (N_{i2} - N_{i1} \tan \thw )
           + N_{i3} (N_{12} - N_{11} \tan \thw ))
           \sin \alpha 
           \nl  +(N_{14} (N_{i2} - N_{i1} \tan \thw )
           + N_{i4} (N_{12} - N_{11} \tan \thw ))
             \cos \alpha )	 \nonumber \\
f_{CR}(i) &=&  g_2/2 (
             (N_{i3} (N_{12} - N_{11} \tan \thw )
           + N_{13} (N_{i2} - N_{i1} \tan \thw ))
           \sin \alpha 
           \nl  +(N_{i4} (N_{12} - N_{11} \tan \thw )
           + N_{14} (N_{i2} - N_{i1} \tan \thw ))
             \cos \alpha )	 \nonumber \\
{\cal T}_{\rm I}\!\!\times\!\!{\cal T}_{\rm I} &=& 1/2(f_A(1,1)(2 f_B(1,1) 
     \mchi \mtau (\mchi^2 + \mst^2 
         + 3 \mtau^2 + s - t - 
          u) \nl + f_A(1,1) (\mchi^4 - \mst^4 + \mtau^4 + 3 \mtau^2 s - 
                \mtau^2 t + \mchi^2 (5 \mtau^2 - t - u) - s u \nl + 
                \mst^2 (-4 \mtau^2 + s + t + u))) + 
       f_B(1,1) (2 f_A(1,1) \mchi \mtau (\mchi^2 + \mst^2 + 3 \mtau^2 + s \nl 
       - t - u) + f_B(1,1) (\mchi^4 - \mst^4 + \mtau^4 + 3 \mtau^2 s - 
                \mtau^2 t + \mchi^2 (5 \mtau^2 - t - u) - s u \nl + 
                \mst^2 (-4 \mtau^2 + s + t + u)))) 
         /(s - \mtau^2)^2 \nonumber \\	        
{\cal T}_{\rm II}\!\!\times\!\!{\cal T}_{\rm II} &=& 1/2 (f_A(i,1) (2 f_B(j,1) 
        \mchi \mtau + f_A(j,1) (\mchi^2 + \mtau^2 
        - t)) \nl + f_B(i,1) (2 f_A(j,1) \mchi \mtau + f_B(j,1) (\mchi^2 
	 + \mtau^2 - t)))/((t - \msti^2) (t - \mstj^2))   \nonumber \\
{\cal T}_{\rm III}\!\!\times\!\!{\cal T}_{\rm III} &=& 1/2 
       (f_A(1,i) (f_B(1,j) \mtau (f_{CR}(i) \nevalsi (2 f_{CL}(j) \mchi 
       \nevalsj + f_{CR}(j) (\mchi^2 - m_{h[H]}^2  + u)) \nl + 
         f_{CL}(i) (2 f_{CR}(j) \mchi u + 
         f_{CL}(j) \nevalsj (\mchi^2 - m_{h[H]}^2 + u))) \nl + 
     f_A(1,j) (f_{CR}(i) \nevalsi (f_{CR}(j) \nevalsj (\mchi^2 + \mtau^2 - t) + 
         f_{CL}(j) \mchi (\mchi^2 + m_{h[H]}^2 \nl + 2 \mtau^2 - s - t)) + 
         f_{CL}(i) (f_{CR}(j) \mchi \nevalsj (\mchi^2 + m_{h[H]}^2 + 
                    2 \mtau^2 - s - t) \nl + 
         f_{CL}(j) (\mchi^4 - m_{h[H]}^4 + \mchi^2 (2 \mtau^2 - s - t) + 
          (\mtau^2 - s) u + m_{h[H]}^2 (-2 \mtau^2 + s + t \nl + u))))) + 
       f_B(1,i) (f_A(1,j) \mtau (f_{CL}(i) \nevalsi (2 f_{CR}(j) \mchi 
       \nevalsj + 
         f_{CL}(j) (\mchi^2 - m_{h[H]}^2 \nl + u)) + 
         f_{CR}(i) (2 f_{CL}(j) \mchi u + 
         f_{CR}(j) \nevalsj (\mchi^2 - m_{h[H]}^2 + u))) \nl + 
       f_B(1,j) (f_{CL}(i) \nevalsi (f_{CL}(j) \nevalsj (\mchi^2 + \mtau^2 
       - t) + 
         f_{CR}(j) \mchi (\mchi^2 + m_{h[H]}^2 + 2 \mtau^2 \nl - s - t)) + 
         f_{CR}(i) (f_{CL}(j) \mchi \nevalsj (\mchi^2 + m_{h[H]}^2 + 
                    2 \mtau^2 - s - t) + 
         f_{CR}(j) (\mchi^4 \nl - m_{h[H]}^4 + \mchi^2 (2 \mtau^2 - s - t) + 
           (\mtau^2 - s) u + m_{h[H]}^2 (-2 \mtau^2 + s + t + u))))))
       \nl   /((u - \nevalsi^2) (u-\nevalsj^2)) \nonumber \\
{\cal T}_{\rm I}\!\!\times\!\!{\cal T}_{\rm II} &=& 1/2 
       (f_B(1,1) (f_B(i,1) \mtau (2 \mchi^2 - \mst^2 + \mtau^2 + s - t) 
       + f_A(i,1) \mchi (\mchi^2 + \mst^2 \nl + 4 \mtau^2 - t - u)) + 
        f_A(1,1) (f_A(i,1) \mtau (2 \mchi^2 - \mst^2 + \mtau^2 + s - t) \nl + 
         f_B(i,1) \mchi (\mchi^2 + \mst^2 + 4 \mtau^2 - t - u))) 
           /((s - \mtau^2) (t - \msti^2))	\nonumber \\ 
{\cal T}_{\rm I}\!\!\times\!\!{\cal T}_{\rm III} &=& 1/2 ((1/2)
        (f_A(1,i)(2 f_A(1,1) \mtau (f_{CR}(i) \nevalsi (2 \mchi^2 
         - \mst^2 + \mtau^2 + s - t) \nl + 
       f_{CL}(i) \mchi (\mchi^2 - m_{h[H]}^2 - 2 \mst^2 + 2 \mtau^2 + u)) + 
       f_B(1,1) (2 f_{CR}(i) \mchi \nevalsi (\mchi^2 \nl + \mst^2 + 4 \mtau^2 
       - t - u) 
       + f_{CL}(i) (2 \mchi^4 + \mst^2 s + \mtau^2 s - s^2 - 
        \mst^2 t - \mtau^2 t + t^2 \nl + \mchi^2 (2 m_{h[H]}^2 
        + 2 \mst^2 + 6 \mtau^2 - s - 3 t - u) + \mst^2 u 
        + 3 \mtau^2 u - u^2 + m_{h[H]}^2 (-2 \mst^2 \nl - 2 \mtau^2 
        + s - t + u)))) + 
       f_B(1,i) (2 f_B(1,1) \mtau (f_{CL}(i) \nevalsi 
       (2 \mchi^2 - \mst^2 + \mtau^2 + s \nl - t) + 
        f_{CR}(i) \mchi (\mchi^2 - m_{h[H]}^2 - 2 \mst^2 + 2 \mtau^2 + u)) 
	\nl + 
       f_A(1,1) (2 f_{CL}(i) \mchi \nevalsi (\mchi^2  + \mst^2 + 4 \mtau^2 
       - t - u) 
        + f_{CR}(i) (2 \mchi^4 + \mst^2 s \nl + \mtau^2 s - s^2 - \mst^2 t 
         - \mtau^2 t + t^2  + \mchi^2 (2 m_{h[H]}^2 + 2 \mst^2 
          + 6 \mtau^2 - s - 3 t - u) \nl + \mst^2 u + 3 \mtau^2 u - 
           u^2 + m_{h[H]}^2 (-2 \mst^2  - 2 \mtau^2 + s - t + u)))))) 
        \nl  /((s - \mtau^2) ( u - \nevalsi^2))	\nonumber \\ 
{\cal T}_{\rm II}\!\!\times\!\!{\cal T}_{\rm III} &=& 1/2 (f_B(j,1) (f_B(1,i) 
       (f_{CL}(i) \nevalsi (\mchi^2 + \mtau^2 - t) + 
       f_{CR}(i) \mchi (\mchi^2 + m_{h[H]}^2 \nl + 2 \mtau^2  - s - t)) + 
       f_A(1,i) \mtau (2 f_{CR}(i) \mchi \nevalsi + 
                f_{CL}(i) (\mchi^2 - m_{h[H]}^2 + u))) \nl + 
       f_A(j,1) (f_A(1,i) (f_{CR}(i) \nevalsi (\mchi^2 + \mtau^2 - t) + 
          f_{CL}(i) \mchi (\mchi^2 + m_{h[H]}^2 + 2 \mtau^2 \nl - s - t)) + 
          f_B(1,i) \mtau (2 f_{CL}(i) \mchi \nevalsi + 
                f_{CR}(i) (\mchi^2 - m_{h[H]}^2 + u))))  
         \nl   /((t - \mstj^2) (u - \nevalsi^2)) \nonumber \\	       
\tsq &=&  f_1^2 {\cal T}_{\rm I}\!\!\times\!\!{\cal T}_{\rm I} 
   + \sum_{i,j} f_2(i) f_2(j) {\cal T}_{\rm II}\!\!\times\!\!{\cal T}_{\rm II} 
   + \sum_{i,j}  {\cal T}_{\rm III}\!\!\times\!\!{\cal T}_{\rm III}
    + 2 \sum_i f_1 f_2(i) {\cal T}_{\rm I}\!\!\times\!\!{\cal T}_{\rm II}
  \nl  + 2 \sum_i f_1 {\cal T}_{\rm I}\!\!\times\!\!{\cal T}_{\rm III}
    + 2 \sum_{i,j} f_2(j) {\cal T}_{\rm II}\!\!\times\!\!{\cal T}_{\rm III} 
\end{eqnarray}

\subsection*{$\slept_1 \chi \longrightarrow   \ell A  $}
 I. $s$-channel $\ell$ annihilation  \hfill \\
 II. $t$-channel $\slept_2$ exchange  \hfill \\ 
 III. $u$-channel $\chi_{1,2,3,4}$ exchange  
\begin{eqnarray}
f_A(1,i) &=& - g_2/ \sqrt{2} (\cos \thf (-N_{i2} - \tan \thw 
            N_{i1} ) + \sin \thf \mtau/(\mw \cos \beta ) 
            N_{i3} ) \nonumber \\
f_A(2,i) &=&  - g_2/\sqrt{2} (- \sin \thf (- N_{i2} - \tan \thw 
            N_{i1} ) + \cos \thf \mtau/(\mw \cos \beta ) 
            N_{i3} )\nonumber \\
f_B(1,i) &=& - g_2/\sqrt{2} (\cos \thf \mtau/(\mw \cos \beta ) 
            N_{i3}  + \sin \thf (2 \tan \thw 
            N_{i1}  ) ) \nonumber \\
f_B(2,i) &=& - g_2/\sqrt{2} (- \sin \thf \mtau/(\mw \cos \beta ) 
            N_{i3}  + \cos \thf (2 \tan \thw 
            N_{i1}  ) ) \nonumber \\
f_1 &=& -(g_2 \mtau \tan \beta )/(2 \mw) \nonumber \\
f_2 &=& -g_2 \mtau/(2 \mw) (\Atau \tan \beta - \mu) \nonumber \\
f_{CL}(i) &=& g_2/2(
            (N_{i3} (N_{12} - N_{11} \tan \thw )
           + N_{13} (N_{i2} - N_{i1} \tan \thw ))
           \sin \beta \nl
             - (N_{i4} (N_{12} - N_{11} \tan \thw )
           + N_{14} (N_{i2} - N_{i1} \tan \thw ))
             \cos \beta ) \nonumber \\
f_{CR}(i) &=& -g_2/2 (
             (N_{i3} (N_{12} - N_{11} \tan \thw )
           + N_{13} (N_{i2} - N_{i1} \tan \thw ))
           \sin \beta  \nl
             -(N_{i4} (N_{12} - N_{11} \tan \thw )
           + N_{14} (N_{i2} - N_{i1} \tan \thw ))
             \cos \beta ) \nonumber \\
{\cal T}_{\rm I}\!\!\times\!\!{\cal T}_{\rm I} &=& 1/2 (f_A(1,) (2 f_B(1,1) 
         \mchi \mtau (-\mchi^2 - \mst^2 
           - \mtau^2 + s + t + u) \nl + 
             f_A(1,1) (-\mchi^4 + \mst^4 - \mtau^4 + \mtau^2 s + 
                \mtau^2 t + s u + \mchi^2 (-\mtau^2 + t + u) \nl - 
                \mst^2 (s + t + u))) + 
            f_B(1,1) (2 f_A(1,1) \mchi \mtau (-\mchi^2 - \mst^2 - \mtau^2 
               + s + t + u) \nl + 
         f_B(1,1) (-\mchi^4 + \mst^4 - \mtau^4 + \mtau^2 s + 
                \mtau^2 t + s u + \mchi^2 (-\mtau^2 + t + u) \nl - 
                \mst^2 (s + t + u)))) 
         /(s - \mtau^2)^2 \nonumber \\
{\cal T}_{\rm II}\!\!\times\!\!{\cal T}_{\rm II} &=& 1/2 (f_A(2,1) (2 f_B(2,1) 
       \mchi \mtau + 
                  f_A(2,1) (\mchi^2 + \mtau^2 - t)) \nl + 
                  f_B(2,1) (2 f_A(2,1) \mchi \mtau + 
                 f_B(2,1) (\mchi^2 + \mtau^2 - t)))
           /((t - \mstwo^2)^2) \nonumber \\
{\cal T}_{\rm III}\!\!\times\!\!{\cal T}_{\rm III} &=& 1/2 (f_A(1,i) (f_B(1,j) 
        \mtau (f_{CR}(i) \nevalsi (
       2 f_{CL}(j) \mchi \nevalsj + f_{CR}(j) (-\mA^2 + \mchi^2 + u)) \nl + 
       f_{CL}(i) (2 f_{CR}(j) \mchi u + f_{CL}(j) \nevalsj (-\mA^2 + \mchi^2 
       + u))) \nl + 
       f_A(1,j) (f_{CR}(i) \nevalsi (f_{CR}(j) \nevalsj (\mchi^2 + \mtau^2 
       - t) + 
       f_{CL}(j) \mchi (\mA^2 + \mchi^2 + 2 \mtau^2 \nl - s - t)) + 
      f_{CL}(i) (f_{CR}(j) \mchi \nevalsj (\mA^2 + \mchi^2 + 2 \mtau^2 - s - t) 
       + f_{CL}(j) (-\mA^4 \nl + \mchi^4 + 
               \mchi^2 (2 \mtau^2 - s - t) + (\mtau^2 - s) u + 
                \mA^2 (-2 \mtau^2 + s + t + u))))) \nl + 
       f_B(1,i) (f_A(1,j) \mtau (f_{CL}(i) \nevalsi (2 f_{CR}(j) \mchi 
       \nevalsj + 
          f_{CL}(j) (-\mA^2 + \mchi^2 + u)) \nl + 
          f_{CR}(i) (2 f_{CL}(j) \mchi u + 
          f_{CR}(j) \nevalsj (-\mA^2 + \mchi^2 + u))) \nl + 
     f_B(1,j) (f_{CL}(i) \nevalsi (f_{CL}(j) \nevalsj (\mchi^2 + \mtau^2 - t) + 
          f_{CR}(j) \mchi (\mA^2 + \mchi^2 + 2 \mtau^2 \nl - s - t)) + 
          f_{CR}(i) (f_{CL}(j) \mchi \nevalsj (\mA^2 + \mchi^2 + 2 \mtau^2 
              - s - t) + 
          f_{CR}(j) (-\mA^4 \nl + \mchi^4 + \mchi^2 (2 \mtau^2 - s - t) + 
          (\mtau^2 - s) u + \mA^2 (-2 \mtau^2 + s + t + u))))))
        \nl  /((u - \nevalsi^2) (u-\nevalsj^2)) \nonumber \\
{\cal T}_{\rm I}\!\!\times\!\!{\cal T}_{\rm II} &=& 1/2 (f_A(1,1) (f_A(2,1) 
      \mtau (-\mst^2 - \mtau^2 + s + t) + 
          f_B(2,1) \mchi (\mchi^2 + \mst^2 - t - u)) \nl + 
          f_B(1,1) (f_B(2,1) \mtau (\mst^2 + \mtau^2 - s - t) + 
          f_A(2,1) \mchi (-\mchi^2 - \mst^2 + t + u))) 
         \nl  /((s - \mtau^2) (t - \mstwo^2)) \nonumber \\
{\cal T}_{\rm I}\!\!\times\!\!{\cal T}_{\rm III} &=& 1/2 ((1/2) (f_B(1,i) (2
       f_B(1,1) \mtau (f_{CL}(i) \nevalsi 
               (\mst^2 + \mtau^2 - s - t) \nl + 
          f_{CR}(i) \mchi (3 \mA^2 + \mchi^2 + 2 \mst^2 + 
                          2 \mtau^2 - 2 s - 2 t - u)) \nl + 
          f_A(1,1) (2 f_{CL}(i) \mchi \nevalsi (\mchi^2 + \mst^2 - t - u) + 
             f_{CR}(i) (2 \mchi^4 + \mst^2 s + \mtau^2 s \nl - s^2 - 
                          \mst^2 t - \mtau^2 t + t^2 + 
             \mchi^2 (2 \mst^2 + 2 \mtau^2 - s - 3 t - u) + 
     	     \mst^2 u - \mtau^2 u - u^2 \nl + 
             \mA^2 (2 \mchi^2 - 2 \mst^2 + 2 \mtau^2 + s - 
                                t + u)))) \nl + 
        f_A(1,i) (-2 f_A(1,1) \mtau (f_{CR}(i) \nevalsi (\mst^2 + \mtau^2 
	- s - t) + 
           f_{CL}(i) \mchi (3 \mA^2 \nl + \mchi^2 + 2 \mst^2 + 
                2 \mtau^2 - 2 s - 2 t - u)) + 
              f_B(1,1) (2 f_{CR}(i) \mchi \nevalsi (-\mchi^2 - \mst^2 
	    \nl  + t + u) + 
              f_{CL}(i) (-2 \mchi^4 - \mst^2 s - \mtau^2 s + s^2 + 
                \mst^2 t + \mtau^2 t - t^2 - \mst^2 u + 
                \mtau^2 u \nl + u^2 - 
                \mA^2 (2 \mchi^2 - 2 \mst^2 + 2 \mtau^2 + s - 
                                t + u) + 
                \mchi^2 (-2 \mst^2 - 2 \mtau^2 + s + 
                                3 t  \nl + u)))))) 
          /((s - \mtau^2) ( u - \nevalsi^2)) \nonumber \\
{\cal T}_{\rm II}\!\!\times\!\!{\cal T}_{\rm III} &=& 1/2 (f_B(2,1) (f_B(1,i) 
      (f_{CL}(i) \nevalsi (\mchi^2 + \mtau^2 - t) + 
            f_{CR}(i) \mchi (\mA^2 + \mchi^2 + 2 \mtau^2 \nl - s - t)) + 
          f_A(1,i) \mtau (2 f_{CR}(i) \mchi \nevalsi + 
            f_{CL}(i) (-\mA^2 + \mchi^2 + u))) \nl + 
        f_A(2,1) (f_A(1,i) (f_{CR}(i) \nevalsi (\mchi^2 + \mtau^2 - t) + 
          f_{CL}(i) \mchi (\mA^2 + \mchi^2 + 2 \mtau^2 \nl - s - t)) + 
          f_B(1,i) \mtau (2 f_{CL}(i) \mchi \nevalsi + 
            f_{CR}(i) (-\mA^2 + \mchi^2 + u))))  
         \nl    /((t - \mstwo^2) (u - \nevalsi^2)) \nonumber \\
\tsq &=&  f_1^2 {\cal T}_{\rm I}\!\!\times\!\!{\cal T}_{\rm
           I} + f_2^2 {\cal T}_{\rm II}\!\!\times\!\!{\cal T}_{\rm
           II} + 
   \sum_{i,j}  {\cal T}_{\rm III}\!\!\times\!\!{\cal T}_{\rm III}
    + 2 f_1 f_2 {\cal T}_{\rm I}\!\!\times\!\!{\cal T}_{\rm II}
    + 2 \sum_i f_1 {\cal T}_{\rm I}\!\!\times\!\!{\cal T}_{\rm III}
  \nl  + 2 \sum_i f_2 {\cal T}_{\rm II}\!\!\times\!\!{\cal T}_{\rm III}    
\end{eqnarray}
 
\vskip 0.2in

\setcounter{equation}{0}
\renewcommand{\theequation}{C\arabic{equation}}
\section*{Appendix C: Chargino-Slepton Coannihilation}

Below is the list of the amplitudes squared for chargino-slepton coannihilation.
Note that, for identical-particle final states,
one needs to divide them by two when performing the 
momentum integrations.

\subsection*{$\slept_1 \schi^+_1 \longrightarrow \nu_{\ss \ell} Z$}
 I. $t$-channel $\schi^-_{(1,2)}$ exchange \hfill \\
 II. $u$-channel $\slept_{(1,2)}$ exchange 
\begin{eqnarray}
f_{1LL}(i) &=& C_{\slept_1-\nu_{\ss \ell}-\schi^-_i}^{L} \;
C_{\schi^+_1-\schi^+_i-Z}^{L} \nonumber \\
f_{1LR}(i) &=& C_{\slept_1-\nu_{\ss \ell}-\schi^-_i}^{L} \;
C_{\schi^+_1-\schi^+_i-Z}^{R} \nonumber \\
f_{1RL}(i) &=& C_{\slept_1-\nu_{\ss \ell}-\schi^-_i}^{R} \;
C_{\schi^+_1-\schi^+_i-Z}^{L} \nonumber \\
f_{1RR}(i) &=& C_{\slept_1-\nu_{\ss \ell}-\schi^-_i}^{R} \;
C_{\schi^+_1-\schi^+_i-Z}^{R} \nonumber \\
f_{2L}(i) &=& C_{\slept_1-Z-\slept^-_i} \; C_{\schi^-_1-\nu_{\ss
\ell}-\slept_i}^{L} \nonumber \\
f_{2R}(i) &=& C_{\slept_1-Z-\slept^-_i} \; C_{\schi^-_1-\nu_{\ss
\ell}-\slept_i}^{R} \nonumber \\
{\cal T}_{\rm I}\!\!\times\!\!{\cal T}_{\rm I} &=&  ( \mxi ( 3 f_{1LL}(j)
   (f_{1LR}(i) + f_{1RL}(i)) \mchar \mz^2 ( \mstau^2 - 
                    t) \nl + \mxj (f_{1RL}(i) f_{1RL}(j) (\mz^4 + 
                          s (-\mchar^2 + t) - 
                          \mz^2 (s + t - u)) \nl + 
                    f_{1LR}(i) f_{1LR}(j) (\mz^4 + s (-\mchar^2 + t) - 
                          \mz^2 (s + t - u)))) \nl + 
        f_{1LL}(i) (3 (f_{1LR}(j) +  f_{1RL}(j)) \mchar \mxj \mz^2 (\mstau^2 -
	t) \nl - 
              2 f_{1LL}(j) (-\mstau^4 t - \mz^4 t - \mz^2 s t + 
                    \mz^2 t^2 + \mz^2 t u - t^2 u \nl + 
                    \mstau^2 (2 \mz^4 - 2 \mz^2 t + s t + 
                          t^2 + t u) + 
                    \mchar^2 (\mstau^4 + t u - 
                          \mstau^2 (s + t + 
                                u))))) \nl /(\mz^2 (\mxi^2 - 
            t) (-\mxj^2 + t)) \nonumber \\     
{\cal T}_{\rm II}\!\!\times\!\!{\cal T}_{\rm II} &=&  ((f_{2L}(i) f_{2L}(j) +
f_{2R}(i) f_{2R}(j))(\mchar^2 - 
            u) (\mstau^4 + (\mz^2 - u)^2 \nl - 
            2 \mstau^2 (\mz^2 + u)))/(\mz^2 (-m_{\slept_i}^2 + 
            u) (-m_{\slept_j}^2 + u))  \nonumber \\
{\cal T}_{\rm I}\!\!\times\!\!{\cal T}_{\rm II} &=&   (-2 (f_{1LR}(j) f_{2L}(i)
+ f_{1RL}(j) f_{2R}(i)) \mchar \mxj (\mz^4 + 
              \mstau^2 (3 \mz^2 - s) + s u \nl - 
              \mz^2 (s + 2 t + u)) + 
        f_{1LL}(j) (f_{2L}(i) + 
              f_{2R}(i)) (\mstau^4 (s - t - u) \nl + (\mz^2 - 
                    u) (-s^2 + t^2 + \mz^2 (s - t - u) + 
                    u^2) \nl + 
              \mstau^2 (-s^2 + t^2 - s u + t u + 2 u^2 + 
                    2 \mz^2 (s - t + u)) \nl + 
              \mchar^2 (2 \mstau^4 - (\mz^2 - u) (-s + t + 
                          u) - 
                    \mstau^2 (2 \mz^2 - s + t + 
                          3 u)))) \nl /(2 \mz^2 (\mxj^2 - 
            t) (m_{\slept_i}^2 - u))   \nonumber \\
\tsq &=&   \sum_{i,j} \left( {\cal T}_{\rm I}\!\!\times\!\!{\cal T}_{\rm I} 
+  {\cal T}_{\rm II}\!\!\times\!\!{\cal T}_{\rm II} 
+ 2 {\cal T}_{\rm I}\!\!\times\!\!{\cal T}_{\rm II} \right)    
\end{eqnarray}

\subsection*{$\slept_1 \schi^+_1 \longrightarrow \nu_{\ss \ell} \gamma$}
 I. $t$-channel $\schi_1$ exchange \hfill\\
 II. $u$-channel $\slept_1$ exchange 
\begin{eqnarray}
f_{1L} &=&  C_{\slept_1-\nu_{\ss \ell}-\schi^-_1}^{L} \;
C_{\schi^+_1-\schi^+_1-\gamma}  \nonumber \\
f_{1R} &=&  C_{\slept_1-\nu_{\ss \ell}-\schi^-_1}^{R} \;
C_{\schi^+_1-\schi^+_1-\gamma}  \nonumber \\
f_{2L} &=&  C_{\slept_1-\slept_1-\gamma} \; C_{\schi^-_1-\nu_{\ss
\ell}-\slept_i}^{L}  \nonumber \\
f_{2R} &=&  C_{\slept_1-\slept_1-\gamma} \; C_{\schi^-_1-\nu_{\ss
\ell}-\slept_i}^{R}  \nonumber \\
{\cal T}_{\rm I}\!\!\times\!\!{\cal T}_{\rm I} &=&  (2 (f_{1L}^2 + f_{1R}^2)
(\mchar^4 - 
          \mstau^4 - s t - \mchar^2 (6 \mstau^2 - 5 t + u) + 
          \mstau^2 (s + t + u))) \nl /(\mchar^2 - t)^2   \nonumber \\
{\cal T}_{\rm II}\!\!\times\!\!{\cal T}_{\rm II} &=&  -2 (f_{2L}^2 + f_{2R}^2)
(\mchar^2 - u) (\mstau^2 + 
              u)/(\mstau^2 - u)^2   \nonumber \\
{\cal T}_{\rm I}\!\!\times\!\!{\cal T}_{\rm II} &=&   ((f_{1L} f_{2L} + f_{1R}
f_{2R}) (-s^2 + t^2 + u^2 - 
            \mchar^2 (6 \mstau^2 - 3 s - 3 t + u) \nl + 
            \mstau^2 (s - t + 3 u)))/(2 (\mchar^2 - 
            t) (\mstau^2 - u))    \nonumber \\  
\tsq &=&    {\cal T}_{\rm I}\!\!\times\!\!{\cal T}_{\rm I} 
+   {\cal T}_{\rm II}\!\!\times\!\!{\cal T}_{\rm II} 
+ 2  {\cal T}_{\rm I}\!\!\times\!\!{\cal T}_{\rm II}    
\end{eqnarray}

\subsection*{$\slept_1 \schi^+_1 \longrightarrow \nu_{\ss \ell} h \quad [\nu_{\ss
\ell} H ] $}
 I. $t$-channel $\schi_{1,2}$ exchange \hfill \\
 II. $u$-channel $\stau_{1,2}$ exchange  
\begin{eqnarray}
f_{1LL}(i) &=& C_{\slept_1-\nu_{\ss \ell}-\schi^-_i}^{L} \;
C_{\schi^+_1-\schi^+_i-h}^{L} \quad \quad [C_{\slept_1-\nu_{\ss \ell}-\schi^-_i}^{L} \;
C_{\schi^+_1-\schi^+_i-H}^{L}] \nonumber \\
f_{1LR}(i) &=& C_{\slept_1-\nu_{\ss \ell}-\schi^-_i}^{L} \;
C_{\schi^+_1-\schi^+_i-h}^{R} \quad \quad [C_{\slept_1-\nu_{\ss \ell}-\schi^-_i}^{L} \;
C_{\schi^+_1-\schi^+_i-H}^{R}] \nonumber \\
f_{1RL}(i) &=& C_{\slept_1-\nu_{\ss \ell}-\schi^-_i}^{R} \;
C_{\schi^+_1-\schi^+_i-h}^{L} \quad \quad [C_{\slept_1-\nu_{\ss \ell}-\schi^-_i}^{R} \;
C_{\schi^+_1-\schi^+_i-H}^{L}] \nonumber \\
f_{1RR}(i) &=& C_{\slept_1-\nu_{\ss \ell}-\schi^-_i}^{R} \;
C_{\schi^+_1-\schi^+_i-h}^{R} \quad \quad [C_{\slept_1-\nu_{\ss \ell}-\schi^-_i}^{R} \;
C_{\schi^+_1-\schi^+_i-H}^{R}] \nonumber \\
f_{2L}(i) &=& C_{\slept_1-h-\slept^-_i} \; C_{\schi^-_1-\nu_{\ss
\ell}-\slept_i}^{L} \quad \quad [C_{\slept_1-H-\slept^-_i} \; C_{\schi^-_1-\nu_{\ss
\ell}-\slept_i}^{L}] \nonumber \\
f_{2R}(i) &=& C_{\slept_1-h-\slept^-_i} \; C_{\schi^-_1-\nu_{\ss
\ell}-\slept_i}^{R} \quad \quad [C_{\slept_1-H-\slept^-_i} \; C_{\schi^-_1-\nu_{\ss
\ell}-\slept_i}^{R}] \nonumber \\
{\cal T}_{\rm I}\!\!\times\!\!{\cal T}_{\rm I} &=&  (\mxi (f_{1LL}(i)
(f_{1LR}(j) \mchar (-\mstau^2 + t) + 
                    f_{1LL}(j) \mxj (-\mchar^2 + u)) \nl + 
              f_{1RR}(i) (f_{1RL}(j) \mchar (-\mstau^2 + t) + 
                    f_{1RR}(j) \mxj (-\mchar^2 + u))) \nl + 
        f_{1LR}(i) (f_{1LL}(j) \mchar \mxj (-\mstau^2 + t) + 
              f_{1LR}(j) (\mstau^4 + \mchar^2 (2 \mstau^2 - t) \nl + s t - 
                    \mstau^2 (s + t + u))) + 
        f_{1RL}(i) (f_{1RR}(j) \mchar \mxj (-\mstau^2 + t) \nl + 
              f_{1RL}(j) (\mstau^4 + \mchar^2 (2 \mstau^2 - t) + s t - 
                    \mstau^2 (s + t + u)))) \nl /((\mxi^2 - 
            t) (-\mxj^2 + t))  \nonumber \\
{\cal T}_{\rm II}\!\!\times\!\!{\cal T}_{\rm II} &=&  (f_{2L}(i) f_{2L}(j) +
f_{2R}(i) f_{2R}(j)) (\mchar^2 - u)/(-m_{\slept_i}^2 
+ u) (-m_{\slept_j}^2 + u)  \nonumber \\
{\cal T}_{\rm I}\!\!\times\!\!{\cal T}_{\rm II} &=&  (f_{1LR}(i) f_{2L}(j)
\mchar (\mstau^2 - t) + 
        f_{1RL}(i) f_{2R}(j) \mchar (\mstau^2 - t) \nl + (f_{1LL}(i) f_{2L}(j) + 
              f_{1RR}(i) f_{2R}(j)) \mxi (\mchar^2 - u))/((\mxi^2 - 
            t) (m_{\slept_j}^2 - u))  \nonumber \\
\tsq &=&   \sum_{i,j}  \left( {\cal T}_{\rm I}\!\!\times\!\!{\cal T}_{\rm I}
+   {\cal T}_{\rm II}\!\!\times\!\!{\cal T}_{\rm II}
+  2  {\cal T}_{\rm I}\!\!\times\!\!{\cal T}_{\rm II} \right)
\end{eqnarray}

\subsection*{$\slept_1 \schi^+_1 \longrightarrow \nu_{\ss \ell} A$}
 I. $t$-channel $\schi_{1,2}$ exchange \hfill\\
 II. $u$-channel $\slept_2$  exchange  
\begin{eqnarray}
f_{1LL}(i) &=& C_{\slept_1-\nu_{\ss \ell}-\schi^-_i}^{L} \;
C_{\schi^+_1-\schi^+_i-A}^{L} \nonumber \\
f_{1LR}(i) &=& C_{\slept_1-\nu_{\ss \ell}-\schi^-_i}^{L} \;
C_{\schi^+_1-\schi^+_i-A}^{R} \nonumber \\
f_{1RL}(i) &=& C_{\slept_1-\nu_{\ss \ell}-\schi^-_i}^{R} \;
C_{\schi^+_1-\schi^+_i-A}^{L} \nonumber \\
f_{1RR}(i) &=& C_{\slept_1-\nu_{\ss \ell}-\schi^-_i}^{R} \;
C_{\schi^+_1-\schi^+_i-A}^{R} \nonumber \\
f_{2L} &=& C_{\slept_1-A-\slept^-_2} \; C_{\schi^-_1-\nu_{\ss
\ell}-\slept_2}^{L} \nonumber \\
f_{2R} &=& C_{\slept_1-A-\slept^-_2} \; C_{\schi^-_1-\nu_{\ss
\ell}-\slept_2}^{R} \nonumber \\
{\cal T}_{\rm I}\!\!\times\!\!{\cal T}_{\rm I} &=& (\mxi (f_{1LL}(i)
(f_{1LR}(j) \mchar (-\mstau^2 + t) + 
                    f_{1LL}(j) \mxj (-\mchar^2 + u)) \nl + 
              f_{1RR}(i) (f_{1RL}(j) \mchar (-\mstau^2 + t) + 
                    f_{1RR}(j) \mxj (-\mchar^2 + u))) \nl + 
        f_{1LR}(i) (f_{1LL}(j) \mchar \mxj (-\mstau^2 + t) + 
              f_{1LR}(j) (\mstau^4 + \mchar^2 (2 \mstau^2 - t) \nl + s t - 
                    \mstau^2 (s + t + u))) + 
        f_{1RL}(i) (f_{1RR}(j) \mchar \mxj (-\mstau^2 + t) \nl + 
              f_{1RL}(j) (\mstau^4 + \mchar^2 (2 \mstau^2 - t) + s t - 
                    \mstau^2 (s + t + u)))) \nl /((\mxi^2 - 
            t) (-\mxj^2 + t))   \nonumber \\
{\cal T}_{\rm II}\!\!\times\!\!{\cal T}_{\rm II} &=&  (f_{2L}^2 + f_{2R}^2)
(\mchar^2 - u)/(m_{\slept_2}^2 - u)^2  \nonumber \\
{\cal T}_{\rm I}\!\!\times\!\!{\cal T}_{\rm II} &=&   (f_{1LR}(i) f_{2L} \mchar
(\mstau^2 - t) + 
        f_{1RL}(i) f_{2R} \mchar (\mstau^2 - t) + (f_{1LL}(i) f_{2L} \nl + 
              f_{1RR}(i) f_{2R}) \mxi (\mchar^2 - u))/((\mxi^2 - 
            t) (m_{\slept_2}^2 - u))    \nonumber \\ 
\tsq &=&   \sum_{i,j} {\cal T}_{\rm I}\!\!\times\!\!{\cal T}_{\rm I} 
+  {\cal T}_{\rm II}\!\!\times\!\!{\cal T}_{\rm II} 
+ 2  \sum_i  {\cal T}_{\rm I}\!\!\times\!\!{\cal T}_{\rm II} 
\end{eqnarray}

\subsection*{$\slept_1 \schi^+_1 \longrightarrow \ell W^+$}
 I. $s$-channel $\nu_{\ss \ell}$ annihilation \hfill\\
 II. $t$-channel $\schi^0_{(1,2,3,4)}$  exchange \hfill\\
Note the L-R switch for the neutralino couplings below.
\begin{eqnarray}
f_{1L} &=&  C_{\slept_1-\nu_{\ss \ell}-\schi^-_1}^{L} \; C_{\nu_{\ss
\ell}-\ell-W^+} \nonumber \\
f_{1R} &=&  C_{\slept_1-\nu_{\ss \ell}-\schi^-_1}^{R} \; C_{\nu_{\ss
\ell}-\ell-W^+} \nonumber \\
f_{2LL}(i) &=&  C_{\slept_1-\ell-\schi^0_i}^{R} \;
C_{W-\schi^-_1-\schi^0_i}^{L}  \nonumber \\
f_{2LR}(i) &=&  C_{\slept_1-\ell-\schi^0_i}^{R} \;
C_{W-\schi^-_1-\schi^0_i}^{R} \nonumber \\
f_{2RL}(i) &=&  C_{\slept_1-\ell-\schi^0_i}^{L} \;
C_{W-\schi^-_1-\schi^0_i}^{L} \nonumber \\
f_{2RR}(i) &=&  C_{\slept_1-\ell-\schi^0_i}^{L} \;
C_{W-\schi^-_1-\schi^0_i}^{R} \nonumber \\
% Note the L-R switch for ttn
{\cal T}_{\rm I}\!\!\times\!\!{\cal T}_{\rm I} &=&  1/(\mw^2 s^2)(f_{1R}^2
(\mstau^4 (\mtau^2 - s) + 
          \mchar^4 (-\mtau^2 + s) + 
          s (-\mw^4 + (\mtau^2 - s) u \nl + 
                \mw^2 (s - t + u)) + 
          \mchar^2 (-2 \mw^4 + 
                2 \mw^2 s + (\mtau^2 - s) (t + u)) \nl + 
          \mstau^2 (2 \mw^4 - 
                2 \mw^2 s - (\mtau^2 - s) (s + t + u))))  \nonumber \\
{\cal T}_{\rm II}\!\!\times\!\!{\cal T}_{\rm II} &=&  (\nevalsi (f_{2RL}(i) (3
f_{2RR}(j) \mchar \mw^2 (\mstau^2 - \mtau^2 - 
                          t) \nl + 
                    \nevalsj (6 f_{2LR}(j) \mchar \mtau \mw^2 + 
                          f_{2RL}(j) (\mw^4 + 
                                \mchar^2 (\mtau^2 - 
                                      s) + (-\mtau^2 + s) t \nl - 
                                \mw^2 (s + t - u)))) + 
              f_{2LL}(j) (3 f_{2LR}(i) \mchar \mw^2 (\mstau^2 - \mtau^2 - 
                          t) \nl + 
                    f_{2RL}(i) \mtau (-2 \mw^4 + 2 \mw^2 t + 
                          t (\mstau^2 + \mtau^2 - s - u) + 
                          \mchar^2 (-\mstau^2 - \mtau^2 \nl + s + 
                                u))) + 
              f_{2LR}(i) (\nevalsj (6 f_{2RL}(j) \mchar \mtau \mw^2 + 
                          f_{2LR}(j) (\mw^4 + 
                                \mchar^2 (\mtau^2 - 
                                      s) \nl + (-\mtau^2 + s) t - 
                                \mw^2 (s + t - u))) + 
                    f_{2RR}(j) \mtau (-2 \mw^4 + 2 \mw^2 t + 
                          t (\mstau^2 + \mtau^2 \nl - s - u) + 
                          \mchar^2 (-\mstau^2 - \mtau^2 + s + 
                                u)))) + 
      f_{2LL}(i) (3 \mchar \mw^2 (f_{2LR}(j) \nevalsj (\mstau^2 \nl - \mtau^2 - 
                          t) + 2 f_{2RR}(j) \mtau t) + 
              f_{2RL}(j) \mtau \nevalsj (-2 \mw^4 + 2 \mw^2 t + 
                    t (\mstau^2 + \mtau^2 \nl - s - u) + 
                    \mchar^2 (-\mstau^2 - \mtau^2 + s + u)) + 
              f_{2LL}(j) (2 \mtau^2 \mw^4 + \mstau^4 t - \mtau^4 t \nl - 
                    2 \mtau^2 \mw^2 t + \mw^4 t + \mtau^2 s t + 
                    \mw^2 s t - \mw^2 t^2 + \mtau^2 t u - \mw^2 t u + 
                    t^2 u \nl + 
                    \mchar^2 (-\mstau^4 + \mtau^4 - t u - 
                          \mtau^2 (s + u) + 
                          \mstau^2 (s + t + u)) - 
                    \mstau^2 (2 \mw^4 \nl - 2 \mw^2 t + 
                          t (s + t + u)))) + 
        f_{2RR}(i) (3 \mchar \mw^2 (f_{2RL}(j) \nevalsj (\mstau^2 - \mtau^2 - 
                          t) \nl + 2 f_{2LL}(j) \mtau t) + 
              f_{2LR}(j) \mtau \nevalsj (-2 \mw^4 + 2 \mw^2 t + 
                    t (\mstau^2 + \mtau^2 - s - u) \nl + 
                    \mchar^2 (-\mstau^2 - \mtau^2 + s + u)) + 
              f_{2RR}(j) (2 \mtau^2 \mw^4 + \mstau^4 t - \mtau^4 t - 
                    2 \mtau^2 \mw^2 t \nl + \mw^4 t + \mtau^2 s t + 
                    \mw^2 s t - \mw^2 t^2 + \mtau^2 t u - \mw^2 t u + 
                    t^2 u \nl + 
                    \mchar^2 (-\mstau^4 + \mtau^4 - t u - 
                          \mtau^2 (s + u) + 
                          \mstau^2 (s + t + u)) \nl - 
                    \mstau^2 (2 \mw^4 - 2 \mw^2 t + 
                          t (s + t + 
                                u))))) \nl /(\mw^2 (\nevalsi^2 - 
            t) (-\nevalsj^2 + t))  \nonumber \\
{\cal T}_{\rm I}\!\!\times\!\!{\cal T}_{\rm II} &=&    (f_{1R} (-2 f_{2LL}(j)
\mchar \mtau (\mstau^4 + 4 \mw^4 - 
                  \mw^2 s - \mtau^2 t - \mw^2 t + s t \nl + 
                  \mstau^2 (\mtau^2 + 4 \mw^2 - s - t - 2 u) + 
                  \mchar^2 (\mstau^2 + \mtau^2 - s - u) - \mtau^2 u \nl - 
                  3 \mw^2 u + s u + t u + u^2) + 
            2 \nevalsj (3 f_{2LR}(j) \mtau \mw^2 (\mchar^2 - \mstau^2 + 
                        s) \nl + 
                  f_{2RL}(j) \mchar (2 \mw^4 + \mchar^2 (\mtau^2 - s) + 
                        \mstau^2 (\mtau^2 - s) - 2 \mw^2 s - 
                        \mtau^2 t + s t \nl - \mtau^2 u + s u)) + 
            f_{2RR}(j) (-5 \mtau^2 \mw^2 s - \mw^4 s + \mw^2 s^2 - 
                  \mtau^2 \mw^2 t - \mw^4 t \nl - 2 \mw^2 s t + 
                  \mw^2 t^2 - \mstau^4 (s + t - u) + 
                  \mtau^2 \mw^2 u + \mw^4 u + \mtau^2 s u + 
                  \mw^2 s u \nl - s^2 u + \mtau^2 t u + \mw^2 t u - 
                  t^2 u - \mtau^2 u^2 - 2 \mw^2 u^2 + u^3 + 
                  \mchar^2 ((-2 \mtau^2 + s \nl + t - u) u + 
                        \mw^2 (-s - 5 t + u) + 
                        \mstau^2 (4 \mtau^2 + 8 \mw^2 - 3 s - t + 
                              u)) \nl + 
                  \mstau^2 (4 \mw^4 + s^2 + 2 s t + t^2 + s u + 
                        t u - 2 u^2 + 
                        \mtau^2 (8 \mw^2 - s - 3 t + u) \nl - 
                        2 \mw^2 (s + t + 
                              u)))))/(2 \mw^2 s (\nevalsj^2 - t))  \nonumber \\ 
\tsq &=&    {\cal T}_{\rm I}\!\!\times\!\!{\cal T}_{\rm I} 
+  \sum_{i,j} {\cal T}_{\rm II}\!\!\times\!\!{\cal T}_{\rm II} 
+ 2  \sum_i  {\cal T}_{\rm I}\!\!\times\!\!{\cal T}_{\rm II} 
\end{eqnarray}

\subsection*{$\slept_1 \schi^+_1 \longrightarrow \ell H^+$}
 I. $s$-channel $\nu_{\ss \ell}$ annihilation \hfill\\
 II. $t$-channel $\schi^0_{(1,2,3,4)}$  exchange \hfill\\ 
Note the L-R switch for the neutralino couplings below.
\begin{eqnarray}
f_{1L} &=&  C_{\slept_1-\nu_{\ss \ell}-\schi^-_1}^{L} \; C_{\nu_{\ss
\ell}-\ell-H^+}  \nonumber \\
f_{1R} &=&  C_{\slept_1-\nu_{\ss \ell}-\schi^-_1}^{R} \; C_{\nu_{\ss
\ell}-\ell-H^+} \nonumber \\
f_{2LL}(i) &=& C_{\slept_1-\ell-\schi^0_i}^{R} \;
C_{H^+-\schi^-_1-\schi^0_i}^{L}   \nonumber \\
f_{2LR}(i) &=&  C_{\slept_1-\ell-\schi^0_i}^{R} \;
C_{H^+-\schi^-_1-\schi^0_i}^{R} \nonumber \\
f_{2RL}(i) &=&  C_{\slept_1-\ell-\schi^0_i}^{L} \;
C_{H^+-\schi^-_1-\schi^0_i}^{L} \nonumber \\
f_{2RR}(i) &=&  C_{\slept_1-\ell-\schi^0_i}^{L} \;
C_{H^+-\schi^-_1-\schi^0_i}^{R} \nonumber \\
%   Note the L-R switch for ttn
{\cal T}_{\rm I}\!\!\times\!\!{\cal T}_{\rm I} &=&  (f_{1R}^2 (\mchar^4 -
\mstau^4 + s (\mtau^2 - t) + 
          \mchar^2 (2 \mtau^2 - t - u) \nl + 
          \mstau^2 (-2 \mtau^2 + s + t + u)))/s^2  \nonumber \\
{\cal T}_{\rm II}\!\!\times\!\!{\cal T}_{\rm II} &=&  (\nevalsi (f_{2RR}(i)
(f_{2RL}(j) \mchar (-\mstau^2 + \mtau^2 + 
                          t) - 
                    \nevalsj (2 f_{2LL}(j) \mchar \mtau \nl + 
                          f_{2RR}(j) (\mchar^2 + \mtau^2 - u)) + 
                    f_{2LR}(j) \mtau (2 \mchar^2 + \mstau^2 + \mtau^2 - s - 
                          u)) \nl + 
              f_{2LL}(i) (f_{2LR}(j) \mchar (-\mstau^2 + \mtau^2 + t) - 
                    \nevalsj (2 f_{2RR}(j) \mchar \mtau  \nl + 
                          f_{2LL}(j) (\mchar^2 + \mtau^2 - u)) + 
                    f_{2RL}(j) \mtau (2 \mchar^2 + \mstau^2 + \mtau^2 - s - 
                          u))) \nl + 
        f_{2LR}(i) (f_{2LL}(j) \mchar \nevalsj (-\mstau^2 + \mtau^2 + t) + 
              \mtau (-2 f_{2RL}(j) \mchar t \nl + 
                    f_{2RR}(j) \nevalsj (2 \mchar^2 + \mstau^2 + \mtau^2 - s - 
                          u)) + 
              f_{2LR}(j) (\mstau^4 - \mtau^4 + \mtau^2 s \nl + 
                    \mchar^2 (2 \mstau^2 - 2 \mtau^2 - t) + s t + 
                    \mtau^2 u - \mstau^2 (s + t + u))) \nl + 
        f_{2RL}(i) (f_{2RR}(j) \mchar \nevalsj (-\mstau^2 + \mtau^2 + t) + 
              \mtau (-2 f_{2LR}(j) \mchar t \nl + 
                    f_{2LL}(j) \nevalsj (2 \mchar^2 + \mstau^2 + \mtau^2 - s - 
                          u)) + 
              f_{2RL}(j) (\mstau^4 - \mtau^4 + \mtau^2 s \nl + 
                    \mchar^2 (2 \mstau^2 - 2 \mtau^2 - t) + s t + 
                    \mtau^2 u - 
                    \mstau^2 (s + t + u)))) \nl /((\nevalsi^2 - 
            t) (-\nevalsj^2 + t))  \nonumber \\
{\cal T}_{\rm I}\!\!\times\!\!{\cal T}_{\rm II} &=&   -(f_{1R} (\nevalsj
(f_{2RR}(j) \mtau (\mchar^2 - \mstau^2 + s) + 
               f_{2LL}(j) \mchar (\mchar^2 + \mstau^2 \nl + 2 \mtau^2 - t - 
                          u)) - 
              f_{2LR}(j) (\mstau^4 + \mchar^2 (\mstau^2 + \mtau^2) + 
                    s t \nl + \mstau^2 (\mtau^2 - s - t - u)) + 
              f_{2RL}(j) \mchar \mtau (-2 \mchar^2 - 2 \mtau^2 + s + t + 
                    u))) \nl /(s (\nevalsj^2 - t))    \nonumber \\ 
\tsq &=&    {\cal T}_{\rm I}\!\!\times\!\!{\cal T}_{\rm I} 
+  \sum_{i,j} {\cal T}_{\rm II}\!\!\times\!\!{\cal T}_{\rm II} 
+ 2  \sum_i  {\cal T}_{\rm I}\!\!\times\!\!{\cal T}_{\rm II} 
\end{eqnarray}

\subsection*{$\slept_1 \schi^+_1 \longrightarrow W^- \ell^+$}
 I. $t$-channel $\widetilde{\nu}_{\ss \ell}$ exchange 
\begin{eqnarray}
f_{1L} &=&  C_{\ell-{\widetilde \nu}_{\ss \ell}-\schi^-_1}^{L} \; C_{{\widetilde
\nu}_{\ss \ell}-\slept_1-W}  \nonumber \\
f_{1R} &=&  C_{\ell-{\widetilde \nu}_{\ss \ell}-\schi^-_1}^{R} \; C_{{\widetilde
\nu}_{\ss \ell}-\slept_1-W} \nonumber \\
{\cal T}_{\rm I}\!\!\times\!\!{\cal T}_{\rm I} &=&  ((-4 f_{1L} f_{1R} \mchar 
\mtau + f_{1L}^2 (\mchar^2 + \mtau^2 - t) + 
          f_{1R}^2 (\mchar^2 + \mtau^2 - 
                t)) (\mstau^4 \nl + (\mw^2 - t)^2 - 
          2 \mstau^2 (\mw^2 + t)))/(\mw^2 (\msn^2 - t)^2)  \nonumber \\ 
\tsq &=&    {\cal T}_{\rm I}\!\!\times\!\!{\cal T}_{\rm I}  
\end{eqnarray}

\subsection*{$\slept_1 \schi^+_1 \longrightarrow H^- \ell^+$}
 I. $t$-channel $\widetilde{\nu}_{\ss \ell}$ exchange 
\begin{eqnarray}
f_{1L} &=&  C_{\ell-{\widetilde \nu}_{\ss \ell}-\schi^-_1}^{L} \; C_{{\widetilde
\nu}_{\ss \ell}-\slept_1-H^+} \nonumber \\
f_{1R} &=&  C_{\ell-{\widetilde \nu}_{\ss \ell}-\schi^-_1}^{R} \; C_{{\widetilde
\nu}_{\ss \ell}-\slept_1-H^+} \nonumber \\
{\cal T}_{\rm I}\!\!\times\!\!{\cal T}_{\rm I} &=&  (-4 f_{1L} f_{1R} \mchar
     \mtau + f_{1L}^2 (\mchar^2 + \mtau^2 - t) + 
      f_{1R}^2 (\mchar^2 + \mtau^2 - t))/(t-\msn^2)^2  \nonumber \\ 
\tsq &=&    {\cal T}_{\rm I}\!\!\times\!\!{\cal T}_{\rm I}  
\end{eqnarray}

\subsection*{$\slept_1 \schi^-_1 \longrightarrow \ell W^-$}
 I. $t$-channel $\schi^0_{(1,2,3,4)}$  exchange \hfill\\ 
 II. $u$-channel $\nu_{\ss \ell}$  exchange \hfill\\
Note the L-R switch for the couplings below.
\begin{eqnarray}
f_{1LL}(i) &=&  C_{\slept_1-\ell-\schi^0_i}^{R} \;
C_{W-\schi^-_1-\schi^0_i}^{L}  \nonumber \\
f_{1LR}(i) &=&  C_{\slept_1-\ell-\schi^0_i}^{R} \;
C_{W-\schi^-_1-\schi^0_i}^{R} \nonumber \\
f_{1RL}(i) &=&  C_{\slept_1-\ell-\schi^0_i}^{L} \;
C_{W-\schi^-_1-\schi^0_i}^{L} \nonumber \\
f_{1RR}(i) &=&  C_{\slept_1-\ell-\schi^0_i}^{L} \;
C_{W-\schi^-_1-\schi^0_i}^{R} \nonumber \\
%   Note the L-R switch for ttn
f_{2L} &=&  C_{\ell-{\widetilde \nu}_{\ss \ell}-\schi^-_1}^{R} \; C_{{\widetilde
\nu}_{\ss \ell}-\slept_1-W}   \nonumber \\
f_{2R} &=&  C_{\ell-{\widetilde \nu}_{\ss \ell}-\schi^-_1}^{L} \; C_{{\widetilde
\nu}_{\ss \ell}-\slept_1-W}  \nonumber \\
%   Note the L-R switch 
{\cal T}_{\rm I}\!\!\times\!\!{\cal T}_{\rm I} &=&  (\nevalsi (f_{1RL}(i) (3
      f_{1RR}(j) \mchar \mw^2 (\mstau^2 - \mtau^2 - 
                          t) \nl + 
                    \nevalsj (6 f_{1LR}(j) \mchar \mtau \mw^2 + 
                          f_{1RL}(j) (\mw^4 + 
                                \mchar^2 (\mtau^2 - 
                                      s) \nl + (-\mtau^2 + s) t - 
                                \mw^2 (s + t - u)))) + 
              f_{1LL}(j) (3 f_{1LR}(i) \mchar \mw^2 (\mstau^2 - \mtau^2 - 
                          t) \nl + 
                    f_{1RL}(i) \mtau (-2 \mw^4 + 2 \mw^2 t + 
                          t (\mstau^2 + \mtau^2 - s - u) \nl + 
                          \mchar^2 (-\mstau^2 - \mtau^2 + s + 
                                u))) + 
              f_{1LR}(i) (\nevalsj (6 f_{1RL}(j) \mchar \mtau \mw^2 \nl + 
                          f_{1LR}(j) (\mw^4 + 
                                \mchar^2 (\mtau^2 - 
                                      s) + (-\mtau^2 + s) t - 
                                \mw^2 (s + t - u))) \nl + 
                    f_{1RR}(j) \mtau (-2 \mw^4 + 2 \mw^2 t + 
                          t (\mstau^2 + \mtau^2 - s - u) \nl + 
                          \mchar^2 (-\mstau^2 - \mtau^2 + s + 
                                u)))) + 
        f_{1LL}(i) (3 \mchar \mw^2 (f_{1LR}(j) \nevalsj (\mstau^2 - \mtau^2 \nl 
	-  t) + 2 f_{1RR}(j) \mtau t) + 
              f_{1RL}(j) \mtau \nevalsj (-2 \mw^4 + 2 \mw^2 t + 
                    t (\mstau^2 + \mtau^2 - s - u) \nl + 
                    \mchar^2 (-\mstau^2 - \mtau^2 + s + u)) + 
              f_{1LL}(j) (2 \mtau^2 \mw^4 + \mstau^4 t - \mtau^4 t \nl - 
                    2 \mtau^2 \mw^2 t + \mw^4 t + \mtau^2 s t + 
                    \mw^2 s t - \mw^2 t^2 + \mtau^2 t u - \mw^2 t u + 
                    t^2 u \nl + 
                    \mchar^2 (-\mstau^4 + \mtau^4 - t u - 
                          \mtau^2 (s + u) + 
                          \mstau^2 (s + t + u)) - 
                    \mstau^2 (2 \mw^4 \nl - 2 \mw^2 t + 
                          t (s + t + u)))) + 
        f_{1RR}(i) (3 \mchar \mw^2 (f_{1RL}(j) \nevalsj (\mstau^2 - \mtau^2 - 
                          t) \nl + 2 f_{1LL}(j) \mtau t) + 
              f_{1LR}(j) \mtau \nevalsj (-2 \mw^4 + 2 \mw^2 t + 
                    t (\mstau^2 + \mtau^2 - s - u) \nl + 
                    \mchar^2 (-\mstau^2 - \mtau^2 + s + u)) + 
              f_{1RR}(j) (2 \mtau^2 \mw^4 + \mstau^4 t - \mtau^4 t \nl - 
                    2 \mtau^2 \mw^2 t + \mw^4 t + \mtau^2 s t + 
                    \mw^2 s t - \mw^2 t^2 + \mtau^2 t u - \mw^2 t u + 
                    t^2 u \nl + 
                    \mchar^2 (-\mstau^4 + \mtau^4 - t u - 
                          \mtau^2 (s + u) + 
                          \mstau^2 (s + t + u)) - 
                    \mstau^2 (2 \mw^4 \nl - 2 \mw^2 t + 
                          t (s + t + 
                                u)))))/(\mw^2 (\nevalsi^2 - 
            t) (-\nevalsj^2 + t))  \nonumber \\
{\cal T}_{\rm II}\!\!\times\!\!{\cal T}_{\rm II} &=&  ((4 f_{2L} f_{2R} \mchar
        \mtau + 
          f_{2L}^2 (\mchar^2 + \mtau^2 - u) + 
          f_{2R}^2 (\mchar^2 + \mtau^2 - 
                u)) (\mstau^4 \nl + (\mw^2 - u)^2 - 
          2 \mstau^2 (\mw^2 + u)))/(\mw^2 (\msn^2 - u)^2)  \nonumber \\
{\cal T}_{\rm I}\!\!\times\!\!{\cal T}_{\rm II} &=&     (2 \nevalsi (f_{1LR}(i)
      (f_{2R} \mtau (\mw^4 - 2 \mw^2 s + 
                          \mstau^2 (3 \mw^2 - t) - \mw^2 t \nl + 
                          \mchar^2 (\mstau^2 + 3 \mw^2 - u) - 
                          \mw^2 u + t u) + 
                    f_{2L} \mchar (-\mw^4 + \mw^2 s \nl + 
                          \mstau^2 (-\mtau^2 - 3 \mw^2 + s) + 
                          2 \mw^2 t + \mw^2 u - s u + 
                          \mtau^2 (-3 \mw^2 + u))) \nl + 
              f_{1RL}(i) (f_{2L} \mtau (\mw^4 - 2 \mw^2 s + 
                          \mstau^2 (3 \mw^2 - t) - \mw^2 t + 
                          \mchar^2 (\mstau^2 + 3 \mw^2 \nl - u) - 
                          \mw^2 u + t u) + 
                    f_{2R} \mchar (-\mw^4 + \mw^2 s + 
                          \mstau^2 (-\mtau^2 - 3 \mw^2 + s) \nl + 
                          2 \mw^2 t + \mw^2 u - s u + 
                          \mtau^2 (-3 \mw^2 + u)))) + 
        f_{1LL}(i) (2 f_{2R} \mchar \mtau (\mstau^4 \nl + 2 \mw^4 - \mw^2 s - 
                    2 \mw^2 t + 
                    \mstau^2 (\mtau^2 + \mw^2 - s - 2 u) + 
                    \mtau^2 (3 \mw^2 - u) \nl - 3 \mw^2 u + s u + 
                    u^2) + 
              f_{2L} (5 \mtau^2 \mw^2 s + \mw^4 s - \mw^2 s^2 + 
                    \mtau^2 \mw^2 t - \mw^4 t \nl + \mw^2 t^2 + 
                    \mstau^4 (s - t - u) - \mtau^2 \mw^2 u - 
                    \mw^4 u - \mtau^2 s u - \mw^2 s u + s^2 u \nl - 
                    \mtau^2 t u + \mw^2 t u - t^2 u + \mtau^2 u^2 + 
                    2 \mw^2 u^2 - u^3 + 
                    \mstau^2 (-s^2 + t^2 \nl + 
                          \mtau^2 (-8 \mw^2 + s + t - u) - s u + 
                          t u + 2 u^2 + 2 \mw^2 (s - t + u)) \nl + 
                    \mchar^2 (2 \mstau^4 - (\mw^2 - u) (-s + 
                                t + u) + 
                          \mtau^2 (-6 \mw^2 + 2 u) \nl - 
                          \mstau^2 (2 \mtau^2 + 2 \mw^2 - s + t + 
                                3 u)))) + 
        f_{1RR}(i) (2 f_{2L} \mchar \mtau (\mstau^4 + 2 \mw^4 \nl - \mw^2 s - 
                    2 \mw^2 t + 
                    \mstau^2 (\mtau^2 + \mw^2 - s - 2 u) + 
                    \mtau^2 (3 \mw^2 - u) - 3 \mw^2 u + s u \nl + 
                    u^2) + 
              f_{2R} (5 \mtau^2 \mw^2 s + \mw^4 s - \mw^2 s^2 + 
                    \mtau^2 \mw^2 t - \mw^4 t + \mw^2 t^2 \nl + 
                    \mstau^4 (s - t - u) - \mtau^2 \mw^2 u - 
                    \mw^4 u - \mtau^2 s u - \mw^2 s u + s^2 u - 
                    \mtau^2 t u \nl + \mw^2 t u - t^2 u + \mtau^2 u^2 + 
                    2 \mw^2 u^2 - u^3 + 
                    \mstau^2 (-s^2 + t^2 \nl + 
                          \mtau^2 (-8 \mw^2 + s + t - u) - s u + 
                          t u + 2 u^2 + 2 \mw^2 (s - t + u)) \nl + 
                    \mchar^2 (2 \mstau^4 - (\mw^2 - u) (-s + 
                                t + u) + 
                          \mtau^2 (-6 \mw^2 + 2 u) \nl - 
                          \mstau^2 (2 \mtau^2 + 2 \mw^2 - s + t + 
                                3 u)))))/(2 \mw^2 
         (\nevalsi^2 - t) (\msn^2 - u))  \nonumber \\ 
\tsq &=&  \sum_{i,j}  {\cal T}_{\rm I}\!\!\times\!\!{\cal T}_{\rm I} 
+   {\cal T}_{\rm II}\!\!\times\!\!{\cal T}_{\rm II} 
+ 2  \sum_i  {\cal T}_{\rm I}\!\!\times\!\!{\cal T}_{\rm II} 
\end{eqnarray}

\subsection*{$\slept_1 \schi^-_1 \longrightarrow \ell H^-$}
 I. $t$-channel $\schi^0_{(1,2,3,4)}$  exchange \hfill\\ 
 II. $u$-channel $\nu_{\ss \ell}$  exchange \hfill\\
Note the L-R switch for the couplings below.
\begin{eqnarray}
f_{1LL}(i) &=& C_{\slept_1-\ell-\schi^0_i}^{R} \;
C_{H^+-\schi^-_1-\schi^0_i}^{L}   \nonumber \\
f_{1LR}(i) &=&  C_{\slept_1-\ell-\schi^0_i}^{R} \;
C_{H^+-\schi^-_1-\schi^0_i}^{R} \nonumber \\
f_{1RL}(i) &=&  C_{\slept_1-\ell-\schi^0_i}^{L} \;
C_{H^+-\schi^-_1-\schi^0_i}^{L} \nonumber \\
f_{1RR}(i) &=&  C_{\slept_1-\ell-\schi^0_i}^{L} \;
C_{H^+-\schi^-_1-\schi^0_i}^{R} \nonumber \\
%   Note the L-R switch for ttn
f_{2L} &=&  C_{\ell-{\widetilde \nu}_{\ss \ell}-\schi^-_1}^{R} \; C_{{\widetilde
\nu}_{\ss \ell}-\slept_1-H^+}  \nonumber \\
f_{2R} &=&  C_{\ell-{\widetilde \nu}_{\ss \ell}-\schi^-_1}^{L} \; C_{{\widetilde
\nu}_{\ss \ell}-\slept_1-H^+}  \nonumber \\
%  (Note the L-R switch) 
{\cal T}_{\rm I}\!\!\times\!\!{\cal T}_{\rm I} &=&  (\nevalsi (-f_{1RL}(i)
          (f_{1RR}(j) \mchar (\mstau^2 - \mtau^2 - 
                          t) + 
                    \nevalsj (2 f_{1LR}(j) \mchar \mtau \nl + 
                          f_{1RL}(j) (\mchar^2 + \mtau^2 - u))) + 
              f_{1LL}(j) (f_{1LR}(i) \mchar (-\mstau^2 + \mtau^2 + t) \nl + 
                    f_{1RL}(i) \mtau (2 \mchar^2 + \mstau^2 + \mtau^2 - s - 
                          u)) - 
              f_{1LR}(i) (\nevalsj (2 f_{1RL}(j) \mchar \mtau \nl + 
                          f_{1LR}(j) (\mchar^2 + \mtau^2 - u)) - 
                    f_{1RR}(j) \mtau (2 \mchar^2 + \mstau^2 + \mtau^2 - s - 
                          u))) \nl + 
        f_{1LL}(i) (f_{1LR}(j) \mchar \nevalsj (-\mstau^2 + \mtau^2 + t) + 
              \mtau (-2 f_{1RR}(j) \mchar t \nl + 
                    f_{1RL}(j) \nevalsj (2 \mchar^2 + \mstau^2 + \mtau^2 - s - 
                          u)) + 
              f_{1LL}(j) (\mstau^4 - \mtau^4 + \mtau^2 s \nl + 
                    \mchar^2 (2 \mstau^2 - 2 \mtau^2 - t) + s t + 
                    \mtau^2 u - \mstau^2 (s + t + u))) \nl + 
        f_{1RR}(i) (f_{1RL}(j) \mchar \nevalsj (-\mstau^2 + \mtau^2 + t) + 
              \mtau (-2 f_{1LL}(j) \mchar t \nl + 
                    f_{1LR}(j) \nevalsj (2 \mchar^2 + \mstau^2 + \mtau^2 - s - 
                          u)) + 
              f_{1RR}(j) (\mstau^4 - \mtau^4 + \mtau^2 s \nl + 
                    \mchar^2 (2 \mstau^2 - 2 \mtau^2 - t) + s t + 
                    \mtau^2 u - 
                    \mstau^2 (s + t + u)))) \nl /((\nevalsi^2 - 
            t) (-\nevalsj^2 + t))  \nonumber \\
{\cal T}_{\rm II}\!\!\times\!\!{\cal T}_{\rm II} &=&  (4 f_{2L} f_{2R} \mchar
      \mtau + 
      f_{2L}^2 (\mchar^2 + \mtau^2 - u) + 
      f_{2R}^2 (\mchar^2 + \mtau^2 - u))/(\msn^2 - u)^2  \nonumber \\
{\cal T}_{\rm I}\!\!\times\!\!{\cal T}_{\rm II} &=&   (\nevalsi (f_{1LR}(i) (2
      f_{2R} \mchar \mtau + 
                    f_{2L} (\mchar^2 + \mtau^2 - u)) + 
              f_{1RL}(i) (2 f_{2L} \mchar \mtau \nl + 
                    f_{2R} (\mchar^2 + \mtau^2 - u))) + 
        f_{1RR}(i) (f_{2R} \mchar (\mstau^2 - \mtau^2 - t) \nl + 
              f_{2L} \mtau (-2 \mchar^2 - \mstau^2 - \mtau^2 + s + 
                    u)) + 
        f_{1LL}(i) (f_{2L} \mchar (\mstau^2 - \mtau^2 - t) \nl + 
              f_{2R} \mtau (-2 \mchar^2 - \mstau^2 - \mtau^2 + s + 
                    u))) \nl /((\nevalsi^2 - t) (\msn^2 - u))    \nonumber \\ 
\tsq &=&  \sum_{i,j}  {\cal T}_{\rm I}\!\!\times\!\!{\cal T}_{\rm I} 
+   {\cal T}_{\rm II}\!\!\times\!\!{\cal T}_{\rm II} 
+ 2  \sum_i  {\cal T}_{\rm I}\!\!\times\!\!{\cal T}_{\rm II} 
\end{eqnarray}

\setcounter{equation}{0}
\renewcommand{\theequation}{D\arabic{equation}} 
\section*{Appendix D: Neutralino-Sneutrino Coannihilation} 

Below is the list of the amplitudes squared for neutralino-sneutrino
coannihilation. Note that, for identical-particle final states,
one needs to divide them by two when performing the 
momentum integrations. Below, $\widetilde \nu$ refers to $\widetilde
\nu_{e,\mu}$. 

\subsection*{$\snu \snu^\ast \longrightarrow W W$}
 I. $s$-channel $H$ annihilation \hfill \\
 II. $s$-channel $h$ annihilation \hfill \\ 
 III. $u$-channel $\sel_L$ exchange \hfill \\
 IV. point interaction  \hfill \\
 V. $s$-channel $Z$ annihilation
\begin{eqnarray}
f_1 &=& C_{H-W-W} \,  C_{\snu-\snu-H} \nonumber \\
f_2 &=& C_{h-W-W} \,  C_{\snu-\snu-h} \nonumber \\
f_3 &=& (C_{\snu-\sel -W})^2  \nonumber \\
f_4 &=&  C_{\snu-\snu-W-W}  \nonumber \\
f_5 &=& C_{Z-W-W} \, C_{\snu-\snu-Z}  \nonumber \\ 
{\cal T}_{\rm I}\!\!\times\!\!{\cal T}_{\rm I} &=&  (12 \mw^4 - 4 \mw^2 s +
     s^2)/(4 \mw^4 (s - m_H^2)^2) \nonumber \\     
{\cal T}_{\rm II}\!\!\times\!\!{\cal T}_{\rm II} &=&  (12 \mw^4 - 4 \mw^2 s + 
     s^2)/(4 \mw^4 (s - m_h^2)^2)   \nonumber \\
{\cal T}_{\rm III}\!\!\times\!\!{\cal T}_{\rm III} &=&  (\msnu^4 + (\mw^2 - u)^2
     - 2 \msnu^2 (\mw^2 + u))^2
          /(\mw^4 (u - \mselL^2)^2)   \nonumber \\
{\cal T}_{\rm IV}\!\!\times\!\!{\cal T}_{\rm IV} &=&  (12 \mw^4 - 4 \mw^2 s +
      s^2)/(4 \mw^4)   \nonumber \\
{\cal T}_{\rm V}\!\!\times\!\!{\cal T}_{\rm V} &=&  (-32 \msnu^6 \mw^2 - 24
      \mw^6 s + (t^2 - u^2)^2 - 
             8 \mw^4 (s^2 - 2 (t - u)^2 - s (t + u)) \nl + 
             2 \mw^2 (s^3 - 2 s^2 (t + u) - 
            2 (t - u)^2 (t + u) - 
            s (t^2 - 6 t u + u^2)) \nl + 
          4 \msnu^4 (16 \mw^4 + (t - u)^2 + 
            \mw^2 (-6 s + 8 (t + u))) \nl + 
            4 \msnu^2 (24 \mw^6 - (t - u)^2 (t + u) + 
            4 \mw^4 (s - 2 (t + u)) + 
            2 \mw^2 (-4 t u \nl + s (t + u))))
             /(4 \mw^4 (s - \mz^2)^2)   \nonumber \\
{\cal T}_{\rm I}\!\!\times\!\!{\cal T}_{\rm II} &=& (12 \mw^4 - 4 \mw^2 s + s^2)
      /(4 \mw^4 (s - m_H^2) (s - m_h^2)) \nonumber \\
{\cal T}_{\rm I}\!\!\times\!\!{\cal T}_{\rm III} &=& (6 \mw^6 + \msnu^4 (6 \mw^2
        - s) - s u^2 + 2 \mw^2 u (s + 2 t + u) - \mw^4 (5 s + 4 t + 8 u) \nl + 
         2 \msnu^2 (10 \mw^4 + s u - \mw^2 (s + 2 t + 4 u)))
        /(2 \mw^4 (s - m_H^2) (u - \mselL^2)) \nonumber \\
{\cal T}_{\rm I}\!\!\times\!\!{\cal T}_{\rm IV} &=& -(12 \mw^4 - 4 \mw^2 s + s^2
       )/(4 \mw^4 (s - m_H^2)) \nonumber \\
{\cal T}_{\rm I}\!\!\times\!\!{\cal T}_{\rm V} &=& -((t - u) (16 \mw^4 + \msnu^2
    (4 \mw^2 - 2 s) + s (t + u) - 2 \mw^2 (2 s + t + u)))
        \nl /(4 \mw^4 (s - m_H^2) (s - \mz^2)) \nonumber \\
{\cal T}_{\rm II}\!\!\times\!\!{\cal T}_{\rm III} &=& (6 \mw^6 + \msnu^4 (6
        \mw^2  - s) - s u^2 + 
         2 \mw^2 u (s + 2 t + u) - \mw^4 (5 s + 4 t + 8 u) \nl + 
         2 \msnu^2 (10 \mw^4 + s u - \mw^2 (s + 2 t + 4 u)))
        /(2 \mw^4 (s - m_h^2)(u - \mselL^2)) \nonumber \\
{\cal T}_{\rm II}\!\!\times\!\!{\cal T}_{\rm IV} &=& -(12 \mw^4 - 4 \mw^2 s +
   s^2) /(4 \mw^4 (s - m_h^2)) \nonumber \\
{\cal T}_{\rm II}\!\!\times\!\!{\cal T}_{\rm V} &=& -((t - u) (16 \mw^4 +
      \msnu^2 (4 \mw^2 - 2 s) 
         + s (t + u) - 2 \mw^2 (2 s + t + u)))
       \nl  /(4 \mw^4 (s - m_h^2) (s - \mz^2)) \nonumber \\
{\cal T}_{\rm III}\!\!\times\!\!{\cal T}_{\rm IV} &=& - (6 \mw^6 + \msnu^4 (6
       \mw^2 - s) - s u^2 + 
            2 \mw^2 u (s + 2 t + u) - \mw^4 (5 s + 4 t + 8 u) \nl + 
           2 \msnu^2 (10 \mw^4 + s u - \mw^2 (s + 2 t + 4 u)))
           /(2 \mw^4 (u- \mselL^2)) \nonumber \\
{\cal T}_{\rm III}\!\!\times\!\!{\cal T}_{\rm V} &=& (t^2 u^2 - u^4 + 2 \msnu^6 
         (8 \mw^2 - t + u) 
         + 4 \mw^6 (s - t + u) + \mw^4 (2 s^2 + 3 t^2 \nl + 
          4 s (t - u) + 4 t u - 7 u^2) + 
        2 \mw^2 u (s^2 - 2 s t - 2 t^2 + 2 u^2) - 
        2 \msnu^2 (8 \mw^6 \nl + \mw^4 (4 s + 5 t - 5 u) + 
              u (t^2 + t u - 2 u^2) + 
              \mw^2 (s^2 - 2 s t - 2 t^2 + 2 s u - 8 t u \nl + 
           2 u^2)) + \msnu^4 (t^2 + 4 t u - 5 u^2 + 
              4 \mw^2 (s - 4 (t + u))))
           \nl /(2 \mw^4 (s- \mz^2) (u - \mselL^2)) \nonumber \\
{\cal T}_{\rm IV}\!\!\times\!\!{\cal T}_{\rm V} &=& ((t - u) (16 \mw^4 +
      \msnu^2 (4 \mw^2 - 2 s) 
         + s (t + u) - 2 \mw^2 (2 s + t + u)))
       \nl  /(4 \mw^4 (s - \mz^2)) \nonumber \\
\tsq &=&   f_1^2 {\cal T}_{\rm I}\!\!\times\!\!{\cal T}_{\rm I} 
+  f_2^2 {\cal T}_{\rm II}\!\!\times\!\!{\cal T}_{\rm II}
+  f_3^2 {\cal T}_{\rm III}\!\!\times\!\!{\cal T}_{\rm III}
+  f_4^2 {\cal T}_{\rm IV}\!\!\times\!\!{\cal T}_{\rm IV}
+  f_5^2 {\cal T}_{\rm V}\!\!\times\!\!{\cal T}_{\rm V}
+ 2 f_1 f_2 {\cal T}_{\rm I}\!\!\times\!\!{\cal T}_{\rm II}
\nl + 2 f_1 f_3 {\cal T}_{\rm I}\!\!\times\!\!{\cal T}_{\rm III}
+ 2 f_1 f_4 {\cal T}_{\rm I}\!\!\times\!\!{\cal T}_{\rm IV}
+ 2 f_1 f_5 {\cal T}_{\rm I}\!\!\times\!\!{\cal T}_{\rm V}
+ 2 f_2 f_3 {\cal T}_{\rm II}\!\!\times\!\!{\cal T}_{\rm III}
+ 2 f_2 f_4 {\cal T}_{\rm II}\!\!\times\!\!{\cal T}_{\rm IV}
\nl + 2 f_2 f_5 {\cal T}_{\rm II}\!\!\times\!\!{\cal T}_{\rm V}
+ 2 f_3 f_4 {\cal T}_{\rm III}\!\!\times\!\!{\cal T}_{\rm IV}
+ 2 f_3 f_5 {\cal T}_{\rm III}\!\!\times\!\!{\cal T}_{\rm V}
+ 2 f_4 f_5 {\cal T}_{\rm IV}\!\!\times\!\!{\cal T}_{\rm V}    
\end{eqnarray}

\subsection*{$\snu \snu^\ast \longrightarrow Z Z$}
 I. $s$-channel $H$ annihilation \hfill \\
 II. $s$-channel $h$ annihilation \hfill \\ 
 III. $u$-channel $\snu$ exchange \hfill \\
 IV. $t$-channel $\snu$ exchange \hfill \\ 
 V. point interaction  
\begin{eqnarray}
f_1 &=& C_{H-Z-Z} \,  C_{\snu-\snu-H} \nonumber \\
f_2 &=& C_{h-Z-Z} \,  C_{\snu-\snu-h} \nonumber \\
f_3 &=& (C_{\snu-\snu -Z})^2  \nonumber \\
f_4 &=& (C_{\snu-\snu -Z})^2  \nonumber \\
f_5 &=&  C_{\snu-\snu-Z-Z}  \nonumber \\
{\cal T}_{\rm I}\!\!\times\!\!{\cal T}_{\rm I} &=&  (12 \mz^4 - 4 \mz^2 s + s^2)
        /(4 \mz^4 (s - m_H^2)^2)\nonumber \\
{\cal T}_{\rm II}\!\!\times\!\!{\cal T}_{\rm II} &=& (12 \mz^4 - 4 \mz^2 s 
       + s^2)/(4 \mz^4 (s - m_h^2)^2) \nonumber \\
{\cal T}_{\rm III}\!\!\times\!\!{\cal T}_{\rm III} &=& (\msnu^4 + (\mz^2 - u)^2 
        - 2 \msnu^2 (\mz^2 + u))^2
          /(\mz^4 (u - \msnu^2)^2) \nonumber \\
{\cal T}_{\rm IV}\!\!\times\!\!{\cal T}_{\rm IV} &=& (\msnu^4 + (\mz^2 - t)^2 
      - 2 \msnu^2 (\mz^2 + t))^2
          /(\mz^4 (t - \msnu^2)^2) \nonumber \\
{\cal T}_{\rm V}\!\!\times\!\!{\cal T}_{\rm V} &=& (12 \mz^4 - 4 \mz^2 s + s^2)
     /(4 \mz^4) \nonumber \\
{\cal T}_{\rm I}\!\!\times\!\!{\cal T}_{\rm II} &=& (12 \mz^4 - 4 \mz^2 s + s^2)
      /(4 \mz^4 (s - m_H^2)(s - m_h^2)) \nonumber \\
{\cal T}_{\rm I}\!\!\times\!\!{\cal T}_{\rm III} &=& (6 \mz^6 + \msnu^4 (6 
      \mz^2 - s) - s u^2 \nl + 
         2 \mz^2 u (s + 2 t + u) - \mz^4 (5 s + 4 t + 8 u) + 
         2 \msnu^2 (10 \mz^4 + s u - \mz^2 (s + 2 t \nl + 4 u)))
         /(2 \mz^4 (s - m_H^2) (u - \msnu^2)) \nonumber \\
{\cal T}_{\rm I}\!\!\times\!\!{\cal T}_{\rm IV} &=& (6 \mz^6 + \msnu^4 (6 \mz^2 
      - s) - s t^2 + 
         2 \mz^2 t (s + 2 u + t) - \mz^4 (5 s + 4 u + 8 t) \nl + 
         2 \msnu^2 (10 \mz^4 + s t - \mz^2 (s + 2 u + 4 t)))
        /(2 \mz^4 (s - m_H^2)(t - \msnu^2)) \nonumber \\
{\cal T}_{\rm I}\!\!\times\!\!{\cal T}_{\rm V} &=& -(12 \mz^4 - 4 \mz^2 s + s^2)
        /(4 \mz^4 (s - m_H^2)) \nonumber \\
{\cal T}_{\rm II}\!\!\times\!\!{\cal T}_{\rm III} &=& (6 \mz^6 + \msnu^4 (6
       \mz^2 - s) - s u^2 + 
         2 \mz^2 u (s + 2 t + u) - \mz^4 (5 s + 4 t + 8 u) \nl + 
         2 \msnu^2 (10 \mz^4 + s u - \mz^2 (s + 2 t + 4 u)))
        /(2 \mz^4 (s - m_h^2) (u - \msnu^2)) \nonumber \\
{\cal T}_{\rm II}\!\!\times\!\!{\cal T}_{\rm IV} &=& (6 \mz^6 + \msnu^4 (6 
       \mz^2 - s) - s t^2 + 
         2 \mz^2 t (s + 2 u + t) - \mz^4 (5 s + 4 u + 8 t) \nl + 
         2 \msnu^2 (10 \mz^4 + s t - \mz^2 (s + 2 u + 4 t)))
        /(2 \mz^4 (s - m_h^2)(t - \msnu^2)) \nonumber \\
{\cal T}_{\rm II}\!\!\times\!\!{\cal T}_{\rm V} &=& -(12 \mz^4 - 4 \mz^2 s 
      + s^2)/(4 \mz^4 (s - m_h^2)) \nonumber \\
{\cal T}_{\rm III}\!\!\times\!\!{\cal T}_{\rm IV} &=& (\msnu^4 + \mz^4 + \msnu^2
       (6 \mz^2 - t - u) + t u 
           - \mz^2 (2 s + t + u))^2
            \nl /(\mz^4 (t - \msnu^2) (u - \msnu^2)) \nonumber \\
{\cal T}_{\rm III}\!\!\times\!\!{\cal T}_{\rm V} &=& - (6 \mz^6 + \msnu^4 (6 
        \mz^2 - s) - s u^2 + 
            2 \mz^2 u (s + 2 t + u) - \mz^4 (5 s + 4 t + 8 u) \nl + 
           2 \msnu^2 (10 \mz^4 + s u - \mz^2 (s + 2 t + 4 u)))
           /(2 \mz^4 (u- \msnu^2)) \nonumber \\
{\cal T}_{\rm IV}\!\!\times\!\!{\cal T}_{\rm V} &=& - (6 \mz^6 + \msnu^4 
      (6 \mz^2 - s) - s t^2 + 
            2 \mz^2 t (s + 2 u + t) - \mz^4 (5 s + 4 u + 8 t) \nl + 
           2 \msnu^2 (10 \mz^4 + s t - \mz^2 (s + 2 u + 4 t)))
           /(2 \mz^4 (t- \msnu^2)) \nonumber \\
\tsq &=&   f_1^2 {\cal T}_{\rm I}\!\!\times\!\!{\cal T}_{\rm I} 
+  f_2^2 {\cal T}_{\rm II}\!\!\times\!\!{\cal T}_{\rm II}
+  f_3^2 {\cal T}_{\rm III}\!\!\times\!\!{\cal T}_{\rm III}
+  f_4^2 {\cal T}_{\rm IV}\!\!\times\!\!{\cal T}_{\rm IV}
+  f_5^2 {\cal T}_{\rm V}\!\!\times\!\!{\cal T}_{\rm V}
+ 2 f_1 f_2 {\cal T}_{\rm I}\!\!\times\!\!{\cal T}_{\rm II}
\nl + 2 f_1 f_3 {\cal T}_{\rm I}\!\!\times\!\!{\cal T}_{\rm III}
+ 2 f_1 f_4 {\cal T}_{\rm I}\!\!\times\!\!{\cal T}_{\rm IV}
+ 2 f_1 f_5 {\cal T}_{\rm I}\!\!\times\!\!{\cal T}_{\rm V}
+ 2 f_2 f_3 {\cal T}_{\rm II}\!\!\times\!\!{\cal T}_{\rm III}
+ 2 f_2 f_4 {\cal T}_{\rm II}\!\!\times\!\!{\cal T}_{\rm IV}
\nl + 2 f_2 f_5 {\cal T}_{\rm II}\!\!\times\!\!{\cal T}_{\rm V}
+ 2 f_3 f_4 {\cal T}_{\rm III}\!\!\times\!\!{\cal T}_{\rm IV}
+ 2 f_3 f_5 {\cal T}_{\rm III}\!\!\times\!\!{\cal T}_{\rm V}
+ 2 f_4 f_5 {\cal T}_{\rm IV}\!\!\times\!\!{\cal T}_{\rm V} 
\end{eqnarray}

\subsection*{$\snu \snu^\ast \longrightarrow Z h \quad [ZH]$}
 I. $s$-channel $Z$ annihilation \hfill \\
 II.  $t$-channel $\snu$ exchange \hfill \\ 
 III. $u$-channel $\snu$ exchange   
\begin{eqnarray}
f_1 &=& C_{\snu-\snu-Z} \,  C_{h-Z-Z} \quad \quad [C_{\snu-\snu-Z} \,  
        C_{H-Z-Z}] \nonumber \\
f_2 &=& C_{\snu-\snu-h} \,  C_{\snu-\snu-Z} \quad \quad [C_{\snu-\snu-H} \,  
       C_{\snu-\snu-Z}] \nonumber \\
f_3 &=& C_{\snu-\snu-h} \,  C_{\snu-\snu-Z} \quad \quad [C_{\snu-\snu-H} \,  
       C_{\snu-\snu-Z}] \nonumber \\
{\cal T}_{\rm I}\!\!\times\!\!{\cal T}_{\rm I} &=& (-16 \msnu^2 \mz^2 + 4 \mz^2
    s + (t - u)^2)
          /(4 \mz^2 (s-\mz^2)^2) \nonumber \\
{\cal T}_{\rm II}\!\!\times\!\!{\cal T}_{\rm II} &=& (\msnu^4 + (\mz^2 - t)^2 
          - 2 \msnu^2 (\mz^2 + t))
          /(\mz^2 (t-\msnu^2)^2) \nonumber \\
{\cal T}_{\rm III}\!\!\times\!\!{\cal T}_{\rm III} &=& (\msnu^4 + (\mz^2 - u)^2 
      - 2 \msnu^2 (\mz^2 + u))
          /(\mz^2 (u-\msnu^2)^2) \nonumber \\
{\cal T}_{\rm I}\!\!\times\!\!{\cal T}_{\rm II} &=& -(t(t - u) + \msnu^2(-8 
         \mz^2 - t + u) 
         + \mz^2 (2 s - t + u))
           /(2 \mz^2 (s- \mz^2) (t-\msnu^2)) \nonumber \\
{\cal T}_{\rm I}\!\!\times\!\!{\cal T}_{\rm III} &=& -(u (u - t) + \msnu^2 (-8 
       \mz^2 - u + t) 
         + \mz^2 (2 s - u + t))
           /(2 \mz^2 (s- \mz^2) (u-\msnu^2)) \nonumber \\
{\cal T}_{\rm II}\!\!\times\!\!{\cal T}_{\rm III} &=& (-\msnu^4 + \mz^4 + t u 
     - \msnu^2 (-6 \mz^2 + t + u) 
         - \mz^2 (2 s + t + u))
         \nl  /(\mz^2 (u- \msnu^2) (t-\msnu^2)) \nonumber \\
\tsq &=&   f_1^2 {\cal T}_{\rm I}\!\!\times\!\!{\cal T}_{\rm I} 
+  f_2^2 {\cal T}_{\rm II}\!\!\times\!\!{\cal T}_{\rm II}
+  f_3^2 {\cal T}_{\rm III}\!\!\times\!\!{\cal T}_{\rm III} 
+ 2 f_1 f_2 {\cal T}_{\rm I}\!\!\times\!\!{\cal T}_{\rm II}
+ 2 f_1 f_3 {\cal T}_{\rm I}\!\!\times\!\!{\cal T}_{\rm III}
\nl + 2 f_2 f_3 {\cal T}_{\rm II}\!\!\times\!\!{\cal T}_{\rm III} 
\end{eqnarray}

\subsection*{$\snu \snu^\ast \longrightarrow A h \quad [AH]$}
 I. $s$-channel $Z$ annihilation 
\begin{eqnarray}
f_1 &=& C_{\snu-\snu-Z} \, C_{Z-h-A} \quad \quad [C_{\snu-\snu-Z} \, 
         C_{Z-H-A}] \nonumber \\
{\cal T}_{\rm I}\!\!\times\!\!{\cal T}_{\rm I} &=& (t-u)^2/(s - \mz^2)^2 
\nonumber \\
\tsq &=&   f_1^2 {\cal T}_{\rm I}\!\!\times\!\!{\cal T}_{\rm I} 
\end{eqnarray}

\subsection*{$\snu \snu^\ast \longrightarrow f \bar{f}$}
 I. $s$-channel $Z$ annihilation 
\begin{eqnarray}
f_L &=& C_{\snu-\snu-Z} \, C_{Z-f-f}^L \nonumber \\
f_R &=& C_{\snu-\snu-Z} \, C_{Z-f-f}^R  \nonumber \\
{\cal T}_{\rm I}\!\!\times\!\!{\cal T}_{\rm I} &=& -(f_L^2 + f_R^2) 
      (4 \msnu^2 s - s^2 + (t - u)^2)
               /(s - \mz^2)^2 \nonumber \\
\tsq &=&    {\cal T}_{\rm I}\!\!\times\!\!{\cal T}_{\rm I} 
\end{eqnarray}
For quarks $\tsq$ is multiplied by 3 for color.

\subsection*{$\snu \snu^\ast \longrightarrow t \bar{t}$}
 I. $s$-channel $Z$ annihilation \hfill \\
 II. $s$-channel $h$ annihilation \hfill \\ 
 III. $s$-channel $H$ annihilation 
\begin{eqnarray}
f_{1L} &=& C_{\snu-\snu-Z} \, C_{Z-t-t}^L \nonumber \\
f_{1R} &=& C_{\snu-\snu-Z} \, C_{Z-t-t}^R  \nonumber \\
f_2 &=& C_{\snu-\snu-h} \, C_{h-t-t} \nonumber \\
f_3 &=& C_{\snu-\snu-H} \, C_{H-t-t} \nonumber \\
{\cal T}_{\rm I}\!\!\times\!\!{\cal T}_{\rm I} &=& (-2 \mt^2 (f_{1L} 
        - f_{1R})^2 s 
         + 4 \msnu^2 (2 \mt^2 (f_{1L} - f_{1R})^2 - (f_{1L}^2 + f_{1R}^2) s) 
         \nl + (f_{1L}^2 + f_{1R}^2) (s^2 - (t - u)^2))
         /(s - \mz^2)^2 \nonumber \\
{\cal T}_{\rm II}\!\!\times\!\!{\cal T}_{\rm II} &=& 2 (s- 4 \mt^2)
         /(s- m_h^2)^2 \nonumber \\
{\cal T}_{\rm III}\!\!\times\!\!{\cal T}_{\rm III} &=& 2 (s - 4 \mt^2)
        /(s-m_H^2)^2 \nonumber \\
{\cal T}_{\rm I}\!\!\times\!\!{\cal T}_{\rm II} &=& -((f_{1L} + f_{1R}) 
       2 \mt (t-u))
       /((s - \mz^2) (s-m_h^2)) \nonumber \\
{\cal T}_{\rm I}\!\!\times\!\!{\cal T}_{\rm III} &=& -((f_{1L} + f_{1R}) 
       2 \mt (t-u))
        /((s- \mz^2) (s- m_H^2)) \nonumber \\
{\cal T}_{\rm II}\!\!\times\!\!{\cal T}_{\rm III} &=& 2 (s - 4 \mt^2)
        /((s-m_H^2) (s-m_h^2)) \nonumber \\
\tsq &=&   3 ( {\cal T}_{\rm I}\!\!\times\!\!{\cal T}_{\rm I} 
+  f_2^2 {\cal T}_{\rm II}\!\!\times\!\!{\cal T}_{\rm II} 
+  f_3^2 {\cal T}_{\rm II}\!\!\times\!\!{\cal T}_{\rm II}
+ 2 f_2 {\cal T}_{\rm I}\!\!\times\!\!{\cal T}_{\rm II} 
+ 2 f_3 {\cal T}_{\rm I}\!\!\times\!\!{\cal T}_{\rm III} 
\nl  
+ 2 f_2 f_3 {\cal T}_{\rm II}\!\!\times\!\!{\cal T}_{\rm III} 
)
\end{eqnarray}

\subsection*{$\snu \snu^\ast \longrightarrow e \bar{e}$}
 I. $s$-channel $Z$ annihilation \hfill \\
 II. $t$-channel charginos exchange 
\begin{eqnarray}
f_{1L} &=& C_{\snu-\snu-Z} \, C_{Z-e-e}^L \nonumber \\
f_{1R} &=& C_{\snu-\snu-Z} \, C_{Z-e-e}^R \nonumber \\
f_2(i) &=& (C_{\snu-\schi^+_i -e})^2 \nonumber \\
{\cal T}_{\rm I}\!\!\times\!\!{\cal T}_{\rm I} &=& (-(f_{1L}^2 + f_{1R}^2)
      (4 \msnu^2 s - s^2 + (t - u)^2))
          /(s- \mz^2)^2 \nonumber \\
{\cal T}_{\rm II}\!\!\times\!\!{\cal T}_{\rm II} &=& f_{2}(i) f_2(j) 
       ( t u - \msnu^4)
               /((t- \mxi^2) (t- \mxj^2)) \nonumber \\
{\cal T}_{\rm I}\!\!\times\!\!{\cal T}_{\rm II} &=& ((1/2) f_2(j) f_{1L} 
      (4 \msnu^2 s - s^2 + (t - u)^2))
                /((s- \mz^2) (t- \mxj^2)) \nonumber \\
\tsq &=&    {\cal T}_{\rm I}\!\!\times\!\!{\cal T}_{\rm I} 
+ \sum_{i,j} {\cal T}_{\rm II}\!\!\times\!\!{\cal T}_{\rm II} 
+ 2 \sum_j {\cal T}_{\rm I}\!\!\times\!\!{\cal T}_{\rm II} 
\end{eqnarray}

\subsection*{$\snu \snu^\ast \longrightarrow \nu \bar{\nu}$}
 I. $s$-channel $Z$ annihilation \hfill \\
 II. $t$-channel neutralinos exchange 
\begin{eqnarray}
f_1 &=& C_{\snu - \snu - Z} \, C_{Z - \nu - \nu} \nnl
f_2(i) &=& (C_{\snu - \schi^0_i - \nu})^2 \nnl
{\cal T}_{\rm I}\!\!\times\!\!{\cal T}_{\rm I} &=& (- 4 \msnu^2 s + s^2 - 
      (t - u)^2)
            /(s - \mz^2)^2 \nonumber \\
{\cal T}_{\rm II}\!\!\times\!\!{\cal T}_{\rm II} &=& (- 2 \msnu^4 
        + 2 t u)
             /((t - \nevalsi^2) (t - \nevalsj^2)) \nonumber \\
{\cal T}_{\rm I}\!\!\times\!\!{\cal T}_{\rm II} &=& - (1/2)(- 4 
       \msnu^2 s + s^2 - ( t - u)^2)
             /((s - \mz^2) (t - \nevalsi^2)) \nonumber \\
\tsq &=&   f_1^2 {\cal T}_{\rm I}\!\!\times\!\!{\cal T}_{\rm I} 
+  \sum_{i,j} f_2(i) f_2(j) {\cal T}_{\rm II}\!\!\times\!\!{\cal T}_{\rm II} 
+ 2 \sum_i f_1 f_2(i){\cal T}_{\rm I}\!\!\times\!\!{\cal T}_{\rm II} 
\end{eqnarray}

\subsection*{$\snu \snu^\ast \longrightarrow W^+ H^-$}
 I. $s$-channel $H$ annihilation \hfill \\
 II. $s$-channel $h$ annihilation \hfill \\ 
 III. $t$-channel $\sel_L$ exchange 
\begin{eqnarray}
f_1 &=& C_{H - W^+ - H^-} \, C_{\snu - \snu - H} \nnl
f_2 &=& C_{h - W^+ - H^-} \, C_{\snu - \snu - h} \nnl
f_3 &=& C_{\snu - \sel_L - W} \, C_{\snu - \sel_L - H^+} \nnl
{\cal T}_{\rm I}\!\!\times\!\!{\cal T}_{\rm I} &=& (\mHp^4 - 2 \mHp^2 (2 \msnu^2
     + 7 \mw^2 + s - t - u) 
          + (-2 \msnu^2 + \mw^2 - s + t + u)^2)
        \nl   /(4 \mw^2 (s- m_H^2)^2) \nonumber \\
{\cal T}_{\rm II}\!\!\times\!\!{\cal T}_{\rm II} &=& (\mHp^4 - 2 \mHp^2 (2 
         \msnu^2 + 7 \mw^2 + s - t - u) 
          + (-2 \msn^2 + \mw^2 - s + t + u)^2)
        \nl  /(4 \mw^2 (s-m_h^2)^2) \nonumber \\
{\cal T}_{\rm III}\!\!\times\!\!{\cal T}_{\rm III} &=& (\msnu^4 + (\mw^2 - t)^2 
       - 2 \msnu^2 (\mw^2 + t))
          /(\mw^2 (t- \mselL^2)^2) \nonumber \\
{\cal T}_{\rm I}\!\!\times\!\!{\cal T}_{\rm II} &=& (\mHp^4 - 2 \mHp^2 (2 
        \msnu^2 + 7 \mw^2 + s - t - u) 
          + (-2 \msnu^2 + \mw^2 - s + t + u)^2)
        \nl  /(4 \mw^2 (s- m_H^2) (s-m_h^2)) \nonumber \\
{\cal T}_{\rm I}\!\!\times\!\!{\cal T}_{\rm III} &=& (2 \msnu^4 + \mw^4 - 
        \mw^2 s - 2 \mw^2 t - 
             s t + t^2 + \mHp^2 (-\msnu^2 - 3 \mw^2 + t) \nl + 
             \msnu^2 (\mw^2 + s - 3 t - u) + \mw^2 u + t u)
           /(2 \mw^2 (s-m_H^2)(t- \mselL^2)) \nonumber \\
{\cal T}_{\rm II}\!\!\times\!\!{\cal T}_{\rm III} &=& (2 \msnu^4 + \mw^4 - 
       \mw^2 s - 2 \mw^2 t - 
             s t + t^2 + \mHp^2 (-\msnu^2 - 3 \mw^2 + t) \nl + 
             \msnu^2 (\mw^2 + s - 3 t - u) + \mw^2 u + t u)
           /(2 \mw^2 (s-m_h^2) (t- \mselL^2)) \nonumber \\
\tsq &=&   f_1^2 {\cal T}_{\rm I}\!\!\times\!\!{\cal T}_{\rm I} 
+  f_2^2 {\cal T}_{\rm II}\!\!\times\!\!{\cal T}_{\rm II} 
+  f_3^2 {\cal T}_{\rm III}\!\!\times\!\!{\cal T}_{\rm III}
+ 2 f_1 f_2 {\cal T}_{\rm I}\!\!\times\!\!{\cal T}_{\rm II} 
+ 2 f_1 f_3 {\cal T}_{\rm I}\!\!\times\!\!{\cal T}_{\rm III}
\nl + 2 f_2 f_3 {\cal T}_{\rm II}\!\!\times\!\!{\cal T}_{\rm III}
\end{eqnarray}

\subsection*{$\snu \snu^\ast \longrightarrow H^+ H^-$}
 I. $s$-channel $H$ annihilation \hfill \\
 II. $s$-channel $h$ annihilation \hfill \\ 
 III. $t$-channel $\sel_L$ exchange \hfill \\
 IV.  point interaction  \hfill \\
 V. $s$-channel $Z$ annihilation
\begin{eqnarray}
f_1 &=& C_{H - H^+ - H^-} \, C_{\snu - \snu - H} \nnl
f_2 &=& C_{h - H^+ - H^-} \, C_{\snu - \snu - h}  \nnl
f_3 &=& (C_{\snu - \sel_L - H^+})^2 \nnl
f_4 &=& C_{\snu-\snu-H^+-H^-} \nnl
f_5 &=& C_{\snu - \snu - H} \, C_{Z - H^+ - H^-} \nnl
{\cal T}_{\rm I}\!\!\times\!\!{\cal T}_{\rm I} &=& 1/(s-m_H^2)^2 \nonumber \\
{\cal T}_{\rm II}\!\!\times\!\!{\cal T}_{\rm II} &=& 1/(s-m_h^2)^2 \nonumber \\
{\cal T}_{\rm III}\!\!\times\!\!{\cal T}_{\rm III} &=& 1/(t- \mselL^2)^2 
\nonumber \\
{\cal T}_{\rm IV}\!\!\times\!\!{\cal T}_{\rm IV} &=& 1 \nonumber \\
{\cal T}_{\rm V}\!\!\times\!\!{\cal T}_{\rm V} &=& (u-t)^2/(s-\mz^2)^2 
\nonumber \\
{\cal T}_{\rm I}\!\!\times\!\!{\cal T}_{\rm II} &=& 1/((s-m_H^2)(s-m_h^2)) 
\nonumber \\
{\cal T}_{\rm I}\!\!\times\!\!{\cal T}_{\rm III} &=& 1/((s-m_H^2)(t-\mselL^2)) 
\nonumber \\
{\cal T}_{\rm I}\!\!\times\!\!{\cal T}_{\rm IV} &=& -1/(s-m_H^2) \nonumber \\
{\cal T}_{\rm I}\!\!\times\!\!{\cal T}_{\rm V} &=& -(u-t)/((s-m_H^2)(s-\mz^2)) 
\nonumber \\
{\cal T}_{\rm II}\!\!\times\!\!{\cal T}_{\rm III} &=& 1/((s-m_h^2)(t-\mselL^2)) 
\nonumber \\
{\cal T}_{\rm II}\!\!\times\!\!{\cal T}_{\rm IV} &=& -1/(s-m_h^2) \nonumber \\
{\cal T}_{\rm II}\!\!\times\!\!{\cal T}_{\rm V} &=& -(u-t)/((s-m_h^2)(s-\mz^2)) 
\nonumber \\
{\cal T}_{\rm III}\!\!\times\!\!{\cal T}_{\rm IV} &=& -1/(t- \mselL^2) 
\nonumber \\
{\cal T}_{\rm III}\!\!\times\!\!{\cal T}_{\rm V} &=&
-(u-t)/((t-\mselL^2)(s-\mz^2))  \nonumber \\
{\cal T}_{\rm IV}\!\!\times\!\!{\cal T}_{\rm V} &=& (u-t)/(s-\mz^2) \nonumber \\
\tsq &=&    f_1^2 {\cal T}_{\rm I}\!\!\times\!\!{\cal T}_{\rm I} 
+  f_2^2 {\cal T}_{\rm II}\!\!\times\!\!{\cal T}_{\rm II} 
+  f_3^2 {\cal T}_{\rm III}\!\!\times\!\!{\cal T}_{\rm III}
+  f_4^2 {\cal T}_{\rm IV}\!\!\times\!\!{\cal T}_{\rm IV}
+  f_5^2 {\cal T}_{\rm V}\!\!\times\!\!{\cal T}_{\rm V}
+ 2 f_1 f_2 {\cal T}_{\rm I}\!\!\times\!\!{\cal T}_{\rm II} 
\nl + 2 f_1 f_3 {\cal T}_{\rm I}\!\!\times\!\!{\cal T}_{\rm III}
+ 2 f_1 f_4 {\cal T}_{\rm I}\!\!\times\!\!{\cal T}_{\rm IV}
+ 2 f_1 f_5 {\cal T}_{\rm I}\!\!\times\!\!{\cal T}_{\rm V}
+ 2 f_2 f_3 {\cal T}_{\rm II}\!\!\times\!\!{\cal T}_{\rm III}
+ 2 f_2 f_4 {\cal T}_{\rm II}\!\!\times\!\!{\cal T}_{\rm IV}
\nl + 2 f_2 f_5 {\cal T}_{\rm II}\!\!\times\!\!{\cal T}_{\rm V}
+ 2 f_3 f_4 {\cal T}_{\rm III}\!\!\times\!\!{\cal T}_{\rm IV}
+ 2 f_3 f_5 {\cal T}_{\rm III}\!\!\times\!\!{\cal T}_{\rm V}
+ 2 f_4 f_5 {\cal T}_{\rm IV}\!\!\times\!\!{\cal T}_{\rm V}
\end{eqnarray}

\subsection*{$\snu \snu^\ast \longrightarrow H H \quad [h h] \quad [h H]$}
 I. $s$-channel $H$ annihilation \hfill \\
 II. $s$-channel $h$ annihilation \hfill \\ 
 III. $t$-channel $\snu$ exchange \hfill \\ 
 IV. point interaction  
\begin{eqnarray}
f_1 &=& C_{H-H-H} \, C_{\snu - \snu -H} \quad \quad [C_{H-h-h} \, 
      C_{\snu - \snu -H} ] \quad [C_{H-H-h} \, C_{\snu - \snu -H} ] \nnl
f_2 &=&  C_{H-H-h} \, C_{\snu - \snu -h} \quad \quad [C_{h-h-h} \, 
      C_{\snu - \snu -h} ] \quad [C_{H-h-h} \, C_{\snu - \snu -h} ] \nnl
f_3 &=& (C_{\snu - \snu -H})^2 \quad \quad [(C_{\snu - \snu -h})^2] \quad
         [C_{\snu - \snu -h} \,C_{\snu - \snu -H} ] \nnl
f_4 &=& C_{\snu - \snu -H-H } \quad \quad [C_{\snu - \snu -h-h }] \quad 
    [C_{\snu - \snu -H-h }] \nnl
{\cal T}_{\rm I}\!\!\times\!\!{\cal T}_{\rm I} &=& 1/(s - m_H^2)^2 \nonumber \\
{\cal T}_{\rm II}\!\!\times\!\!{\cal T}_{\rm II} &=& 1/(s-m_h^2)^2 \nonumber \\
{\cal T}_{\rm III}\!\!\times\!\!{\cal T}_{\rm III} &=& 1/(t-\msnu^2)^2 
\nonumber \\
{\cal T}_{\rm IV}\!\!\times\!\!{\cal T}_{\rm IV} &=& 1 \nonumber \\
{\cal T}_{\rm I}\!\!\times\!\!{\cal T}_{\rm II} &=& 1/((s-m_H^2)(s-m_h^2)) 
\nonumber \\
{\cal T}_{\rm I}\!\!\times\!\!{\cal T}_{\rm III} &=& 1/((s-m_H^2)(t-\msnu^2)) 
\nonumber \\
{\cal T}_{\rm I}\!\!\times\!\!{\cal T}_{\rm IV} &=& -1/(s-m_H^2) \nonumber \\
{\cal T}_{\rm II}\!\!\times\!\!{\cal T}_{\rm III} &=& 1/((s-m_h^2)(t-\msnu^2)) 
\nonumber \\
{\cal T}_{\rm II}\!\!\times\!\!{\cal T}_{\rm IV} &=& -1/(s-m_h^2) \nonumber \\
{\cal T}_{\rm III}\!\!\times\!\!{\cal T}_{\rm IV} &=&  -1/(t-\msnu^2) 
\nonumber \\
\tsq &=&   f_1^2 {\cal T}_{\rm I}\!\!\times\!\!{\cal T}_{\rm I} 
+  f_2^2 {\cal T}_{\rm II}\!\!\times\!\!{\cal T}_{\rm II} 
+  f_3^2 {\cal T}_{\rm III}\!\!\times\!\!{\cal T}_{\rm III}
+  f_4^2 {\cal T}_{\rm IV}\!\!\times\!\!{\cal T}_{\rm IV}
+ 2 f_1 f_2 {\cal T}_{\rm I}\!\!\times\!\!{\cal T}_{\rm II} 
\nl + 2 f_1 f_3 {\cal T}_{\rm I}\!\!\times\!\!{\cal T}_{\rm III}
+ 2 f_1 f_4 {\cal T}_{\rm I}\!\!\times\!\!{\cal T}_{\rm IV}
+ 2 f_2 f_3 {\cal T}_{\rm II}\!\!\times\!\!{\cal T}_{\rm III}
+ 2 f_2 f_4 {\cal T}_{\rm II}\!\!\times\!\!{\cal T}_{\rm IV}
\nl + 2 f_3 f_4 {\cal T}_{\rm III}\!\!\times\!\!{\cal T}_{\rm IV}
\end{eqnarray}

\subsection*{$\snu \snu^\ast \longrightarrow A A$}
 I. $s$-channel $H$ annihilation \hfill \\
 II. $s$-channel $h$ annihilation \hfill \\ 
 III. point interaction  
\begin{eqnarray}
f_1 &=& C_{\snu - \snu - H } \, C_{H-A-A} \nnl
f_2 &=&  C_{\snu - \snu - h } \, C_{h-A-A} \nnl
f_3 &=&   C_{\snu - \snu - A-A }\nnl
{\cal T}_{\rm I}\!\!\times\!\!{\cal T}_{\rm I} &=& 1/(s-m_H^2)^2 \nonumber \\
{\cal T}_{\rm II}\!\!\times\!\!{\cal T}_{\rm II} &=& 1/(s-m_h^2)^2 \nonumber \\
{\cal T}_{\rm III}\!\!\times\!\!{\cal T}_{\rm III} &=& 1 \nonumber \\
{\cal T}_{\rm I}\!\!\times\!\!{\cal T}_{\rm II} &=&
1/((s-m_H^2)(s-m_h^2)) 
\nonumber \\
{\cal T}_{\rm I}\!\!\times\!\!{\cal T}_{\rm III} &=& -1/(s-m_H^2) \nonumber \\
{\cal T}_{\rm II}\!\!\times\!\!{\cal T}_{\rm III} &=& -1/(s-m_h^2) \nonumber \\
\tsq &=&   f_1^2 {\cal T}_{\rm I}\!\!\times\!\!{\cal T}_{\rm I} 
+  f_2^2 {\cal T}_{\rm II}\!\!\times\!\!{\cal T}_{\rm II} 
+  f_3^2 {\cal T}_{\rm III}\!\!\times\!\!{\cal T}_{\rm III}
+ 2 f_1 f_2 {\cal T}_{\rm I}\!\!\times\!\!{\cal T}_{\rm II} 
 + 2 f_1 f_3 {\cal T}_{\rm I}\!\!\times\!\!{\cal T}_{\rm III}
\nl + 2 f_2 f_3 {\cal T}_{\rm II}\!\!\times\!\!{\cal T}_{\rm III}
\end{eqnarray}

\subsection*{$\snu \snu^\ast \longrightarrow A Z$}
 I. $s$-channel $H$ annihilation \hfill \\
 II. $s$-channel $h$ annihilation   
\begin{eqnarray}
f_1 &=& C_{\snu-\snu-H} \, C_{H-Z-A} \nnl
f_2 &=& C_{\snu-\snu-h} \, C_{h-Z-A}  \nnl
{\cal T}_{\rm I}\!\!\times\!\!{\cal T}_{\rm I} &=& (\mA^4 - 2 \mA^2 (2 \msnu^2 
      + 7 \mz^2 + s - t - u) 
           + (-2 \msnu^2 + \mz^2 - s + t + u)^2)
         \nl  /(4 \mz^2 (s- m_H^2)^2) \nonumber \\
{\cal T}_{\rm II}\!\!\times\!\!{\cal T}_{\rm II} &=& (\mA^4 - 2 \mA^2 (2 \msnu^2
        + 7 \mz^2 + s - t - u) 
           + (-2 \msnu^2 + \mz^2 - s + t + u)^2)
         \nl  /(4 \mz^2 (s- m_h^2)^2) \nonumber \\
{\cal T}_{\rm I}\!\!\times\!\!{\cal T}_{\rm II} &=& (\mA^4 - 2 \mA^2 (2 \msnu^2 
      + 7 \mz^2 + s - t - u) 
           + (-2 \msnu^2 + \mz^2 - s + t + u)^2)
         \nl  /(4 \mz^2 (s- m_H^2) (s-m_h^2))  \nonumber \\
\tsq &=&   f_1^2 {\cal T}_{\rm I}\!\!\times\!\!{\cal T}_{\rm I} 
+  f_2^2 {\cal T}_{\rm II}\!\!\times\!\!{\cal T}_{\rm II} 
+ 2 f_1 f_2 {\cal T}_{\rm I}\!\!\times\!\!{\cal T}_{\rm II} 
\end{eqnarray}

\subsection*{$\snu \snu \longrightarrow \nu \nu $}
 I. $t$-channel neutralino exchange \hfill \\
 II. $u$-channel neutralino exchange  
\begin{eqnarray}
f(i) &=& (C_{\snu-\schi^0_i -\nu})^2 \nnl
{\cal T}_{\rm I}\!\!\times\!\!{\cal T}_{\rm I} &=&  (2 s \, \nevalsi
\nevalsj)/((t -      \nevalsi^2)(t - \nevalsj^2)) \nonumber \\
{\cal T}_{\rm II}\!\!\times\!\!{\cal T}_{\rm II} &=&  (2 s \,\nevalsi
        \nevalsj)/((u - \nevalsi^2)(u - \nevalsj^2)) \nonumber \\
{\cal T}_{\rm I}\!\!\times\!\!{\cal T}_{\rm II} &=&  (2 s \,\nevalsi
          \nevalsj)/((t - \nevalsi^2)(u - \nevalsj^2)) \nonumber \\
\tsq &=&   \sum_{i,j} f(i) f(j) 
\left( {\cal T}_{\rm I}\!\!\times\!\!{\cal T}_{\rm I} 
+  {\cal T}_{\rm II}\!\!\times\!\!{\cal T}_{\rm II} 
+ 2 {\cal T}_{\rm I}\!\!\times\!\!{\cal T}_{\rm II} \right)
\end{eqnarray}

\subsection*{$\chi \snu \longrightarrow \nu Z$}
 I. $s$-channel $\nu$ annihilation \hfill \\
 II. $t$-channel $\snu$ exchange \hfill \\ 
 III. $u$-channel neutralino exchange  
\begin{eqnarray}
f_1 &=& C_{\snu-\schi^0_1 -\nu} \, C_{Z-\nu-\nu}  \nnl
f_2 &=& C_{\snu-\snu-Z} \, C_{\snu-\schi^0_1 -\nu} \nnl
f_{3L}(i) &=& C_{\snu-\schi^0_i -\nu} \, C_{\schi^0_1-\schi^0_i - Z}^L \nnl
f_{3R}(i) &=& C_{\snu-\schi^0_i -\nu} \, C_{\schi^0_1-\schi^0_i - Z}^R \nnl
{\cal T}_{\rm I}\!\!\times\!\!{\cal T}_{\rm I} &=& (1/2) (1/\mz^2) 
          (\mchi^4 s - \msnu^4 s + 
            s (-\mz^4 - s t + \mz^2 (s + t - u)) \nl - 
            \mchi^2 (2 \mz^4 - 2 \mz^2 s + s (t + u)) + 
          \msnu^2 (2 \mz^4 - 2 \mz^2 s + s (s + t + u)))/(s)^2 
	  \nonumber \\
{\cal T}_{\rm II}\!\!\times\!\!{\cal T}_{\rm II} &=& (1/2) (\mchi^2 - t) 
         (\msnu^4 + (\mz^2 - t)^2 
            - 2 \msnu^2 (\mz^2 + t))/(\mz^2 (t - \msnu^2 )^2) \nonumber \\
{\cal T}_{\rm III}\!\!\times\!\!{\cal T}_{\rm III} &=& (1/2) (-1) (\nevalsi 
       f_{3L}(i) (3 (-\mchi) \mz^2 f_{3R}(j) 
           (-\mchi^2 - 
          \mz^2 + s + t) \nl + \nevalsj f_{3L}(j) (\mz^4 + s (-\mchi^2 + u) - 
  \mz^2 (s - t + u))) \nl + f_{3R}(i) (-3 (-\mchi^3) \nevalsj \mz^2 f_{3L}(j) - 
 \mchi^6 f_{3R}(j) \nl + 3 (-\mchi) \nevalsj \mz^2 f_{3L}(j) (-\mz^2 + s + t) + 
              \mchi^4 f_{3R}(j) (-2 \mz^2 + s + t + 2 u) \nl + 
              f_{3R}(j) (2 \mz^6 + 2 \mz^2 s u + t u^2 - 2 \mz^4 (s + 
     	      t + u)) \nl + \mchi^2 f_{3R}(j) (\mz^4 + \mz^2 (s + t + u) - 
              u (s + 2 t + u))))\nl /(\mz^2 (u - \nevalsi^2) (u - \nevalsj^2)) 
	      \nonumber \\
{\cal T}_{\rm I}\!\!\times\!\!{\cal T}_{\rm II} &=& (1/2) (-1/(2 \mz^2)) 
          (\msnu^4 (s + t - u) + 
           (\mz^2 - t) (-s^2 - t^2 + \mz^2 (s + t - u) + u^2) \nl + 
          \msnu^2 (-s^2 - s t - 2 t^2 + 2 \mz^2 (s - t - u) + 
            t u + u^2) + \mchi^2 (\mz^2 (-s + t - 5 u) \nl + 
              \msnu^2 (8 \mz^2 - s + t - u) + 
              t (s - t + u)))/(s (t - \msnu^2 )) \nonumber \\
{\cal T}_{\rm I}\!\!\times\!\!{\cal T}_{\rm III} &=& (1/2) (1/(2 \mz^2)) 
         (-2 (-\mchi^3) \nevalsj f_{3L}(j) s - 
  \mchi^4 f_{3R}(j) (s + t - u) \nl + 2 (-\mchi) \nevalsj f_{3L}(j) (2 \mz^4 - 
            2 \mz^2 s + s (-\msnu^2 + t + u)) \nl + 
         \mchi^2 f_{3R}(j) (s^2 + 2 s t + t^2 + s u + t u - 
           2 u^2 + \msnu^2 (8 \mz^2 - 3 s - t + u) \nl - 
            \mz^2 (5 s + t + 3 u)) + f_{3R}(j) (-4 \mz^4 s + 
           \msnu^2 (4 \mz^4 - 4 \mz^2 (s + t) + (s + t - u) u) \nl - 
           u (s^2 + t^2 - u^2) + \mz^2 (4 s^2 + (t - u) u + 
           s (4 t + u))))/(s (u - \nevalsj^2)) \nonumber \\
{\cal T}_{\rm II}\!\!\times\!\!{\cal T}_{\rm III} &=& (1/2) (1/\mz^2) 
          (\mchi^4 f_{3R}(j) (\msnu^2 + 3 \mz^2 - t) 
      + (-\mchi) \nevalsj f_{3L}(j) (\mz^4 + \msnu^2 (3 \mz^2 - s) \nl + s t - 
            \mz^2 (s + t + 2 u)) + 
        \mchi^2 f_{3R}(j) (3 \mz^4 + \msnu^2 (5 \mz^2 - s - t - u) \nl + 
            t (s + t + u) - \mz^2 (4 s + 5 t + 2 u)) + 
         f_{3R}(j) (-\mz^4 (s + t - u) - t^2 u \nl + 
            \mz^2 (s^2 + 2 s t + t^2 + t u - u^2) + 
            \msnu^2 (t u + \mz^2 (-s - 3 t + u))))
            \nl /((t - \msnu^2) (u - \nevalsj^2)) \nonumber \\
\tsq &=&   f_1^2 {\cal T}_{\rm I}\!\!\times\!\!{\cal T}_{\rm I} 
+  f_2^2 {\cal T}_{\rm II}\!\!\times\!\!{\cal T}_{\rm II}
+ \sum_{i,j} {\cal T}_{\rm III}\!\!\times\!\!{\cal T}_{\rm III}
+ 2 f_1 f_2 {\cal T}_{\rm I}\!\!\times\!\!{\cal T}_{\rm II} 
\nl + 2 \sum_j \left( f_1 {\cal T}_{\rm I}\!\!\times\!\!{\cal T}_{\rm III} 
+ f_2 {\cal T}_{\rm II}\!\!\times\!\!{\cal T}_{\rm III} \right)
\end{eqnarray}

\subsection*{$\chi \snu \longrightarrow e W^+$}
 I. $s$-channel $\nu$ annihilation \hfill \\
 II. $t$-channel $\sel$ exchange \hfill \\ 
 III. $u$-channel chargino exchange  
\begin{eqnarray}
f_1 &=& C_{\nu-e-W} \, C_{\snu-\schi^0_1 -\nu} \nnl
f_{2L}(i) &=& C_{\snu-\sel-W} \, C_{\sel_i-\schi^0_1 - e}^L \nnl
f_{2R}(i) &=& C_{\snu-\sel-W} \, C_{\sel_i-\schi^0_1 - e}^R \nnl
f_{3L}(i) &=& C_{\snu-\schi^+_i -e} \, C_{\schi^0_1 - \schi^+_i - W}^L \nnl
f_{3R}(i) &=& C_{\snu-\schi^+_i -e} \, C_{\schi^0_1 - \schi^+_i - W}^R  \nnl
{\cal T}_{\rm I}\!\!\times\!\!{\cal T}_{\rm I} &=&  (1/2) (\mchi^4 s 
         - \msnu^4 s + s (-\mw^4 - s t 
          + \mw^2 (s + t - u))  - \mchi^2 (2 \mw^4 - 2 \mw^2 s 
        \nl  + s (t + u)) + \msnu^2 (2 \mw^4 - 2 \mw^2 s 
          + s (s + t + u)))/(\mw^2 s^2) \nonumber \\
{\cal T}_{\rm II}\!\!\times\!\!{\cal T}_{\rm II} &=&  (1/2) 
       ((f_{2R}(i) f_{2R}(j) + f_{2L}(i) f_{2L}(j) ) (\mchi^2 - t)
           (\msnu^4 + (\mw^2 - t)^2 \nl - 2 \msnu^2 (\mw^2 + t)))
             /(\mw^2 (t-\msei^2) (t-\msej^2)) \nonumber \\
{\cal T}_{\rm III}\!\!\times\!\!{\cal T}_{\rm III} &=& (-1/2) ( 
           (f_{3L}(i) \mxi (3 (-\mchi) \mw^2 f_{3R}(j)  
            (-\mchi^2 - \mw^2 + s + t) \nl + \mxj f_{3L}(j) (\mw^4 
           + s (-\mchi^2 + u) - \mw^2 (s - t + u))) \nl + 
            f_{3R}(i) (-3 (-\mchi)^3 \mxj \mw^2 f_{3L}(j) - \mchi^6 f_{3R}(j) 
	    \nl + 
                  3 (-\mchi) \mxj \mw^2 f_{3L}(j) (-\mw^2 + s + t) + 
                  \mchi^4 f_{3R}(j) (-2 \mw^2 + s + t + 2 u) \nl + 
                  f_{3R}(j) (2 \mw^6 + 2 \mw^2 s u + t u^2 - 
          2 \mw^4 (s + t + u)) \nl + \mchi^2 f_{3R}(j) (\mw^4 + 
          \mw^2 (s + t + u) - u (s + 2 t + u)))))
         \nl  /(\mw^2 (u-\mxi^2) (u-\mxj^2)) \nonumber \\
{\cal T}_{\rm I}\!\!\times\!\!{\cal T}_{\rm II} &=&  (-1/2) (f_{2R}(j) 
      (\msnu^4 (s + t - u) + (\mw^2 - 
         t) (-s^2 - t^2 + \mw^2 (s + t - u) + u^2) \nl + 
         \msnu^2 (-s^2 - s t - 2 t^2 + 2 \mw^2 (s - t - u) 
          + t u + u^2) + \mchi^2 (\mw^2 (-s + t - 5 u) \nl + 
          \msnu^2 (8 \mw^2 - s + t - u) + t (s - t + u))))
         /(2 \mw^2 s (t-\msej^2)) \nonumber \\
{\cal T}_{\rm I}\!\!\times\!\!{\cal T}_{\rm III} &=&  (1/2) ( 
    (-2 (-\mchi)^3 \mxj f_{3L}(j) s - 
        \mchi^4 f_{3R}(j) (s + t - u) \nl + 2 (-\mchi) \mxj f_{3L}(j) (2 \mw^4 
        - 2 \mw^2 s + s (-\msnu^2 + t + u)) \nl + 
         \mchi^2 f_{3R}(j) (s^2 + 2 s t + t^2 + s u + t u - 2 u^2 + 
         \msnu^2 (8 \mw^2 - 3 s - t + u) \nl - \mw^2 (5 s + t + 3 u)) 
         + fprj (-4 \mw^4 s + \msnu^2 (4 \mw^4 - 4 \mw^2 (s + t) 
         \nl + (s + t - u) u) - u (s^2 + t^2 - u^2) + 
         \mw^2 (4 s^2 + (t - u) u + s (4 t + u)))))
       \nl   /(2 \mw^2 s (u-\mxj^2)) \nonumber \\
{\cal T}_{\rm II}\!\!\times\!\!{\cal T}_{\rm III} &=&  (1/2) (f_{2R}(i)  
      (\mchi^4 f_{3R}(j) (\msnu^2 + 3 \mw^2 
          - t) + (-\mchi) \mxj f_{3L}(j) (\mw^4  + \msnu^2 (3 \mw^2 \nl - s) 
           + s t - 
          \mw^2 (s + t + 2 u)) + \mchi^2 f_{3R}(j) (3 \mw^4 
          + \msnu^2 (5 \mw^2 - s - t - u) \nl + t (s + t + u) 
          - \mw^2 (4 s + 5 t + 2 u)) + f_{3R}(j) (-\mw^4 (s + t - u) 
          - t^2 u \nl + \mw^2 (s^2 + 2 s t + t^2 + t u - u^2) + 
          \msnu^2 (t u + \mw^2 (-s - 3 t + u)))))
        \nl  /(\mw^2 (t-\msei^2) (u-\mxj^2)) \nonumber \\
\tsq &=&   f_1^2 {\cal T}_{\rm I}\!\!\times\!\!{\cal T}_{\rm I} 
+  \sum_{i,j} {\cal T}_{\rm II}\!\!\times\!\!{\cal T}_{\rm II} 
+  \sum_{i,j} {\cal T}_{\rm III}\!\!\times\!\!{\cal T}_{\rm III} 
+ 2 f_1 \sum_j {\cal T}_{\rm I}\!\!\times\!\!{\cal T}_{\rm II} 
+ 2 f_1 \sum_j {\cal T}_{\rm I}\!\!\times\!\!{\cal T}_{\rm III}
\nl + 2 \sum_{i,j} {\cal T}_{\rm II}\!\!\times\!\!{\cal T}_{\rm III}
\end{eqnarray}

\subsection*{$\chi \snu \longrightarrow h \nu \quad [H \nu]$}
 I. $t$-channel neutralino exchange \hfill \\ 
 II. $u$-channel $\snu$ exchange  
\begin{eqnarray}
f_{1L}(i) &=& C_{\snu - \schi^0_1 - \nu} \, C_{\schi^0_1 - \schi^0_i - h}^L 
\quad
\quad [ C_{\snu - \schi^0_1 - \nu} \, C_{\schi^0_1 - \schi^0_i - H}^L] \nnl
f_{1R}(i) &=& C_{\snu - \schi^0_1 - \nu} \, C_{\schi^0_1 - \schi^0_i - h}^R 
\quad
\quad [ C_{\snu - \schi^0_1 - \nu} \, C_{\schi^0_1 - \schi^0_i - H}^R]\nnl
f_2 &=& C_{\snu - \schi^0_1 - \nu} \, C_{\snu - \snu -h } \quad \quad [C_{\snu -
\schi^0_1 - \nu} \, C_{\snu - \snu -H }] \nnl
{\cal T}_{\rm I}\!\!\times\!\!{\cal T}_{\rm I} &=& (1/2) (f_{1R}(i) 
       \nevalsi (f_{1R}(j) \nevalsj (\mchi^2 - u) + 
           f_{1L}(j) (-\mchi) (-\mchi^2 - m_{h[H]}^2 + s + u)) \nl + 
   f_{1L}(i) (f_{1R}(j) (-\mchi) \nevalsj (-\mchi^2 - m_{h[H]}^2 + s + u) \nl + 
           f_{1L}(j) (\mchi^4 - m_{h[H]}^4 - s t - \mchi^2 (s + u) + 
           m_{h[H]}^2 (s + t + u))))
         \nl   /((t-\nevalsi^2) (t-\nevalsj^2)) \nonumber \\
{\cal T}_{\rm II}\!\!\times\!\!{\cal T}_{\rm II} &=& (1/2) 
     (\mchi^2-u)/(u-\msnu^2)^2 \nonumber \\
{\cal T}_{\rm I}\!\!\times\!\!{\cal T}_{\rm II} &=& (1/2) (f_{1R}(i) \nevalsi 
       (\mchi^2 - u) + 
             f_{1L}(i) (-\mchi) (-\mchi^2 - m_{h[H]}^2 + s + u))
      	  \nl   /((t-\nevalsi^2) (u-\msnu^2)) \nonumber \\
\tsq &=&   \sum_{i,j}  {\cal T}_{\rm I}\!\!\times\!\!{\cal T}_{\rm I} 
+ f_2^2 {\cal T}_{\rm II}\!\!\times\!\!{\cal T}_{\rm II} 
+ 2 \sum_i f_2 {\cal T}_{\rm I}\!\!\times\!\!{\cal T}_{\rm II} 
\end{eqnarray}

\subsection*{$\chi \snu \longrightarrow A \nu$}
 I. $t$-channel neutralino exchange 
\begin{eqnarray}
f_L(i) &=& C_{\snu - \schi^0_1 - \nu} \, C_{\schi^0_1 - \schi^0_i - A}^L \nnl
f_R(i) &=&  C_{\snu - \schi^0_1 - \nu} \, C_{\schi^0_1 - \schi^0_i - A}^R \nnl
{\cal T}_{\rm I}\!\!\times\!\!{\cal T}_{\rm I} &=& (1/2) (f_R(i) \nevalsi 
         (f_R(j) \nevalsj (\mchi^2 - u) + 
           f_L(j) (-\mchi) (-\mchi^2 - \mA^2 + s + u)) \nl + 
           f_L(i) (f_R(j) (-\mchi) \nevalsj (-\mchi^2 - \mA^2 + s + u) + 
           f_L(j) (\mchi^4 - \mA^4 - s t - \mchi^2 (s + u) \nl + 
           \mA^2 (s + t + u))))
           /((t- \nevalsi^2) (t- \nevalsj^2)) \nonumber \\
\tsq &=&   \sum_{i,j} {\cal T}_{\rm I}\!\!\times\!\!{\cal T}_{\rm I} 
\end{eqnarray}

\subsection*{$\chi \snu \longrightarrow e H^+$}
 I. $t$-channel chargino exchange \hfill \\
 II. $u$-channel $\sel_L$ exchange 
\begin{eqnarray}
f_{1aL}(i) &=& C_{\schi^0_1 - \schi^-_i - H^+}^L \nnl
f_{1aR}(i) &=& C_{\schi^0_1 - \schi^-_i - H^+}^R \nnl
f_{1bL}(i) &=& C_{\snu - \schi^+_i - e}^L \nnl
f_{1bR}(i) &=& C_{\snu - \schi^+_i - e}^R \nnl
f_2 &=& C_{\sel - \schi^0_1 - e} \, C_{\snu - \sel - H^+} \nnl
{\cal T}_{\rm I}\!\!\times\!\!{\cal T}_{\rm I} &=& (1/2) (f_{1bR}(i) f_{1bR}(j) 
      (\mchi^4 f_{1aL}(i) f_{1aL}(j) - \mHp^4 f_{1aL}(i) f_{1aL}(j) \nl - 
    (-\mchi)^3 (\mxj f_{1aL}(i) f_{1aR}(j)  + \mxi f_{1aL}(j) f_{1aR}(i)) 
    - f_{1aL}(i) f_{1aL}(j) s t \nl - 
        (-\mchi) (\mxj f_{1aL}(i) f_{1aR}(j)  + \mxi f_{1aL}(j) f_{1aR}(i)) 
	(\mHp^2 - s - u) \nl - 
        \mxi \mxj f_{1aR}(j) f_{1aR}(i) u + \mHp^2 f_{1aL}(i) f_{1aL}(j) 
	(s + t + u) \nl + 
        \mchi^2 (\mxi \mxj f_{1aR}(j) f_{1aR}(i) - f_{1aL}(i) f_{1aL}(j) 
	(s + u))))
          /((t-\mxi^2) (t-\mxj^2)) \nonumber \\
{\cal T}_{\rm II}\!\!\times\!\!{\cal T}_{\rm II} &=& (1/2) 
        (\mchi^2-u)/(u-\mselL^2)^2 \nonumber \\
{\cal T}_{\rm I}\!\!\times\!\!{\cal T}_{\rm II} &=& (1/2) (f_{1bR}(i) 
       (-(-\mchi)^3 f_{1aL}(i) 
        + \mchi^2 \mxi f_{1aR}(i) - 
          \mxi f_{1aR}(i) u \nl + (-\mchi) f_{1aL}(i) (-\mHp^2 + s + u)))
          /((t-\mxi^2) (u-\mselL^2)) \nonumber \\
\tsq &=&   \sum_{i,j}  {\cal T}_{\rm I}\!\!\times\!\!{\cal T}_{\rm I} 
+  f_2^2 {\cal T}_{\rm II}\!\!\times\!\!{\cal T}_{\rm II} 
+ 2 f_2 \sum_i {\cal T}_{\rm I}\!\!\times\!\!{\cal T}_{\rm II} 
\end{eqnarray}

\subsection*{$\snu_e \snu_\mu^\ast \longrightarrow \nu_e \bar{\nu}_\mu$}
 I. $t$-channel neutralino exchange 
\begin{eqnarray}
f(i) &=& (C_{\snu -\schi^0_i - \nu})^2 \nnl
{\cal T}_{\rm I}\!\!\times\!\!{\cal T}_{\rm I} &=& f(i) f(j) (- 2 \msnu^4 
       + 2 t u)
           /((t - \nevalsi^2) (t - \nevalsj^2)) \nonumber \\
\tsq &=&   \sum_{i,j} {\cal T}_{\rm I}\!\!\times\!\!{\cal T}_{\rm I} 
\end{eqnarray}

\subsection*{$\snu_e \snu_\mu^\ast \longrightarrow e \bar{\mu}$}
 I. $t$-channel charginos exchange 
\begin{eqnarray}
f(i) &=& (C_{\snu -\schi^+_i - e})^2 \nnl
{\cal T}_{\rm I}\!\!\times\!\!{\cal T}_{\rm I} &=& f(i) f(j) 
     (t u - \msnu^4)/((t-
        \mxi^2) (t- \mxj^2)) \nonumber \\
\tsq &=&  \sum_{i,j}  {\cal T}_{\rm I}\!\!\times\!\!{\cal T}_{\rm I} 
\end{eqnarray}

\subsection*{$\snu_e \snu_\mu \longrightarrow \nu_e \nu_\mu$}
 I. $t$-channel neutralino exchange 
\begin{eqnarray}
f(i) &=& (C_{\snu -\schi^0_i - \nu})^2 \nnl
{\cal T}_{\rm I}\!\!\times\!\!{\cal T}_{\rm I} &=& f(i) f(j) (2 s \, \nevalsi 
       \nevalsj)
         /((t - \nevalsi^2) (t - \nevalsj^2)) \nonumber \\
\tsq &=&    {\cal T}_{\rm I}\!\!\times\!\!{\cal T}_{\rm I} 
\end{eqnarray}

\end{document}